\documentclass[12pt]{article}
\pdfoutput=1
\usepackage[square,comma,numbers,sort&compress]{natbib}
\usepackage{graphicx,epstopdf,amssymb,amsfonts,amsmath,amsthm,array,
mathrsfs,amscd}
\usepackage{enumerate}
\usepackage{float}
\usepackage{xcolor}
\usepackage{appendix}
\usepackage{hyperref}
\newcommand{\pbs}[1]{\let\temp=\\#1\let\\=\temp}
\numberwithin{equation}{section}
\newcommand{\be}{\begin{equation}}
\newcommand{\ee}{\end{equation}}

\def\npb#1#2#3{{\it Nucl.\ Phys.\ }{\bf B #1} (#2) #3}

\def\prl#1#2#3{{\it Phys.\ Rev.\ Lett.\ }{\bf #1} (#2) #3}
\def\jhep#1#2#3{{\it J. High Energy Phys.\ }{\bf #1} (#2) #3}
\def\prd#1#2#3{{\it Phys.\ Rev.\ }{\bf D #1} (#2) #3}
\def\prb#1#2#3{{\it Phys.\ Rev.\ }{\bf B #1} (#2) #3}

\def\cmp#1#2#3{{\it Comm.\ Math.\ Phys.\ }{\bf #1} (#2) #3}
\def\pr#1#2#3{{\it Phys.\ Rep.\ }{\bf #1} (#2) #3}

\def\imath#1#2#3{{\it Invent math }{\bf #1} (#2) #3}

\newcommand{\N}{n}
\newcommand{\D}{d}
\newcommand{\dof}{N}
\newcommand{\maxk}{k_{\text{UV}}}

\def\tr{\mathop{\text{tr}}\nolimits}\def\Tr{\mathop{\text{Tr}}\nolimits}
\def\im{\mathop{\text{Im}}\nolimits}
\def\re{\mathop{\text{Re}}\nolimits}

\def\cvp{\raise 2pt\hbox{,}} \def\sign{\mathop{\text{sign}}\nolimits}
 \def\tr{\mathop{\text{tr}}\nolimits}
\def\im{\mathop{\text{Im}}\nolimits}
\def\re{\mathop{\text{Re}}\nolimits}  

 \def\d{{\rm d}} 
\def\la{\lambda}
\def\uN{\text{U}(N)}

\def\oD{\text{O}(\D)}
\def\uNL{\text{U}(\N)_{\text L}}
\def\uNR{\text{U}(\N)_{\text R}}
\def\uN{\text{U}(\N)}
\def\udof{\text{U}(\dof)}
\def\uoneQ{\text{U}(1)_{Q}}
\def\slR{\text{SL}(2,\mathbb R)}

\newcommand{\modelone}{\uN^2}
\newcommand{\modeltwo}{\uN}
\newcommand{\modone}{$\uN^2$}
\newcommand{\modtwo}{$\uN$}

\def\la{\lambda}

\def\mc{m_{\text c}}\def\Tc{T_{\text c}}
\def\Qc{Q_{\text c}}
\def\lyap{\eta_{\text{L}}}

\newcommand{\TcValue}{0.067_1}
\newcommand{\mcValue}{0.34_1}
\newcommand{\mStarOne}{0.213_2}
\newcommand{\mStarTwo}{0.44_1}

\def\Fock{|0\rangle}

\def\nF{n_{\text F}}

\def\sG{\mathsf G}\def\sS{\mathsf\Sigma}\def\sSeff{\mathsf{S}_{\text{eff}}}

\def\sShat{\hat{\mathsf{S}}_{\text{eff}}}
\def\Seff{S_{\text{eff}}}\def\Shat{\hat S_{\text{eff}}}
\def\meff{m_{\text{eff}}}

\def\Gr{G_{\text r}}\def\Ga{G_{\text a}}\def\chir{\chi_{\text r}}\def\chia{\chi_{\text a}}
\def\GWp{G_{\text{W}+}}\def\GWm{G_{\text{W}-}}

\def\nuqn{\nu_{\text{qn}}}\def\tqn{t_{\text{qn}}}

\begin{document}
%
%
{\pagestyle{empty}
\parskip 0in
\

\vfill
\begin{center}
{\LARGE\sffamily Phases Of Melonic Quantum Mechanics}



\vspace{0.4in}


Frank F{\scshape errari}${}^1$ and Fidel I. S{\scshape chaposnik} M{\scshape assolo}${}^{1,2}$\\

\medskip
${}^1${\it Service de Physique Th\'eorique et Math\'ematique\\
Universit\'e Libre de Bruxelles (ULB) and International Solvay Institutes\\
Campus de la Plaine, CP 231, B-1050 Bruxelles, Belgique}

\medskip
${}^2${\it Institut des Hautes \'Etudes Scientifiques\\
35 route de Chartres, 91440 Bures-sur-Yvette, France}
{\renewcommand{\thefootnote}{$\!\!\!\!\!\dagger$}
\footnote{Present address.}}

\medskip
{\tt frank.ferrari@ulb.ac.be, fidel.s@gmail.com}
\end{center}
\vfill\noindent

We explore in detail the properties of two melonic quantum mechanical theories which can be formulated either as fermionic matrix quantum mechanics in the new large $D$ limit, or as disordered models. Both models have a mass parameter $m$ and the transition from the perturbative large $m$ region to the strongly coupled ``black-hole'' small $m$ region is associated with several interesting phenomena. One model, with \modone\ symmetry and equivalent to complex SYK, has a line of first-order phase transitions terminating, for a strictly positive temperature, at a critical point having non-trivial, non-mean-field critical exponents for standard thermodynamical quantities. Quasi-normal frequencies, as well as Lyapunov exponents associated with out-of-time-ordered four-point functions, are also singular at the critical point, leading to interesting new critical exponents. The other model, with reduced $\modeltwo$ symmetry, has a quantum critical point at strictly zero temperature and positive critical mass $m_*$. For $0<m<m_*$, it flows to a new gapless IR fixed point, for which the standard scale invariance is spontaneously broken by the appearance of distinct scaling dimensions $\Delta_{+}$ and $\Delta_{-}$ for the Euclidean two-point function when $t\rightarrow +\infty$ and $t\rightarrow -\infty$ respectively. We provide several detailed and pedagogical derivations, including rigorous proofs or simplified arguments for some results that were already known in the literature.

\vfill

\medskip
%
\begin{flushleft}
\today
\end{flushleft}
\newpage\pagestyle{plain}
\baselineskip 16pt
\setcounter{footnote}{0}

}

\tableofcontents\clearpage

\section{Introduction and discussion}

Melonic theories are models for which the perturbative expansion is dominated, in some well-defined limit, by a special class of Feynman graphs called \emph{melonic}. These graphs are relevant in a surprisingly wide range of problems. They were discovered for tensor models in the context of the discretized approach to quantum gravity \cite{1/N} but also appeared implicitly in the study of disordered condensed matter systems \cite{SachdevYe}. They dominate a suitably defined large $d$ limit of planar graphs \cite{ferra1,ferra2,ferra3} and thus provide an interesting non-perturbative approximation to the large $\N$ limit of matrix models. Moreover, melonic graphs can be recursively constructed and explicitly summed over using the technique of Schwinger-Dyson equations, allowing for considerable analytical progress to be made in the study of such models.

Strong interest in melonic theories, mostly fermionic quantum models, has been rekindled in recent times by the Sachdev-Ye-Kitaev (SYK) model \cite{SachdevYe,Kitaev} and various generalizations thereof \cite{SusySYK,FlavorSYK}, see also \cite{Maldacena:2016hyu} and the review \cite{Rosenhaus:2018dtp}. Within a program that has seen the convergence of very diverse approaches, new and interesting relations have been found between condensed-matter disordered systems \cite{reparacondmat,condmat,zeroTcondmat,Sachdev:2010um}, tensor models \cite{witten1,klebanov1,TensorSYK} and random matrix theory \cite{HanadaShenker,RmtSYK}. The emerging general picture is that melonic theories provide a quantum mechanical description of the near-horizon, low energy limit of near-extremal black holes \cite{Kitaev,Maldacena:2016hyu}. The $\text{AdS}_{2}$ factor in the near-horizon geometry is at the origin of universal properties that are also found in the infrared limit of the melonic models \cite{Sachdev:2015efa,Davison:2016ngz,Sachdev:2019bjn}. A crucial ingredient is the emergence of an approximate time reparameterization symmetry. This symmetry is spontaneously broken down to $\slR$, and the associated pseudo-Goldstone modes are governed by a Schwarzian action which matches the gravitational dynamics in nearly $\text{AdS}_{2}$ space-time \cite{holomod,BulkSYK}. The low-temperature thermodynamical properties of the gravitational system are reproduced, together with some correlation functions and Lyapunov chaos exponents \cite{ChaosSYK}.

Melonic quantum models are also extremely interesting in their own right \cite{1/N,TanasaMelonic,klebanov1,Melonic,Dartois:2018kfy}. The class of melonic graphs is situated in between the very simple trees of bubbles which dominate the large $n$ limit of vector models, and the intractable set of planar graphs which are relevant for the matrix models. This ``midpoint'' has striking properties: the models remain simple enough to be amenable to analytical study; yet they display non-perturbative features typically associated with matrix models, including the emergence of a continuous spectrum and chaos. It is even plausible that the melonic large $d$ approximation to matrix models \cite{ferra1} is able to capture the correct qualitative physics associated with the full sum over planar diagrams.

The aim of this paper is to present an extensive study of two fermionic quantum mechanical models in the melonic limit, motivated by the line of research initiated in \cite{letter}. The models can be formulated either as matrix quantum mechanics, or as disordered models. The random Hamiltonians in the disordered formulation read
\be\label{h1intro}
H= N m\, \chi^\dagger_i \chi^i + \sqrt{N}\la^{ij}_{kl}\,\chi^\dagger_i \chi^\dagger_j \chi^k \chi^l
\ee
and
\be\label{h2intro}
H'= N m\, \chi^\dagger_i \chi^i + \sqrt{N}\bigl( \xi^{i}_{jkl}\, \chi^{\dagger}_{i}\chi^{j}\chi^{k}\chi^{l} + \xi^{ijk}_{l}\, \chi^{\dagger}_{i}\chi^{\dagger}_{j}\chi^{\dagger}_{k}\chi^{l}\bigr) \,,
\ee
where the $\chi^{i}$ are complex fermions for $1\leq i\leq \dof$, $\la^{ij}_{kl}$ and $\xi^{i}_{jkl}$ are random couplings and $m$ is a mass parameter. The model based on $H$ is often called the ``complex SYK model'' and has already been considered in the literature, see \emph{e.g.}\ \cite{Sachdev:2015efa,Davison:2016ngz,Cai:2017vyk,Bulycheva:2017uqj,Bhattacharya:2018nrw}. It preserves the fermion number symmetry and its matrix model version, discussed in Sec.~\ref{PresentationSec}, has a full $\uN\times\uN\times\oD$ symmetry. For this reason, we call this model the ``\modone\ model.'' The model based on $H'$ has to the best of our knowledge never been studied before. It breaks the fermion number symmetry and its matrix model version has a reduced $\uN\times\oD$ symmetry; we therefore call it the ``\modtwo\ model.'' The Hamiltonians \eqref{h1intro} and \eqref{h2intro} look very similar, but this is misleading. We shall see that they yield qualitatively very different physics.

In the natural units for which the quartic coupling strengths are set to one, the models depend on two parameters, the mass $m$ and the temperature $T$. Both the large $T$, fixed $m$ and large $m$, fixed $T$ regions of the $(m,T)$-plane are weakly coupled. The matrix model formulation makes it clear that the large mass regime describes a situation akin to a system of weakly interacting and well-separated D-particles governed by open string perturbation theory. At small masses, on the other hand, the models are strongly coupled and share properties with black holes. The introduction of the mass parameter thus allows us to study in detail the interesting transition between a weakly coupled ``D-brane'' regime and a strongly coupled ``black-hole'' regime \cite{Balasubramanian:2016ids}. We are going to discuss at length both the physics of Euclidean time (two-point and four-point functions, thermodynamics and phase diagrams) and real time (two-point and out-of-time-ordered four-point functions, quasi-normal frequencies, chaos exponents).

\subsubsection*{Salient features and main new results for the \modone\ model}

A fundamental property of the \modone\ model, first observed in \cite{letter}, is the presence of a line of first order phase transitions in the strongly coupled region of the $(m,T)$-plane, see Fig.\ \ref{fig:M1PhaseDiagram} in Sec.\ \ref{PhaseMod1Sec}. The first order phase transition exists because the Schwinger-Dyson equations resumming the melonic diagrams of the theory have two distinct solutions within a finite region of the $(m,T)$-plane. One solution is SYK-like and describes black-hole physics, whereas the other is an analytic continuation of the perturbative solution at large mass. The first order phase transition occurs when the free energies of the two solutions match. A crucial feature is that the line of first order phase transitions terminates at a critical point $(m=\mc,T=\Tc)$. Standard thermodynamical critical exponents were evaluated numerically in \cite{letter} and found to be non-mean-field and asymmetric; for instance, at $m=\mc$, different exponents are obtained depending on whether $T\to\Tc$ from above or from below.

In the present paper, we provide full derivations of the results presented in \cite{letter} and extend the analysis in two directions.
\begin{enumerate}[i)]
	\item First, we study the phase diagram of the $q$-generalizations of the model \eqref{h1intro}, for which the quartic interaction is replaced by a fermion number preserving term of even degree $q$. The physics is similar for all values of $q\geq 4$, with a line of first order phase transitions terminating at a critical point $(\mc^{(q)},\Tc^{(q)})$. We find that the values of some of the critical exponents do depend on $q$, see Table \ref{tab:CriticalExponents} in Sec.\ \ref{Pha1exponentsqSec}. The variations with $q$ seem to be well above numerical error bars, especially for low values of $q$, providing evidence that the critical points found for different values of $q$ are inequivalent. Moreover, we can also define ``multi-$q$'' models for which interactions of various degrees are included simultaneoulsy. We have initiated the study of such models, whose phase diagrams depend on the relative strengths of the couplings, see Fig.\ \ref{fig:CriticalLine} in Sec.~\ref{Pha1continuousexpSec} for the case where the quartic and sextic couplings are both turned on. We find in this case evidence for a line of continuously varying non-trivial critical points, interpolating between the pure $q=4$ and $q=6$ cases.
	\item Second, we study real time correlation functions, from which we extract the leading quasi-normal frequency $\nuqn$ governing the exponential decay of two-point functions, and the Lyapunov chaos exponent $\lyap$ governing the exponential growth of the connected piece of out-of-time-ordered four-point functions. We find that the chaos bound of \cite{Maldacena:2015waa} is saturated in the zero-temperature limit for all values of the mass in the black-hole, SYK-like phase. This is true for both the particular four-point function studied in \cite{Maldacena:2015waa}, but also for the physical OTOC, for which the proof of the bound presented in \cite{Maldacena:2015waa} does not a apply. An interesting new result comes about when one studies the behaviour of $\nuqn$ and $\lyap$ near the critical point $(\mc,\Tc)$, see Fig.\ \ref{fig:M1QNFCritExp} and \ref{fig:LyapunovCriticalExp} in Sec.\ \ref{quasifreqSec} and \ref{LyapAppSec}, respectively. We find that both $\nuqn$ and $\lyap$ have a singular behaviour at the critical point. This singular behaviour leads to a new set of critical exponents, tied to the real time physics of dissipation and chaos and characterizing the critical point.
\end{enumerate}
\subsubsection*{Main results for the \modtwo\ model}

In contrast with the situation in the \modone\ model, for any strictly positive temperature the \modtwo\ model undergoes a smooth crossover from the weakly coupled large mass regime to the strongly coupled small mass regime. In particular, we find only one physically acceptable solution of the Schwinger-Dyson equations for each set of values of the parameters $m$ and $T$. Interesting physics thus occurs at zero temperature. The mass gap $\meff$ is found to decrease from its perturbative value $\meff\simeq m$ at large mass to zero for a strictly positive critical mass $m=m_{*}>0$, see Fig.\ \ref{fig:M2MassGap} in Sec.\ \ref{PhaseMod2Sec}. Below the quantum critical point at $m_{*}$ the model is in an interesting new gapless phase.

The most novel and important feature of the model has to do with the infrared behaviour of this gapless phase found for $0<m<m_{*}$. Instead of the standard conformally invariant solution, we find that \emph{the conformal symmetry $\slR$, including the usual scale transformations, is spontaneously broken.} This occurs because the power-law decay of the zero-temperature Euclidean two-point function is governed by two \emph{distinct} exponents $\Delta_{+}$ and $\Delta_{-}$,
\be\label{Gasintro}
G(\tau)\sim
\begin{cases}
\displaystyle\frac{b_{+}}{|\tau|^{2\Delta_{+}}} &\quad\text{if $\tau\rightarrow +\infty$}\\
\displaystyle
-\frac{b_{-}}{|\tau|^{2\Delta_{-}}} &\quad\text{if $\tau\rightarrow -\infty$\, ,}
\end{cases}
\ee
with 
\be\label{Delineqintro} \Delta_{-}>\Delta_{+}\, .\ee
This non-standard behaviour, with the strict inequality \eqref{Delineqintro}, is  established numerically in Sec.~\ref{EucSolmod2Sec} with a very high degree of confidence.

The low-temperature thermodynamical properties of the exotic gapless phase are similar to the standard behaviour also found in SYK or in the \modone\ model. The entropy can be expanded as $S/N = \sigma + c T + o(T)$, with the zero-temperature entropy $\sigma$ being strictly positive for all $0\leq m< m_{*}$ and vanishing at $m=m_{*}$. The coefficient $c$ varies smoothly in the interval $0\leq m< m_{*}$ and jumps discontinuously to zero at $m=m_{*}$, see Fig.\ \ref{fig:M2Thermo} in Sec.~\ref{PhaseMod2Sec}. On the other hand, we find convincing numerical evidence supporting the fact that the Lyapunov exponent does not saturate the bound $\lyap\leq 2\pi T$ when $T\to 0$ (compare Fig.\ \ref{fig:LyapunovExponentsa} for the \modone\ model with Fig.\ \ref{fig:LyapunovExponentsb} for the \modtwo\ model in Sec.\ \ref{LyapAppSec}). Note that there is no inconsistency here. The standard methods \cite{Maldacena:2016hyu} to compute $\lyap$ in the zero-temperature limit  rely on the $\slR$ symmetry; it is a non-trivial open problem to generalize these methods to a phase satisfying \eqref{Gasintro}, \eqref{Delineqintro} in which the $\slR$ symmetry is spontaneously broken.

\subsubsection*{Plan of the paper}

In Sec.~\ref{PresentationSec}, we introduce and motivate the \modone\ and \modtwo\ models in detail. We explain that, to leading melonic order, they can be formulated equivalently either as large $\D$ planar matrix quantum mechanics with fundamental degrees of freedom given by $\D$ complex fermionic matrices $\psi^{a}_{\mu\, b}$, $1\leq a,b\leq n$, $1\leq\mu\leq \D$ (Sec.\ \ref{matformSec}), or as large $\dof$ disordered models with fundamental degrees of freedom $\chi^{i}$, $1\leq i\leq \dof$ (Sec.\ \ref{DisGenSec}). We emphasize the matrix formulation because it provides a genuine quantum mechanical theory even when the number of degrees of freedom is finite. We explain that these models are the most general single-trace matrix quantum mechanics based on the variables $\psi^{a}_{\mu\, b}$, with quartic interaction terms and the additional condition that the fermion number violating two-point functions $\langle\tr\psi_{\mu}\psi_{\mu}\rangle$ vanish at leading melonic order. We also comment in Sec.\ \ref{technicalsubsec} on the relation with tensor models and the different large $\dof$ limits that can be defined in these models; on subtleties associated with the case where the symmetry $\uN^{2}\times\oD$ is broken down to $\uN\times\oD$, as in the \modtwo\ model; and on gauging.

In Sec.~\ref{EuclideSec} we introduce the Euclidean time technology which is heavily used in the rest of the paper. In Sec.~\ref{SDeq} we recall, in a detailed and pedagogical way, the fundamental properties of Euclidean two-point functions, the Schwinger-Dyson equations they satisfy, some important properties of the space of solutions of these equations (non-trivial monodromies), their reparameterization invariance and the standard low energy ansatz. We also introduce and briefly discuss the fundamental new low energy ansatz \eqref{Gasintro} in Sec.~\ref{NewansatzSec}. In Sec.~\ref{SeffSec} we derive explicit formulas for all the thermodynamical potentials (free energy, energy and entropy). These derivations are well known in the case of the disordered models, for which an effective action can be derived through a straightforward path integral argument based on the introduction of a suitable bilocal auxiliary field. We focus on the case of the matrix (or tensor) models, for which such a simple argument does not exist. We clarify the meaning of the effective actions for these models and provide rigorous and elementary derivations of all the required identities. In Sec.~\ref{NumSolEucSec} we introduce the numerical techniques used to solve the Euclidean Schwinger-Dyson equations and provide extensive tests of their precision by accurately reproducing well-known properties of the IR limit of the \modone\ model in the SYK-like phase. We then apply the same methods to the \modtwo\ model and discover that its low energy limit, for low enough mass $m$, is governed by the generalized ansatz \eqref{Gasintro}, establishing unambiguously the inequality \eqref{Delineqintro}. Finally, in Sec.\ \ref{EuclidefourSec} we discuss the Euclidean four-point functions. As usual, their connected pieces are obtained by resumming a geometric series associated with a set of ladder diagrams. We provide a diagrammatic derivation of the associated kernels, together with a fully rigorous and general algebraic proof (see also App.~\ref{algApp}).

Sec.~\ref{PhaseSec} is devoted to the study of the phase diagrams of the models, \emph{i.e.}\ their properties in the $(m,T)$-plane. A general physical discussion is given in Sec.~\ref{phasemotivSec}, where the main features are explained and motivated in simple terms. In Sec.~\ref{PhaseMod1Sec}, we focus on the \modone\ model. After recalling and extending the main results already obtained in \cite{letter} (line of first order phase transition, critical point, physics at fixed fermion number, low temperature behaviour), we study the $q$-generalizations of the model (see also App.~\ref{qGenApp}). This includes the derivation of the critical exponents for the $q$-generalizations of the critical points in Sec.~\ref{Pha1exponentsqSec}, and the discussion of multi-$q$ models with their associated domain wall solutions and line of critical points in Sec.~\ref{Pha1continuousexpSec}. The \modtwo\ model is discussed in \ref{PhaseMod2Sec}. The main feature of its phase diagram is the vanishing of the mass gap at a quantum critical point $m=m_{*}>0$. For $0<m<m_{*}$, the model is in the exotic gapless phase with a spontaneously broken scale symmetry. We find strong evidence that the zero-temperature entropy is non-vanishing in this phase and also compute the energy and heat capacity at low temperature.

Real time physics is discussed in Sec.~\ref{RealTimeSec}. We begin in Sec.~\ref{defrealtimeSec} with a discussion of the general properties of real time two-point functions, including Carlson's theorem. We then explain in Sec.~\ref{realSDSec} how the Schwinger-Keldysh formalism can be used to constrain the two-point functions in the melonic limit, and how the resulting real time Kadanov-Baym equations can be solved by an elegant argument based on Carlson's theorem, apparently yet unpublished. This provides a rigorous path from the Euclidean to the real time solution of the models. The numerical algorithm used to solve the real time equations, which is an extension of the algorithm used for the Euclidean equations, is presented in Sec.~\ref{realnum2ptSec} and used to compute some spectral functions. We then apply our results on two-point functions to the computation of quasi-normal frequencies in Sec.~\ref{quasifreqSec}, finding an interesting singular behaviour at the critical point of the \modone\ model, with associated non-trivial critical exponents. The problem of the real time four-point functions is tackled in Sec.~\ref{real4ptSec}. General properties (symmetries, spectral functions, KMS condition) are reviewed in Sec.~\ref{real4ptgenSec} and the Schwinger-Keldysh formalism in the melonic case is set up in Sec.~\ref{realSD4ptSec}. We then discuss the expected behaviour of the real time four-point functions, which strongly depend on the time ordering of the operators, and introduce the chaos Lyapunov exponent $\lyap$ associated with the out-of-time-ordered correlators in Sec.~\ref{OTOCSec}. We carefully explain in Sec.~\ref{OTOCSec2} how $\lyap$ can be extracted by simplifying the exact Schwinger-Keldysh equations written down in Sec.~\ref{realSD4ptSec}. This yields the so-called real time ladder kernels. In Sec.~\ref{LyapnumSec}, we present two methods to extract $\lyap$ numerically from the ladder kernels. The first method is the most direct and rigorous and does not rely on the assumption of any particular ansatz for the correlators. It amounts to inverting a large matrix of discretized real time ladder kernels. The second method is due to Kitaev \cite{Kitaev} and is based on a natural ansatz for the long time behaviour of the correlators. It reduces the problem to the computation of the largest eigenvalue of a one-dimensional kernel operator. Both methods are found to yield perfectly consistent results and are applied in Sec.~\ref{LyapAppSec} to the explicit computation of $\lyap$ in both models. The most interesting new results we obtain concern the non-saturation of the chaos bound in the gapless phase of the \modtwo\ model, and the singular behaviour of $\lyap$ near the critical point in the \modone\ model.

\section{\label{PresentationSec}Presentation of the models}

The models we shall study can be formulated in two different ways, either as matrix or disordered models. In the matrix models there are $\dof=\N^2\D$ degrees of freedom, organized as a vector of complex fermionic matrices $\psi^{a}_{\mu\, b}$ with $\uN$ indices $a,b$ and an $\oD$ index $\mu$. On the other hand, the disordered model variables form a vector of $\dof$ complex fermions $\chi^{i}$ with $i = 1,2,\dots,\dof$. For the purposes of the present work, we restrict ourselves to leading order in a suitably defined $\dof\rightarrow\infty$ limit \cite{ferra1,ferra2,ferra3}. It is a remarkable fact that in this limit both formulations turn out to be \emph{strictly equivalent}, although this equivalence breaks down if one were to consider $1/\dof$ corrections.

The advantage of the disordered models is that they are \emph{a priori} technically much simpler. Being vector models, they are straightforwardly solved in the case of annealed disorder using the standard auxiliary field techniques. The main and crucial novelty, compared to the old-fashioned vector models \cite{ZJvector}, is that the auxiliary field is now bilocal \cite{BrayMoore1980}. By examining the leading graphs it is also easy to show that the quenched and annealed disorder averages coincide at leading large $\dof$ order, see \emph{e.g.}\ \cite{Gurau:2017xhf} for a detailed discussion.

The advantage of the matrix models is that they make the link with possible applications in holography much more explicit \cite{ferra1}, since they are standard quantum mechanical matrix models with the usual string theory interpretation. Indeed, in retrospect one may see the startling equivalence with the disordered models at leading order in $\dof$ as an explanation for why disordered models \emph{\`a la} SYK could be related to black-hole physics.\footnote{Of course, only a special subclass of matrix models may be expected to yield good models of quantum black holes. For instance, as discussed in Sec.\ \ref{real4ptSec}, the Lyapunov exponent in our $\modeltwo$ model does not seem to agree with the universal value obtained for Einstein gravity black holes.}

The leading order equivalence between the two points of view deserves to be carefully justified. This essentially amounts to devising proofs that never use the auxiliary field technique, which is not applicable in the matrix case. This will be explained and exemplified in detail along the rest of the paper. 

The models we consider are, in some sense to be clarified below, the most general ones based on complex fermions with quartic interaction terms and some simple additional conditions. In this Section we provide a general presentation and exposition of their basic properties.

\subsection{\label{matformSec}Large $d$ matrix models}

\subsubsection{General setup and the large $\dof$ limit}

For convenience, we work with fermionic matrices $\psi^a_{\mu\, b}$ normalized in such a way that the canonical anticommutator reads
\be\label{anticom1}
\bigl\{\psi_{\mu\, b}^{\, a},\psi_{\nu\, d}^{\dagger c}\bigr\} = \frac{1}{\N\D}\,\delta_{\mu\nu}\delta^{a}_{d}\delta^{c}_{b}\,.
\ee
The fermion number operator and fermion number are respectively defined by
\be\label{Qdef1}
\hat Q = \frac{1}{n}\tr\psi_{\mu}^{\dagger}\psi_{\mu}\, ,\quad Q=\bigl\langle\hat Q\bigr\rangle\,,
\ee
where we use matrix notation for the $\uN$ indices. With the normalization \eqref{anticom1}, we have that $0\leq Q\leq 1$.

We wish to consider the most general single-trace Hamiltonian invariant under $\uN\times\oD$ with a quadratic mass term and quartic interactions. It is important to distinguish two classes of interaction terms, according to the cyclic order of the $\oD$ indices in the traces. We will say that interaction terms with alternating indices as in $\mu\nu\mu\nu$ are of type I or \emph{tetrahedric}, while those with repeated consecutive indices such as $\mu\mu\nu\nu$ are of type II or \emph{pillow}.\footnote{A possibly confusing terminology associated to similar terms in the tensor model literature is discussed in Sec.~\ref{technicalsubsec}.} For example, $\tr\psi^{\dagger}_{\mu}\psi_{\nu}\psi_{\mu}\psi_{\nu}$ is tetrahedric whereas $\tr\psi^{\dagger}_{\mu}\psi_{\mu}\psi^{\dagger}_{\nu}\psi_{\nu}$ and $\tr\psi^{\dagger}_{\nu}\psi_{\mu}\psi^{\dagger}_{\mu}\psi_{\nu}$ are pillows. Packing up terms of each class in $h_{\text I}$ and $h_{\text{II}}$ respectively, we can write the Hamiltonian as
\be\label{Hgeneral} H= nd\Bigl( m \tr\psi^{\dagger}_{\mu}\psi_{\mu} + \sqrt{d}\,  h_{\text{I}}+ h_{\text{II}}\Bigr)\, .\ee
Note that the most general quadratic term depends on a unique mass parameter $m$, because expressions like $\tr\psi_{\mu}\psi_{\mu}$ identically vanish.

We then consider the large $\N$ and large $\D$ limits defined in the following way \cite{ferra1,ferra2,ferra3}:
\begin{enumerate}[i)]
	\item First, we take $\N\rightarrow\infty$ keeping $\D$ and the couplings in $h_{\text{I}}$ and $h_{\text{II}}$ fixed. This is the usual 't~Hooft limit \cite{tHooft}, so for example the free energy has a large $\N$ expansion of the form
\be\label{Freegenusexp}
F = \sum_{g\geq 0}n^{2-2g}F_{g}\,.
\ee
Here $F_{g}$ is the sum over genus $g$ vacuum Feynman graphs. We will focus on the leading order $F_{0}$ which selects the planar Feynman graphs.
\item Next, we take $\D\rightarrow\infty$ again keeping the couplings in $h_{\text{I}}$ and $h_{\text{II}}$ fixed. Note that the $\sqrt{d}$ enhancement of the tetrahedric (type I) terms in \eqref{Hgeneral} is absolutely crucial here. Without it, the $\D\rightarrow\infty$ limit would exist but would be trivially vector-like \cite{ferra1,ferra2}. This second limit generates a $1/\sqrt\D$ expansion of the sums over planar graphs, so for example for the leading-order free energy
\be\label{Freelargedexp} F_{0} = \sum_{\ell\geq 0}d^{1-\ell/2}F_{0,\ell}\, .\ee
Once more, in what follows we focus on the leading contribution, \emph{i.e.}\ $F_{0,0}$ for the free energy and similarly for other quantities.
\end{enumerate}

In the limiting procedure above, the order in which one takes the limits $\N\to\infty$ and $\D\to\infty$ is crucial. At fixed $\N$, due to the $\sqrt\D$ enhancement of the tetrahedric couplings in \eqref{Hgeneral}, it is easy to find graphs having arbitrarily high powers of $\D$. The large $\D$ limit at fixed $\N$ is thus ill-defined. On the other hand, taking the large $\N$ limit first restricts us to consider only planar graphs, and these have a power of $\D$ which is bounded above. Thus the large $\D$ limit of the sum over planar graphs is well defined even with the enhanced couplings.

For conciseness, recalling that $\dof = \N^2\D$ we shall refer to the above successive large $\N$ and large $\D$ limits as ``the large $\dof$ limit'' and call the leading contributions in this limit ``the leading large $\dof$ contributions.'' In this way, the terminology is also appropriate for the discussion of the disordered models presented in Sec.~\ref{DisGenSec}, so it is possible to switch back and forth between both equivalent formulations.

\subsubsection{Inequivalent models}

It is straightforward to check that all the single-trace interaction terms corresponding to a given cyclic ordering of the fields in the trace are equivalent at leading order in the large $\dof$ limit described above, up to a possible redefinition of the mass parameter $m$ and a c-number shift of the Hamiltonian. For example, 
\be\label{cyclicequivex1} n d^{3/2}\tr\psi_{\mu}\psi^{\dagger}_{\nu}\psi_{\mu}\psi^{\dagger}_{\nu}=-n d^{3/2}\tr\psi^{\dagger}_{\mu}\psi_{\nu}\psi^{\dagger}_{\mu}\psi_{\nu} +n^{2}\sqrt{d}\, , \ \text{\emph{etc}.}\ee
Note that in some instances a double-trace term $\tr\psi_{\mu}\tr\psi^{\dagger}_{\mu}$ may be produced, but this is also sub-leading at large $\N$. Alternatively, such terms can be altogether eliminated by imposing a tracelessness condition $\tr\psi_{\mu}=0$, which does not play any role at leading order.\footnote{The tracelessness condition is useful if one wants to study the models at all orders, see Sec.\ \ref{technicalsubsec}.}

This remark greatly reduces the number of inequivalent terms we may consider at leading order. For the tetrahedric interactions, noting that $\tr\psi_{\mu}\psi_{\nu}\psi_{\mu}\psi_{\nu}=0$, we are left with three possibilities
\be\label{hIterms} h_{\text{I},1}= \frac{\la}{2}\tr\psi^{\dagger}_{\mu}\psi_{\nu}\psi^{\dagger}_{\mu}\psi_{\nu}\, ,\quad h_{\text{I},2}= \frac{\la'}{2}\tr\psi^{\dagger}_{\mu}\psi^{\dagger}_{\nu}\psi_{\mu}\psi_{\nu}\, ,\quad
h_{\text{I},3}= \frac{\xi}{2}\tr\psi^{\dagger}_{\mu}\psi_{\nu}\psi_{\mu}\psi_{\nu} + \text{H.c.}\ee
There are a few more possibilities for pillow interactions, but those that contain the combination $\psi_{\mu}\psi_{\mu}$ or its complex conjugate cannot contribute to leading order graphs. The simplest proof of this uses the auxiliary field technique, which works for pillow interactions but not for tetrahedric interactions (an explicit example will be given in Sec.\ \ref{EuclideSec}). Taking this into account, we are also left with the three pillows
\be\label{hIIterms} h_{\text{II},1}= \frac{\kappa}{2}\tr\psi^{\dagger}_{\mu}\psi_{\mu}\psi^{\dagger}_{\nu}\psi_{\nu}\, ,\quad h_{\text{II},2}= \frac{\kappa'}{2}\tr\psi_{\mu}\psi^{\dagger}_{\mu}\psi_{\nu}\psi^{\dagger}_{\nu}\, ,\quad
h_{\text{II},3}= \frac{\kappa''}{2}\tr\psi^{\dagger}_{\mu}\psi_{\mu}\psi_{\nu}\psi^{\dagger}_{\nu} + \text{H.c.}\ee

Treating these three terms by introducing auxiliary fields $\psi_{\mu}\psi^{\dagger}_{\mu}$ and $\psi^{\dagger}_{\mu}\psi_{\mu}$, one may show that their effect is to renormalize the mass and to shift the energy by a simple function of the fermion number $Q$. These changes are rather trivial and  can be taken into account straightforwardly. For this reason, we shall discard the pillow terms in most of the following discussion, except for a few specific remarks.

~

The action of $\uNL\times\uNR\times\oD$ is defined in matrix notation by
\be\label{uNLRdef} \psi'_{\mu} = R_{\mu\nu}\, U_{\text L}\psi_{\nu}U_{\text R}^{-1}\, .\ee
The diagonal subgroup for which $U_{\text L}=U_{\text R}$ is simply denoted by $\uN$, and the fermion number symmetry is the $\uoneQ$ subgroup acting by a phase as $\psi_\mu'=e^{-i\alpha}\psi_\mu$.

All the terms we are considering are $\oD$-symmetric. The terms $h_{{\rm I},1}$, $h_{{\rm II},1}$ and $h_{{\rm II},2}$ are $\uNL\times\uNR$-symmetric and thus also conserve the fermion number. The terms $h_{{\rm I},2}$ and $h_{{\rm II},3}$ violate $\uNL\times\uNR$ but preserve the fermion number, and the term $h_{{\rm I},3}$ violates both $\uNL\times\uNR$ and $\uoneQ$.

One can also define an anti-unitary time reversal operator $\hat T$, acting like the identity on the Fock basis obtained by the action of the creation operators $\psi^{\dagger}$ on the Fock vacuum $\Fock$ defined by $\psi^a_{\mu b} \Fock = 0$. The hermiticity of the Hamiltonian implies that the couplings $\la$, $\la'$, $\kappa$ and $\kappa'$ must be real, and the corresponding terms are then automatically T-invariant. If complex, the couplings $\xi$ and $\kappa''$ violate T; see, however, the discussion in Sec.~\ref{genmelonsstruc}.

\subsubsection{\label{genmelonsstruc} Generalized melons}

There are many ways to depict the Feynman graphs of these models. For example, we can use the standard 't~Hooft ribbon graph representation, with an additional middle strand associated to the $\oD$ index. In the case of $\uNL\times\uNR$-invariant models the two strands making up the ribbons are associated with distinct unitary groups and can thus be colored, say in green and red. The faces of the graphs then respect such coloring. In this case a so-called ``colored graph representation'' is also possible, and is used in the tensor model literature (see \emph{e.g.}\ \cite{ferra1,ferra2}). For more general models, in which the $\uNL\times\uNR$ symmetry can be broken down to $\uN$, a more economic and convenient representation is as follows. 

\begin{figure}[h!]
\centering
\def\svgwidth{4.5in}
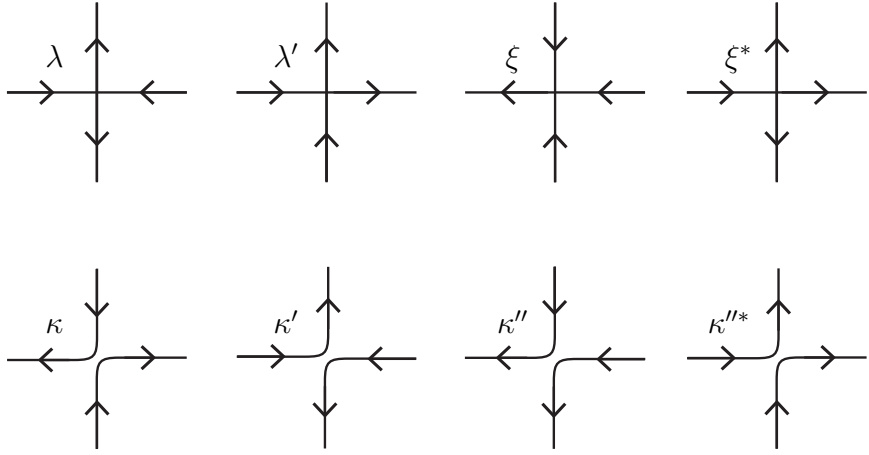
\caption{\label{fig1}The tetrahedric (upper) and pillow (lower) single-trace interaction vertices \eqref{hIterms} and \eqref{hIIterms}, respectively.}
\end{figure}

The propagator is represented by a single oriented line. We choose conventionally the orientation to represent the flow of $\uoneQ$ charge, with $\psi^{\dagger}_\mu$ creating and $\psi_\mu$ destroying one unit of fermion number. At the vertices, the cyclic ordering of the lines is important: going around a vertex clockwise corresponds to the positions of operators in a single trace from left to right. Moreover, the line in the propagator may be seen as being associated with the $\oD$ index. Tetrahedric and pillow interaction vertices are thus distinguished by the fact that the lines cross or not, respectively. The vertices associated with the terms in \eqref{hIterms} and \eqref{hIIterms} are depicted in Fig.\ \ref{fig1}.

Following the terminology of \cite{ferra2}, the graphs that contribute at leading order in $\dof$ are called \emph{generalized melons}. More detailed comments on these leading graphs will be given in Sec.~\ref{technicalsubsec}. Here we limit ourselves to describing them.

\begin{figure}[h!]
\centering
\def\svgwidth{5in}
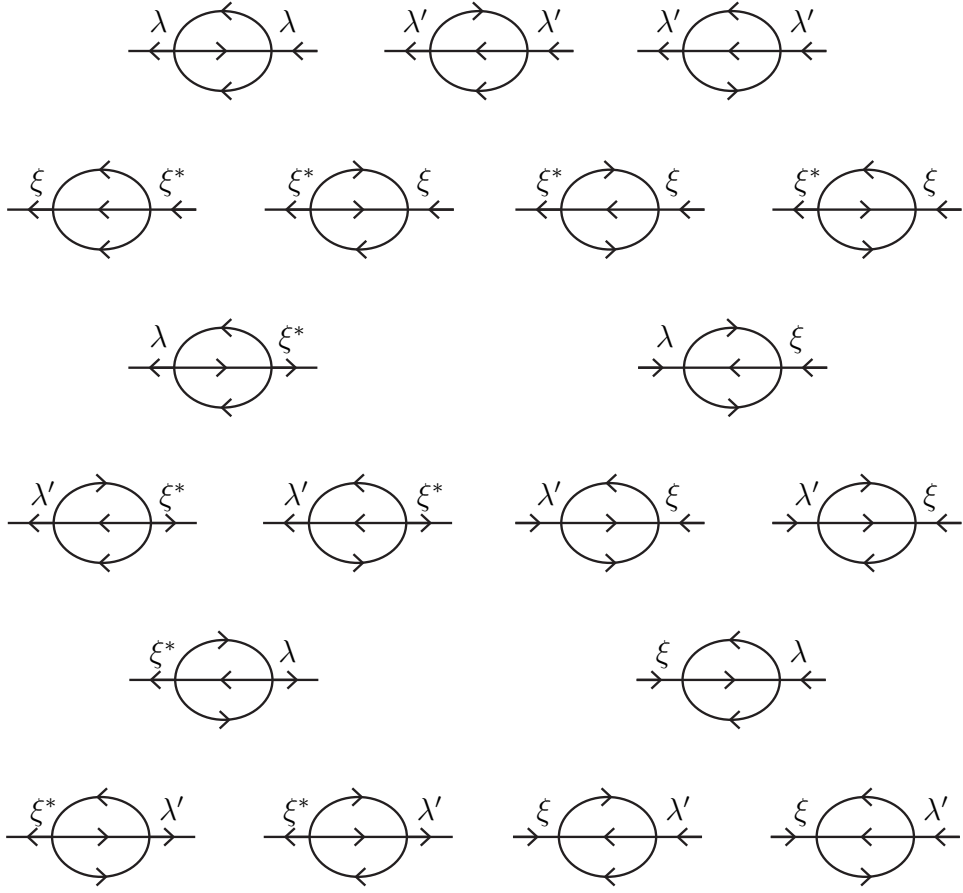
\caption{Generalized elementary melons built from tetrahedric (type I) vertices.\label{fig2}}
\end{figure}

\begin{figure}[h!]
\centering
\def\svgwidth{5in}
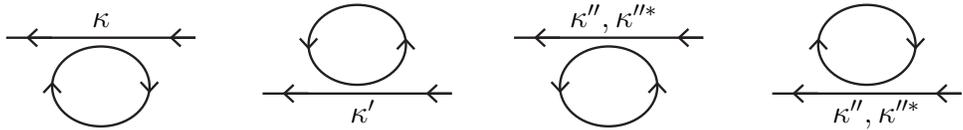
\caption{Elementary melons built from pillow (type II) vertices.\label{fig3}}
\end{figure}

\begin{figure}[h!]
\centering
\def\svgwidth{4in}
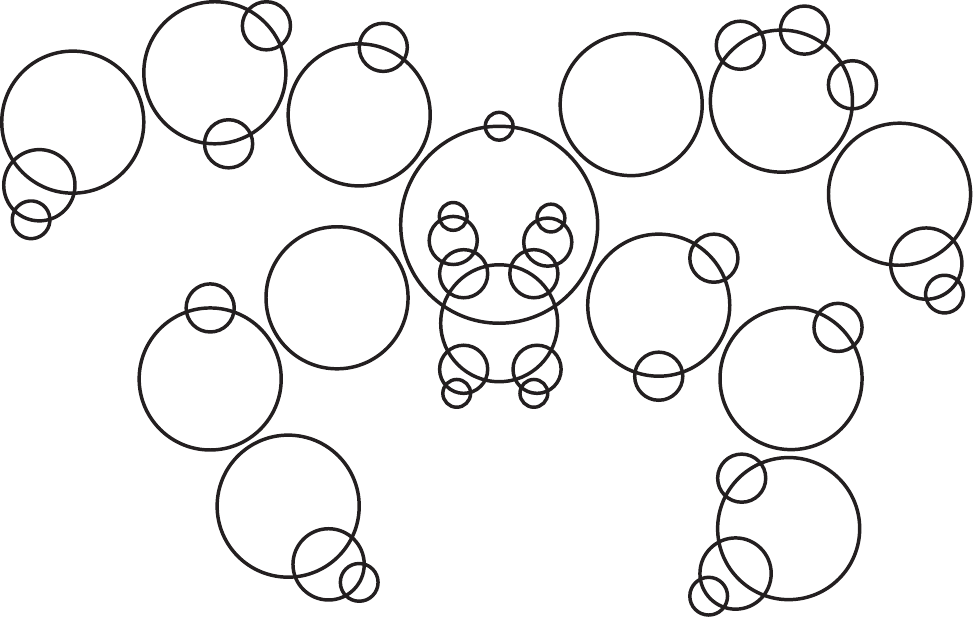
\caption{\label{fig4}Sketch of a leading vacuum Feynman graph built from both tetrahedric and pillow vertices.}
\end{figure}

When only tetrahedric interactions are used, the leading order graphs contributing to the two-point functions are obtained by inserting on bare propagators the generalized elementary melons depicted in Fig.\ \ref{fig2}, and repeating this process recursively. Note that on top of the ``symmetric'' elementary melons $(\la,\la)$, $(\la',\la')$ and $(\xi,\xi^{*})$, there exist mixed structures like $(\la,\xi^{*})$, \emph{etc}. These mixed structures, inserted once, can only contribute to the fermion number violating two-point functions $\langle\tr\psi_{\mu}\psi_{\mu}\rangle$ and $\langle\tr\psi_{\mu}^{\dagger}\psi_{\mu}^{\dagger}\rangle$, but multiple insertions do contribute to $\langle\tr\psi_{\mu}\psi_{\mu}^{\dagger}\rangle$ as well. When pillow interactions are included, we can also insert recursively on bare propagators the elementary melonic structures depicted in Fig.\ \ref{fig3}. A generic leading order Feynman graph thus has a tree-like structure built out of bubbles attached to each other by a pillow vertex, on which arbitrary recursive insertions of generalized elementary melons may be added. Leading order graphs contributing to the vacuum amplitude, or free energy, are obtained by gluing the external legs of a graph contributing to $\langle\tr\psi_{\mu}\psi^{\dagger}_{\mu}\rangle$ at leading order. A sketch of such a leading vacuum graph is depicted in Fig.\ \ref{fig4}. 


Further simplifications result from restricting to the models for which the fermion number violating two-point functions vanish at leading order, $\langle\tr\psi_{\mu}\psi_{\mu}\rangle=\langle\tr\psi_{\mu}^{\dagger}\psi_{\mu}^{\dagger}\rangle=0$. This yields one model for which we can keep the interactions $h_{{\rm I},1}$ and $h_{{\rm I},2}$ and another model based on $h_{{\rm I},3}$. But the structure of the elementary melons implies that, at leading order in $\dof$, the only effect of adding $h_{{\rm I},2}$ on top of $h_{{\rm I},1}$ is to renormalize the coupling $\la \to \la+2\la'$. It is thus enough to keep only the term $h_{{\rm I},1}$. We are left with two basic models,
\begin{align}\label{Hone}
H_{\modelone} &= nd\Bigl( m \tr\psi^{\dagger}_{\mu}\psi_{\mu} + \frac{\la}{2}\sqrt{d}\, \tr\psi^{\dagger}_{\mu}\psi_{\nu}\psi^{\dagger}_{\mu}\psi_{\nu}\Bigr)\, ,\\
\label{Htwo}
H_{\modeltwo} &= nd\Bigl( m \tr\psi^{\dagger}_{\mu}\psi_{\mu}
+ \frac{\xi}{2}\sqrt{d}\tr\psi^{\dagger}_{\mu}\psi_{\nu}\psi_{\mu}\psi_{\nu} + 
\frac{\xi^{*}}{2}\sqrt{d}\tr\psi^{\dagger}_{\mu}\psi^{\dagger}_{\nu}\psi^{\dagger}_{\mu}\psi_{\nu}\Bigr)\, .
\end{align}
As already noted, pillow interactions may also be added to either of these rather trivially. The models \eqref{Hone} and \eqref{Htwo} are those on which we shall focus in the rest of the paper. In spite of very similar-looking Hamiltonians, their physics, in particular their phase diagrams and infrared behaviours, will turn out to be drastically different. 

Note that the model \eqref{Hone} has the full $\uNL\times\uNR$ symmetry and in particular conserves the fermion number. The model \eqref{Htwo}, on the other hand, only has the diagonal $\uN$ symmetry and violates both $\uoneQ$ and time-reversal invariance (for a general complex coupling $\xi$). However, at leading order the symmetric structure of the elementary melons implies that time reversal invariance is actually restored, because only the combination $\xi\xi^{*}=|\xi|^{2}$ can enter. Moreover, for the same reasons a $\mathbb Z_{4}$ subgroup of $\uoneQ$ acting as $\psi\mapsto i\psi$ is also preserved at leading order, explaining why $\langle\tr\psi_{\mu}\psi_{\mu}\rangle$ and $\langle\tr\psi_{\mu}^{\dagger}\psi_{\mu}^{\dagger}\rangle$ vanish.

Because of their symmetry properties, we shall refer to the models with Hamiltonians \eqref{Hone} and \eqref{Htwo} as the \modone\ and \modtwo\ models, respectively. Note that the \modone\ model is the matrix version of the complex tensor model mentioned in \cite{klebanov1}. The \modtwo\ model does not seem to have been introduced before, either in tensor or matrix versions.

There are also interesting generalizations involving higher order interaction terms, similar to the $q$-body interacting SYK model introduced in \cite{Maldacena:2016hyu} and based on the complete interaction bubble studied in \cite{ferra2}. We shall briefly comment on such a generalization for \eqref{Hone} in the following sections, see also App.~\ref{qGenApp}.

\subsection{\label{DisGenSec}Disordered models}

In the disordered model framework, the degrees of freedom are $\dof$ complex fermions conveniently normalized to satisfy the anti-commutation relations
\be\label{anticom2}
\bigl\{ \chi^i, \chi_j^\dagger \bigr\} = \frac{1}{\dof} \delta^i_j \,,
\ee
and we consider a general Hamiltonian of the form
\begin{multline}\label{Hdis} H = N\Bigl( m\, \chi^\dagger_i \chi^i +\frac{1}{2} \kappa \,\chi^\dagger_i \chi^i \chi^\dagger_j \chi^j\Bigr) + \sqrt{N}\Bigl(\la^{ij}_{kl}\,\chi^\dagger_i \chi^\dagger_j \chi^k \chi^l\\ + \xi^{i}_{jkl}\, \chi^{\dagger}_{i}\chi^{j}\chi^{k}\chi^{l} + \xi^{ijk}_{l}\, \chi^{\dagger}_{i}\chi^{\dagger}_{j}\chi^{\dagger}_{k}\chi^{l} + 
\zeta_{ijkl}\, \chi^{i}\chi^{j}\chi^{k}\chi^{l} + \zeta^{ijkl}\, \chi^{\dagger}_{i}\chi^{\dagger}_{j}\chi^{\dagger}_{k}\chi^{\dagger}_{l}\Bigr)\, .
\end{multline}
The coupling $\kappa$ mimics the pillow interactions of the matrix models. The other couplings satisfy the hermiticity conditions
\be\label{rancouplingcond1}
\bigl(\la^{ij}_{kl}\bigr)^{*}=\la^{lk}_{ji}\, ,\quad \bigl(\xi^{i}_{jkl}\bigr)^{*}=\xi^{lkj}_{i}\, ,\quad \bigl(\zeta_{ijkl}\bigr)^{*}=\zeta^{lkji}\,,
\ee
and the symmetry conditions
\be\label{rancouplingcond2} \la^{ij}_{kl}=\la^{[ij]}_{[kl]}\, ,\quad \xi^{i}_{jkl}=\xi^{i}_{[jkl]}\, ,\quad \zeta_{ijkl}=\zeta_{[ijkl]}\, .\ee
They are drawn randomly from a Gaussian distribution with weight
\be\label{randomweight} \exp\Bigl[-\sum_{ijkl}\Bigl(|\la^{ij}_{kl}|^{2}/\la^{2}-4|\xi^{i}_{jkl}|^{2}/\xi^{2}-4|\zeta_{ijkl}|^{2}/\zeta^{2}\Bigr)\Bigr]\, .\ee
The large $\dof$ limit is defined by taking the real couplings $\kappa$, $\la$, $\xi$ and $\zeta$ fixed. In this limit, the so-called self-averaging observables, which include the thermodynamical potentials and the two- and four-point functions studied later in the paper, coincide with their average over disorder. This quenched average is usually performed by introducing replicas, but at leading large $\dof$ order this turns out to be unnecessary. Indeed, the solution is given by the diagonal replica ansatz, which means that the quenched average coincides with the annealed average that is obtained by directly integrating over the random couplings as if they were dynamical variables themselves. The models can then be solved using the standard auxiliary field technique, as will be briefly reviewed in Sec.~\ref{EuclideSec}.

It is easy to check that at leading order in $\dof$, and for all the purposes of the present paper, the random Hamiltonian \eqref{Hdis} is equivalent to \eqref{Hone} or \eqref{Htwo} when turning on the coupling $\la$ or $\xi$, respectively. The effect of the matrix model pillow terms are also correctly mimicked by the coupling $\kappa$ in \eqref{Hdis}. 

Note that differences do exist between the matrix model and the disordered model formulations. For instance, the coupling $\zeta$ in \eqref{Hdis} has no analogue in the matrix case, because a term like $\tr\psi_{\mu}\psi_{\nu}\psi_{\mu}\psi_{\nu}$ identically vanishes. We will therefore take $\zeta = 0$ in the rest of this paper. Conversely,  some features of the matrix model, such as non-vanishing fermion number violating correlation functions, cannot be reproduced by \eqref{Hdis} because the annealed average automatically yields an emergent $\udof$ symmetry which implies fermion number conservation. Equivalently, the integration over the random couplings can produce ``symmetric'' elementary melons, similar to the ones depicted in the first two rows of Fig.\ \ref{fig2}, but cannot yield the other asymmetric structures. To achieve this, one needs to generalize the disordered model. A simple possibility is to consider two species of fermionic vectors $\chi^{i}$ and $\tilde\chi_{i}$ transforming in complex conjugate representations of $\udof$, using a Gaussian integration measure over random couplings that includes terms mixing different couplings. An explicit example could be a term $\smash{\la^{ijk}_{l}\chi^{\dagger}_{i}\chi^{\dagger}_{j}\tilde\chi_{k}\chi^{l}+\xi^{i}_{jkl}\chi^{\dagger}_{i}\chi^{j}\chi^{k}\chi^{l}+ \text{H.c.}}$ in the Hamiltonian with a contribution of the form $\smash{\la^{ijk}_{l}\xi^{l}_{ijk} + \text{H.c.}}$ in the Gaussian weight for the couplings. This does generate asymmetric melonic structures and non-zero fermion number violating two-point functions of the form $\langle\chi^{i}\tilde\chi_{i}\rangle$, but will not be considered further in this paper.

\medskip

To summarize, the \modone\ matrix model is mimicked at leading large $\dof$ order by the random Hamiltonian\footnote{One should not confuse this model with the so-called ``mass-deformed SYK model,'' which combines random quadratic and quartic terms (usually with Majorana fermions), see for example \cite{Garcia-Garcia:2017bkg,Nosaka:2018iat}.}
\be\label{Honerandom}
H= N m\, \chi^\dagger_i \chi^i + \sqrt{N}\la^{ij}_{kl}\,\chi^\dagger_i \chi^\dagger_j \chi^k \chi^l\,,
\ee
which has already been considered in the literature and is often called the ``complex SYK model,'' see \emph{e.g.}\ \cite{Sachdev:2015efa,Davison:2016ngz,Cai:2017vyk,Bulycheva:2017uqj,Bhattacharya:2018nrw}. As for the \modtwo\ model, it is on the other hand mimicked by
\be\label{Htworandom}
H'= N m\, \chi^\dagger_i \chi^i + \sqrt{N}\bigl( \xi^{i}_{jkl}\, \chi^{\dagger}_{i}\chi^{j}\chi^{k}\chi^{l} + \xi^{ijk}_{l}\, \chi^{\dagger}_{i}\chi^{\dagger}_{j}\chi^{\dagger}_{k}\chi^{l}\bigr) \,,
\ee
and we are not aware of any prior work on it, even though it has obvious similarities with \eqref{Honerandom} and might be expected to also be relevant for condensed-matter purposes.

\subsection{\label{technicalsubsec}Discussion and relation to other tensor models}

This subsection provides details and comments on the matrix model point of view that are not strictly necessary to follow the main new results of the paper, as presented in the next sections. It may thus be skipped upon first reading. The aim is to provide further motivation for our work, as well as to place it in the wider context of recent developments in tensor models. The reader can refer to \cite{Guraubook,Klebanov:2018fzb,Delporte:2018iyf} for reviews on the latter.

We shall first clarify the nature of the double large $\N$ and large $\D$ expansions we are considering, since some confusion seems to exist in the literature about certain terminology. We will discuss models with the full $\uNL\times\uNR$ symmetry in Sec.~\ref{MelonicSection}, and consider models with only one $\uN$ symmetry in Sec.~\ref{TracelessSection}. Finally, in Sec.~\ref{MatrixTensorComparison} we elaborate on the string-theory inspired interpretation of the models we are studying, as well as the relation between these and other tensor models considered in the literature.

\subsubsection{\label{MelonicSection}Degree expansion, index expansion, melonic and non-melonic interactions}

We have already mentioned in Sec.~\ref{genmelonsstruc} that diagrams for models with $\uNL\times\uNR$ symmetry can be ``colored''. As reviewed in detail for instance in \cite{ferra2}, both interaction vertices and Feynman graphs have a representation in terms of regular colored graphs, the so-called ``bubbles.'' To any bubble we can associate a Gurau degree \cite{Guraubook,1/N,colored} which is in general a non-negative half integer. The original and standard definition of a \emph{melon} is to be a \emph{regular colored graph of Gurau degree zero.} 

Since interaction vertices are in one-to-one correspondence with regular colored graphs, they can be classified into two groups: the melonic and non-melonic interactions. In our case, the tetrahedric term $h_{{\rm I},1}$ in \eqref{hIterms} is non-melonic, whereas the pillow terms $h_{{\rm II},1}$ and $h_{{\rm II},2}$ in \eqref{hIIterms} are melonic.

To define a large $\dof$ limit, one should choose the way the coupling constants scale with $\dof$. In the case of matrix models, there is a unique interesting choice first discussed by 't~Hooft \cite{tHooft}, which yields the genus expansion. In the case of tensor models, the situation is much more complex and interesting. Several inequivalent and non-trivial scalings can \emph{a priori} be defined. In any given model, the \emph{optimal} scaling of a coupling constant is the one for which it is impossible to further enhance that coupling and still have a well-defined expansion. Computing the optimal scaling in tensor models is in general a difficult technical problem, see \emph{e.g.}\ \cite{optimalscaling}.

There are two general procedures known to define scalings with well-defined large $\dof$ limits for arbitrary interactions in tensor models.\footnote{This applies to the general ``uncolored'' models, in tensor model terminology.} 

One is the so-called BGR scaling \cite{BGR}, which was used originally in the tensor model literature but can also be straightforwardly generalized to the matrix-tensors by distinguishing two sets of indices with $\D\neq \N$, \cite{ferra2}. This scaling has two fundamental properties: 
\begin{enumerate}[i)]
\item The Feynman graphs in the large $\dof$ expansion of the tensor models are classified according to their Gurau degree. In particular, the leading graphs are the standard melons, whose structure was elucidated in \cite{melonstruc}. Subleading orders may be studied using the results in \cite{GurS}.

\item Only melonic interaction vertices can contribute to leading order graphs. In particular, the tetrahedric terms in \eqref{hIterms} do \emph{not} contribute at leading order in this framework. This is due to the fact that the $\sqrt{d}$ enhancement that we have introduced in \eqref{Hgeneral} is \emph{not} present in the BGR scaling.
\end{enumerate}

The physics of quantum mechanical models based on the BGR scaling is similar to the physics of the large $\dof$ vector models \cite{ZJvector} and in particular cannot describe black-hole physics (for instance, there is no continuous spectrum at large $\dof$ in this scheme). Note also that, in the BGR scaling, if we distinguish $\uN$ and $\oD$ indices the large $\N$ and large $\D$ limits commute.

The other scaling was introduced in \cite{ferra2}, generalizing to any tensor (or matrix-tensor) model the scaling used by Carrozza and Tanasa for a $\text{O}(n)^{3}$ tensor model in their seminal work \cite{CT}. The scaling of \cite{ferra2} is actually optimal for a large class of interactions, called maximally single-trace, which generalize the concept of single-trace matrix model interactions to general tensor or matrix-tensor models. This new scaling is the one used in the present paper, and is such that:
\begin{enumerate}[i)]
\item The Feynman graphs in the large $N$ expansion of the tensor models are classified according to their \emph{index}, which is a non-negative half-integer naturally associated to the colored graph. In particular, the leading graphs have index zero and are called \emph{generalized melons} in \cite{ferra2}. Note that a graph of Gurau degree zero automatically has index zero, but the converse is not true: there are many more generalized melons than melons, and their full classification is at the moment unknown. Some special cases, which include the interactions used in the present paper, have been treated in \cite{CT,ferra2}.
\item Both melonic and some non-melonic interactions can appear in the leading order graphs. In particular, both the tetrahedric $h_{{\rm I},1}$ vertex in \eqref{hIterms} and the pillows $h_{{\rm II},1}$ and $h_{{\rm II},2}$ in \eqref{hIIterms} contribute at leading order, which makes the physics much richer and more interesting than that of the BGR scaling.
\end{enumerate}

Even though the generalized melons built out of the tetrahedric interaction vertex $h_{{\rm I},1}$ in \eqref{hIterms} have a strictly positive Gurau degree, they share the basic fundamental recursive structure associated with standard melons: they are built by arbitrary repeated insertions of generalized elementary melons, depicted in Fig.\ \ref{fig2}, and the sum over such graphs can be evaluated using Schwinger-Dyson equations of ``melonic'' type. This simple but non-trivial result is known to be valid for the interactions used in the present paper, as well as for some generalizations thereof \cite{ferra2,ferra3}. In more general cases, it is still usually very easy to show that graphs built in such a way will contribute at leading order. However, let us emphasize that the hard part of the proof is to demonstrate that no other graph can contribute; the simple ``melonic'' recursive structure is not \emph{a priori} required. This hard part is often overlooked in the modern literature. The full classification of generalized melons remains a very interesting open problem.

\subsubsection{\label{TracelessSection}$\udof$-symmetric models and the tracelessness condition}

Almost all the tensor model literature, including \cite{ferra2}, relies on the fact that each distinct index of the tensors (or of the matrix-tensors) is associated to a distinct symmetry group. This simplifies the analysis of the Feynman graphs considerably, because different indices of a given tensor cannot mix in a given face. 

For a long time it was believed that this simplifying property was necessary to define a consistent large $\dof$ limit. The fact that this is actually not the case was first pointed out in \cite{ferra1}. A simple argument, based on the possibility to color the faces of the relevant planar graphs with only two colors, implies that the large $d$ limit with the enhanced scaling of \cite{CT, ferra2} used in the present paper is consistent and that the graphs contributing at leading order are again the generalized melons. The models containing terms that break $\uNL\times\uNR$ down to $\uN$, like \eqref{Htwo}, can thus have a well-defined large $\dof$ limit. 

An obvious question to ask is what happens beyond the planar graphs.  To some extent this question may be purely technical or academic, because for the most important physical applications we are aware of, only the planar graphs are needed. Moreover, in the matrix interpretation the higher genus terms are also subleading to all the $d^{1-\ell/2}$, $\ell>0$, corrections that already contribute at the planar level. These planar corrections are themselves out of reach with present technology, except for the leading piece in $\sqrt{d}$ which is easy to include \cite{CT}.\footnote{The physics at finite $\N$ and the resulting mechanisms that restore unitarity are of course of great interest, see \emph{e.g.}\ \cite{HanadaShenker}. However, these are non-perturbative effects in the coupling $1/\N^2$ and cannot be studied by computing to higher genera in the $1/\N^2$ expansion.} 

This being said, it is an extremely interesting problem from the graph-theoretic point of view, and is not limited to the matrix-tensor models. From the purely tensorial point of view, it is very natural to impose symmetry properties on the indices by considering, for instance, completely symmetric or antisymmetric tensors. Such symmetry conditions have the same effect as the $\uNL\times\uNR$ symmetry breaking terms in the models we study, the result being that distinct indices mix up in faces and the usual tensor model technology based on colored graphs does not apply.

It was very soon realised after \cite{ferra1} that in these models there exists a specific class of non-planar graphs that makes the new large $\D$ limit inconsistent, see \cite{ferra3} for explicit examples. Similar graphs spoil the large $\dof$ limit of other (anti)symmetric tensor models. An elegant way to remove these graphs, which have a very specific structure, is to impose tracelessness conditions \cite{klebtrace}. This was shown to be correct in the case of symmetric tensors in \cite{tensorhardproof}, using a technical tour de force based on a set of local moves simplifying the graphs \cite{guraumoves}. 

The problem remains open in the case of the matrix-tensors. The additional difficulty is that we do not have complete symmetry between all the indices in this case, and in particular the tracelessness condition is imposed only on the matrix indices.\footnote{We would like the thank Sylvain Carrozza, Adrian Tanasa and Guillaume Valette for thorough discussions on this problem.} A precise conjecture for the models of interests in the present paper is as follows.

\noindent\emph{Conjecture}: For the models \eqref{Hgeneral} for which we impose the condition $\tr\psi_{\mu}=0$, the genus $g$ free energy (and similarly any other observable) has a well-defined large $d$ expansion of the form
\be\label{Freelargedexpgen} F_{g} = \sum_{\ell\geq 0}d^{\eta_{g}-\ell/2}F_{g,\ell}\, ,\ee
where $\eta_{g}$ is an upper bound on the highest possible power of $d$ for a graph of fixed genus $g$. From \cite{ferra1}, we already know that $\eta_{0}=1$. Generalizing results proven in \cite{ferra1,ferra2} for the case of $\uNL\times\uNR$ models at higher genera, for which the tracelessness condition is not required, we may further conjecture that $\eta_{g}=1+g$.

\subsubsection{\label{MatrixTensorComparison}Matrix models vs.\ tensor models, brane picture and gauging}

As discussed above, the models we are considering are close cousins of the fermionic tensor models introduced in \cite{witten1,klebanov1}. To make the relation more explicit, one can set $\D=\N$ and treat the $\psi_{\mu\, b}^{a}$ as a three-index tensor to take the large $\N$ limit. 

In the present context, this procedure seems rather artificial because the indices $a,b$ and $\mu$ are associated with different symmetry groups. We believe it is more physical and interesting to keep the matrix model point of view. In the matrix model picture, one can always interpret the indices $a,b$ associated with the unitary symmetry as Chan-Paton factors of oriented open strings ending on some kind of D-particle. The additional orthogonal $\oD$ symmetry is then naturally interpreted as corresponding to the rotational invariance in the directions transverse to the D-particle worldline. This picture allows interesting physical interpretations of the results, and makes the relation with string theory, holography and black-hole physics much clearer. The large $\D$ limit might also have an interesting relation with the large dimension limit of general relativity studied in \cite{emparanetal}.

At subleading order, the difference between the matrix model and the tensor model interpretations becomes quantitative. 

One reason is that the corrections are organized differently in the two pictures. In the matrix models, planar contributions are always dominant over non-planar contributions, even if we consider higher order corrections in $1/\sqrt{d}$. On the other hand, if $\D=\N$ as in the tensor model interpretation, a given order in the large $\N$ expansion can mix up matrix model graphs having different genera.

Another reason is that the gauging of the global symmetries is done in a different way in the matrix models and in the tensor models. In the former, in line with the open string picture, we gauge only the $\uN$ or $\uNL\times\uNR$ symmetries that act on the matrix indices. The $\oD$ symmetry remains ungauged and is associated with the conservation of angular momentum. One then gets a standard string-like spectrum of operators. On the other hand, in the tensor model picture it is more natural to gauge the full global symmetry group, $\uN^{2}\times\text{O}(n)$ in our case, or $\text{O}(n)^{3}$ in well-studied examples \cite{gaugingtensor}. 

Note, however, that both in large $\D$ matrix models and in tensor models the gauging is relevant only at next-to-next-to leading order, because the dimension of the gauge group is of order $n^{2}$. Any contribution coming from the gauging, for instance a gauge-fixing action and a Fadeev-Popov determinant, will thus contribute at order $n^{2}$, whereas the leading and next-to-leading orders are $n^{2}d$ and $n^{2}\sqrt{d}$ (or $n^{3}$ and $n^{5/2}$ in tensor models). For instance, any Hagedorn-like phase transition at weak coupling will be pushed to zero temperature at leading order (in large $d$ matrix models, it is easy to check that the weak coupling Hagedorn temperature is of order $1/\ln d$).

\section{\label{EuclideSec} Euclidean time picture}

We now proceed to study the models \eqref{Hone} and \eqref{Htwo} in the Euclidean formalism, leaving for Sec.~\ref{RealTimeSec} the corresponding analysis in real time. Detailed derivations of various formulas are provided for thermodynamic quantities, as well as for two- and four-point functions. The low energy solutions to the relevant Schwinger-Dyson equations are spelled out, and we also discuss the numerical methods used to solve these equations at finite coupling, exemplifying their application. 

An important point we discuss is the possibility to have non-trivial monodromies in parameter space for the two-point functions. This means that the Schwinger-Dyson equations can have several distinct and consistent solutions for fixed values of their parameters. This possibility was first discovered in \cite{letter} and later also observed in \cite{Maldacena:2018lmt} for a related system with two coupled SYK models. It will play a major role in Sec.~\ref{PhaseSec} when we study the phase diagram of the \modone\ model \eqref{Hone}.

An innovative aspect of our discussion is the introduction of a new ansatz for the infrared behaviour of two-point functions. We assume that their power-law decay is governed in general by two \emph{distinct} exponents $\Delta_{+}$ and $\Delta_{-}$ when the Euclidean time goes to $+\infty$ and $-\infty$ respectively. The usual ansatz used so far in the literature corresponds to $\Delta_{+}=\Delta_{-}$, but this is not strictly required in certain cases. The possibility to have $\Delta_{+}\not=\Delta_{-}$ will play a crucial role in elucidating the physics of the \modtwo\ model \eqref{Htwo}.

We also provide a careful justification of the use of effective actions for the standard bilocal fields $G(\tau_1,\tau_2)$ and $\Sigma(\tau_1,\tau_2)$ in the matrix  (or tensor) models. Most of the literature has focused on disordered models, in which this effective action can be straightforwardly derived by a path integral argument, see \emph{e.g.} \cite{Das:2017eiw}, but this is not applicable in the case of matrix or tensor models.

Finally, we derive a general formula \emph{\`a la} Bethe-Salpeter \eqref{Kalggen} for the kernel used in the computation of the four-point functions, which is applicable in a wide class of models. Its analytic continuation to real time will be performed in Sec.~\ref{RealTimeSec}, allowing us to study the strength of quantum chaos as diagnosed by the Lyapunov exponent.

\subsection{\label{SDeq} Euclidean time two-point function}

\subsubsection{\label{genpropEucfunc}General definitions and properties}

We define the time-ordered finite temperature Euclidean two-point function by
\be\label{GEucliddef} G(\tau) =
\begin{cases}
\frac{1}{n}\bigl\langle\text{T}\tr\psi_{\mu\text{E}}(\tau)\psi_{\mu}^{\dagger}\bigr\rangle_{\beta} & \text{(matrix models)}\\
\bigl\langle\text{T}\chi^{i}_{\text E}(\tau)\chi_{i}^{\dagger}\bigr\rangle_{\beta} & \text{(disordered models)}\, .
\end{cases}
\ee
The Euclidean time evolution of any operator is defined as usual by
\be\label{Euctimeevolution} \mathscr O_{\text E}(\tau) = e^{H \tau}\mathscr O e^{-H\tau}\, .\ee
The equations \eqref{GEucliddef} are valid for $-\beta<\tau<\beta$. The correlators satisfy the KMS condition and can be extended to all Euclidean times by antiperiodicity of period $\beta=1/T$,
\be\label{Ganti}
G(\tau+\beta) = -G(\tau)\,.
\ee
In particular, we use the Fourier decomposition
\be\label{GFourier} G(\tau)=\frac{1}{\beta}\sum_{k\in\mathbb Z+\frac{1}{2}}G_{k}e^{-i\nu_{k}\tau}\, ,\ee
with Matsubara-Fourier coefficients $G_{k}$ and Matsubara frequencies
\be\label{matsubara}
\nu_k = \frac{2\pi}{\beta} k\,.
\ee
The tree-level Matsubara coefficients are given by
\be\label{Gk0def}
G_k^{(0)} = \frac{1}{m - i\nu_k}\, \cvp
\ee
yielding the tree-level two-point function
\be\label{Gtree}
G^{(0)}(\tau;m) =
\begin{cases} 
\nF(m)e^{m(\beta - \tau)}&\text{if}\quad 0<\tau<\beta\\
-\nF(m)e^{-m\tau}& \text{if}\quad -\beta<\tau<0\, ,
\end{cases}
\ee
where we have introduced the Fermi distribution function
\be\label{Fermidef} \nF(\omega) = \frac{1}{e^{\beta\omega}+1} \,\cdotp\ee

On top of \eqref{Ganti}, the Euclidean two-point function has a set of interesting general properties.\footnote{See \emph{e.g.}\ \cite{ferraRG} for a thorough discussion  in the bosonic case, which can be straightforwardly generalised to the fermionic case relevant for the present paper.} To understand these properties, it is very useful to introduce the so-called spectral function, defined by
\be\label{specdef}\rho(\omega) = \frac{1}{Z}\sum_{p,q}\bigl(e^{-\beta E_{p}}+e^{-\beta E_{q}}\bigr) |\alpha_{pq}|^{2}\delta(\omega + E_{p}-E_{q})\,,
\ee
where
\be\label{partfunctiondef} Z=\Tr e^{-\beta H}\ee
is the partition function, $E_{p}$ are the eigenvalues of the Hamiltonian, the double sum in \eqref{specdef} is over all the eigenstates $|p\rangle$ and $|q\rangle$ of the Hamiltonian\footnote{Note that in all of the models we consider the finite $\dof$ spectrum is discrete.} and 
\be\label{alphapqdef}
|\alpha_{pq}|^{2} =
\begin{cases} \sum_{\mu,a,b}|\langle p|\psi^{a}_{\mu\, b}|q\rangle|^{2} &
\text{(matrix models)}\\
\sum_{i}|\langle p|\chi^{i}|q\rangle|^{2} &
\text{(disordered models)}\, .
\end{cases}
\ee
The two-point function has the following spectral decomposition,
\be\label{Gspecdec} G(\tau) = \int_{-\infty}^{+\infty}G^{(0)}(\tau;\omega)\rho(\omega)\,\d\omega\, ,\ee
which is equivalent to
\be\label{Gkspecdec} G_{k}=\int_{-\infty}^{+\infty}\frac{\rho(\omega)}{\omega-i\nu_{k}}\,\d\omega\, .\ee

The spectral function $\rho$ is manifestly real and positive. This yields
\be\label{Gkreality}
G_{k}^{*}=G_{-k}\,,
\ee
and from \eqref{Gspecdec} and \eqref{Gkspecdec},
\begin{gather}\label{Gsign} G(\tau) >0\quad\text{if}\quad 0<\tau<\beta\, ,\quad G(\tau) <0\quad\text{if}\quad -\beta<\tau<0\, ,\\
\label{imGsign} \im G_{k}>0\quad\text{if}\quad k>0\, .
\end{gather}
The moments of the spectral function,
\be\label{rhomomentdef} \rho_{n} = \int_{-\infty}^{+\infty}\omega^{n}\rho(\omega)\,\d\omega\, ,\ee
govern the UV behaviour of the Matsubara-Fourier coefficients \eqref{Gkspecdec},
\be\label{GkUV} 
G_{k} = \sum_{n=0}^{r-1}(-1)^{n}\Bigl(\frac{i}{\nu_{k}}\Bigr)^{n+1}\rho_{n} + \mathcal{O}\bigl(1/k^{r+1}\bigr)\,,
\ee
and through \eqref{Gspecdec} they fix the discontinuities of the two-point function and its derivatives,
\be\label{disc} G^{(n)}(0^{+}) - G^{(n)}(0^{-}) = (-1)^{n}\rho_{n}\, .\ee
In particular, the canonical commutation relations \eqref{anticom1} or \eqref{anticom2} yield
\be\label{rhozero}G(0^{+})-G(0^{-}) = \rho_{0} = \int_{-\infty}^{+\infty}\rho(\omega)\,\d\omega = 1\, .\ee
Moreover, as will be demonstrated in the following (see \ref{matrixthermoSec}), the first moment $\rho_{1}$ is exactly given by a one-loop calculation, and the one-loop contribution is subleading in the $\dof\to\infty$ limit. At leading large $\dof$ order, $\rho_{1}$ thus coincides with the tree-level result
\be\label{rhoone}
\dot G(0^{-})-\dot G (0^{+})= \rho_{1} = \int_{-\infty}^{+\infty}\omega \rho(\omega)\,\d\omega = m\quad\text{at leading large $\dof$ order.}
\ee

A last very useful quantity that we need to introduce is the so-called resolvent $R$, defined by
\be\label{resolventdef} R(z) = \int_{-\infty}^{+\infty}\frac{\rho(\omega)}{z-\omega}\,\d\omega\, .\ee
The resolvent is a holomorphic function for $\im z>0$ and $\im z<0$. Eq.\ \eqref{Gkspecdec} is equivalent to
\be\label{GkRrel} R(i\nu_{k})=-G_{k}\, .\ee
The large $z$ expansion of $R$ is governed by the moments \eqref{rhomomentdef} and in particular
\be\label{Rasymp} R(z)\underset{z\rightarrow\infty}{=}\frac{1}{z}+\frac{m}{z^{2}} + \mathcal{O}(1/z^{3})\, .\ee
When the spectrum of the Hamiltonian is discrete, which is the case in these models at finite $\dof$, the only singularities of $R$ are simple poles on the real axis at the Bohr frequencies. At infinite $\dof$, the spectrum becomes continuous and $R$ develops a discontinuity across the real axis. The spectral function can be read off from this discontinuity as
\be\label{rhofromR} \rho(\omega) = \frac{i}{2\pi}\bigl( R(\omega + i0^{+}) - R(\omega - i0^{+})\bigr)\, .\ee

\subsubsection{Schwinger-Dyson equations}

Let us introduce the self-energy $\Sigma(\tau)$, which is an antiperiodic function with Matsu\-bara coefficients
\be\label{Sigmadef} \Sigma_{k}=\frac{1}{G_{k}}-\frac{1}{G_{k}^{(0)}}= \frac{1}{G_{k}}-m + i\nu_{k}\, .\ee
This equation is equivalent to 
\be\label{Sigmadef2} \Bigl(\frac{\d}{\d\tau} + m\Bigr)G(\tau-\tau') +\int_{0}^{\beta}\Sigma(\tau-\tilde\tau)G(\tilde\tau-\tau')\,\d\tilde\tau = \delta_{\beta}(\tau-\tau')\, ,\ee
where we have introduced the notation $\delta_{\beta}$ for the antiperiodic $\delta$-function
\be\label{deltabetadef}\delta_{\beta}(\tau) = \sum_{n\in\mathbb Z}(-1)^{n}\delta(\tau+n\beta)\, .\ee
The self-energy is computed by resumming the one-particle irreducible (1PI) two-point graphs from which the full two-point function is obtained by the usual geometric series,
\be\label{GkfromSigmak}
G_{k}=\frac{1}{m-i\nu_{k}+\Sigma_{k}} = \frac{1}{m-i\nu_{k}} -\frac{1}{m-i\nu_{k}}\Sigma_{k}\frac{1}{m-i\nu_{k}} + \cdots \,.
\ee
The recursive structure of the leading order graphs discussed in Sec.~\ref{genmelonsstruc} implies simple self-consistency conditions, called Schwinger-Dyson equations, on the sum of the 1PI two-point graphs. For instance, for the \modone\ model \eqref{Hone} and the \modtwo\ model \eqref{Htwo}, we get
\begin{align}
\label{SDone} \Sigma(\tau) &=\la^{2}G(\tau)^{2}G(-\tau) &&\text{(\modone\ model)}\,,\\
\label{SDtwo} \Sigma(\tau) &= -\frac{|\xi|^{2}}{4}\bigl(G(\tau)^{3}+3 G(\tau)G(-\tau)^{2}\bigr) &&\text{(\modtwo\ model)}\,.
\end{align}
The contributions for the \modone\ model come from the recursive insertions of the generalized elementary melon depicted on the top-left corner of Fig.\ \ref{fig2}. The contributions for the \modtwo\ model come from the melons on the second line of Fig.\ \ref{fig2}: the one on the left yields the term $G(\tau)^{3}$ and the other three the term $3 G(\tau)G(-\tau)^{2}$.

\begin{figure}
\centering
\def\svgwidth{5in}
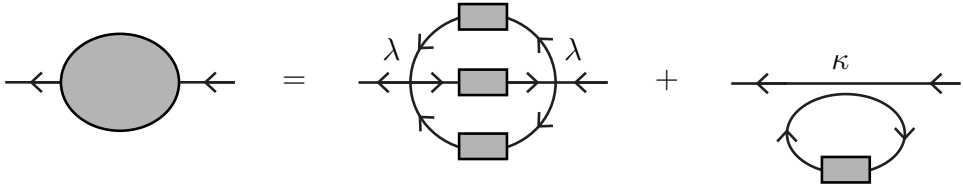
\caption{The Schwinger-Dyson equation for the \modone\ model with the pillow vertex $h_{{\rm II},1}$ turned on. The disk represents the self-energy $\Sigma$ and the rectangle the full two-point function $G$. In this case, the self-energy gets contributions from the structures depicted on the upper-left corner of Fig.\ \ref{fig2} and on the left of Fig.\ \ref{fig3}.\label{fig5}}
\end{figure}

Including contributions from the pillow vertices is very easy. For instance, if we turn on the $h_{\text{II},1}$ term of \eqref{hIIterms} in the \modone\ model, the right-hand side of \eqref{SDone} gets an additional contribution $-\kappa G(0^{-})\delta_{\beta}(\tau)=\kappa Q\delta_{\beta}(\tau)$ due to the possibility of inserting the structure on the left of Fig.\ \ref{fig3}. The full Schwinger-Dyson equation resulting from this addition is depicted schematically in Fig.\ \ref{fig5}. The same term would be added if $h_{\text{II},1}$ were to be turned on in the \modtwo\ model. The term $h_{\text{II},2}$ produces a contribution $-\kappa' G(0^{+})\delta_{\beta}(t)=\kappa' (Q-1)\delta_{\beta}(\tau)$ and the term $h_{\text{II},3}$ a contribution $(\re\kappa'') (G(0^{+})+G(0^{-}))\delta_{\beta}(\tau)=(\re\kappa'')(1-2Q)\delta_{\beta}(\tau)$, consistently with the other structures in Fig.\ \ref{fig3}.

From the point of view of the disordered models, one does not need to analyze the leading graphs to derive the Schwinger-Dyson equations. One can instead apply a purely algebraic reasoning using the standard auxiliary field method, based on the saddle-point equations of the effective action derived from the path integral.

\subsubsection{\label{MonodromySec}On the solutions of Schwinger-Dyson equations and monodromies}

The Schwinger-Dyson equations are perturbative in nature, since they are associated with the summation of a class of Feynman graphs. However, they also contain non-perturbative information because the sum of Feynman graphs they compute has a finite radius of convergence. Performing the analytic continuation of the perturbative series to strong coupling can yield extremely interesting and non-perturbative physics.

Morally speaking, this is very similar to what happens, for example, in some supersymmetric gauge theories \cite{SW}, where one resums weakly coupled semiclassical contributions that have a finite radius of convergence to obtain non-trivial results at strong coupling via analytic continuation. In the models studied in \cite{SW}, these are instanton contributions, not perturbative contributions, but the spirit is essentially the same.

In both the \modone\ and \modtwo\ models, the perturbative series in $\la^{2}$ or $|\xi|^{2}$ can be straightforwardly computed from the Schwinger-Dyson equations. For instance, in the \modone\ model we may plug \eqref{Gtree} into \eqref{SDone} to obtain the first non-trivial order of $\Sigma$, which is $\mathcal{O}(\la^{2})$. Inserting this approximate solution into \eqref{GkfromSigmak} to get a correction for $G_{k}$, we find
\be\label{Gkmod1order1} G_{k}=\frac{1}{m-i\nu_{k}}\Bigl[1+\frac{e^{\beta m}}{(e^{\beta m}+1)^{2}}\frac{\la^{2}}{(m-i\nu_{k})^{2}}+\mathcal \mathcal{O}(\la^{4})\Bigr]\, .\ee
This procedure can be iterated to any desired order in $\la$. In particular, it leads to Eq.\ \eqref{rhoone} since all the perturbative corrections to the tree-level result decay at least as fast as $1/k^{3}$ when $k\rightarrow\infty$.

An important feature of the models we are studying is that the perturbative expansion converges in a region of parameter space that includes two physically very different regimes.   
\begin{enumerate}[i)]
	\item One regime is the high temperature limit at fixed mass and coupling, $\beta m\ll 1$ and $\beta\la\ll 1$ (in the \modone\ model) or $\beta |\xi|\ll 1$ (in the \modtwo\ model). This regime is perturbative because the fermions get a large effective thermal mass of order $\pi T$. In this limit, the leading order solution is $G_{k}=i/\nu_{k}$, which is equivalent to $G(\tau)=\frac{1}{2}\sign (\tau)$. At the level of the states, this is an expansion around the maximally entropic state associated with the density matrix $2^{-N}\mathbb I$.
	\item Another regime is the high mass limit $\la/m\ll 1$ (in the \modone\ model) or ${|\xi|/m\ll1}$ (in the \modtwo\ model). This is the standard perturbative regime in particle physics, whose leading order is governed by the usual harmonic oscillator solution \eqref{Gtree}. In this limit, the models remain perturbative even at zero temperature and the ground state is simply the Fock vacuum $\Fock$.
\end{enumerate}
Overall, there is an effective coupling constant $g_{\text{eff}}=\min (\la/T,\la/m)$ in the \modone\ model, or $g_{\text{eff}}=\min(|\xi|/T,|\xi|/m)$ in the \modtwo\ model. The perturbative series converges when $g_{\text{eff}}$ is small enough.

\begin{figure}[h!]
\centering
\def\svgwidth{4.5in}
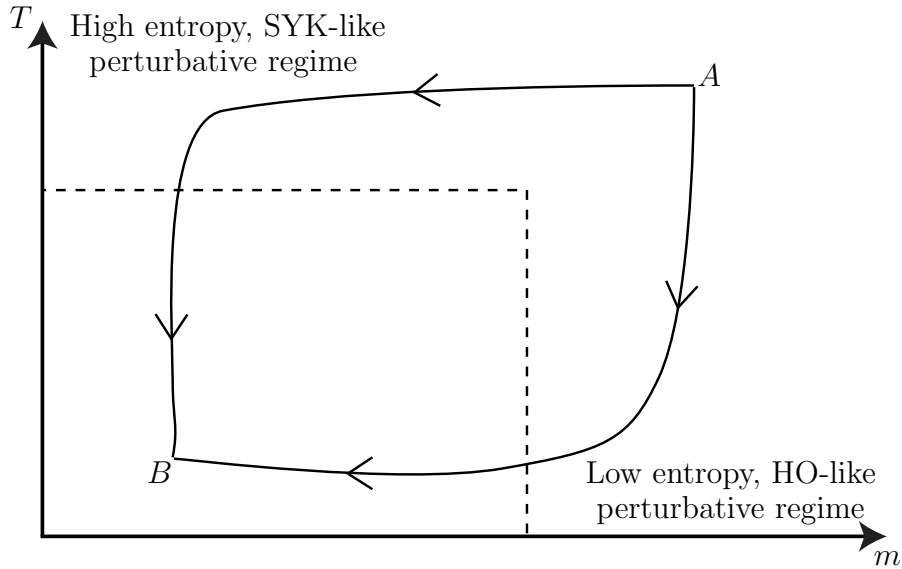
\caption{\label{fig6}Non-trivial monodromy in parameter space associated with two paths along which the weakly coupled solution at $A$ is analytically continued to $B$. Units are such that $\la=|\xi|=1$ and the dashed line represents the separation between the weakly coupled and the strongly coupled regions. When going first through the high $T$, high entropy perturbative state and then to the low $T$, non-perturbative regime (upper path), it is natural to get a solution \emph{\`a la} SYK at point $B$. When going first through the high mass, low $T$, low entropy, harmonic oscillator (HO)-like perturbative state (lower path), one may obtain a completely different, low entropy solution at $B$ \cite{letter}.}\end{figure}

A crucial property of the solutions of the Schwinger-Dyson equations is that they can undergo non-trivial monodromies in parameter space. This possibility was first realized in \cite{letter}, and later also found in other disordered models with quadratic terms \cite{Maldacena:2018lmt}. In other words, depending on the path one uses to perform the analytic continuation from the weakly coupled region $g_{\text{eff}}\ll 1$ where the perturbative series converges to a point in the strongly coupled region beyond the radius of convergence, one may end up with \emph{distinct} solutions. These distinct solutions are \emph{a priori} all physically relevant, in the sense that they satisfy all the general consistency conditions listed in Sec.~\ref{genpropEucfunc}, namely \eqref{Ganti}, \eqref{Gkreality}, \eqref{Gsign}, \eqref{imGsign}, \eqref{rhozero} and \eqref{rhoone}. This property is illustrated in Fig.\ \ref{fig6} and will be abundantly discussed in the following. 

\subsubsection{\label{ReparaSec}Reparameterization invariance and standard low energy ansatz}

\paragraph{The \modone\ model}

Let us look at the \modone\ model in which the term $h_{\text{II},1}$ is being turned on as well (including other pillow vertices would change the discussion only in a trivial way). The Schwinger-Dyson equations read
\begin{align}
\label{SD11}
&\frac{1}{G_{k}}=-i\nu_{k}+m+\Sigma_{k}\,,\\
\label{SD12}
&\Sigma(\tau) = \la^{2} G(\tau)^{2}G(-\tau) -\kappa G(0^{-})\delta_{\beta}(\tau)\, .
\end{align}
If we define
\be\label{sigmahatdef}\hat\Sigma (\tau_{1},\tau_{2}) =\Sigma(\tau_{1}-\tau_{2}) + m\delta_{\beta}(\tau_{1}-\tau_{2})\, ,\quad G(\tau_{1},\tau_{2})=G(\tau_{1}-\tau_{2})\ee
and we neglect the term $i\nu_{k}$ and the ultraviolet $\delta$-function contributions on the right-hand side of \eqref{SD11} and \eqref{SD12}, we can write the low energy (or long time) limit of the equations in the convenient form
\begin{align}\label{SD1le1} &\int_{0}^{\beta}G(\tau_{1},\tau)\hat\Sigma(\tau,\tau_{2})\,\d \tau = \delta_{\beta}(\tau_{1}-\tau_{2})\,,\\\label{SD1le2} 
& \hat\Sigma(\tau_{1},\tau_{2}) = \la^{2} G(\tau_{1},\tau_{2})^{2}G(\tau_{2},\tau_{1})\, .
\end{align}
These equations are covariant under the transformations
\begin{align}
\label{diffeoG1}& G'(\tau'_{1},\tau'_{2}) = 
\Bigl(\frac{\d \tau_{1}}{\d \tau'_{1}}\Bigr)^{\Delta}\Bigl(\frac{\d \tau_{2}}{\d \tau'_{2}}\Bigr)^{\Delta} G(\tau_{1},\tau_{2})\,,\\
\label{diffeoS1}& \hat\Sigma'(\tau'_{1},\tau'_{2}) = \Bigl(\frac{\d \tau_{1}}{\d \tau'_{1}}\Bigr)^{1-\Delta}\Bigl(\frac{\d \tau_{2}}{\d \tau'_{2}}\Bigr)^{1-\Delta}\hat\Sigma (\tau_{1},\tau_{2})\,,
\end{align}
for $\Delta=1/4$ and where $\tau'(\tau)$ is an arbitrary diffeomorphism of $\text{S}^{1}_{\beta}\rightarrow\text{S}^{1}_{\beta'}$,
\be\label{diffcond} \frac{\d \tau'}{\d \tau} >0\, ,\quad \tau'(\tau+\beta)=\tau'(\tau)+\beta'\, .\ee
Invariance of the equations is achieved for diffeomorphisms such that $\beta'=\beta$. The equations are also invariant under the local multiplications
\be\label{localu1} G'(\tau_{1},\tau_{2})=\frac{g(\tau_{2})}{g(\tau_{1})}G(\tau_{1},\tau_{2})\, ,\quad  \hat\Sigma'(\tau_{1},\tau_{2})=\frac{g(\tau_{2})}{g(\tau_{1})}\hat\Sigma(\tau_{1},\tau_{2})\, ,\ee
where $g$ is an arbitrary non-vanishing function of period $\beta$. The emergence of the reparameterization symmetry \eqref{diffeoG1}-\eqref{diffeoS1} is a remarkable feature of the low energy limit of Schwinger-Dyson equations resumming melonic graphs, consistent with a dual gravitational description of the models \cite{Kitaev,Maldacena:2016hyu}. The invariance \eqref{localu1} is a local Euclidean version of the global fermion number symmetry of \eqref{Hone}. The fact that a global symmetry becomes local is also beautifully consistent with holographic ideas.

The full low energy reparameterization symmetry \eqref{diffeoG1}-\eqref{diffeoS1} will always be spontaneously broken. In particular, time translation invariance implies that the physical two-point functions depend only on $\tau_{1}-\tau_{2}$, a property which is of course not invariant under arbitrary reparameterizations. Moreover, Eq.\ \eqref{SD1le1} involves an integral over all time scales. Since it is a low energy equation, a UV cutoff is typically needed to avoid UV divergences. In practice, the reparameterization symmetry is thus also explicitly broken by this cutoff.

A fundamental property of the models, first discussed in \cite{reparacondmat}, is that some solutions can preserve a $\slR$ subgroup of reparameterizations, or more generally a $\slR$ subgroup of the full set of transformations \eqref{diffeoG1}, \eqref{diffeoS1} and \eqref{localu1} \cite{Davison:2016ngz}, at low energy. For the \modone\ model, the standard zero-temperature ansatz is
\be\label{Gasymp}
G(\tau)\sim
\begin{cases}
\displaystyle\frac{b_{+}}{|\tau|^{2\Delta}} & \quad\text{if $\tau\rightarrow +\infty$}\\
\displaystyle
-\frac{b_{-}}{|\tau|^{2\Delta}} & \quad\text{if $\tau\rightarrow -\infty$}\, .
\end{cases}
\ee
Instead of $b_{+}$ and $b_{-}$, it is often convenient to use $C>0$ and $\theta$ defined by
\be\label{Cthetadef} b_{+}=\frac{C}{\sqrt{\pi}}\sin\bigl(\pi/4 + \theta\bigr)\, ,\quad b_{-}=\frac{C}{\sqrt{\pi}}\sin\bigl(\pi/4 - \theta\bigr)\, .\ee
Note that \eqref{Gsign} implies that $b_{\pm}>0$ and thus $|\theta|<\pi/4$. Computing the Fourier transform $\tilde G$ of $G$, Eq.\ \eqref{Gasymp} and \eqref{Cthetadef} can be rewritten as
\be\label{FourierGle} \tilde G(\omega) \underset{\omega\rightarrow 0^{+}}{\sim}\frac{i}{\omega^{1-2\Delta}} Ce^{-i\theta}\, .\ee
In terms of the spectral function introduced in Sec.~\ref{genpropEucfunc}, this is equivalent to
\be\label{rhozeroTmodelone} \rho(\omega)\underset{\omega\rightarrow 0^{\pm}}\sim
\frac{C}{\pi\sqrt{|\omega|}}\sin\bigl(\pi/4\pm\theta\bigr)\, .\ee
Plugging the above ansatz in \eqref{SD1le1} and \eqref{SD1le2} yields in the standard way
\be\label{Deltaform1} \Delta = \frac{1}{4}\,\cvp\ee
and
\be\label{Cthetarel} C^{4}\cos (2\theta) = \frac{\pi}{\la^{2}}\,\cdotp\ee
One low energy parameter remains unconstrained, for example the angle $\theta$. This parameter is naturally related, via a complicated non-perturbative formula, to the mass parameter $m$ in the Hamiltonian, possibly renormalized by the pillow couplings. When $m = 0$ the model recovers a large $\dof$ particle-hole symmetry which yields $G(-\tau)=-G(\tau)$, as in the standard SYK model, and thus $\theta=0$.

Note that \eqref{SD1le1} is equivalent to 
\be\label{Sigtilde1}
\tilde{\hat\Sigma}(\omega) =\frac{1}{\tilde G(\omega)} \underset{\omega\rightarrow 0^{+}}{\sim} -\frac{i e^{i\theta}}{C}\,\omega^{1-2\Delta}\,,
\ee
and thus implies in particular that $\tilde{\hat\Sigma}(0)=0$. Using the first equation in \eqref{sigmahatdef}, we see that the Fourier transform of the self-energy $\Sigma$ at zero frequency is constrained to satisfy
\be\label{Sigtildeconst} \tilde\Sigma (0) = -m\ee
for the $\slR$-preserving solution. Satisfying this condition is therefore an indicator that the solution is indeed conformal in the IR. 

The scaling form of the two-point function at non-zero temperature can then be obtained from the zero-temperature ansatz by using the mapping $\tau=\tan(\pi \tau'/\beta)$ and \eqref{diffeoG1} to go from a solution defined on the real line ($\beta=+\infty$) to a solution defined on the interval $[-\beta/2,\beta/2]$.\footnote{The particular mapping $\tau=\tan(\pi \tau'/\beta)$ from $\text{S}^{1}_{\beta}$ to the real line is singled out by the fact that it is consistent with time translation invariance. We would like to thank Paolo Gregori for providing this argument.} The resulting solution does not satisfy the antiperiodicity $G(-\beta/2)=-G(\beta/2)$, but this can be corrected by performing an additional transformation of the form \eqref{localu1}, with a function $g(\tau)=(b_{+}/b_{-})^{\tau/\beta}$. This yields
\be\label{Gscaling} G(\tau)=\Bigl(\frac{\pi}{\beta}\Bigr)^{1/2}\frac{e^{-2\pi a \tau/\beta}}{\sqrt{\sin\bigl(\pi |\tau|/\beta\bigr)}}\Bigl(b_{+}\Theta(\tau) - b_{-}\Theta(-\tau)\Bigr)\, ,\ee
with a damping factor
\be\label{aform} a = \frac{1}{2\pi}\ln\frac{b_{+}}{b_{-}}=\frac{1}{2\pi}\ln\tan\bigl(\pi/4 + \theta\bigr)\, .\ee
Eq.\ \eqref{Gscaling} provides a good approximation to the two-point function in the scaling regime $\beta\la\gg 1$, with the additional condition $|\tau|\gg 1/\la$ which excises the UV region. By taking the Fourier transform, we get the Matsubara-Fourier coefficients
\be\label{Gkscaling} G_{k}=i \Bigl(\frac{\beta}{2\la}\Bigr)^{1/2}
\frac{e^{-i\theta}}{(\pi\cos(2\theta))^{1/4}}\frac{\Gamma(1/4+k+i a)}{\Gamma(3/4+k+ia)}\,\cvp\ee
which are valid when $\beta\la\gg 1$ and $|\nu_{k}|\ll\la$, as well as
\be\label{rhoscale} \rho(\omega)=\Bigl(\frac{\beta}{2\la}\Bigr)^{1/2}\frac{1}{(\pi^{5}\cos (2\theta))^{1/4}}\re\biggl[e^{-i\theta}\frac{\Gamma\bigl(1/4+ia-i\frac{\beta\omega}{2\pi}\bigr)}{\Gamma\bigl(3/4+ia-i\frac{\beta\omega}{2\pi}\bigr)}\biggr]\, ,\ee
which is valid when $\beta\la\gg 1$ and $|\omega|\ll\la$.

\paragraph{\modtwo\ model}

A very similar analysis can be performed for the \modtwo\ model using \eqref{SDtwo} instead of \eqref{SDone}. At low energy, the reparameterization symmetry \eqref{diffeoG1}, \eqref{diffeoS1} is still valid, but \eqref{localu1} does not hold anymore, consistently with the fact that the fermion number is no longer conserved. The low energy ansatz \eqref{Gasymp}, \eqref{Cthetadef} can still be used, and plugging it into the Schwinger-Dyson equations we find that the only possibility with $b_{\pm}\not = 0$ corresponds to 
\be\label{model2sol1} \Delta=\frac{1}{4}\,\cvp\quad\theta=0\, ,\quad
C^{4}=\frac{\pi}{|\xi|^{2}}\,\cvp\quad b_{+}^{4}=\frac{1}{4\pi |\xi|^{2}}\,
\cdotp\ee

\noindent\textit{Important remark:} The existence of a consistent low energy ansatz solving the low energy Schwinger-Dyson equations does not ensure that there exists a physical solution to the full Schwinger-Dyson equations whose low energy limit matches the low energy ansatz. In other words, it may not always be possible to UV complete solutions that seem to be consistent at low energies. Sometimes this occurs for very elementary reasons. For instance, the low energy ansatz \eqref{Gasymp} can \emph{a priori} solve the low energy Schwinger-Dyson equations for any mass parameter $m$ in the Hamiltonian. But, at least in the perturbative regime $m\gg\la$, we know that such a solution cannot be UV completed: the perturbative series converges in this regime and the solution must be very near the harmonic oscillator which has a trivial IR behaviour. The lack of a UV completion is more subtle and has important consequences in the context of purely bosonic models \cite{ferra3,Giombi:2017dtl,Giombi:2018qgp}; see also the discussion in Sec.\ \ref{EucSolmod2Sec}.

\subsubsection{\label{NewansatzSec}A new low energy ansatz}

The standard ansatz \eqref{Gasymp} for the long time behaviour of the zero-temperature two-point function can be generalized to 
\be\label{Gasympnew}
G(\tau)\sim
\begin{cases}
\displaystyle\frac{b_{+}}{|\tau|^{2\Delta_{+}}} &\quad\text{if $\tau\rightarrow +\infty$}\\
\displaystyle
-\frac{b_{-}}{|\tau|^{2\Delta_{-}}} &\quad\text{if $\tau\rightarrow -\infty$}\, ,
\end{cases}
\ee
where the exponents $\Delta_{+}$ and $\Delta_{-}$ are \emph{a priori} distinct. This possibility may seem exotic and, to the best of our knowledge, has never been considered in the literature so far. However, in models for which the particle-hole symmetry is broken it is a plausible and mathematically consistent possibility. One salient new result of the present paper will be to provide strong numerical evidence that this behaviour is actually realized in the \modtwo\ model.

An important new feature of the $\Delta_{+}\not = \Delta_{-}$ case is that the two terms on the right-hand side of \eqref{SDtwo} scale differently at $\tau=+\infty$ and $\tau=-\infty$. Assuming, for example, that $\Delta_{+}<\Delta_{-}$, Eq.\ \eqref{SDtwo} implies that\footnote{Of course, the case $\Delta_{-}<\Delta_{+}$ can be discussed along the same lines.}
\be\label{Sigmaasympnew} \Sigma(\tau)\sim
\begin{cases}
\displaystyle -\frac{|\xi|^{2}}{4}G(\tau)^{3} & \quad\text{if $\tau\rightarrow +\infty$}\\
\displaystyle
-\frac{3|\xi|^{2}}{4}G(\tau)G(-\tau)^{2} & \quad\text{if $\tau\rightarrow -\infty$\,.}
\end{cases}
\ee
The $\tau\rightarrow +\infty$ term then dominates in the Fourier transform at low energy. Using the Schwinger-Dyson equations \eqref{SD1le1} and \eqref{Sigmaasympnew} as before, and keeping the leading terms at low energy, we get
\be\label{model2sol2} \Delta_{+}=\frac{1}{4}\,\cvp\quad b_{+}^{4}=\frac{2}{\pi |\xi|^{2}}\,\cvp\ee
with no further constraints on $\Delta_{-}$ and $b_{-}$. In passing, we note that the relation \eqref{Sigtildeconst} must also be satisfied by such a solution.

Let us remark that both the standard \eqref{Gasymp} and the more general \eqref{Gasympnew} ans\"atze spontaneously break the reparameterization symmetry \eqref{diffeoG1}-\eqref{diffeoS1}. However, the scale symmetry, which is preserved by \eqref{Gasymp}, is spontaneously broken by \eqref{Gasympnew} when $\Delta_{+} \neq \Delta_{-}$. This is an important qualitative difference between the two classes of solutions. 

It is natural to seek a finite temperature solution generalizing \eqref{Gscaling} using the reparamaterization covariance of the Schwinger-Dyson equations. However,  the usual method does not work. First, the invariance under local multiplications \eqref{localu1}, which is crucially used in the \modone\ model to obtain a finite temperature two-point function with the correct antiperiodicity property, is no longer valid in the \modtwo\ model. Second, and even more crucially, the mapping $\tau=\tan (\pi\tau'/\beta)$ yields a two point function that is not time-translation invariant, except if $\Delta_{+}=\Delta_{-}=1/4$.\footnote{One can actually check that no diffeomorphism from the real line to $\text{S}^{1}_{\beta}$ is compatible with time translation invariance if $\Delta_{+}\not = \Delta_{-}$.} Because of these difficulties, we shall limit ourselves to the zero-temperature formula \eqref{Gasympnew} when $\Delta_{+}\not = \Delta_{-}$.

On the practical side, it is interesting to observe that, even though the exponent $\Delta_{+}$ in the case $\Delta_{+}<\Delta_{-}$ and the exponent $\Delta$ in the case of the standard ansatz $\Delta=\Delta_{+}=\Delta_{-}$ are both equal to $1/4$, the two situations can be sharply distinguished even if we look only at the $\tau\rightarrow +\infty$ asymptotics of the two-point functions. Indeed, there is a factor of $8$ discrepancy for  $b_{+}^{4}$ between \eqref{model2sol1} and \eqref{model2sol2}. 

\subsection{\label{SeffSec} Effective action and thermodynamics}
\subsubsection{Disordered model formulation}

In the case of the disordered models, one can compute the Euclidean finite temperature path integral for the Hamiltonian \eqref{Hdis} straightforwardly. This is a completely standard calculation and thus we shall be brief.

Since the annealed and quenched averages coincide at leading order, as already mentioned in Sec.~\ref{DisGenSec}, we first integrate over the disorder with the Gaussian weight \eqref{randomweight}. This produces a vector model with bilocal interaction terms which can be treated in the standard way \cite{ZJvector}. We introduce two bilocal auxiliary fields $\sG$ and $\sS$, where $\sS$ plays the role of a Lagrange multiplier enforcing the constraint $\sG(\tau_{1},\tau_{2})=\chi^{i}(\tau_{1}) \bar\chi_i(\tau_{2})$. Integrating out the complex fermions yields the effective action
\begin{multline}\label{Seffdis1}\frac{1}{N}\sSeff[\sG,\sS] =-\ln\bigl(1+e^{-\beta m}\bigr) -\ln\frac{\det\mathscr O_{\sS}}{\det\mathscr O_{0}} + \frac{1}{2}\kappa\beta \sG(0^{-})^{2}
+\int_{[0,\beta]^{2}}\Bigl[\sS(\tau_{2},\tau_{1})\sG(\tau_{1},\tau_{2})\\
-\frac{1}{4}\la^{2}\sG(\tau_{1},\tau_{2})^{2}\sG(\tau_{2},\tau_{1})^{2}+\frac{1}{4}\xi^{2}
\sG(\tau_{1},\tau_{2})^{3}\sG(\tau_{2},\tau_{1})\Bigr]\d \tau_{1}\d \tau_{2}\, ,
\end{multline}
where the operators $\mathscr O_{\sS}$ and $\mathscr O_{0}$, which act on antiperiodic functions, are defined by
\be\label{OpSdef}\begin{split} \mathscr O_{\sS}(\tau_{1},\tau_{2}) &= \delta_{\beta}(\tau_{1}-\tau_{2})\Bigl[\frac{\partial}{\partial \tau_{2}}+m\Bigr]+\sS(\tau_{1},\tau_{2})\,,\\
\mathscr O_{0}(\tau_{1},\tau_{2}) &= \delta_{\beta}(\tau_{1}-\tau_{2})\Bigl[\frac{\partial}{\partial \tau_{2}}+m\Bigr]\, .\end{split}\ee
Note that, formally, the tree-level terms $-\ln(1+e^{-\beta m})$ and $\ln\det\mathscr O_{0}$ in \eqref{Seffdis1} cancel each other. However, we prefer to use the rigorous formula \eqref{Seffdis1}, which does not have UV divergences and takes into account that only the ratio of determinants $\det\mathscr O_{\sS}/\det\mathscr O_{0}$ is actually well defined.

The large $\dof$ limit is governed by the saddle points of \eqref{Seffdis1}. Varying this effective action with respect to $\sS$ and $\sG$, denoting the saddle point values of $\sS$ and $\sG$ by $\Sigma$ and $G$ respectively, and using time translation invariance, we get
\begin{align}\label{SDdisgen1}
\frac{1}{G_{k}} &=-i\nu_{k}+m+\Sigma_{k}\,,\\\label{SDdisgen2}
\Sigma(\tau) &= \la^{2} G(\tau)^{2}G(-\tau)-\frac{|\xi|^{2}}{4}\bigl(G(\tau)^{3}+3 G(\tau)G(-\tau)^{2}\bigr)-\kappa G(0^{-})\delta_{\beta}(\tau)\, .
\end{align}
These equations generalize the Schwinger-Dyson equations written down in \eqref{Sigmadef}, \eqref{SDone}, \eqref{SDtwo}, \eqref{SD11} and \eqref{SD12} to the case where the couplings $\la$, $\xi$ and $\kappa$ are turned on simultaneously.\footnote{As explained in Sec.\ \ref{genmelonsstruc}, the Schwinger-Dyson equations are more complicated in the case of the matrix model when $\la$ and $\xi$ are both turned on. In the present subsection, we are dealing with the simpler case of the disordered model \eqref{Hdis}.} In particular, and not surprisingly, the on-shell auxiliary field $G$ coincides with the Euclidean two-point function.

Moreover, by construction the on-shell value of the effective action,
\be\label{OnshellSeff}
\Seff = \sSeff (G,\Sigma)\,,
\ee
yields the free energy $F$
\be\label{SeffFrelation}
\beta F = \Seff\,.
\ee
From the free energy one can then obtain the entropy and the energy by using the standard thermodynamical identities,
\be\label{SEthermo} S = -\frac{\partial F}{\partial T}\, \cvp\qquad E = F+ TS\, .\ee

The above results are very powerful: they imply that the knowledge of the Euclidean two-point function only is sufficient to derive all the thermodynamical properties of the model. In practice, we solve the Schwinger-Dyson equations numerically as described in Sec.~\ref{NumSolEucSec}, and plug the result into \eqref{Seffdis1} to compute $F$, $S$ and $E$ straightforwardly. Very explicit formulas are given in Sec.\ \ref{EucformSec} below.

\subsubsection{Matrix model formulation\label{matrixthermoSec}}

The case of the matrix models is a bit more subtle to deal with, because there is no longer an auxiliary field method available to compute the large $\dof$ path integral. It is thus not immediately obvious how to compute the free energy. In particular, it is unclear whether an effective action like \eqref{Seffdis1} could still be relevant. The aim of this subsection is to present a detailed derivation of the free energy and to carefull justify the use of effective actions like \eqref{Seffdis1} in the context of matrix models.\footnote{The arguments apply to tensor models as well, see also \cite{Benedetti:2018goh}.} 

We explained in Sec.\ \ref{genmelonsstruc} that turning on both couplings $\la$ and $\xi$ in the matrix model yields complications due to non-vanishing fermion-number-violating two-point functions. For simplicity and because we won't need the more general results in the rest of this paper, we are then going to focus on the models \eqref{Hone} and \eqref{Htwo}. Note however that our method can be generalised straightforwardly to the models in which both couplings are turned on simultaneously. Note also that, as usual, pillow couplings may be turned on quite trivially by using the auxiliary field method.

The derivations for \eqref{Hone} and \eqref{Htwo} are extremely similar. We thus present the derivation in parallel for both models.

\paragraph{Step one}

The Heisenberg Euclidean equations of motion follow from \eqref{Euctimeevolution} and read
\be\label{EucHeisenberg} \frac{{\d\mathscr O}_{\text{E}}}{\d\tau}=\dot{\mathscr O}_{\text{E}}=\bigl[H,\mathscr O_{\text{E}}\bigr]\, .\ee
In the \modone\ model, this yields
\be\label{psidot1} \dot\psi_{\mu}=\bigl[H_{\modelone},\psi_{\mu}\bigr]= -\Bigl(m-\frac{\la}{2\sqrt{d}}\Bigr)\psi_{\mu}-\la\sqrt{d}\,\psi_{\nu}\psi_{\mu}^{\dagger}\psi_{\nu}\, .\ee
Contracting with $\psi_{\mu}^{\dagger}$ on both sides and taking the trace we get
\be\label{Geqdiff}
\Bigl(\frac{\d}{\d \tau} + m - \frac{\la}{2\sqrt{d}}\Bigr)G(\tau) = \begin{cases}\displaystyle
-\frac{\la\sqrt{d}}{n}\,\bigl\langle\tr\bigl[(\psi_{\mu}\psi_{\nu}^{\dagger}\psi_{\mu})(\tau)\psi_{\nu}^{\dagger}\bigr]\bigr\rangle_{\beta} &\quad \text{if $\tau>0$}\\[5pt]
\displaystyle
\frac{\la\sqrt{d}}{n}\,\bigl\langle\tr\bigl[\psi_{\mu}^{\dagger}(\psi_{\nu}\psi_{\mu}^{\dagger}\psi_{\nu})(\tau)\bigr]\bigr\rangle_{\beta} &\quad \text{if $\tau<0$}\, .
\end{cases}
\ee
In particular,
\begin{align}\label{Gdotzero1} \dot G(0^{+}) & = -\Bigl(m-\frac{\la}{2\sqrt{d}}\Bigr)G(0^{+})
-\frac{\la\sqrt{d}}{n}\,\bigl\langle\tr\psi_{\mu}\psi_{\nu}^{\dagger}\psi_{\mu}\psi_{\nu}^{\dagger}\bigr\rangle_{\beta}\,,\\\label{Gdotzero2}
\dot G(0^{-}) & = -\Bigl(m-\frac{\la}{2\sqrt{d}}\Bigr)G(0^{-})
+\frac{\la\sqrt{d}}{n}\,\bigl\langle\tr\psi^{\dagger}_{\mu}\psi_{\nu}\psi_{\mu}^{\dagger}\psi_{\nu}\bigr\rangle_{\beta}\, .
\end{align}
Let us note \emph{en passant} that, subtracting the above two equations and using \eqref{rhozero} and the canonical anticommutation relations to evaluate
\be\label{traceeval} \tr\bigl(\psi_{\mu}\psi_{\nu}^{\dagger}\psi_{\mu}\psi_{\nu}^{\dagger} + \psi^{\dagger}_{\mu}\psi_{\nu}\psi_{\mu}^{\dagger}\psi_{\nu}\bigr)
= \frac\N\D\,\cvp\ee
we get
\be\label{deltaGdot} \dot G(0^{-})-\dot G(0^{+}) = m+\frac{\la}{2\sqrt\D}\,\cdotp\ee
This together with \eqref{disc} yields an exact formula for the first moment of the spectral function, generalising \eqref{rhoone} at finite $\N$ and $\D$.

In the \modtwo\ model, we have\footnote{At finite $\N$ and $\D$, we use the tracelessness condition $\tr\psi_{\mu}=0$ to simplify the formulas. This plays no role at infinite $\N$.}
\be\label{psidot2} \dot\psi_{\mu}=\bigl[H_{\modeltwo},\psi_{\mu}\bigr]= -m\psi_{\mu}-\frac{\xi}{2}\sqrt{d}\, \psi_{\nu}\psi_{\mu}\psi_{\nu}-\frac{\xi^{*}}{2}\sqrt{d}\,\bigl(\psi_{\nu}^{\dagger}\psi_{\mu}^{\dagger}\psi_{\nu}- \psi_{\nu}^{\dagger}\psi_{\mu}\psi_{\nu}^{\dagger}+ \psi_{\nu}\psi_{\mu}^{\dagger}\psi_{\nu}^{\dagger}\bigr)\, .\ee
Massaging a bit the traces, we get
\begin{multline}\label{Gdotzero3} \dot G(0^{+})+m G(0^{+})=\dot G(0^{-})+m G(0^{-})\\=
\frac{\xi\sqrt{d}}{2n}\,\bigl\langle\tr\psi_{\mu}^{\dagger}\psi_{\nu}\psi_{\mu}\psi_{\nu}\bigr\rangle_{\beta}
+\frac{3\xi^{*}\sqrt{d}}{2n}\,\bigl\langle\tr\psi_{\mu}^{\dagger}\psi_{\nu}^{\dagger}\psi_{\mu}^{\dagger}\psi_{\nu}\bigr\rangle_{\beta}\, .
\end{multline}
In particular, in the \modtwo\ model
\be\label{deltaGdot2} \dot G(0^{-})-\dot G(0^{+}) = m\,,\ee
and Eq.\ \eqref{rhoone} is valid even at finite $\N$ and $\D$.

\paragraph{Step two} 

The partition function $Z$ and free energy $F$ are defined in general by
\be\label{ZFdef1} Z = \Tr e^{-\beta H}=e^{-\beta F}\,,\ee
where $\Tr$ traces over the Hilbert space of states. Then
\begin{align}\label{dFdm1}\frac{\partial F}{\partial m} &= nd\bigl\langle\tr\psi_{\mu}^{\dagger}\psi_{\mu}\bigr\rangle=-n^{2}d\,  G(0^{-})\,,
\\\label{dfdla1}  \frac{\partial F}{\partial\la} &=\frac{1}{2}nd^{3/2}\bigl\langle
\tr\psi_{\mu}^{\dagger}\psi_{\nu}\psi_{\mu}^{\dagger}\psi_{\nu}\bigr\rangle_{\beta}\,,
\end{align}
in the \modone\ model and
\begin{align}\label{dFdm2}\frac{\partial F}{\partial m} &= nd\bigl\langle\tr\psi_{\mu}^{\dagger}\psi_{\mu}\bigr\rangle=-n^{2}d\, G(0^{-})\,,
\\\label{dfdxi2}  \frac{\partial F}{\partial\xi} &=\frac{1}{2}nd^{3/2}\bigl\langle
\tr\psi_{\mu}^{\dagger}\psi_{\nu}\psi_{\mu}\psi_{\nu}\bigr\rangle_{\beta}\,,
\\\label{dfdxi2b}  \frac{\partial F}{\partial\xi^{*}} &=\frac{1}{2}nd^{3/2}\bigl\langle
\tr\psi_{\mu}^{\dagger}\psi_{\nu}^{\dagger}\psi_{\mu}^{\dagger}\psi_{\nu}\bigr\rangle_{\beta}\,,
\end{align}
in the \modtwo\ model. From \eqref{Gdotzero1} and \eqref{Gdotzero3}, neglecting the subleading $1/\sqrt{d}$ term in the \modone\ model, we thus get
\be\label{GdotFderrel}\dot G(0^{-})+mG(0^{-})=\dot G(0^{+})+mG(0^{+})=
\displaystyle
\begin{cases}\displaystyle \frac{2\la}{n^{2}d}\frac{\partial F}{\partial\la} & \quad\text{(\modone\ model)}\\\\
\displaystyle
\frac{1}{n^{2}d}\Bigl(\xi\frac{\partial F}{\partial\xi}+3\xi^{*}\frac{\partial F}{\partial\xi^{*}}\Bigr) & \quad\text{(\modtwo\ model)}\, .
\end{cases}
\ee
In particular, since $G$ is real, in the \modtwo\ model we must have
\be\label{dfrelxi}\xi\frac{\partial F}{\partial\xi}=\xi^{*}\frac{\partial F}{\partial\xi^{*}}\,\cvp\ee
or, equivalently,
\be\label{Fofxi} F(\xi,\xi^{*}) = F(|\xi|)\, .\ee
This result is obvious from the structure of the generalized melons in the large $\dof$ limit, but we see here that it is valid even at finite $\N$ and $\D$. In the \modtwo\ model, we can thus rewrite \eqref{GdotFderrel} as
\be\label{GdotFderrel2}
\dot G(0^{-})+mG(0^{-})=\dot G(0^{+})+mG(0^{+})=
\frac{4\xi}{n^{2}d}\frac{\partial F}{\partial\xi}=\frac{4\xi^{*}}{n^{2}d}\frac{\partial F}{\partial\xi^{*}}\,\cdotp\ee
Finally, note that, by definition of the self-energy \eqref{Sigmadef}, we can also write
\be\label{GdotSrel} \dot G(0^{-})+mG(0^{-})=\dot G(0^{+})+mG(0^{+})=-\int_{0}^{\beta}G(\tau)\Sigma(-\tau)\,\d \tau\, .\ee
From the point of view of the Fourier transform, the first equality in the above equation follows from the fact that both $G_{k}$ and $\Sigma_{k}$ are proportional to $1/k$ at large $k$ and thus the series $\sum_{k\in\mathbb Z+1/2}G_{k}\Sigma_{k}$ converges.

\paragraph{Step three}

Define the functionals
\begin{align}\label{Seffmat1}
\begin{split}
\frac{1}{n^{2}d}\sSeff^{\modelone}[\sG,\sS] &=-\ln\bigl(1+e^{-\beta m}\bigr) -\ln\frac{\det\mathscr O_{\sS}}{\det\mathscr O_{0}} + \int_{[0,\beta]^{2}}\Bigl[\sS(\tau_{2},\tau_{1})\sG(\tau_{1},\tau_{2})\\ & \hskip 3cm
-\frac{1}{4}\la^{2}\sG(\tau_{1},\tau_{2})^{2}\sG(\tau_{2},\tau_{1})^{2}\Bigr]\d \tau_{1}\d \tau_{2}\, ,
\end{split}
\\\label{Seffmat2}
\begin{split}
\frac{1}{n^{2}d}\sSeff^{\modeltwo}[\sG,\sS] &=-\ln\bigl(1+e^{-\beta m}\bigr)  -\ln\frac{\det\mathscr O_{\sS}}{\det\mathscr O_{0}} +\int_{[0,\beta]^{2}}\Bigl[\sS(\tau_{2},\tau_{1})\sG(\tau_{1},\tau_{2})\\ & \hskip 3cm+\frac{1}{4}|\xi|^{2}\sG(\tau_{1},\tau_{2})^{3}\sG(\tau_{2},\tau_{1})\Bigr]\d \tau_{1}\d \tau_{2}\, ,
\end{split}
\end{align}
where $\sG$ and $\sS$ are antiperiodic functions of their arguments and the operator $\mathscr O_{\sS}$ is defined as in \eqref{OpSdef}. Equations \eqref{Seffmat1} and \eqref{Seffmat2} are, at this point, nothing more than mathematical definitions. They are constructed in such a way that the saddle point equations obtained by varying $\sSeff^{\modelone}$ and $\sSeff^{\modeltwo}$ with respect to $\sS$ and $\sG$ do match the correct Schwinger-Dyson equations \eqref{Sigmadef}, \eqref{SDone} and \eqref{SDtwo} of the \modone\ and \modtwo\ models respectively. There is no reference to a path integral argument here: we simply use the matrix model Schwinger-Dyson equations, which are derived diagrammatically, and the mathematical statement that varying the functionals \eqref{Seffmat1}, \eqref{Seffmat2} yields precisely these equations. 

Let us denote by $\Sigma$ and $G$ the on-shell values of $\sS$ and $\sG$, \emph{i.e.}\ solutions to \eqref{Sigmadef} and \eqref{SDone} or \eqref{SDtwo}. Using the fact that the functionals \eqref{Seffmat1} and \eqref{Seffmat2} are stationary when evaluated at $\sG=G$ and $\sS=\Sigma$, we get
\begin{align}\label{Seffderla1} \frac{\partial\Seff^{\modelone}}{\partial\la} & = -\frac{\la\beta}{2}\int_{0}^{\beta}G(\tau)^{2}G(-\tau)^{2}\,\d \tau\, ,\\
\label{Seffderxi1} \frac{\partial\Seff^{\modeltwo}}{\partial\xi} & = \frac{\xi^{*}\beta}{4}\int_{0}^{\beta}G(\tau)^{3}G(-\tau)\,\d \tau\, ,
\end{align}
which, using the Schwinger-Dyson equations \eqref{SDone} or \eqref{SDtwo}, as well as the antiperiodicity of $G$, can be rewritten as
\begin{align}\label{Seffderla2} \la\frac{\partial\Seff^{\modelone}}{\partial\la} = -\frac{\beta}{2}\int_{0}^{\beta}G(\tau)\Sigma(-\tau)\,\d \tau\, ,\\\label{Seffderxi2}
\xi\frac{\partial\Seff^{\modeltwo}}{\partial\xi} = -\frac{\beta}{4}\int_{0}^{\beta}G(\tau)\Sigma(-\tau)\,\d \tau\, .
\end{align}
Comparing with \eqref{GdotFderrel}, \eqref{GdotFderrel2} and \eqref{GdotSrel}, we get
\be\label{SeffFmatrel}
\frac{\partial\Seff^{\modelone}}{\partial\la}=\beta\frac{\partial F}{\partial\la} \qquad\text{and}\qquad \frac{\partial\Seff^{\modeltwo}}{\partial\xi}=\beta\frac{\partial F}{\partial\xi}\,\cvp
\ee
for the \modone\ and \modtwo\ models, respectively.

\paragraph{Step four}

Integrating \eqref{SeffFmatrel} and fixing the constant of integration by using the fact that at zero coupling both models are simply a collection of $\dof=\N^2\D$ decoupled fermionic oscillators, we obtain \eqref{SeffFrelation}. This is exactly as in the case of the disordered model formulation, where $\Seff$ is replaced by $\smash{\Seff^{\modelone}}$ or $\smash{\Seff^{\modeltwo}}$ for the \modone\ or \modtwo\ models, according to which couplings are turned on.

\subsubsection{Useful formulas\label{EucformSec}}

Numerically, the solution of the Schwinger-Dyson equations is most naturally encoded in the Matsubara-Fourier coefficients $G_{k}$ (or $\Sigma_{k}$). It is thus very convenient to express all the relevant physical quantities, such as fermion number, free energy, energy and entropy, as convergent sums over these coefficients.

\paragraph{Fermion number}

The fermion number \eqref{Qdef1} is given in terms of the two-point function
\be\label{QGrel} Q = -G(0^{-})=1-G(0^{+})\, .\ee
One cannot use immediately the expansion \eqref{GFourier} to evaluate $G(0^{\pm})$ because $\sum_{k}G_{k}$ diverges. This is related to the fact that the function $G$ is discontinuous at $\tau=0$. To go around this problem, one can use Dirichlet's trick of considering $G(\tau)+G(-\tau)$, which is continuous at $\tau=0$ and has a convergent Fourier expansion. This yields
\be\label{Gzeropm}
G(0^{+}) + G(0^{-}) = \frac{2}{\beta}\sum_{k\in\mathbb Z+\frac{1}{2}}\re G_{k}\, .
\ee
Using then \eqref{QGrel}, \eqref{rhozero} and \eqref{Gkreality}, we get
\be\label{QFourier} Q = \frac{1}{2}-\frac{2}{\beta}\sum_{k\in\mathbb N+\frac{1}{2}}\re G_{k}\, .\ee
Subtracting the tree-level result, with tree-level coefficients $G_{k}^{(0)}$ given in \eqref{Gk0def}, we also get
\be\label{QFourier2}
Q = \nF(m) - \frac{2}{\beta} \sum_{k\in\mathbb N+\frac{1}{2}} \re\bigl(G_k - G_k^{(0)}\bigr)\, .
\ee
Formulas \eqref{QFourier} and \eqref{QFourier2} are of course mathematically equivalent, but \eqref{QFourier2} is more convenient to use in numerical evaluations because the infinite series converges more quickly. This is due to the fact that $\re G_{k}$ and $\re (G_k - G_k^{(0)})$ decay like $1/k^{2}$ and $1/k^{4}$ respectively at large $k$. We shall repeatedly use a similar rewriting of the formulas in the following, separating the tree-level result and the quantum corrections.

There is also an interesting formula for the fermion number in terms of the spectral density, which is obtained from \eqref{QGrel} and \eqref{Gspecdec},
\be\label{Qrhorel} Q = \int_{-\infty}^{+\infty}\nF(\omega)\rho(\omega)\,\d\omega\, .\ee
This formula coincides with the formula for a collection of \emph{decoupled} fermionic harmonic oscillators with a frequency distribution given by the spectral density $\rho$. 

\paragraph{Free energy}

The free energy is obtained by evaluating the effective actions \eqref{Seffmat1} or \eqref{Seffmat2} on the solution for the two-point function and using \eqref{SeffFrelation}. It is convenient to use the Schwinger-Dyson equations \eqref{SDone} or \eqref{SDtwo} to simplify the result. We get in this way
\be\label{FreeEnergy}
\frac{F}{\dof} = m - T\ln\bigl(1+e^{\beta m}\bigr)
+T\sum_{k\in\mathbb N+\frac{1}{2}}\biggl[\ln\biggl|\frac{G_{k}}{G_{k}^{(0)}}\biggr|^2 + \frac{3}{2}\Bigl(1-\re\frac{G_k}{G_k^{(0)}}\Bigr)\biggr]\, .
\ee
Note that this formula is valid for both the \modone\ and the \modtwo\ models.

\paragraph{Energy} The energy is simply the expectation value of the Hamiltonian,
\be\label{Edefgen} E = \langle H\rangle_{\beta}\, .\ee
From the definitions \eqref{Hone} and \eqref{Htwo}, Eq.\ \eqref{GdotFderrel}, \eqref{dfrelxi} and \eqref{QGrel}, we get
\be\label{Energya} \frac{E}{\dof} = mQ +\frac{1}{2}\bigl(\dot G(0^{+})+ m G(0^{+})\bigr)\, ,\ee
which is valid for both the \modone\ and the \modtwo\ models. The right-hand side of \eqref{Energya} can be evaluated using \eqref{QFourier} and \eqref{GdotSrel}, yielding
\be\label{Energy} \frac{E}{N} = \frac{m}{2}-\frac{1}{\beta}\sum_{k\in\mathbb N+\frac{1}{2}}
\Bigl(2m\re G_{k} + 1 - \re\frac{G_k}{G_k^{(0)}}\Bigr)\, .\ee
Subtracting the tree-level result yields the equivalent formula
\be\label{Energy2} \frac{E}{N} =m\,\nF(m) -\frac{1}{\beta}\sum_{k\in\mathbb N+\frac{1}{2}}
\Bigl(2m\re\bigl( G_{k} - G_{k}^{0}\bigr) + 1 - \re\frac{G_k}{G_k^{(0)}}\Bigr)\, ,\ee
to be used in the numerical evaluations.

It is interesting to express the energy in terms of the spectral density, using \eqref{Energya} and \eqref{Gspecdec} together with \eqref{rhozero} and \eqref{rhoone}. We obtain
\be\label{Erhorel} \frac{E}{N} = \frac{1}{2}\int_{-\infty}^{+\infty}\bigl(m+\omega)\nF(\omega)\rho(\omega)\,\d\omega\, .\ee
Note that, unlike the case of the fermion number \eqref{Qrhorel}, this formula differs slightly from the formula for a collection of decoupled fermionic harmonic oscillators, by the replacement of the frequency factor $\omega$ by $\frac{1}{2}(m+\omega)$.

\paragraph{Entropy}

The entropy is then given by the standard relation
\be\label{entropy} S = \beta(E-F)\, .\ee
We have also checked explicitly that the result so obtained is consistent with the first relation in \eqref{SEthermo}.

\paragraph{Zero temperature in the \modone\ model}

One can derive very interesting exact formulas at zero temperature in the \modone\ model \cite{zeroTcondmat,Sachdev:2015efa} when the low energy ansatz discussed in Sec.~\ref{ReparaSec} is valid. The charge at zero temperature can be expressed in terms of the low energy parameter $\theta$ defined in \eqref{Cthetadef} as
\be\label{thetaQrel} Q(T=0) = \frac{1}{2}-\frac{\theta}{\pi}-\frac{1}{4}\sin (2\theta)\, .\ee
Moreover, if we denote by $\sigma$ the zero-temperature entropy,
\be\label{zeroTendef} \sigma(m,\la) = \frac{1}{\dof}\lim_{T\rightarrow 0}S(T,m,\la)\, ,\ee
one finds in the case $m=0$
\be\label{zeroTentropySYK}
\sigma(m=0) = \ln 2 - \int_{0}^{\pi/4}\Bigl(1-\frac{2\theta}{\pi}\Bigr)\tan\theta\,\d\theta =\frac{\text{Catalan}}{\pi} + \frac{1}{4}\ln 2 \simeq 0.46\, ,\ee
where
\be\label{Catalandef}\text{Catalan} = \sum_{k=0}^{\infty}\frac{(-1)^{k}}{(2k+1)^{2}}\ee
is the so-called Catalan constant. When $m\not = 0$, it is useful to consider $\sigma$ as a function of the zero-temperature charge $Q(T=0,m)$ instead of $m$. One can then derive the elegant formula
\be\label{dsigdqzeroT}\frac{\partial\sigma}{\partial Q} = \ln\frac{b_{+}}{b_{-}}\,\cdotp\ee
Using \eqref{Cthetadef}, \eqref{thetaQrel} and \eqref{zeroTentropySYK} as an initial condition,\footnote{One may also use $\sigma(Q=0)=0$ as an initial condition to integrate \eqref{dsigdqzeroT}. This yields the correct answer, but is  marred by an important subtlety. Indeed, as explained in \cite{letter} and reviewed in Sec.~\ref{PhaseSec}, the low energy ansatz discussed in Sec.~\ref{ReparaSec}, which is at the basis of the derivation of \eqref{dsigdqzeroT}, seems to be valid only when the zero-temperature charge $Q$ is greater than a strictly positive critical value.} and after a little massaging, this equation can be integrated to 
\be\label{zeroTentropy} \sigma(Q) = N\int_{\theta(Q)}^{\pi/4}\Bigl(\frac{1}{\pi}+\frac{1}{2}\cos(2\theta')\Bigr)\ln\frac{\cos\theta' + \sin\theta'}{\cos\theta'-\sin\theta'}\,\d\theta'\,,\ee
or more explicitly
\begin{align}\label{zeroTentropy2}
\sigma(\theta) &= \left(\frac{1}{2} - \frac{\theta}{\pi} - \frac{1}{4}\sin 2\theta \right) \ln\frac{\cos\theta + \sin\theta}{\cos\theta-\sin\theta} - \frac{i}{2\pi}\left({\rm Li}_2(i e^{2i\theta}) - {\rm Li}_2(-ie^{2i\theta})\right)\\
&\qquad\qquad+ \frac{1}{4}\ln(2\sec 2\theta) -{\rm atanh}\tan\theta - \frac{2i\theta}{\pi}{\rm atan}\,e^{2i\theta}\,.\nonumber
\end{align}
The above zero-temperature formulas will not play a crucial role in our analysis of the model, but will be used in the next subsection to check the precision of our numerical results.

\subsection{\label{NumSolEucSec} Numerical solution of Schwinger-Dyson equations}

In this subsection, and in all numerical computations except otherwise stated, we always choose ``natural units'' such that the couplings are set to one, \emph{i.e.}\
\be\label{unitsnum}
\la = 1 \quad\text{and}\quad |\xi|=1\,.
\ee
We want to solve numerically, for any values of the parameters, the Schwinger-Dyson equations \eqref{GkfromSigmak}, with self-energy \eqref{SDone} or \eqref{SDtwo}, which we rewrite here for convenience in the units \eqref{unitsnum},
\be
\label{SDcomplete}
\begin{split}
G_k &= \frac{1}{m - i\nu_k + \Sigma_k}\,\cvp \\
\Sigma(\tau) &=\begin{cases}
G(\tau)^{2}G(-\tau) & \text{(\modone model)}\\
 -\frac{1}{4}\bigl(G(\tau)^{3}+3 G(\tau)G(-\tau)^{2}\bigr) & \text{(\modtwo model)}\, .
\end{cases}
\end{split}
\ee
We introduce a high-frequency cutoff $\maxk$ and restrict our attention to the $2\maxk$ Matsubara frequencies labeled by $k = -\maxk+\tfrac12,\ldots, \maxk-\tfrac12$. This effectively reduces \eqref{SDcomplete} to a system of $4\maxk$ algebraic equations for the coefficients $G_{k}$ and $\Sigma_{k}$. We solve this system using a standard iterative procedure which, given an approximate solution $(G_k^{[i-1]},\Sigma_{k}^{[i-1]})$ to \eqref{SDcomplete}, generates another one through \cite{Maldacena:2016hyu}
\be\label{nsolveriteration}
G^{[i]}_k = (1-\alpha) G^{[i-1]}_k + \frac{\alpha}{m - i\nu_k + 
\Sigma^{[i-1]}_k}\,\cvp
\ee
with $\Sigma_k^{[i]}$ calculated from $G_k^{[i]}$ through \eqref{SDcomplete}.

The weighting factor $\alpha$ can \emph{a priori} be an arbitrary real number, and this choice will directly affect the convergence of the algorithm. One may refine it by making $\alpha$ frequency-dependent, or updating it from one iteration to the next as in \cite{Maldacena:2016hyu}. For most purposes of the present paper, we found that a fixed, frequency-independent value $\alpha=0.75$ is sufficient. We found no other solutions by using other values of $\alpha$, by introducing a frequency dependence or by updating the weighting factor during the iterations.

The above algorithm is a generalization of the iterative procedure commonly used to solve fixed-point equations of the form $f(x) = x$, of which \eqref{SDcomplete} can be thought of as being a multivariate version. Solutions $x^*$ to $f(x)=x$ can be found from the iteration $x^{[i]}=f(x^{[i-1]})$ as long as the initial guess for the solution is close enough to $x^*$ and $|f'(x^*)| < 1$. When this condition is satisfied, we say that the fixed point is attractive. If $|f'(x^*)| \geq 1$, on the other hand, the fixed point is repulsive and the method cannot converge. This is where the use of a weighting factor $\alpha$ can be helpful. Indeed, all fixed-points of $f(x)$ are also fixed-points of $f_\alpha(x) = \alpha f(x) + (1-\alpha) x$, and it is easy to check that we can always find a non-zero value of $\alpha$ such that $|f_\alpha'(x^*)| < 1$. Thus, a repulsive fixed-point of the original problem $f(x) = x$ may be found as an attractive fixed point of the modified $f_\alpha(x) = x$ problem, for some suitably chosen value of $\alpha$.

An important ingredient of the algorithm is the choice of the initial approximation. It may be taken to be the tree-level coefficients \eqref{Gk0def}, $G_{k}^{[0]}=G_{k}^{(0)}$, or alternatively any other previously found Matsubara-Fourier coefficients deemed to be close to the actual solution. The iteration is then repeated until the equations are satisfied up to a prescribed tolerance $\varepsilon$, \emph{i.e.}\
\be
\frac{1}{\maxk}\sum_{k=-\maxk+\tfrac12}^{\maxk-\tfrac12} \biggl|G^{[i]}_k -\frac{1}{m - i\nu_k + \Sigma^{[i]}_k}\biggr| < \varepsilon\,.
\ee
Since we work with 64-bit processors, we chose the tolerance to be $\varepsilon = 10^{-12}$ in the units \eqref{unitsnum}, safely above machine-precision.

There is no guarantee that this algorithm terminates, but if convergence is achieved in $I_\alpha(\varepsilon)$ steps then the running time is $\mathcal{O}(I_\alpha(\varepsilon) \times \maxk \log \maxk)$. The $\maxk \log \maxk$ factor in this expression comes from performing two Fast Fourier Transforms (FFTs) in each step, first to obtain $G^{[i]}(\tau)$ from $G_k^{[i]}$ and then to obtain $\smash{\Sigma^{[i]}_k}$ from $\smash{\Sigma^{[i]}(\tau)}$. Note that the calculation of $\smash{\Sigma^{[i]}(\tau)}$ is itself $\mathcal{O}(\maxk)$, or more generally $\mathcal{O}(\maxk \log q)$ for the $q$-body generalizations of \eqref{SDcomplete} (see App.~\ref{qGenApp}), so it is always subleading because we shall always have $q \ll \maxk$.

We implemented the above algorithm in {\tt C++}, and also in Mathematica for reference and double-checking purposes. In our {\tt C++} implementation, FFTs are calculated using the Cooley-Tukey algorithm \cite{Cooley:1965zz} optimized to work with $\maxk$ being a power of $2$. In general, lower temperatures require higher values of $\maxk$ to obtain good solutions. The reason is that the highest Matsubara frequency we include must always be much greater than the scale set by the coupling constant, \emph{i.e.}\
\be\label{condonkUV} \nu_{\maxk-\tfrac12} \equiv \frac{2\pi}{\beta}\left(\maxk-\tfrac12\right) \gg 1 \qquad\text{in natural units}\,.\ee
The condition \eqref{condonkUV} ensures that at scale $\nu_{\maxk-1/2}$ we are well into the UV free asymptotic regime of the models, governed by the expansion \eqref{GkUV} with \eqref{rhozero} and \eqref{rhoone}, for which
\be\label{GUVnum}
G_{\maxk-\frac12} = G^{(0)}_{\maxk-\frac{1}{2}} + \mathcal O\bigl(1/\maxk^{3}\bigr)\, .
\ee
After achieving convergence, one should always check that \eqref{GUVnum} is satisfied, as well as the basic consistency conditions \eqref{Gkreality}, \eqref{Gsign} and \eqref{imGsign}.

In practice, the cutoff $\maxk$ also affects the precision with which we can compute thermodynamic quantities through formulas such as \eqref{QFourier2}, \eqref{FreeEnergy} or \eqref{Energy2}. Indeed, introducing a high-frequency cutoff amounts to the truncation of the corresponding series, thus resulting in a systematic error in the computed values. To keep this error under control at temperatures of the order $T \sim 0.1$, one finds that $\maxk = 2^{10}$ is high enough, but for $T \sim 0.001$ the cutoff should be increased at least to $\maxk = 2^{16}$, and for $T \sim 0.0001$ we can reach $\maxk = 2^{21}$. See App.~\ref{NumericalErrorsApp} for a more thorough discussion.

We shall use the above algorithm extensively in the following. For instance, when computing the phase diagrams in Sec.~\ref{PhaseSec} we map the solutions that exist at each point in the space parameterized by $(m,T)$. In order to do this we may begin by finding numerical solutions of the Schwinger-Dyson equations in a region where an approximate solution is known to be valid, \emph{e.g.}\ in one of the perturbative regimes $T \gg 1$ or $m \gg 1$. The algorithm then converges if we use as initial solution the tree-level coefficients \eqref{Gk0def}, $G^{[0]}_{k}=G^{(0)}_k$. Keeping all but one of the parameters fixed, we can then find another solution close to this point using the previously found solution as an initial approximation, successively repeating this process to explore all of parameter space. While doing this, it is very useful to keep track of the thermodynamic properties of the solutions and the convergence of the algorithm, in order to adjust the step sizes accordingly.

\begin{figure}[h!]
\centering
\def\svgwidth{4.5in}
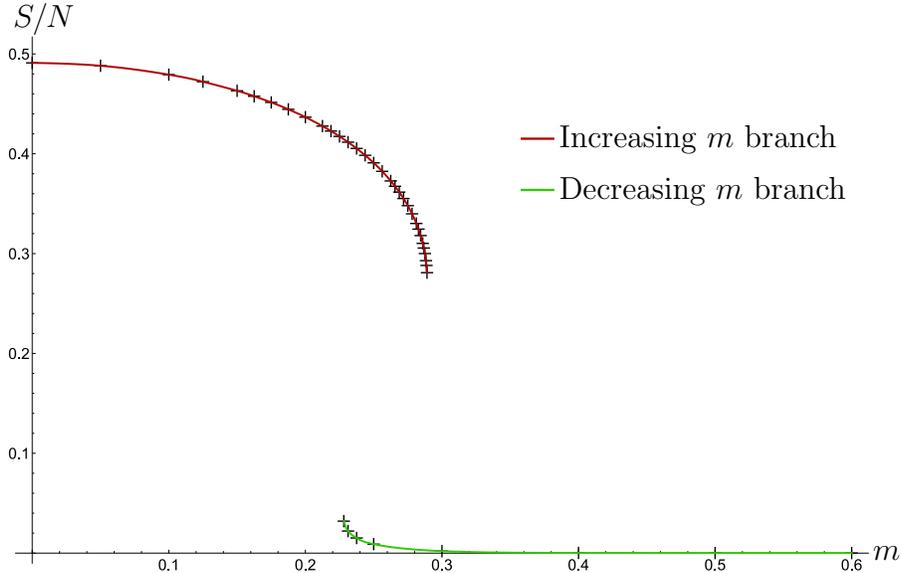
\caption{\label{fig:Hysteresis}Hysteresis curve for the entropy of the \modone\ model as a function of mass, for fixed temperature $T = 0.035$ in natural units (data points shown with crosses for illustrative purposes). The red (dark gray) curve is obtained starting from the solution found at $m = 0$ and gradually increasing the mass, whereas the green (light gray) curve is obtained by decreasing the mass starting from the solution found at $m = 0.6$. In a range of masses around the value $m\sim 0.26$, two distinct solutions to the Schwinger-Dyson equations clearly coexist.}
\end{figure}

One extremely important feature, that we have already mentioned in Sec.~\ref{MonodromySec}, is the possible existence of multiple solutions at a given point $(m,T)$, due to non-trivial monodromies in the space of solutions. This is signaled by the fact that the process described in the paragraph above can sometimes result in hysteresis curves for the thermodynamic functions. A representative example is depicted in Fig.~\ref{fig:Hysteresis}. This means that the solution we find numerically depends not only on the parameters $(m,T)$ explicitly entering the equations \eqref{SDcomplete}, but also on the initial approximation $G_k^{[0]}$ we used to start the iterative procedure. In this way, the solution ``remembers'' the specific sequence of steps in parameter space that was taken to reach it, and through it also the perturbative regime where the whole process typically started. This phenomenon will be discussed extensively in Sec.~\ref{PhaseSec}.

\medskip

\noindent\emph{Note on numerical errors and error bars}\footnote{The remarks below apply also to the the real time numerics presented in Sec.~\ref{RealTimeSec}: } As discussed in detail in App.\ \ref{NumericalErrorsApp}, our numerical results are subject to two main sources of errors: systematic errors due to the introduction of cut-offs or discretizations; and methodological errors related to our fitting methods, \emph{etc}. We do not have rigorous control over systematic errors, which typically may become important at very low temperatures. The data analysis methods are however under better control, and error bars can be associated with them as explained in App.\ \ref{NumericalErrorsApp}. The error bars that appear in the following therefore always refer to the methodological errors, but do not take into account the systematic errors that may appear on top of them. This implies that the actual values of some low-temperature observables may lie outside our error bar intervals. This issue is only relevant in some specific, yet important, instances and will be mentioned and discussed explicitly whenever it occurs.

\subsubsection{\label{EucSolmod1Sec}Basic numerical results for the \modone\ model}

The numerical algorithm described above may be used to study the finite coupling Schwinger-Dyson equations \eqref{SDcomplete} for the \modone\ model, and assess for example the applicability of the low energy ansatz discussed in Sec.~\ref{ReparaSec}. From previous studies of the complex SYK model \cite{Sachdev:2015efa,Davison:2016ngz} we can expect that this should be the case for low enough masses, whereas at large mass the system must be in a gapped perturbative phase (see Sec.\ \ref{PhaseSec} for more details).

\begin{figure}[h!]
\centering
\vskip 0.3cm
\begin{tabular}{cc}
\def\svgwidth{7cm}
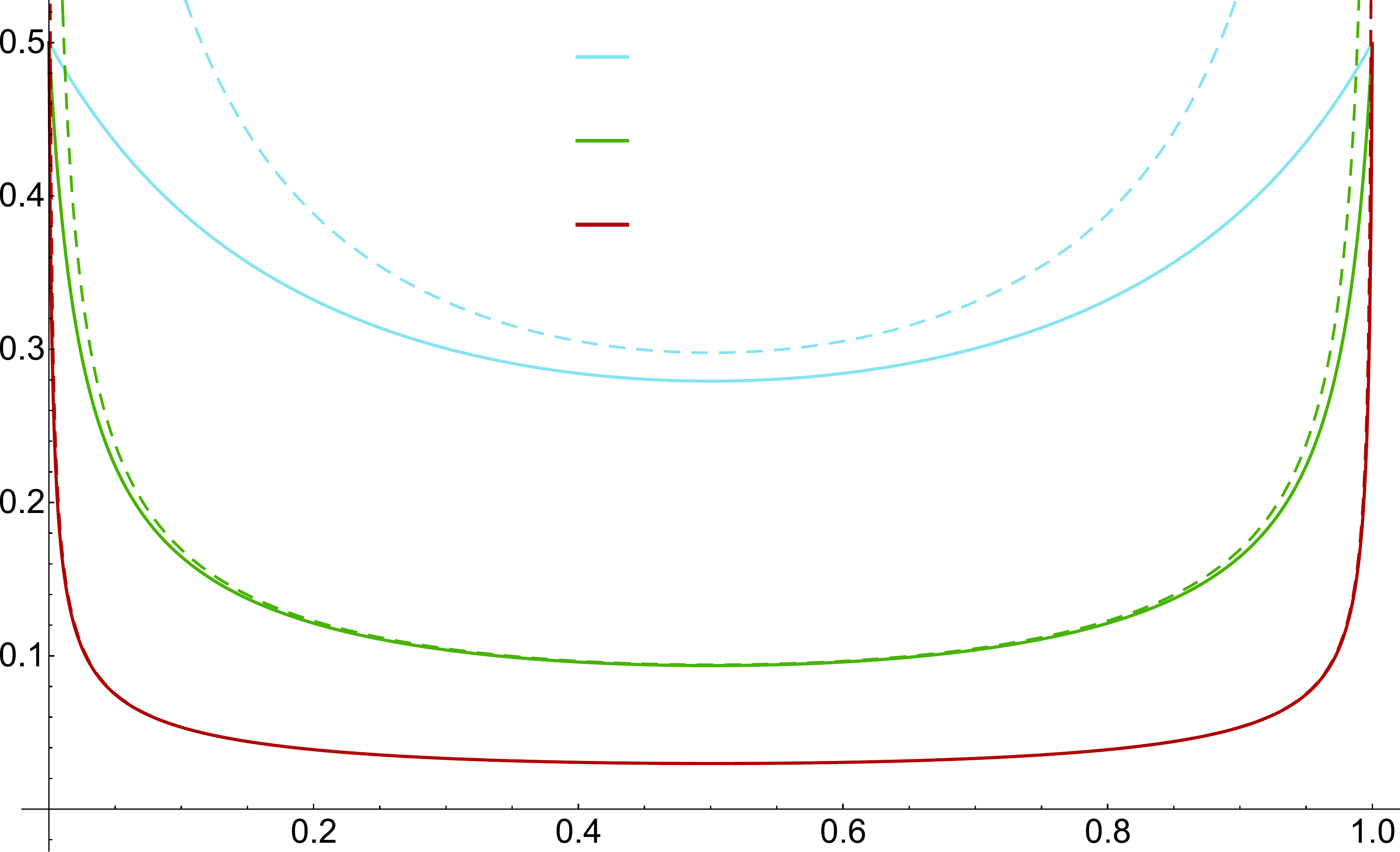
&
\def\svgwidth{7cm}
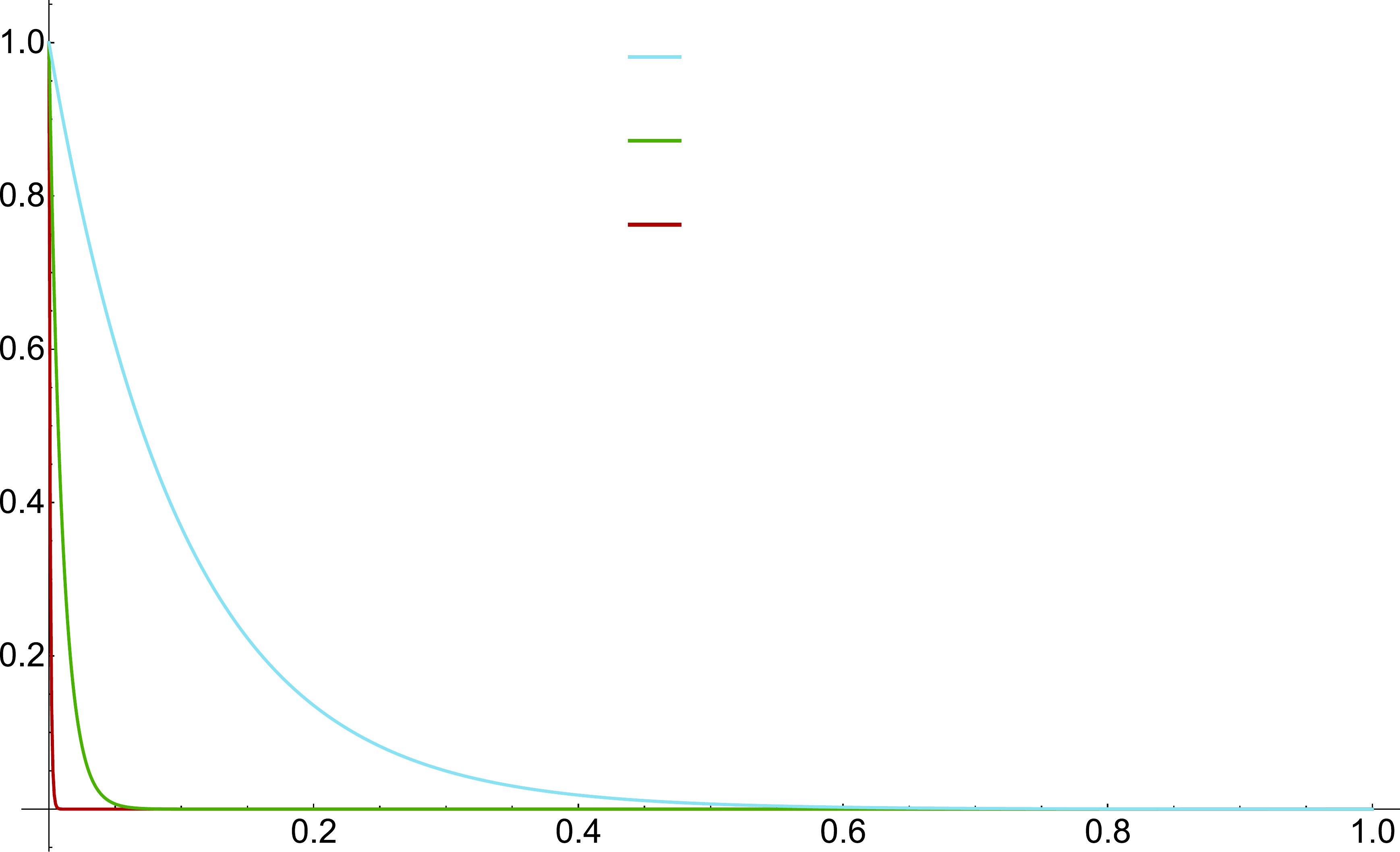 \\
(a) & (b)
\end{tabular}
\caption{\label{fig:M1SolutionExamples}Numerical solution of the Schwinger-Dyson equations for the \modone\ model at various temperatures. (a) Plots for $m = 0$, with the finite-temperature conformal ansatz \eqref{Gscaling} shown in dashed lines for comparison (for $m = 0$, particle-hole symmetry implies $b_+ = b_-$, so we have set $a = \theta = 0$). (b) Plots for $m = 1$, where the exponential decay of the two-point function is clearly visible. Here and elsewhere in this paper, unless otherwise stated, solutions were computed with $\maxk = 2^{16}$, \emph{i.e.}\ approximately $10^5$ frequencies.}
\end{figure}

In Fig.\ \ref{fig:M1SolutionExamples} we plot some solutions found for $m = 0$ and $m = 1$ at various temperatures. The $m=0$ solutions match with the standard SYK model, and we clearly see that the finite-temperature conformal ansatz \eqref{Gscaling} becomes a very good approximation in its range of validity, which in the natural units \eqref{unitsnum} is when $\beta\gg 1$, $\tau\gg 1$ and $\beta-\tau\gg 1$. The $m = 1$ solutions display a distinctly different behaviour, with the two-point function decaying exponentially. An energy gap can be immediately extracted from the log-plot of this type of solutions, see Sec.~\ref{PhaseSec}.

\begin{figure}[h!]
\centering
\def\svgwidth{4.5in}
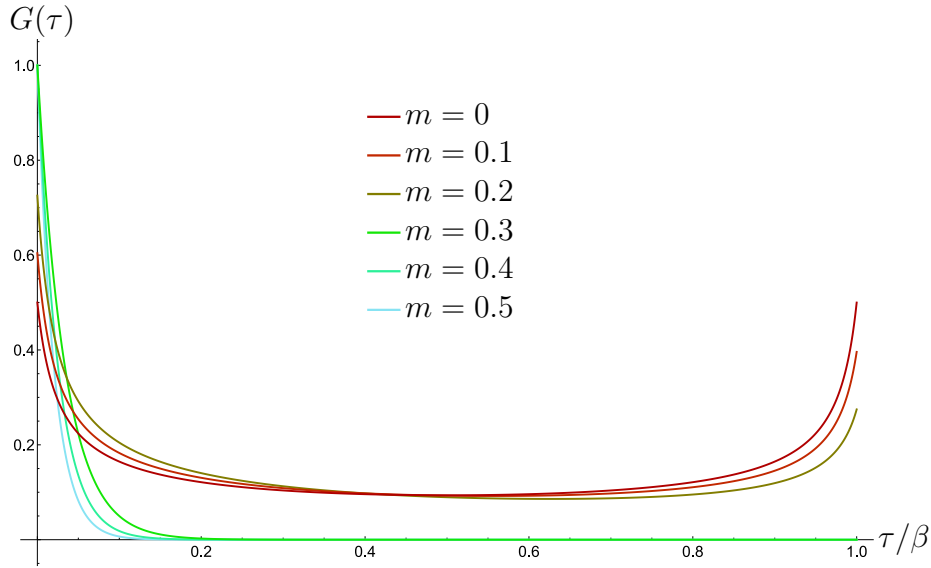
\caption{\label{fig:M1VaryingMass}Numerical solution of the Schwinger-Dyson equations for the \modone\ model at $\beta = 100$ in natural units, for various values of the mass $m$. The loss of the particle-hole symmetry is signaled by the asymmetric shape of $G(\tau)$ about $\tau = \beta/2$. The endpoint values are related to the charge $Q$ through \eqref{QGrel}, so that $Q = G(\beta^-) = 1-G(0^+)$. SYK-like solutions similar to Fig.~\ref{fig:M1SolutionExamples}a and perturbative solutions similar to Fig.~\ref{fig:M1SolutionExamples}b can be immediately distinguished.}
\end{figure}

For $m\neq 0$ the model loses its particle-hole symmetry and thus $G(\beta-\tau)\not = G(\tau)$. In Fig.~\ref{fig:M1VaryingMass} we plot solutions for various intermediate values of the mass at $T=0.01$. The marked change in behaviour around $m\sim0.3$, from power-law decay at low masses to exponential decay at large masses, will be associated in Sec.~\ref{PhaseSec} to a first order phase transition.

A very rough but simple criterion to distinguish the low and large mass solutions is to test the condition \eqref{Sigtildeconst}. Working at finite inverse temperature $\beta$ the lowest frequency we have access to is $\nu_{1/2} = \pi/\beta$, so that we can use $\re \Sigma_{1/2}$ as an approximation to $\tilde{\Sigma}(0)$. For the examples illustrated in Fig.~\ref{fig:M1VaryingMass}, we have $\re\Sigma_{1/2} = 0$, $-0.07$ and $-0.15$ for the solutions with $m = 0$, $0.1$ and $0.2$ respectively, whereas $\Sigma_{1/2} \sim 10^{-4}$ for the solutions with $m = 0.3,0.4,0.5$. These values show that the lowest three masses can satisfy the condition \eqref{Sigtildeconst}, whereas the three largest ones do not.

There are different ways in which we may quantitatively test the low energy ansatz of Sec.~\ref{ReparaSec}, beyond the naked-eye impression provided by Fig.~\ref{fig:M1SolutionExamples} or the very rough testing of \eqref{Sigtildeconst} through $\re\Sigma_{1/2}$. One possible approach starting from the zero-temperature expression \eqref{Gasymp} consists in extracting the scaling dimension from a numerical solution by fitting
\be\label{dGdLog1}
\frac{\d\ln G(\tau)}{\d\ln \tau} \to -2\Delta \qquad\text{as}\qquad \tau\to +\infty\,.
\ee
However, since the numerical evaluations are always performed at finite temperature we cannot use the above equation directly. Instead, Eq.\ \eqref{Gasymp} is approximately valid for values of $\tau$ satisfying $1 \ll \tau\ll \beta$. The precise interval of time we use in practice may be defined in different ways, but fortunately this choice does not greatly affect the fitted value of the scaling dimension $\Delta$ unless obviously unreasonable. Comparing the results obtained with different choices can serve the dual purpose of making sure that the numerical results  make sense, as well as providing an estimation of the error or uncertainty associated to the computed values, see App.~\ref{NumericalErrorsApp} for more details.

One important advantage of the procedure just described is that we can repeat it independently for $\tau\to-\infty$. We then obtain from each numerical solution not one but two values for the scaling dimension, denoted by $\Delta_\pm$ for $\tau\to\pm\infty$ as in \eqref{Gasympnew}. These values could \emph{a priori} be different, but we find that in the \modone\ model they are always very similar, and certainly equivalent within the numerical precision we can achieve. This is in sharp contrast with the case of the \modtwo\ model discussed in the next subsection. Applying the same methods, we will clearly see there that $\Delta_+ \neq \Delta_-$, signaling the fact that the behaviour of the \modtwo\ model indeed follows the non-standard ansatz \eqref{Gasympnew}.

\begin{figure}[h!]
\centering
\vskip 0.4cm
\def\svgwidth{4.5in}
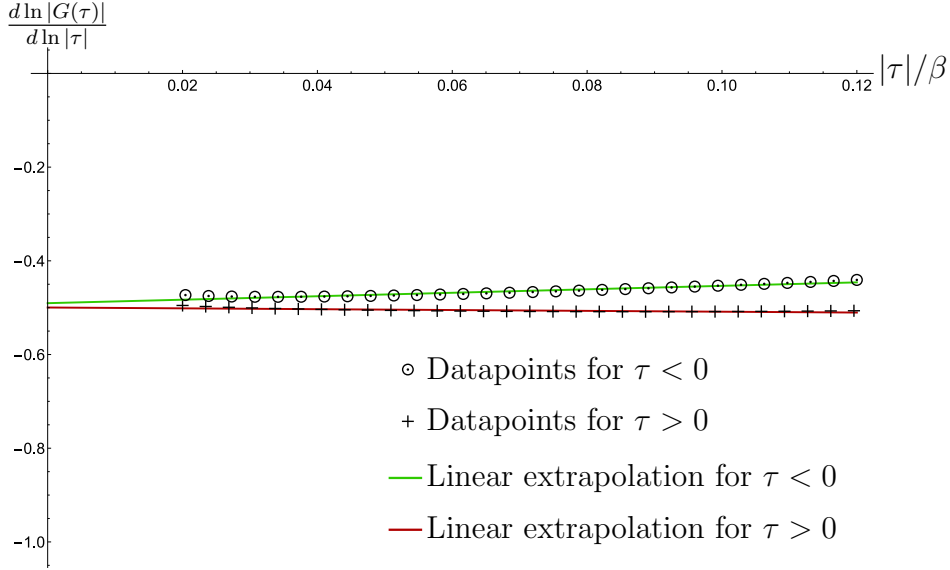
\caption{\label{fig:M1DeltaZeroTa}Extraction of the scaling dimension $\Delta$ and damping factor $a$ for the \modone\ model using \eqref{dGdLog2}, with $m = 0.1$, $\beta = 1000$ and $\la=1$. The linear behaviour of $\smash{\frac{\d\ln |G(\tau)|}{\d\ln |\tau|}}$ in the chosen range of Euclidean times is clearly seen. The scaling dimensions $\Delta_\pm$ correspond to the $y$-intercept of the linear fits, in this case $\Delta_+ = 0.250_3$ (red, dark gray) and $\Delta_- = 0.246_3$ (green, light gray), results consistent with $\Delta_{-}=\Delta_{+}=1/4$ as predicted by the standard low energy ansatz.
From the small slope we get the values $a_{+}=0.02_1$ and $a_{-}=0.06_1$ for the parameter $a$. This result is consistent with $a=a_{+}=a_{-} = 0.04_3$ and through \eqref{aform} yields an angle $\theta=0.12_9$. A more precise determination of $a$ and $\theta$ is obtained in Fig.\ \ref{fig:M1DeltaZeroTb}.
}
\end{figure}

More precisely, the $\beta\rightarrow\infty$ limit of the finite temperature formula \eqref{Gscaling} predicts
\be\label{dGdLog2}
\frac{\d\log |G(\tau)|}{\d\ln |\tau|} \underset{1\ll |\tau|\ll\beta}{\simeq} -2\Delta - \frac{2\pi a}{\beta}\tau \qquad\text{with}\qquad \Delta = \frac14 \,\cdotp
\ee
We thus expect a linear dependence in $\tau$, with a vanishingly small slope in the $\beta\to\infty$ limit. This is illustrated in Fig.~\ref{fig:M1DeltaZeroTa} for $T=10^{-3}$ and $m=0.1$. We indeed find a linear behaviour within the range of Euclidean times considered, both for $\tau\to+\infty$ and $\tau\to-\infty$. The scaling dimensions $\Delta_\pm$ are then obtained from the $\tau$-independent part of the corresponding linear fits, and seen to be consistent with the expected value given by \eqref{Deltaform1}, $\Delta = 1/4$. We can also read the coefficient $a$ from the slope and then derive the angle $\theta$ from \eqref{aform}. Note that this determination uses a subleading term and thus cannot be very accurate.

\begin{figure}[h!]
\centering
\vskip 0.3cm
\def\svgwidth{4.5in}
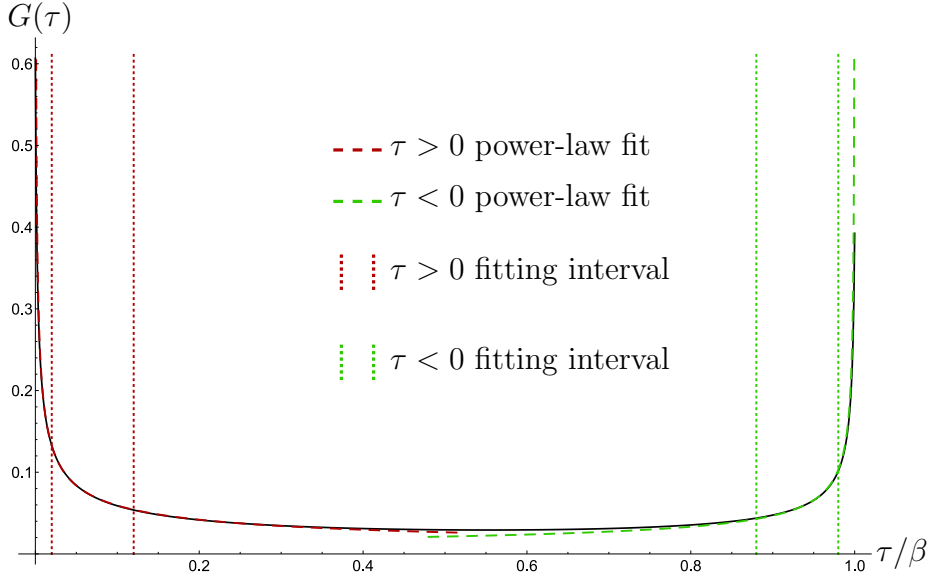
\caption{\label{fig:M1DeltaZeroTb}Extraction of the coefficients $b_{+}$ (red, dark gray) and $b_{-}$ (green, light gray) for the \modone\ model using the zero-temperature ansatz \eqref{Gasymp}, fitting the numerical solution with $m = 0.1$, $\beta = 1000$ and $\la=1$ (solid black curve). The intervals used for fitting are delimited by vertical dotted lines; we obtain $b_+ = 0.59_2$ and $b_- = 0.45_2$, which through the parameterization \eqref{Cthetadef} result in $C = 1.32_3$, $a=0.043_6$ and $\theta = 0.13_2$, satisfying \eqref{Cthetarel} to within $10\%$.}
\end{figure}

We can also use the previously found values for $\Delta_\pm$ as input to fit the coefficients $b_\pm$ directly in $G(\tau)$, again taking the numerical solution restricted to the finite intervals considered before. This is illustrated in Fig.~\ref{fig:M1DeltaZeroTb} for the example with $\beta=1000$ and $m=0.1$, and provides in particular a better determination of the parameters $a$ and $\theta$ through \eqref{aform}. A non-trivial check of the whole procedure is the fact that these satisfy with very good accuracy the relation \eqref{Cthetarel}.

\begin{figure}[h!]
\centering
\begin{tabular}{cc}
\def\svgwidth{7cm}
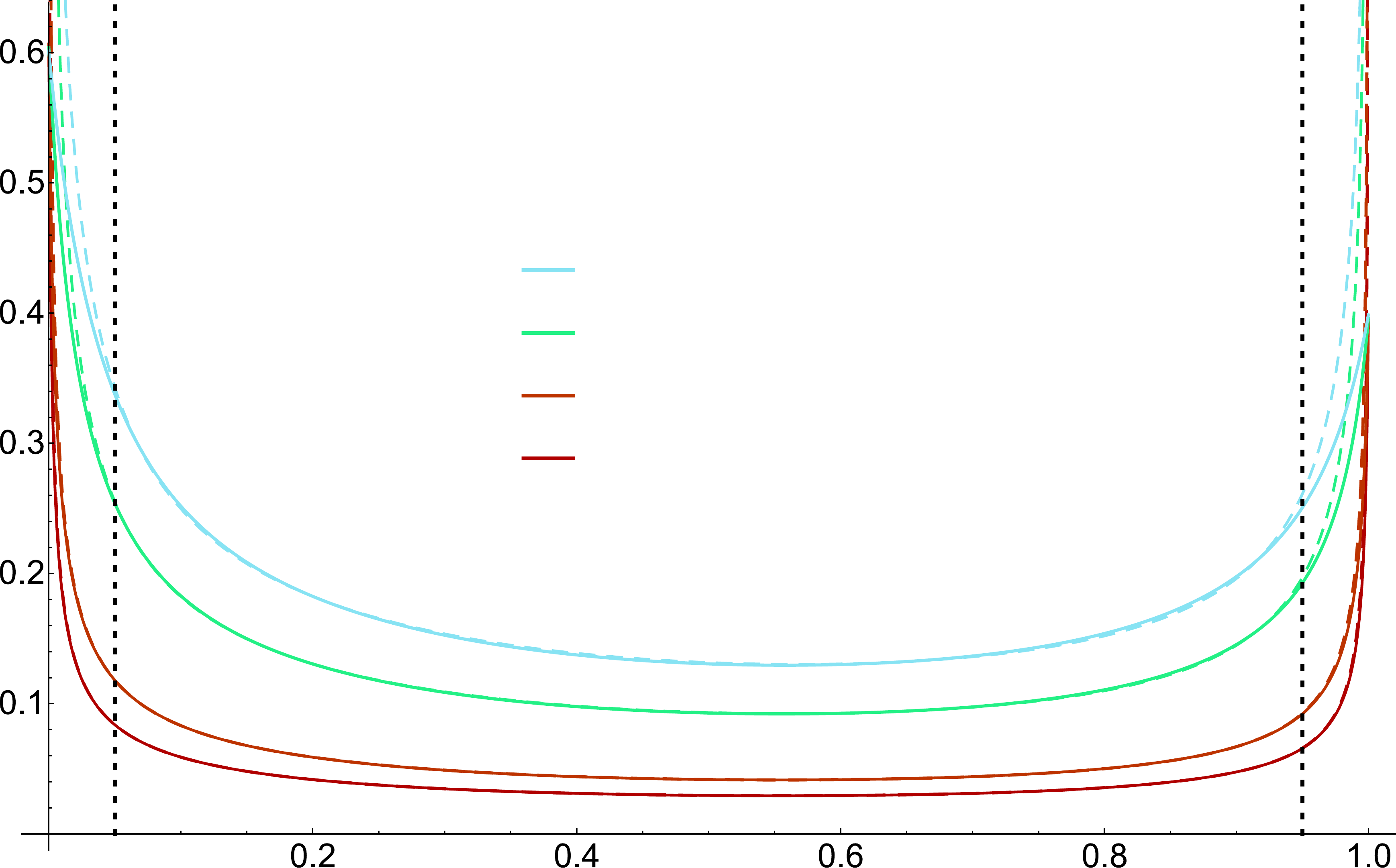
&
\def\svgwidth{7cm}
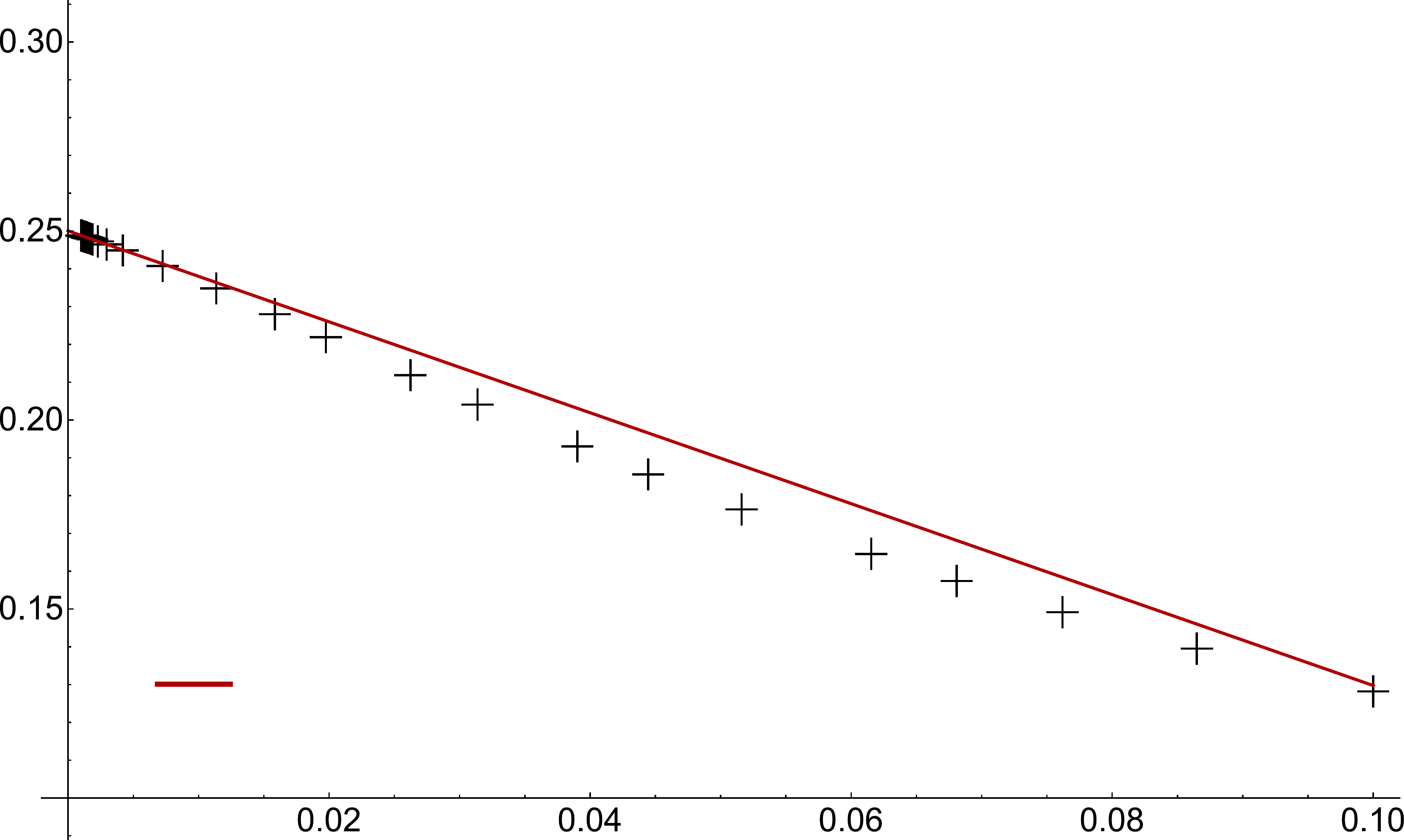\\
(a) & (b)
\end{tabular}
\caption{\label{fig:M1DeltaFiniteT}Extraction of the scaling dimension $\Delta$ and $b_\pm$ coefficients for the \modone\ model, with $m = 0.1$ at various temperatures, using the finite-temperature ansatz \eqref{Gscalingq}. (a) Numerical solutions (solid lines) and the fitted finite temperature scaling ansatz \eqref{Gscalingq} (dashed lines); the dotted vertical lines delimit the region used for fitting. (b) Dependence of the extracted scaling dimension $\Delta$ with the temperature $T$, showing linear convergence to $\Delta = 1/4$ as $T\to0$; the linear fit (red, dark gray) was performed using $\sim 20$ points taken with $\beta \in [500, 1000]$ (shown in inset), and results in extrapolated values $\Delta = 0.250_1$, $b_+ = 0.600_1$, $b_- = 0.459_1$, $2\pi a=0.268_4$, $C = 1.339_3$ and $\theta=0.132_2$, matching \eqref{Deltaform1} and \eqref{Cthetarel} to within $2\%$ precision.}
\end{figure}

Another complementary approach is to fit $\Delta$ and $b_\pm$, with $\smash{a = \frac{1}{2\pi}\ln\frac{b_+}{b_-}}$ as before, by using the natural generalization of \eqref{Gasymp} to arbitrary $\Delta$ given in App.\ \ref{qGenApp}, Eq.\ \eqref{Gscalingq}. This ansatz can be trusted as soon as $|\tau|\gg 1$, without the need to restrict to $|\tau|\ll\beta$ as before. It is therefore not surprising that the results obtained in this way are even more precise. Some examples are shown in Fig.~\ref{fig:M1DeltaFiniteT}, for solutions with $m=0.1$ at various temperatures. The fitting interval is delimited by vertical dotted lines, but once more the results are fairly robust with respect to its variations. Repeating this process at a given fixed mass for various temperatures, we observe that the scaling dimension varies linearly with the temperature for low enough temperatures. Such a linear dependence could certainly be explained by looking at subleading corrections to \eqref{Gasymp}, but we have not tried to do so. We may consider the appearance of such linear behaviour as an indicator that the IR regime has been properly reached, since it is then immediate to extrapolate the finite temperature results to $T \to 0$.

The consistency of the results convincingly demonstrates that the solution to the Schwinger-Dyson equations approaches the low energy ansatz discussed in Sec.~\ref{ReparaSec}. Similar results are obtained for other small masses, with the approximation becoming better and better as the temperature gets lower. Note that the aim of the detailed analysis of some standard cases, as discussed above, is mainly to test our numerical methods before applying the same ideas to the much more interesting \modtwo\ model. In that case the non-standard ansatz \eqref{Gasympnew} will play a crucial role, and we do not have a finite temperature expression analogous to \eqref{Gscaling}. We will therefore not have the luxury of added precision. However, the consistency observed in the \modone\ model between the different methods provides us with a high degree of confidence that the results for the \modtwo\ model that we present in what follows are also very reliable.

\subsubsection{\label{EucSolmod2Sec}Basic numerical results for the \modtwo\ model}

We can perform on the \modtwo\ model a numerical analysis similar to the one done in the previous section for the \modone\ model. Note that for $m=0$ we can use the particle-hole symmetry to look for solutions satisfying $G(-\tau) = -G(\tau)$, exactly as in the \modone\ model. The Schwinger-Dyson equations are then equivalent to the standard SYK model \cite{SachdevYe,Maldacena:2016hyu,Kitaev}, the two-point function following at low temperatures the standard ansatz of Sec.~\ref{ReparaSec}.

\begin{figure}[h!]
\centering
\vskip 0.3cm
\def\svgwidth{4.5in}
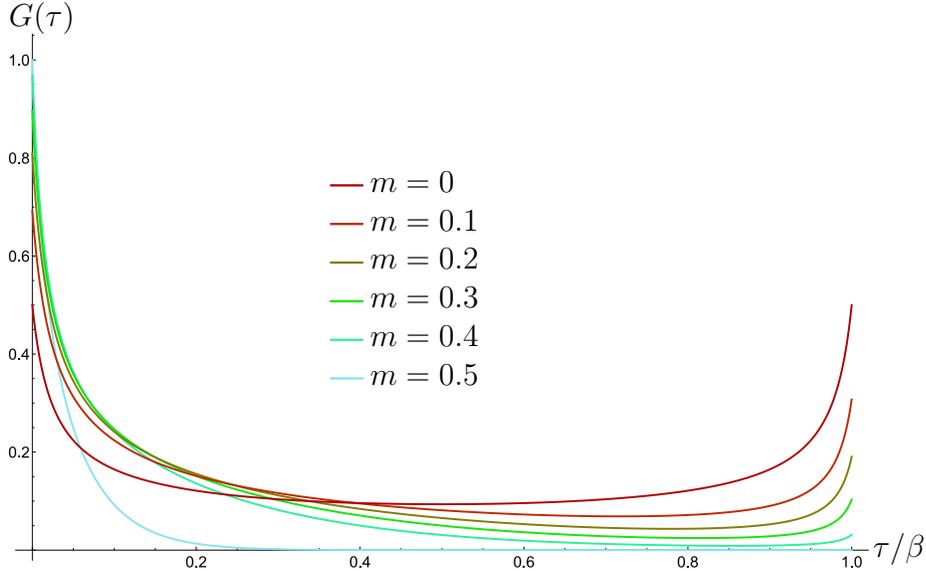
\caption{\label{fig:M2VaryingMass}Numerical solution of the Schwinger-Dyson equations for the \modtwo\ model at $\beta = 100$ and $|\xi|=1$, for various values of the mass $m$. The loss of the particle-hole symmetry is signalled by the asymmetric shape of $G(\tau)$ about $\tau = \beta/2$. The plot is consistent with a smooth change in the long time behaviour of the two-point function, from power-law to exponential decay, to be contrasted with that of the \modone\ model shown in Fig.~\ref{fig:M1VaryingMass}.}
\end{figure}

For $m \neq 0$ the situation is very different already at a qualitative level. To illustrate this, we show in Fig.~\ref{fig:M2VaryingMass} solutions of the Schwinger-Dyson equations found for various masses at $T = 0.01$ in natural units. Again we expect the system to develop an energy gap for large enough values of the mass, and indeed we see that the two-point function is exponentially decaying already for $m = 0.5$. However, we observe in this case a gradual, smooth transition into this regime, contrasting with the sharp change of behaviour seen for the \modone\ model, \emph{e.g.}\ in Fig.~\ref{fig:M1VaryingMass}.

Further evidence in favor of a smooth transition into the perturbative regime is provided by the constraint in \eqref{Sigtildeconst}. Note that this condition is equally valid for the standard and non-standard ans\"atze discussed in Sec.\ \eqref{ReparaSec} and \eqref{NewansatzSec}. For the solutions shown in Fig.~\ref{fig:M2VaryingMass} with $m = 0,0.1,0.2,0.3,0.4$ and $0.5$, we have $\re\Sigma_{1/2} = 0, -0.03,-0.11,-0.20,-0.28$ and $-0.23$, respectively, so that $\re\Sigma_{1/2}$ switches from being decreasing to increasing somewhere in the range $0.3 < m < 0.5$. For the moment, we will concentrate on the numerical methods we can use to characterize the small-mass solutions approximately satisfying \eqref{Sigtildeconst}, deferring until Sec.~\ref{PhaseSec} a more detailed discussion of this issue.

To elucidate in more details the nature of the gapless regime, we can try to extract scaling dimensions from the numerical solutions at finite temperature, as done before for the \modone\ model. Having fixed the mass to some small value, a scaling dimension growing unboundedly as the temperature decreases would be a telltale sign that the IR behaviour of the two-point function is trivial. On the other hand, a linear dependence of $\Delta$ with the temperature as $T \to 0$, as seen in the \modone\ model, would be strong evidence in favor of some non-trivial scaling in the IR.

\begin{figure}[h!!]
\centering
\vskip 0.3cm
\def\svgwidth{4.5in}
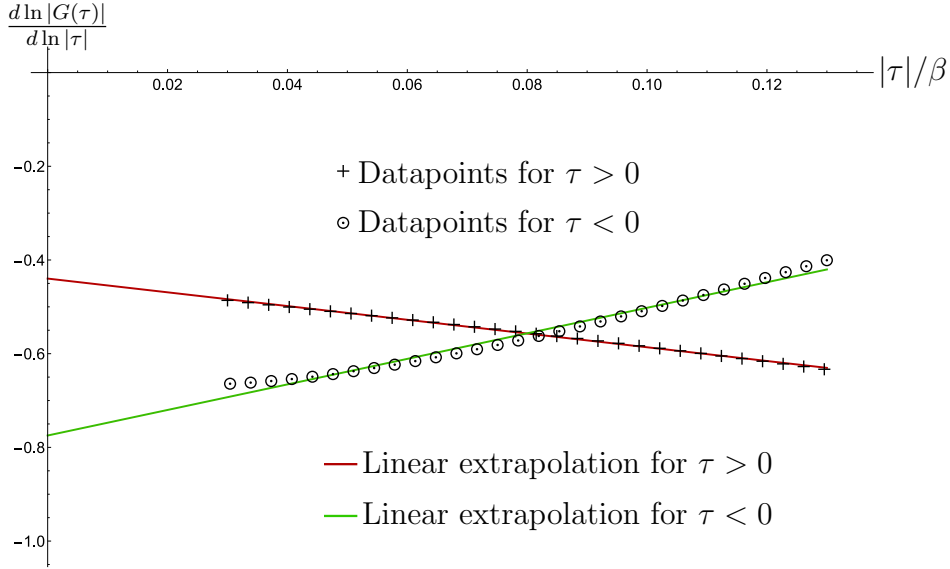 
\caption{\label{fig:M2DeltaZeroTa}Extraction of the scaling dimensions $\Delta_\pm$ for the \modtwo\ model using the zero-temperature ansatz \eqref{Gasympnew}, with $m = 0.1$, $\beta = 1000$ and $|\xi|=1$. We find a clear linear behaviour of $\smash{\frac{\d\ln |G(\tau)|}{\d\ln |\tau|}}$ in the chosen range of Euclidean times when $\tau>0$ (crosses), with a scaling dimension $\Delta_{+}=0.22_2$ corresponding to the $y$-intercept of the linear fit (red, dark gray). The behaviour when $\tau<0$ (circles) is only approximately linear, and does not allow for a very precise determination of $\Delta_{-}$. The linear fit shown on the plot (green, light gray) yields $\Delta_{-} = 0.37_{4}$, but in fact this may be considered a lower bound for the actual value (see discussion in the main text and App.~\ref{NumericalErrorsApp}). In any case, the data does unambiguously show that $\Delta_{-}>\Delta_{+}$.}
\end{figure}

A representative case with $m=0.1$ at $T=10^{-3}$ in natural units is illustrated in Fig.\ \ref{fig:M2DeltaZeroTa}. An exponential decay of the two-point function is clearly ruled out, the results being consistent with a power-law decay both when $\tau\to +\infty$ and when $\tau\to -\infty$. However, as already anticipated the exponents $\Delta_\pm$ relevant at $\tau\to\pm\infty$ must now be distinct, in sharp contrast with the situation observed in the \modone\ model in Sec.~\ref{EucSolmod1Sec} . In other words, the general ansatz \eqref{Gasympnew} appears to be valid, but the standard ansatz \eqref{Gasymp} is unambiguously violated.

More precisely, as in Fig.\ \ref{fig:M1DeltaZeroTa} for the \modone\ model we find an approximately linear behaviour of $\smash{\frac{d\log |G(\tau)|}{d\log|\tau|}}$ when $1\ll|\tau|\ll\beta$ and $\tau>0$. This allows for the determination of an exponent $\Delta_{+}$ with satisfactory precision, the result being consistent with the expected value of $1/4$, see Eq.\eqref{model2sol2}. On the other hand, linearity is not as good when $\tau<0$, suggesting that the inverse temperature $\beta=1000$ is not high enough to properly reach the deep IR regime. This yields a significant uncertainty in the determination of $\Delta_{-}$, but in any case the equality $\Delta_{-}=\Delta_{+}$ is ruled out.\footnote{Note that we have also tried to fit the data to the finite temperature ansatz \eqref{Gscalingq}, which is valid if $\Delta_{+}=\Delta_{-}$. We have found that it is impossible to obtain a good fit, even at very low temperatures, confirming that $\Delta_{+}$ and $\Delta_{-}$ necessarily differ.} We have also proceeded to extract the coefficients $b_\pm$ using the zero-temperature ansatz \eqref{Gasympnew}, as shown for the \modone\ model in Fig.\ \ref{fig:M1DeltaZeroTb}. This yields a rather imprecise determination $b_{+} = 0.64_5$ and $b_{-} = 0.46_7$, again supporting the interpretation that the deep IR regime has not been fully reached.

\begin{figure}[h!]
\centering
\def\svgwidth{4.5in}
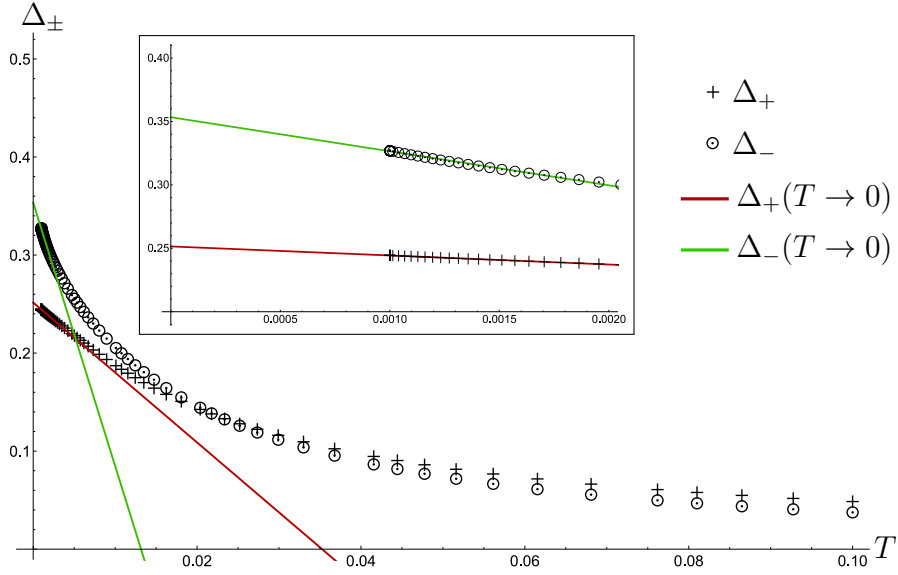\\
\caption{\label{fig:M2DeltaBpmExtrapolationa}Temperature dependence of the scaling dimensions $\Delta_\pm$ obtained for the \modtwo\ model by fitting the zero-temperature ansatz \eqref{Gasympnew}, for $m = 0.1$ in natural units.  Datapoints correspond to scaling dimensions $\Delta_+$ (crosses) and $\Delta_-$ (circles), with the $T\to0$ linear extrapolations shown with red (dark gray) and green (light gray) lines, respectively. The linear fits were performed using $\sim 30$ points taken with $\beta \in [500, 1000]$ (shown in inset), and result in extrapolated values $\Delta_+ = 0.251_1$ (red, dark gray) and $\Delta_- = 0.353_2$ (green, light gray). The low-temperature regime of $\Delta_-$ is not as linear as for $\Delta_+$, so that the indicated extrapolated value should conservatively be interpreted as a lower bound for the actual value of $\Delta_{-}$.}
\end{figure}

\begin{figure}[h!]
\centering
\def\svgwidth{4.5in}
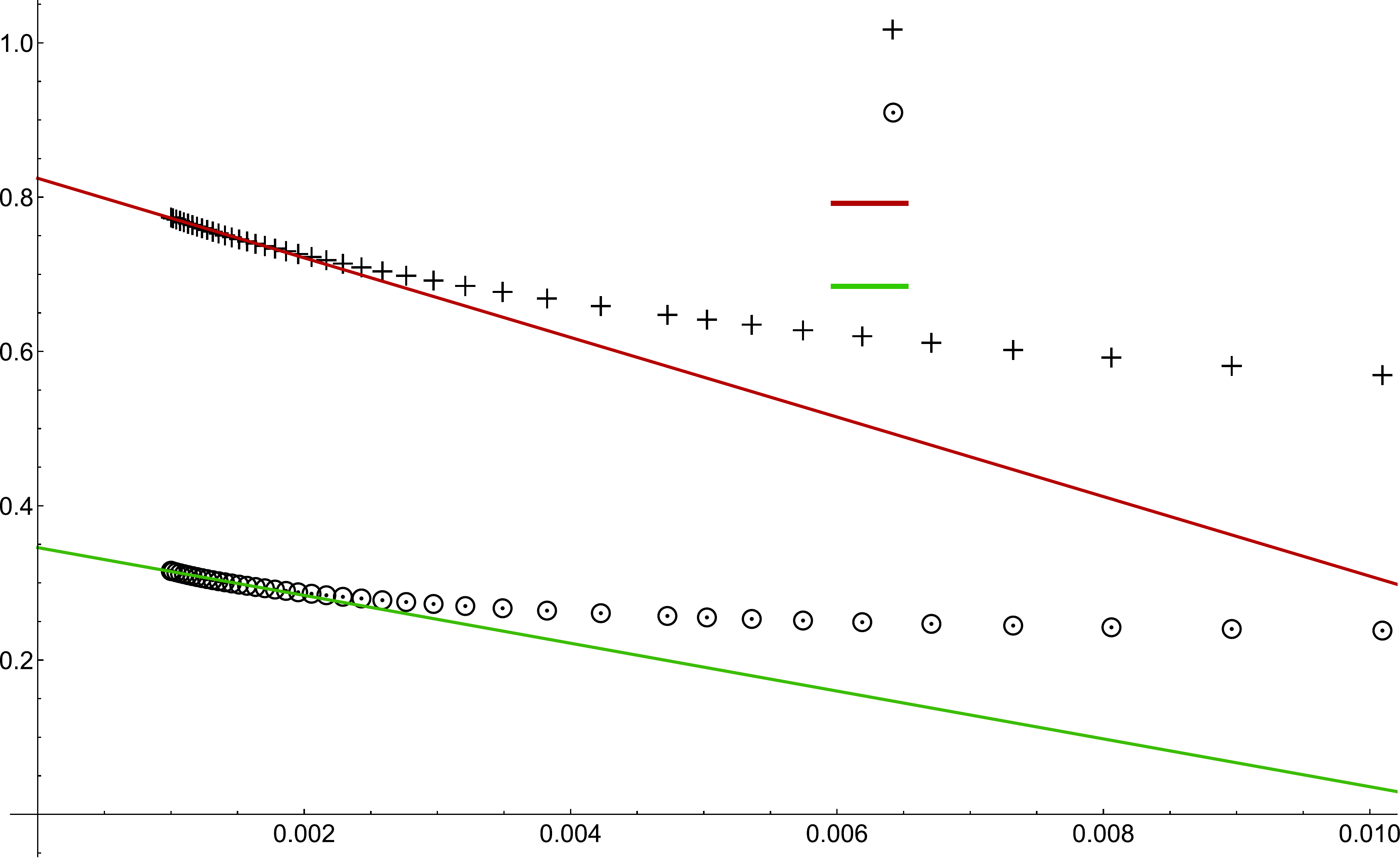
\caption{\label{fig:M2DeltaBpmExtrapolationb}Temperature dependence of the $b_\pm$ coefficients obtained for the \modtwo\ model by fitting the zero-temperature ansatz \eqref{Gasympnew}, for $m = 0.1$ in natural units. Datapoints for the $b_+$ (crosses) and $b_-$ (circles) coefficients, along with the corresponding $T\to0$ linear extrapolations (red and green, or dark and light gray, respectively); linear fits were performed using the same points as in Fig.~\ref{fig:M2DeltaBpmExtrapolationa}, and result in $b_+ = 0.82_4$ and $b_- = 0.35_{5}$. Once more the  linear behaviour is observed to be more accurately followed for $\tau\to+\infty$ than for $\tau\to-\infty$, the determination of $b_{-}$ consequently being more imprecise.}
\end{figure}

Further evidence in favour of the generalized ansatz $\Delta_{+}\not = \Delta_{-}$ is found when analyzing the temperature dependence of the various quantities involved. In Fig.~\ref{fig:M2DeltaBpmExtrapolationa} we illustrate the typical situation at low enough masses by plotting the results obtained for $m = 0.1$. The scaling dimension $\Delta_+$ extracted for $\tau\to+\infty$ is seen to develop a very clear linear behaviour at low temperatures. This allows us to confidently extrapolate its value to $T \to 0$, obtaining $\Delta_+ = 0.251_1$. On the other hand, the linear regime does not seem to be reached for $\Delta_{-}$. Taking into account the sign of the small non-zero second derivative in the plot, our results should then be interpreted as providing a lower bound on $\Delta_{-}$, establishing unambiguously that $\Delta_{-}>\Delta_{+}$. A similar analysis is made in Fig.\ \ref{fig:M2DeltaBpmExtrapolationb} for the coefficients $b_\pm$. The extrapolation of $b_{+}$, which is observed to be quite reliable, is of particular interest. Indeed, it provides an additional criterion to distinguish the standard ansatz $\Delta_{+}=\Delta_{-}$, for which $b_+ = (1/4\pi)^{1/4}\simeq 0.53$ from Eq.\ \eqref{model2sol1}, from the generalized ansatz $\Delta_{+}\neq \Delta_{-}$, for which $b_+ = (2/\pi)^{1/4}\simeq 0.89$ from Eq.\ \eqref{model2sol2}. We find in this case $b_+ = 0.82_4$, a result clearly in favour of $\Delta_{+}\not = \Delta_{-}$ and consistent with our previous discussion of the scaling dimensions.


In conclusion, we have discovered a new IR behaviour in the \modtwo\ model, inconsistent with the standard ansatz \eqref{Gasymp} discussed in Sec.\ \ref{ReparaSec} and used so far in the literature. It is however in good agreement with the more general, non-standard ansatz of Sec.~\ref{NewansatzSec}, at least to the degree in which we can trust the numerical methods involved. More details, in particular a more global picture of the resulting phase diagram, will be provided in Sec.\ \ref{PhaseSec}, whereas the consequences of this new IR behaviour in the real time physics shall be presented in Sec.~\ref{RealTimeSec}.

\subsection{\label{EuclidefourSec} Euclidean time four-point function}

\subsubsection{General definitions and properties}

Let us now introduce the time-ordered finite temperature Euclidean four-point function,
\be\label{PhiEucliddef}
\mathcal \phi(\tau_{1},\tau_{2},\tau_{3},\tau_{4}) = 
\begin{cases}\displaystyle
\frac{1}{n^{2}}\bigl\langle\text{T}\tr\psi_{\mu\text{E}}(\tau_{1})\psi^{\dagger}_{\mu\text{E}}(\tau_{2})\tr\psi_{\nu\text{E}}(\tau_{3})\psi^{\dagger}_{\nu\text{E}}(\tau_{4})\bigr\rangle_{\beta}&\ \text{(matrix models)}\\\displaystyle
\bigl\langle\text{T}\chi^{i}_{\text{E}}(\tau_{1})\chi^{\dagger}_{i\text{E}}(\tau_{2})\chi^{j}_{\text{E}}(\tau_{3})\chi^{\dagger}_{j\text{E}}(\tau_{4})\bigr\rangle_{\beta}&\ \text{(disordered models).}
\end{cases}
\ee
These definitions apply whenever $0<\tau_{i}<\beta$ but the KMS condition allows us to extend them to all values of the Euclidean times by antiperiodicity of period $\beta$,
\be\label{phianti} \phi(\tau_{1}+n_{1}\beta,\tau_{2}+n_{2}\beta,\tau_{3}+n_{3}\beta,\tau_{4}+n_{4}\beta) = (-1)^{n_{1}+n_{2}+n_{3}+n_{4}}\phi (\tau_{1},\tau_{2},\tau_{3},\tau_{4})\, ,\ n_{i}\in\mathbb Z\, .\ee
The four-point function also satisfies the reality and symmetry conditions
\begin{align}
\label{fourptsym1} & \phi (\tau_{1},\tau_{2},\tau_{3},\tau_{4})^{*}=\phi(-\tau_{4},-\tau_{3},-\tau_{2},-\tau_{1})\,,\\
\label{fourptsym2} & \phi (\tau_{1},\tau_{2},\tau_{3},\tau_{4}) = \phi (\tau_{3},\tau_{4},\tau_{1},\tau_{2})\,,\\
\label{fourptsym3} & \phi (\tau_{1},\tau_{2},\tau_{3},\tau_{4})^{*}=\phi (\tau_{1},\tau_{2},\tau_{3},\tau_{4})\, .
\end{align}
The last condition above follows from time-reversal invariance, which is always valid in the leading large $\dof$ melonic limit we are interested in.\footnote{At finite $\N$ and $\D$, Eq.\ \eqref{fourptsym3} is correct only when the couplings are real.} 

It is natural to decompose $\phi$ into a disconnected and a connected piece,
\be\label{PhiFref} \phi(\tau_{1},\tau_{2},\tau_{3},\tau_{4}) = G(\tau_{1},\tau_{2})G(\tau_{3},\tau_{4}) + \frac{1}{N}\mathcal F(\tau_{1},\tau_{2},\tau_{3},\tau_{4})\, .\ee
Our goal is to compute the connected piece $\mathcal F$ at leading order in the large $\dof$ limit. We first present a diagrammatic analysis in the next subsection and then a rigorous and very general algebraic proof in Sec.~\ref{fourptalgargSec}.

\subsubsection{Diagrammatic analysis}

\begin{figure}[h!]
\centering
\def\svgwidth{5in}
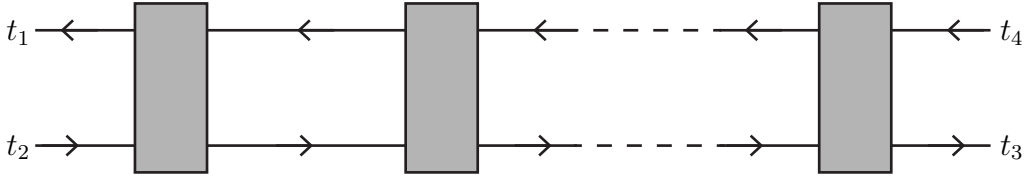
\caption{Ladder graph contributing to the connected four-point function $\mathcal F$. For better readability, we use a simple oriented line to represent the full two-point function (instead of a line with a shaded disk as in Fig.\ \ref{fig5}). The shaded rectangles can be any of the rung structures depicted in Fig.\ \ref{figladder1} for the \modone\ model (see also \cite{TanasaMelonic}), or in Fig.\ \ref{figladder2} for the \modtwo\ model.\label{figladder}}
\end{figure}

Graphs contributing to $\mathcal F$ at leading order are easy to find. One considers ladders of the form depicted in Fig.\ \ref{figladder}, with rungs of various types. These ladder graphs can be easily resummed because they form a geometric series. If we introduce a convenient matrix notation for functions of four variables, with a pair of variables forming a ``matrix index'' and matrix multiplication rule
\be\label{matrixnot} (\mathbb A\cdot\mathbb B)(\tau_{1},\tau_{2},\tau_{3},\tau_{4}) = \int_{[0,\beta]^{2}}\mathcal A(\tau_{1},\tau_{2},\tau,\tau')\mathcal B(\tau,\tau',\tau_{3},\tau_{4})\,\d \tau\d \tau'\, ,\ee
we can write the geometric series for $\mathcal F$ in the form
\be\label{Ffromkernel} \mathbb F = \sum_{n\geq 0}\mathbb K^{n}\cdot\mathbb F_{0}=(\mathbb I-\mathbb K)^{-1}\cdot\mathbb F_{0}\, .\ee
The function
\be\label{F0def}\mathcal F_{0}(\tau_{1},\tau_{2},\tau_{3},\tau_{4})=-G(\tau_{1},\tau_{4}) G(\tau_{3},\tau_{2})\ee
corresponds to the contribution with no rung inserted, whereas the kernel $\mathcal K$ is built from all the possible rung structures that can appear in the ladder. 

\begin{figure}[h!]
\centering
\includegraphics[width=5in]{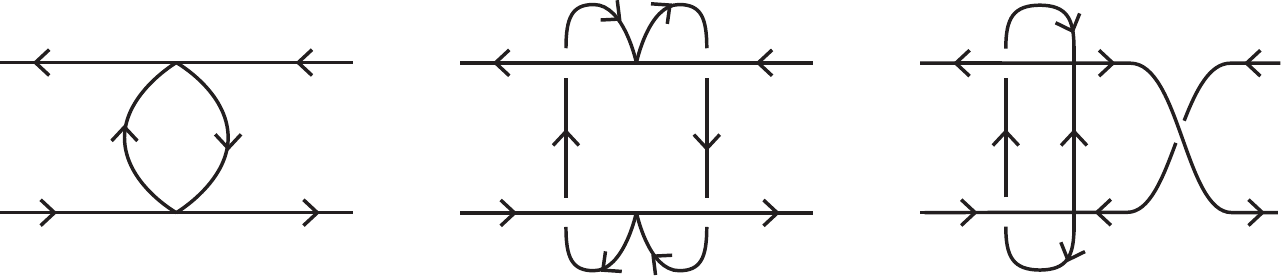}
\caption{\label{figladder1}Rungs serving as building blocks for the ladder diagrams contributing to the leading order of the connected four-point function of the \modone\ model. For better readability, we use a simple oriented line to represent the full two-point function.}
\end{figure}
\begin{figure}[h!]
\centering
\includegraphics[width=5in]{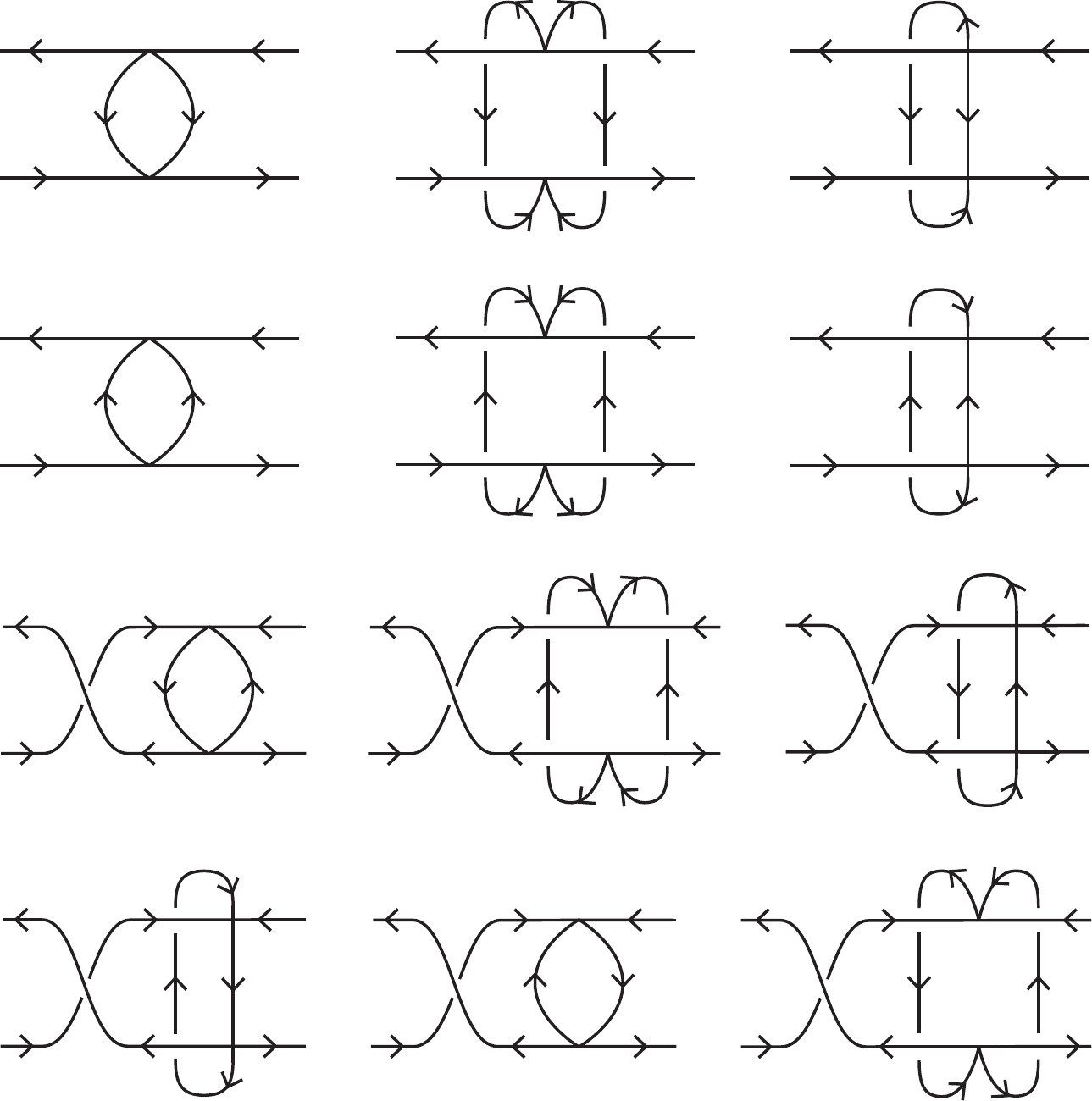}
\caption{\label{figladder2}Rungs serving as building blocks for the ladder diagrams contributing to the leading order of the connected four-point function of the \modtwo\ model. For better readability, we use a simple oriented line to represent the full two-point function.}
\end{figure}

In the \modone\ model, the kernel is given by
\begin{multline}\label{Kone}
\mathcal K^{\modelone}(\tau_{1},\tau_{2},\tau_{3},\tau_{4})=-\la^{2}\Bigl[2G(\tau_{1},\tau_{3})G(\tau_{4},\tau_{2})G(\tau_{3},\tau_{4})G(\tau_{4},\tau_{3})\\
+ G(\tau_{1},\tau_{4})G(\tau_{3},\tau_{2})G(\tau_{4},\tau_{3})^{2}\Bigr]\, .
\end{multline}
The first term on the right-hand side of \eqref{Kone} is associated with the left and central rungs in Fig.\ \ref{figladder1}, while the second term corresponds to the rightmost rung depicted in that figure.\footnote{These rung structures seem to have appeared first in the third reference in \cite{TanasaMelonic}.} 

In the \modtwo\ model, the kernel is similarly given by
\begin{multline}\label{Ktwo}
\mathcal K^{\modeltwo}(\tau_{1},\tau_{2},\tau_{3},\tau_{4})=\frac{|\xi|^{2}}{4}\Bigl[3\bigl(G(\tau_{4},\tau_{3})^{2}+ G(\tau_{3},\tau_{4})^{2}\bigr)G(\tau_{1},\tau_{3})G(\tau_{4},\tau_{2})\\+6G(\tau_{3},\tau_{4})G(\tau_{4},\tau_{3})G(\tau_{1},\tau_{4})G(\tau_{3},\tau_{2})\Bigr]\, .
\end{multline}
The first term on the right-hand side of \eqref{Ktwo} is associated with the three rungs in the first row of Fig.\ \ref{figladder2}; the second term with the three rungs in the second row of Fig.\ \ref{figladder2}; and the third term with the six rungs in the last two rows of Fig.\ \ref{figladder2}.

Even though it is completely straightforward to check that the ladder diagrams do contribute at leading order to $\mathcal F$, it is less immediate to prove diagrammatically that they are the only diagrams having this property. This part of the proof is often omitted in the literature. In the next subsection, we provide a very general algebraic method that permits the rigorous computation of the connected four-point function kernels of a large class of models. Applied to the \modone\ and \modtwo\ models this method yields \eqref{Kone} and \eqref{Ktwo}, respectively, thus proving that the ladder diagrams considered above are indeed the only ones that can contribute at leading large $\dof$ order.

\subsubsection{\label{fourptalgargSec}Algebraic proof}

We shall use the language of matrix models to describe the proof of the formulas \eqref{Kone} and \eqref{Ktwo} for the four-point function kernels, but the disordered model formulation can be treated in exactly the same way. Only some very general features of the models are needed, namely:
\begin{enumerate}[i)]
	\item The large $\dof$ limit is dominated by generalized melonic graphs such that the Schwinger-Dyson equations for the two-point function $G$ and self-energy $\Sigma$ coincide with the saddle-point equations of an effective action functional $\sSeff$, of the general form
\begin{multline}\label{Seffalggen} \frac{1}{\dof}\sSeff[\sG,\sS]=-\ln\bigl(1+e^{-\beta m}\bigr)
-\ln\frac{\det\mathscr O_{\sS}}{\det\mathscr O_{0}}\\+\int_{[0,\beta]^{2}}\sS(\tau_{2},\tau_{1})\sG(\tau_{1},\tau_{2})\,\d \tau_{1}\d \tau_{2} + s[\sG]\,.
\end{multline}
The operators $\mathscr O_{\sS}$ and $\mathscr O_{0}$ are defined in \eqref{OpSdef} and $s$ is an \emph{a priori} arbitrary functional of $\sG$. In particular, $G$ and $\Sigma$ are the on-shell values of $\sG$ and $\sS$.

	\item The partition function $Z$ is given at leading order by the on-shell value of the effective action,
\be\label{ZSeffalg} Z = e^{-\Seff}\, ,\ee
where $\Seff = \sSeff(G,\Sigma)$ as in \eqref{OnshellSeff}.
\end{enumerate}
The Euclidean four-point function \eqref{PhiEucliddef} is then given by \eqref{PhiFref}, \eqref{Ffromkernel} and \eqref{F0def}, with a kernel defined by the general formula
\be\label{Kalggen} {\mathcal K}(\tau_{1},\tau_{2},\tau_{3},\tau_{4}) = \int_{[0,\beta]^{2}}G(\tau_{1},\tau)G(\tau',\tau_{2})\frac{\delta^{2} s}{\delta G(\tau',\tau)\delta G(\tau_{3},\tau_{4})}\,\d \tau\d \tau'\, .\ee

To prove this formula, it is convenient to introduce the source formalism. The partition function $Z$ has a standard path integral representation, with matrix Grassmannian integration variables $\psi_{\mu}$ and $\bar\psi_{\mu}$ and an action
\be\label{Smic1}\begin{split} S &= \N\D\int_{0}^{\beta}\!\d \tau\,\tr\bigl(\bar\psi_{\mu}\dot\psi_{\mu} + m \bar\psi_{\mu}\psi_{\mu}\bigr)+S_{\text{int}}\\ &
=\N\D\int_{[0,\beta]^{2}}\!\d \tau_{1}\d \tau_{2}\,\tr\bar\psi_{\mu}(\tau_{1})\mathscr O_{0}(\tau_{1},\tau_{2})\psi_{\mu}(\tau_{2})+ S_{\text{int}}\, .
\end{split}\ee
The term $S_{\text{int}}$ represents the interactions in the Hamiltonian. We can modify this action by adding a source term for the two-point function. The new action is
\be\label{Smic2}
S[J]=\N\D\int_{[0,\beta]^{2}}\!\d \tau_{1}\d \tau_{2}\,\tr\bar\psi_{\mu}(\tau_{1})\bigl(\mathscr O_{0}(\tau_{1},\tau_{2})+J(\tau_{1},\tau_{2})\bigr)\psi_{\mu}(\tau_{2})+ S_{\text{int}}\,,
\ee
and the associated path integral yields a source-dependent partition function $Z[J]$ and free energy $F[J]=-T\ln Z[J]$. The two-point function \eqref{GEucliddef} and four-point function \eqref{PhiEucliddef} are then given by
\begin{align}
\label{twoptfuncder}G(\tau_{1},\tau_{2}) &= \left.\frac{1}{\dof}\frac{1}{Z}\frac{\delta Z}{\delta J(\tau_{2},\tau_{1})}\right|_{J=0}\,,\\
\label{fourptfuncder} \phi(\tau_{1},\tau_{2},\tau_{3},\tau_{4}) &= 
\left.\frac{1}{\dof^2}\frac{1}{Z}\frac{\delta^{2}Z}{\delta J(\tau_{2},\tau_{1})\delta J(\tau_{4},\tau_{3})}\right|_{J=0}\, \cdotp
\end{align}
Equivalently, we can express $G$ and the connected piece $\mathcal F$ of the four-point function defined in \eqref{PhiFref} in terms of the free energy, so that
\begin{align}
\label{twoptfuncder2} G(t_{1},t_{2}) &= \left.-\frac{\beta}{N}\frac{\delta F}{\delta J(\tau_{2},\tau_{1})}\right|_{J=0}\,,\\
\label{fourptfuncder2} \mathcal F(\tau_{1},\tau_{2},\tau_{3},\tau_{4})&= \left.-\frac{\beta}{N}\frac{\delta^{2}F}{\delta J(\tau_{2},\tau_{1})\delta J(\tau_{4},\tau_{3})}\right|_{J=0}\,\cdotp
\end{align}

Because the source term in \eqref{Smic2} is quadratic in $\psi$ and $\bar\psi$, its addition only modifies the expression of the propagator, \emph{without changing the large $\N$ and large $\D$ diagrammatics of the models.} The results \eqref{Seffalggen} and \eqref{ZSeffalg}, which are \emph{a priori} assumed to be valid without the source term, can thus be immediately generalized to take it into account. This amounts to replacing the mass term $m\delta_{\beta}(\tau_{1}-\tau_{2})$ in $\mathscr O_{\sS}$ and $\mathscr O_{0}$ defined in \eqref{OpSdef} by $m\delta_{\beta}(\tau_{1}-\tau_{2}) + J(\tau_{1},\tau_{2})$. The form \eqref{Seffalggen} of the effective action then implies that this modification yields a new, source-dependent effective action $\sShat[\sG,\sS;J]$ given by
\be\label{SeffJ} \sShat[\sG,\sS;J] = \sSeff[\sG,\sS+J] - \dof
\int_{[0,\beta]^{2}} J(\tau_{2},\tau_{1})\sG(\tau_{1},\tau_{2})\,\d \tau_{1}\d \tau_{2}\, .\ee
The solutions to the source-dependent saddle-point equations
\be\label{saddleJ} \frac{\delta\sShat}{\delta\sG(\tau_{1},\tau_{2})} = 0\,\quad\text{and}\quad
\frac{\delta\sShat}{\delta\sS(\tau_{1},\tau_{2})} = 0\ee
are the on-shell values $G[J]$ and $\Sigma[J]$, and the source-dependent partition function and free energy are given by
\be\label{ZJform} Z[J]=e^{-\beta F[J]}=e^{-\Shat[J]} \qquad\text{with}\qquad \Shat[J]\equiv\sShat[G[J],\Sigma[J];J]\, .\ee
It is then an exercise in partial differentiation to obtain the formula \eqref{Ffromkernel} with a kernel given by \eqref{Kalggen}, starting from \eqref{fourptfuncder2} and using \eqref{SeffJ} and \eqref{ZJform}. The details are provided in App.~\ref{algApp}.

For the particular case of the models we are discussing, one then finds the explicit form of the kernels \eqref{Kone} and \eqref{Ktwo} by plugging in \eqref{Kalggen} the relevant functionals for $s[\sG]$, namely $s_{\modelone}$ and $s_{\modeltwo}$ given by
\begin{align}\label{s1alg} s_{\modelone}[\sG] &= -\frac{\la^{2}}{4}\int_{[0,\beta]^{2}}\sG(\tau_{1},\tau_{2})^{2}\sG(\tau_{2},\tau_{1})^{2}\, \d \tau_{1}\d \tau_{2}\, ,\\ \label{s2alg}
s_{\modeltwo}[\sG] &= \frac{|\xi|^{2}}{4}\int_{[0,\beta]^{2}}\sG(\tau_{1},\tau_{2})^{3}\sG(\tau_{2},\tau_{1})\, \d \tau_{1}\d \tau_{2}\, .
\end{align}
\section{Phase diagrams\label{PhaseSec}}

In this section, unless explicitly stated otherwise, we continue using for convenience the natural units defined in \eqref{unitsnum}. Having set $\lambda = 1$ (for the \modone\ model) or $|\xi| = 1$ (for the \modtwo\ model), we then have two dimensionless parameters, $m$ and $T$, and we are going to study in detail the properties of the models at large $\dof$ in the $(m,T)$-plane. 

We can assume $m\geq 0$ without loss of generality. Indeed, the sign of $m$ may be changed by permuting ``particles'' and ``holes,'' \emph{i.e.}\ by exchanging the role of the operators $\psi$ and $\psi^{\dagger}$ (in the matrix models) or $\chi$ and $\chi^{\dagger}$ (in the disordered models). It is immediate to check that if $G(\tau;m)$ is a solution with mass $m$ of the Schwinger-Dyson equations \eqref{Sigmadef}--\eqref{SDone} (for the \modone\ model) or \eqref{Sigmadef}-\eqref{SDtwo} (for the \modtwo\ model), then $G(\tau;-m)=G(\beta-\tau;m)$ is a solution of the same equations with mass $-m$. 

\subsection{General discussion\label{phasemotivSec}}

\begin{figure}[h!]
\centering
\def\svgwidth{4.5in}
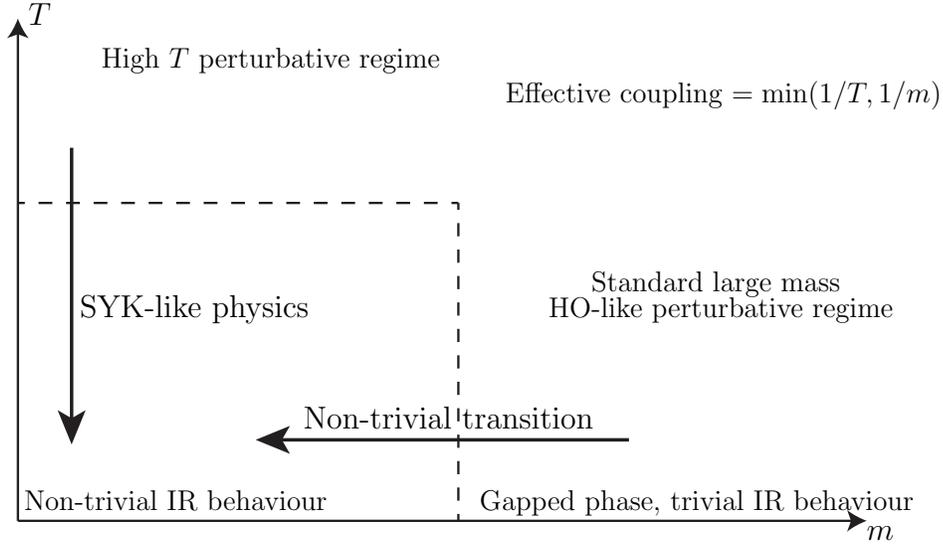
\caption{Qualitative picture of the phase diagram in the $(m,T)$-plane. The physics at small mass is SYK-like, with a flow from a perturbative regime at high $T$ to a strongly coupled conformal phase at low $T$. The physics at large mass is harmonic-oscillator-like, weakly coupled for all values of the temperature. Our main goal below is to elucidate and study the consequences of the non-trivial transition from the weakly coupled low-temperature region at large mass to the strongly coupled low-temperature conformal region at small mass.\label{qualitativephasefig}}
\end{figure}

A qualitative picture of the physics to be expected in the $(m,T)$ plane is illustrated in Fig.\ \ref{qualitativephasefig}. Our goal in this subsection is to elaborate on this picture and elucidate many important features, without resorting to difficult calculations or numerical analysis. Full details are then provided in the following subsections \ref{PhaseMod1Sec} and \ref{PhaseMod2Sec}.

\paragraph{Zero mass}

When the mass strictly vanishes, the models have particle-hole symmetry. Assuming that this symmetry is not spontaneously broken, a possibility that we shall not investigate here, we can seek a solution to the Schwinger-Dyson equations such that $G(-\tau)=-G(\tau)$, and this solution matches the standard solution for the original SYK model studied extensively in the literature \cite{Kitaev,Maldacena:2016hyu,Polchinski:2016xgd,Garcia-Garcia:2016mno,Gross:2017aos}. The only difference with SYK is the doubling of the number of degrees of freedom, because we use complex and not Majorana fermions. This introduces trivial factors of two, for example in the formula for the entropy. The physics at $m=0$ is thus familiar. The models flow from a trivial perturbative behaviour at high temperature, governed by the standard high $T$ perturbation theory for fermionic systems around the maximally entropic state with density matrix $2^{-N}\mathbb I$, to a non-trivial, strongly coupled, conformal behaviour at low temperature, reminiscent of extremal black-hole physics. In particular: the zero-temperature Euclidean two-point function $G$ has a power-law decay $|\tau|^{-2\Delta}$ at large times, with $\Delta=1/4$; there exists a non-vanishing zero-temperature entropy given by \eqref{zeroTentropySYK};  and the chaos exponent, as read-off from real time out-of-time-ordered four-point functions, is $2\pi T$ at low temperatures, saturating the bound discussed in  \cite{Maldacena:2015waa}.

\paragraph{Small non-zero mass}

When the mass is small but non-zero, three qualitatively distinct behaviours could \emph{a priori} be expected. 

The first possibility is that a mass gap is immediately created, as would be required if the physics were weakly coupled. This is what happens, for instance, when two copies of the SYK model are coupled quadratically as in \cite{Maldacena:2018lmt,Garcia-Garcia:2019poj}.

The second possibility is that the physics remains gapless when the mass is turned on.  This is what happens in both of the models we consider, over a finite interval of masses $0\leq m\leq m_{*}$, where $m_{*}$ is a strictly positive critical mass beyond which a mass gap is created. The nature of the deformation induced by the mass term is, however,  qualitatively different in the \modone\ and the \modtwo\ models. This was mentioned in Sec.~\ref{EucSolmod1Sec} and \ref{EucSolmod2Sec}  and will be thoroughly discussed in Sec.~\ref{PhaseMod1Sec} and \ref{PhaseMod2Sec}. More precisely, in the \modone\ model the mass term flows to a marginal deformation in the IR, in the sense that the IR physics smoothly depends on $m$. In the \modtwo\ model, on the other hand, the mass operator is relevant at strictly $m=0$, in the sense that the deep IR physics is governed by two disconnected fixed points at $m=0$ and $m>0$. It becomes marginal for all 
$0<m<m_{*}$.

The third possibility is that the mass term is irrelevant in the IR. This behaviour would actually be quite natural in the \modtwo\ model, since the standard low energy ansatz \eqref{Gasymp}, \eqref{Cthetadef} has no free parameter in this case, see Eq.\ \eqref{model2sol1}. However, our numerical results rule out this possibility in favor of the most general behaviour described in Sec.~\ref{NewansatzSec}.

\paragraph{Large mass}

At very large masses the models are weakly coupled for all values of the temperature, and are governed by the standard perturbation theory around a set of $\dof$ decoupled harmonic oscillators. The physics is then very simple. There is a mass gap $\meff>0$ of order $m$, with the zero-temperature Euclidean two-point function $G$ decaying as $e^{-\meff \tau}$ at large times; the zero-temperature entropy vanishes, the ground state being the unique Fock vacuum; and the chaos exponent is $o(T)$ at low temperatures.

In the \modone\ model, the fermion number is conserved and the mass term can then also be interpreted as (minus) a chemical potential for the fermion number. Under these circumstances, it is straightforward to derive, for instance from the spectral decomposition formula \eqref{Gspecdec}, that in a phase with a mass gap the mass dependence of the zero-temperature two-point function is exactly given by a simple exponential term. More precisely,
\be\label{GzeroTmassgap} G(\tau;m_{1},T=0) = G(\tau;m_{2},T=0)e^{-(m_{1}-m_{2})\tau}
\ee
for $m_{1}$ and $m_{2}$ any two masses in the gapped regime where the ground state is the Fock vacuum. This very general result is realized in the simplest possible way in the \modone\ model. Indeed, the tree-level zero-temperature two-point function,
\be\label{GtreezeroT}
G^{(0)}(\tau;m,T=0) = \Theta(\tau)e^{-m\tau}\,,
\ee
is actually an exact solution of the zero-temperature Schwinger-Dyson equations for any value of the coupling, simply because the right-hand side of \eqref{SDone} involves the product of $G$ at opposite times. 

In the \modtwo\ model this simplification of the large mass physics does not occur. The fermion number is not conserved and the mass is not a chemical potential. Even though the theory is still gapped and the ground state is the Fock vacuum, the simple relation \eqref{GzeroTmassgap} is not valid. Indeed the zero-temperature two-point function can get non-trivial corrections, due to the presence of the $G(\tau)^{3}$ term on the right-hand side of \eqref{SDtwo}. In particular, the physical mass gap $\meff$, which is computed in Sec.~\ref{PhaseMod2Sec}, is a non-trivial function of $m$.

\paragraph{The transition from weak to strong coupling}

The above discussion implies that, when the mass is decreased from large to small values, the behaviour of the models must drastically change from a gapped perturbative regime in which the ground state is the Fock vacuum to a non-perturbative regime with no gap. Therefore, there exists a critical mass $m_{*}$ below which the mass gap closes,
\be\label{meffmstar} \meff >0\ \text{for } m>m_{*}\, ,\quad\meff
= 0 \ \text{for } m<m_{*}\, .\ee
The zero-temperature transition at $m=m_{*}$ is reflected in the zero-temperature energy, which vanishes in the Fock vacuum at $m>m_{*}$ and must be strictly negative at $m<m_{*}$. The transition also comes along with the creation of a non-vanishing zero-temperature entropy. This follows from \eqref{zeroTentropy} for the \modone\ model and will be convincingly demonstrated numerically for the \modtwo\ model in Sec.~\ref{PhaseMod2Sec}.

However, in spite of these similarities between the two models, we can infer from our discussion of the high mass regime that the transition at $m=m_{*}$ must be qualitatively different in the \modone\ and \modtwo\ models.

In the \modone\ model, because the zero-temperature two-point function matches with the tree-level result \eqref{GtreezeroT} in the gapped regime, the mass gap is exactly $\meff=m$ for all $m>m_{*}$. The transition from $m>m_{*}$ to $m<m_{*}$ must then be discontinuous, with in particular a discontinuous jump of the mass gap from $m_{*}$ to zero. Such a first order phase transition was indeed found in \cite{letter}. Its existence is made possible by the non-trivial monodromies in the space of solutions of the Schwinger-Dyson equations that we have already mentioned in Sec.~\ref{MonodromySec}. The first order transition is not limited to zero temperature, but extends up to a critical point at finite temperature \cite{letter}. More details are given in Sec.~\ref{PhaseMod1Sec}.

In the \modtwo\ model, the zero-temperature two-point function gets non-trivial corrections at $m>m_{*}$ and the mass gap is a non-trivial function $\meff(m)$. We shall provide convincing numerical evidence in Sec.~\ref{PhaseMod2Sec} that $\meff$ smoothly vanishes when $m\rightarrow m_{*}$ from above. We thus get in this case a large $\dof$ version of a quantum critical point. In particular, the first order phase transition is replaced by a smooth crossover at finite temperature.

Most of the rest of this paper will focus on providing more details on the transition from weak to strong coupling in both models, using the standard thermodynamical Euclidean point of view in Sec.~\ref{PhaseMod1Sec} and \ref{PhaseMod2Sec} and a more modern and innovative real time point of view in Sec.~\ref{RealTimeSec}.

\paragraph{The gravitational collapse interpretation}

There is an interesting qualitative interpretation of the transition from weak to strong coupling when the mass is varied at a fixed (low) temperature in the models we are studying, based on the compelling analogy between black-hole physics and the strongly coupled conformal phase on the one hand and the matrix model formulation of the models on the other hand.

As already mentioned in Sec.~\ref{MatrixTensorComparison}, quantum mechanical matrix models can always be interpreted in terms of open strings and D-particles. In this point of view, the operators $\psi_{\mu\, b}^{\dagger a}$ create open strings whose endpoints are attached to the D-particles labelled by the $\uN$ indices $a$ and $b$. Since the mass of a string is proportional to its length, the mass parameter $m$ represents the distance between the D-particles joined by the string.

At large mass, we thus have an assembly of D-particles which are very far away from each other and interact very weakly. Their dynamics is governed by standard matrix model perturbation theory. In the usual terminology, this is the regime where the ``brane picture'' is reliable. As the mass is decreased, the D-particles are brought closer and closer together and the matrix model becomes more and more strongly coupled. Eventually, we expect the system to undergo gravitational collapse and form a black hole. It then enters a regime in which the ``brane picture'' is no longer suitable, and should be replaced by a ``gravity picture'' in which the branes are effectively described by a non-trivial gravitational background containing a black hole.

It is very natural to identify this phenomenon with the expected non-trivial transition at $m=m_{*}$ discussed above. This is perfectly consistent with the appearance of a macroscopic zero-temperature entropy for $m<m_{*}$, signaling the formation of a black-hole horizon, as well as with other properties of the strongly coupled regime of both models we consider.

Of course, the matrix models under consideration are still quite far from being truly realistic models of black holes. For instance, they do not contain fundamental bosonic matrix degrees of freedom, which are most naturally associated with transverse dimensions of space-time. Moreover, the strongly coupled phases of the models are ``black holes'' only in the same sense as the SYK model is one: there is a very suggestive matching of various physical properties at the qualitative level (entropy, quasi-normal behaviour, continuum spectrum, chaos) and some non-trivial quantitative agreement (saturation of the bound on the Lyapunov exponent). However, it is also clear that they can at most represent ``string size'' black holes for which a purely Einsteinian description cannot be fully adequate. In particular, the models contain an infinite tower of states in the bulk that cannot be decoupled from the low mass modes \cite{Maldacena:2016hyu}. Moreover, the strongly coupled phase of the \modtwo\ model is quite exotic and its interpretation as a ``black hole'' is then even more questionable than in the case of the \modone\ model (see Sec.~\ref{PhaseMod2Sec}).

Keeping in mind these important caveats, we nevertheless expect that the transition from weak to strong coupling in these models does capture important features of the transition between the ``brane description'' and the ``gravitational description'' of a collection of D-branes, \emph{i.e.}\ of the phenomenon of gravitational collapse. To the best of our knowledge, such a description has never before been given in a fully quantum mechanical framework, even in a very simplified set-up such as the one presented here.

\subsection{\label{PhaseMod1Sec}Phase diagram of the \modone\ model}

\begin{figure}[h!]
\centering
\def\svgwidth{4.5in}
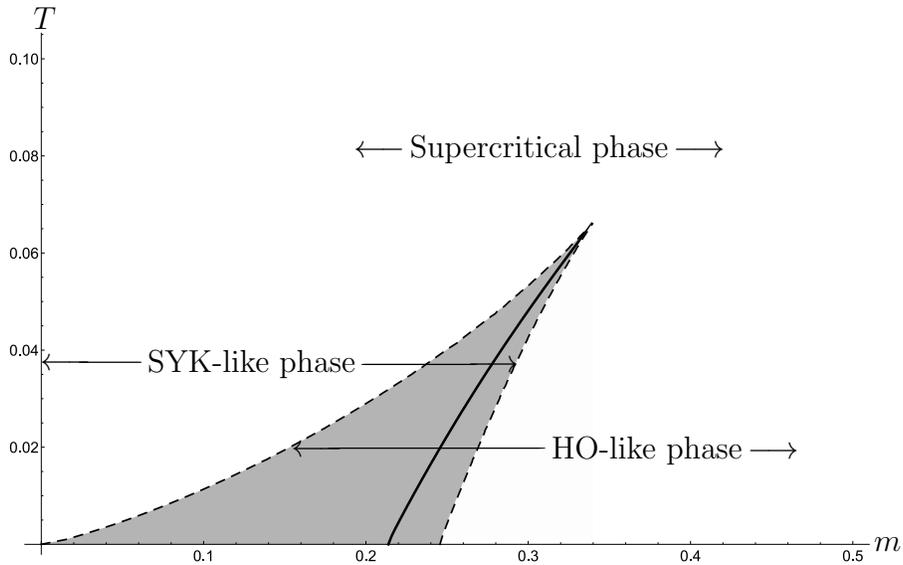
\caption{\label{fig:M1PhaseDiagram}Phase diagram of the \modone\ model in the $(m,T)$ plane, also in \cite{letter}. The dashed lines correspond to $m_-(T)$ (left) and $m_+(T)$ (right), with the triangle-shaped region they delimit (shaded in gray) being where HO-like and SYK-like solutions coexist. The solid line corresponds to $m_{\rm t}(T)$, the locus where the free energies of the two solutions coincide. It starts at $(T=0,m=m_{*} = \mStarOne)$ and terminates at a non-trivial critical point $(T = \Tc = \TcValue, m = \mc  = \mcValue, Q=\Qc= 0.050_{1})$. All along this line the system undergoes a first order phase transition.
}
\end{figure}

The phase diagram of the \modone\ model was first described in \cite{letter}, and is depicted in Fig.\ \ref{fig:M1PhaseDiagram}. We will briefly review here its main features to offer a self-contained discussion, and present additional details about its construction, generalizations and most important properties.

The phase diagram's structure is governed by the existence of a critical temperature $\Tc$ below which two distinct and consistent solutions to the Schwinger-Dyson equations may coexist. The critical temperature itself can be determined numerically using binary search, taking as division criterion the appearance of the hysteresis phenomenon already discussed in Sec.~\ref{MonodromySec} and exemplified in Fig.~\ref{fig:Hysteresis}. The resulting value $\Tc = \TcValue$ is small but finite, and is compatible with theoretical back-of-the-envelope estimates. Having determined the critical temperature, the phase diagram of Fig.~\ref{fig:M1PhaseDiagram} was constructed by taking $\sim130$ fixed temperatures in the range $T\in[10^{-3},\Tc]$, and sweeping the $(m,T)$-plane from small to large masses and vice-versa, as explained in Sec.~\ref{NumSolEucSec}.

One of the solutions found below the critical temperature $\Tc$ is the analytic continuation of the weakly coupled solution at large masses. We shall call this solution ``harmonic oscillator-like'', or simply HO-like solution for short. It can be analytically continued down to a minimum mass $m_-(T)$, but ceases to exist for any mass $m < m_-(T)$. As discussed in Sec.~\ref{phasemotivSec}, at strictly zero temperature this solution matches with the tree-level result, which solves the Schwinger-Dyson equations for all masses, so that $m_-(T = 0) = 0$. At finite temperatures, however, we always find $m_-(T) > 0$.

A different solution existing for $T<\Tc$ is the SYK solution at $m = 0$, which in turn can be analytically continued to higher masses up to the maximum value $m_+(T)$. This solution is conformal in the IR and retains many of the familiar properties of the SYK model, such as non-vanishing zero-temperature entropy, so we shall call it the SYK-like solution.

\begin{figure}[h!]
\centering
\begin{tabular}{cc}
\def\svgwidth{7cm}
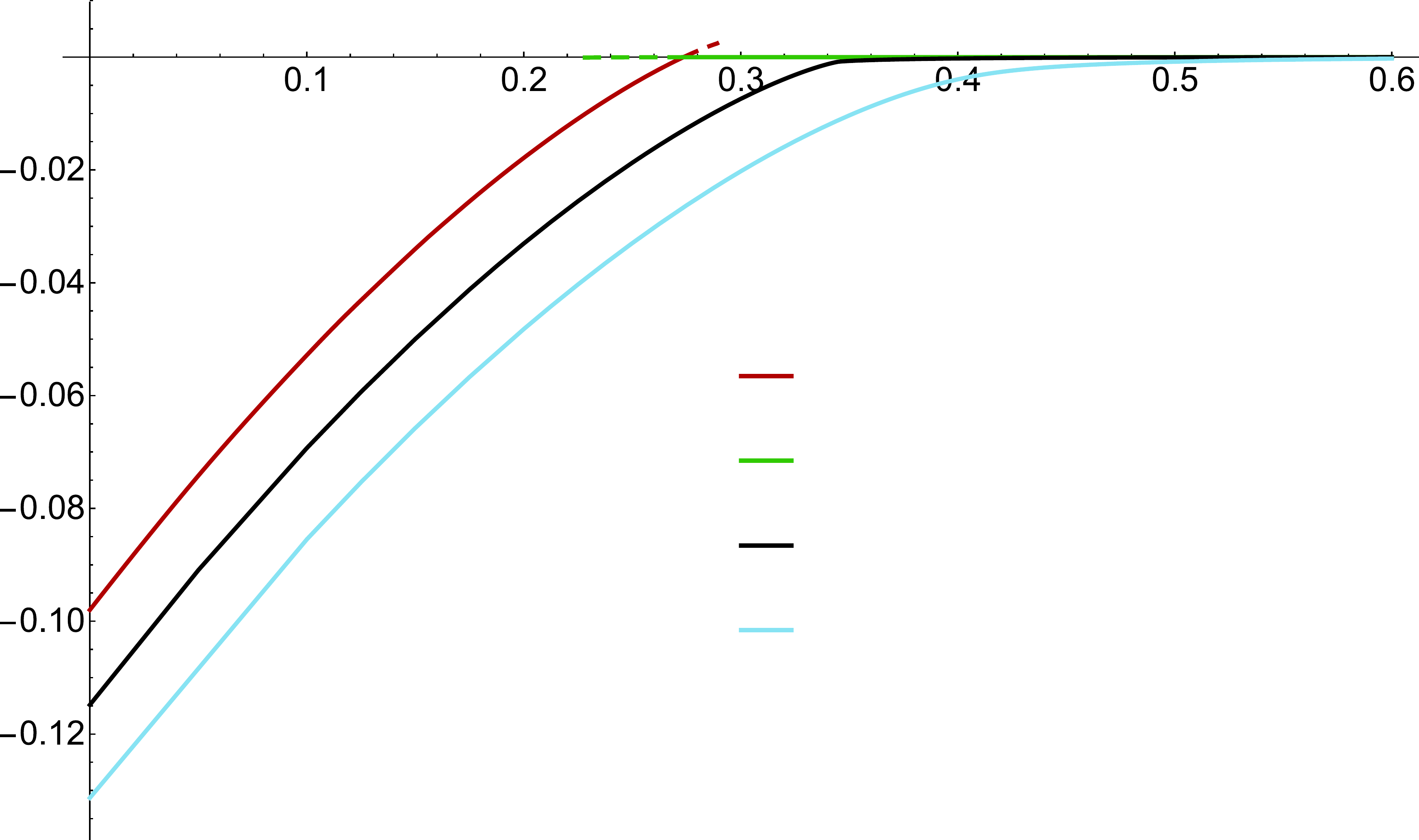 &
\def\svgwidth{7cm}
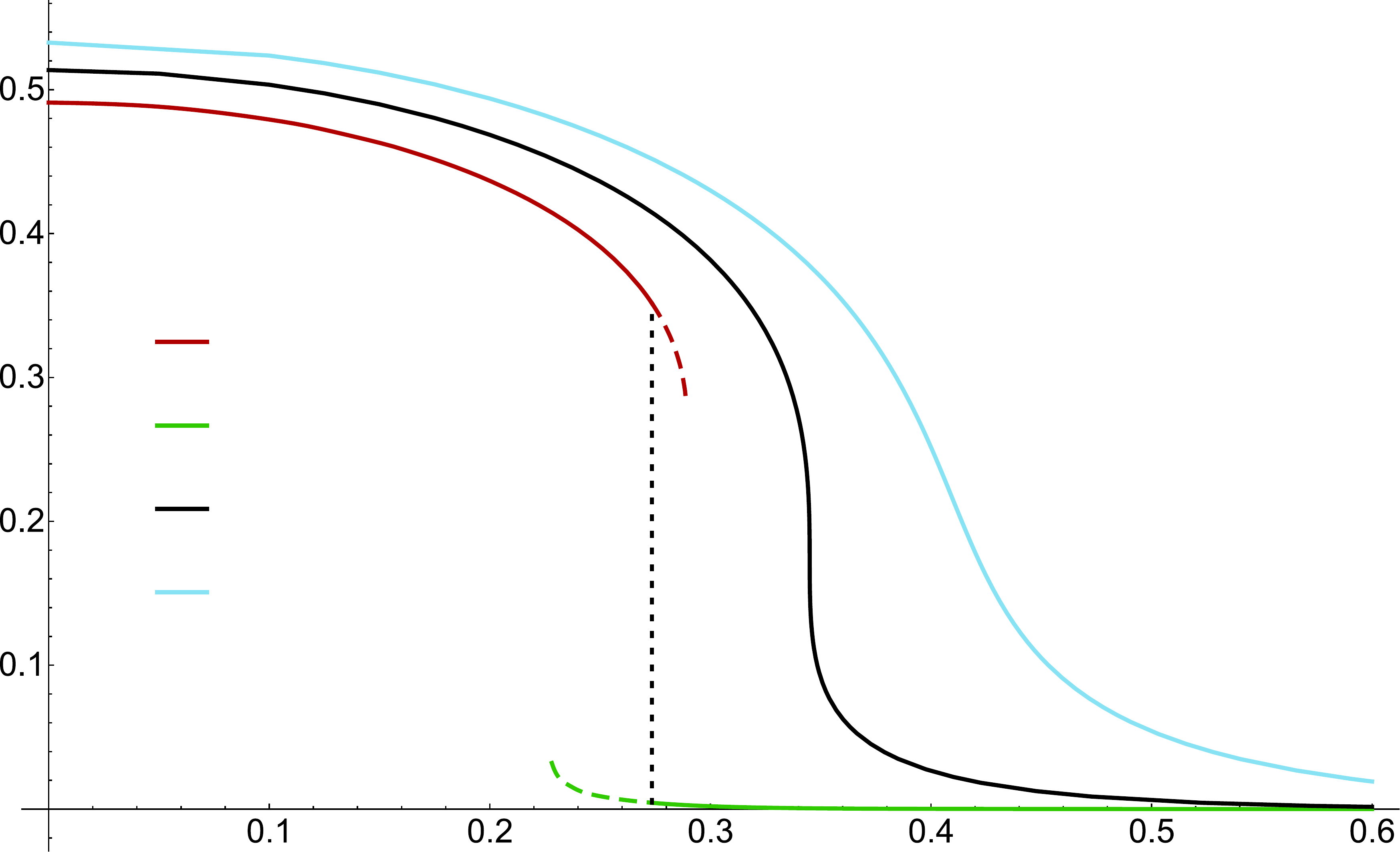 \\
(a) & (b)\\
\def\svgwidth{7cm}
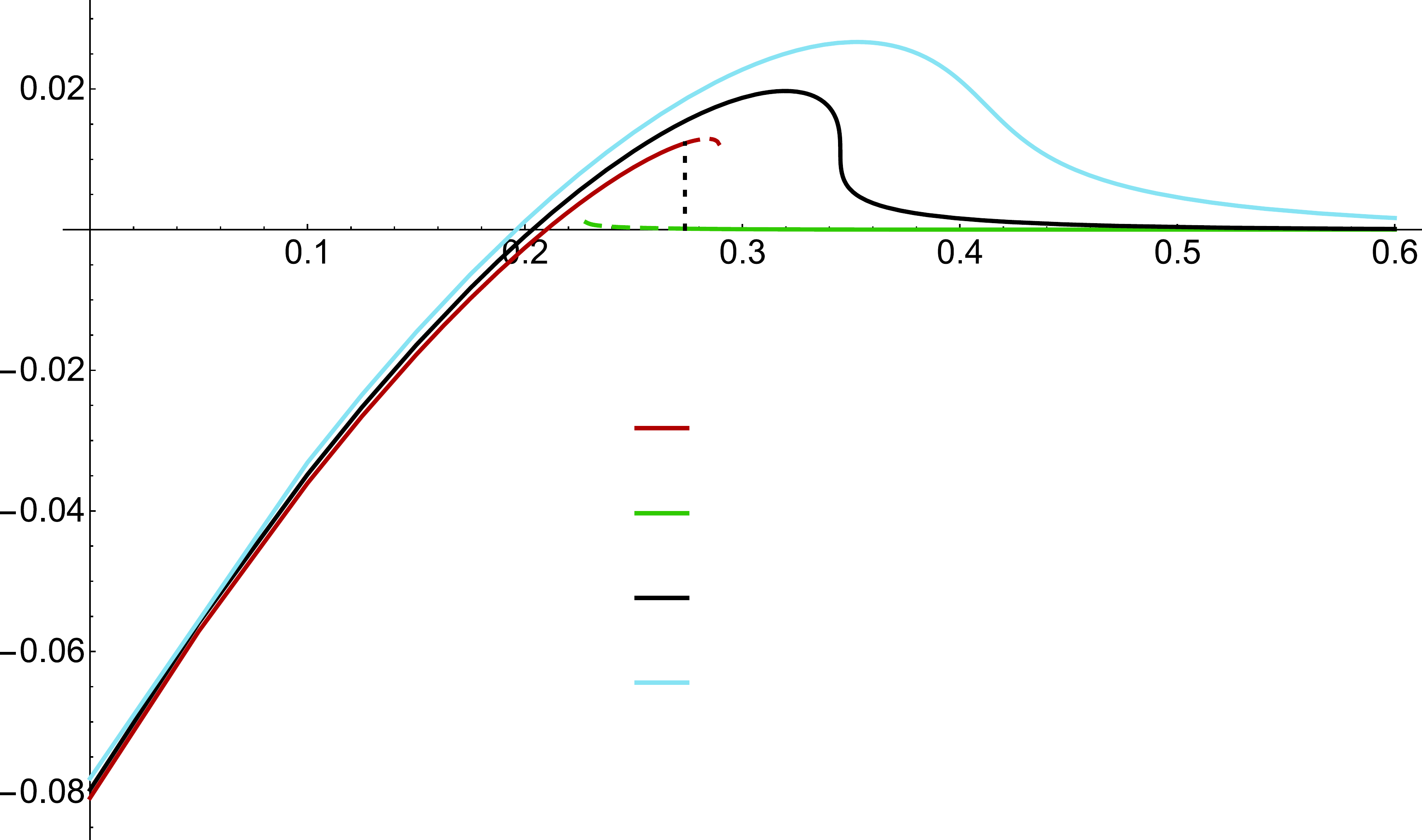 &
\def\svgwidth{7cm}
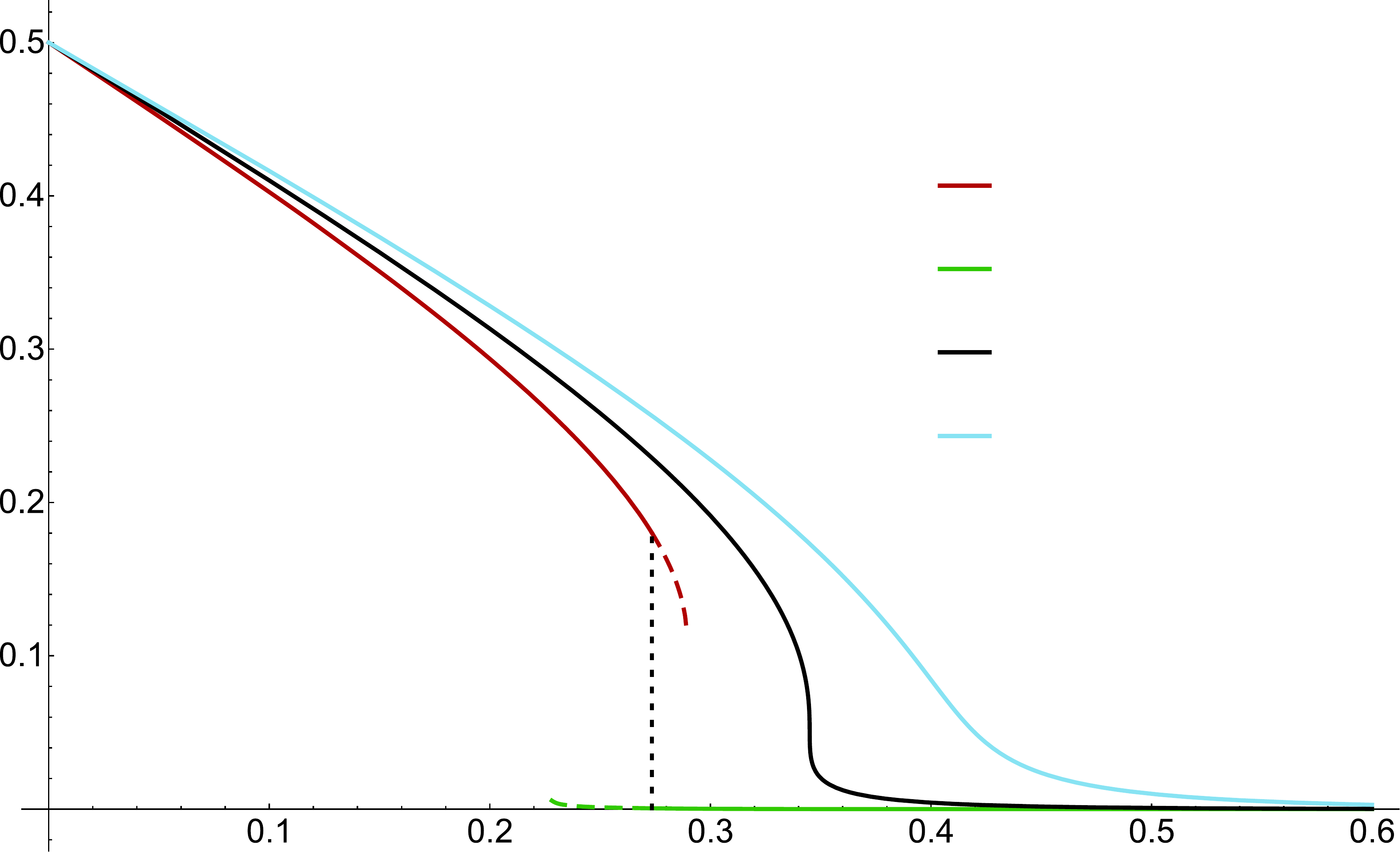 \\
(c) & (d)
\end{tabular}
\caption{\label{M1FreeEnergyEntropy}Free energy (a), entropy (b), energy (c) and charge (d) of the \modone\ model, as functions of the mass, for $\la=1$. Solid lines correspond to the actual physical values associated with the most stable solution, dashed lines to metastable values. For $T<\Tc$ (here $T = 0.035$) there are two distinct solutions, one SYK-like (red, dark gray) and another HO-like (green, gray), with thermodynamic quantities experiencing finite jumps where their free energies cross (black dotted lines at $m_{\text t} = 0.274_1$). For $T = \Tc = \TcValue$ there is a single solution (solid black curve), with a singular behaviour at $m = \mc$. For the supercritical phase at $T > \Tc$ (blue or light gray curve, for $T=0.1$ here) all the thermodynamical potentials are smooth.
}
\end{figure}

The curves $m_-(T)$ and $m_+(T)$ are drawn in Fig.~\ref{fig:M1PhaseDiagram} with dashed lines. For any fixed temperature $T<\Tc$, there is an interval of masses $m_{-}(T) < m < m_{+}(T)$ for which both solutions coexist. Their free energies coincide at a transition mass $m_{\rm t}(T)$, the HO-like and SYK-like solutions being more stable for $m>m_{\rm t}(T)$ and $m<m_{\rm t}(T)$ respectively. The system thus undergoes a phase transition when crossing the line $m_{\rm t}(T)$, the jump from one solution to the other being discontinuous. The corresponding phase transition is therefore of first order. This is illustrated in Fig.\ \ref{M1FreeEnergyEntropy}, where the thermodynamical potentials are plotted as a function of mass for temperatures $T<\Tc$, $T=\Tc$ and $T>\Tc$.

As we have mentioned at the end of Sec.\ \ref{phasemotivSec}, the first order phase transition observed by decreasing $m$ at fixed $T$, for $T<\Tc$, is reminiscent of the phenomenon of gravitational collapse. On the other hand, if we start from very low $T$ in the mass interval $m_{*}<m<\mc$ and increase $T$, then we also cross the line of first order phase transitions. In this case, the phenomenon is somehow reminiscent of the Hawking-Page phase transition, where the thermal AdS space becomes thermodynamically unstable compared with the large Schwarzschild-AdS black hole solution \cite{HawPage}.

A very important feature of the phase diagram is that the mass interval in which the two solutions coexist shrinks as the temperature is increased, and eventually reduces to the point $m=\mc$ when $T=\Tc$. At this point, the two solutions coincide and the transition becomes of second order. Remarkable properties of this critical point and its generalizations will be discussed below in Sec.~\ref{Pha1exponentsqSec} and \ref{Pha1continuousexpSec}. For $T>\Tc$ there is only one solution and the continuation from large to small mass is completely smooth.

\medskip

In \cite{Maldacena:2018lmt}, Maldacena and Qi studied a system consisting of two copies of the SYK model coupled via a quadratic ``mass'' term. The resulting phase diagrams, depicted in Fig.\ 11 and 18 of Ref.\ \cite{Maldacena:2018lmt}, have both similarities and differences with the one described here.

An important similarity is that there are two competing solutions in a region of the $(m,T)$-plane, with an associated first order phase transition. Moreover, it seems that the line of first order phase transitions terminates at a critical point (see in particular Fig.~18 in \cite{Maldacena:2018lmt}), but  this critical point is not explicitly mentioned or analyzed in \cite{Maldacena:2018lmt}. A detailed discussion of the properties of the critical point found in the \modone\ model will be provided below in Sec.~\ref{Pha1exponentsqSec}, \ref{Pha1continuousexpSec} and \ref{RealTimeSec}.

On the other hand, a crucial difference is that the mass parameter $m$ in Maldacena-Qi is \emph{relevant}: a mass gap is created as soon as the mass is turned on. All the solutions for the two-point functions in their model are thus gapped when $m\not = 0$, in parallel with our HO-like solutions; in other words, they do not have SYK-like solutions with a non-trivial IR behaviour when $m\not = 0$. 

Another interesting aspect of the quadratically coupled SYK model is that it can be fruitfully studied in the large $q$ limit, where $q$ is the order of the interactions. In this limit a third branch of solutions is found, and allows to interpolate smoothly between the two other types of solutions responsible for the existence of the first order phase transition that we have discussed above. Solutions in this third branch have a negative specific heat. Their existence is irrelevant in the canonical ensemble, but Maldacena and Qi argued that they could smoothen the transition in the microcanonical ensemble.\footnote{A non-trivial issue is that solutions with negative specific heat are thermodynamically unstable in standard set-ups with local interactions and a notion of extensivity; what really happens in the Maldacena-Qi model, which is a disordered model with all-to-all interactions, or in equivalent matrix models, remains an open question.} Note that this novel class of solutions was not found by solving numerically the Schwinger-Dyson equations at finite $q$, so it might be an artifact of the large $q$ limit, even though this seems unlikely.

The \modone\ model we are describing here can also be generalized to arbitrary $q$-body interactions, see App.\ \ref{qGenApp} and the results in Sec.\ \ref{Pha1exponentsqSec} and \ref{Pha1continuousexpSec}. However, the large $q$ limit, when carried out in a standard way along the lines of \cite{Maldacena:2016hyu} and \cite{Maldacena:2018lmt}, turns out to be rather trivial. In particular, the phase transition is never seen. The reason for this seems to be that the critical masses $m_{*}$ or $\mc$, or the critical temperature $\Tc$, become exponentially small with $q$ at large $q$, instead of being governed by a simple power law. For this reason, we shall not discuss the large $q$ limit any further here. Note also that we have not been able to find numerically any different solution with negative specific heat at finite $q$ either, although we can of course not guarantee it does not exist. 

\subsubsection{\label{Pha1Qm}The $(m,Q)$ plane and physics at fixed $Q$}

\begin{figure}[h!]
\centering
\def\svgwidth{4.5in}
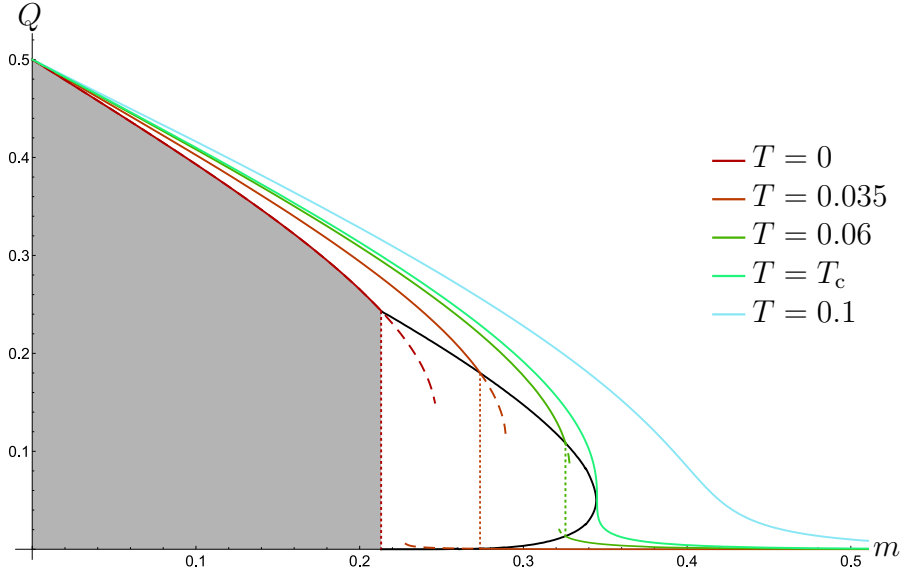
\caption{\label{fig:M1MQPhaseDiagram}Phase diagram of the \modone\ model in the $(m,Q)$ plane, also in \cite{letter}. We have indicated several isothermal curves, including one for the $T\to0$ extrapolation of finite-temperature results (see Sec.\ \ref{Pha1basic} below). Solid lines correspond to thermodynamically stable solutions; dashed lines are used for metastable ones. For $T<\Tc$ these curves go through the transition region delimited by the black line, where both phases coexist. The area shaded in gray is a forbidden region.}
\end{figure}

The phase diagram in the $(m,Q)$ plane is depicted in Fig.\ \ref{fig:M1MQPhaseDiagram}. One notable feature is the existence of a large forbidden region, shaded in grey. Another notable feature is that for temperatures $T<\Tc$ there is an interval of charges that can only be realized if the two phases coexist (vertical dotted lines on the figure).\footnote{Since the model we are considering does not have a notion of space or local interactions, it is unclear how such coexistence might be realized.}

In the context of most condensed matter work, it is natural to interpret the mass term as minus a chemical potential, which is adjusted to fix the fermion number $Q$ to any desired value. This is not the point of view adopted in the present paper, where we always consider the mass to be the most natural variable to be fixed. An interesting consequence of our analysis is that for any given temperature $T<\Tc$, there is a maximum mass $m_{+}(T)$, associated with the dashed line on the right in Fig.\ \ref{fig:M1PhaseDiagram}, above which the system cannot exist in the SYK-like phase. 

The existence of this maximum mass, and thus of a minimum chemical potential $-m_{+}(T)$, is a very well established fact. Independently of the numerical results, we can provide the following simple argument. Let us assume that the $m_{+}(T)$ bound does not exist, so that the SYK-like solution can then be analytically continued anywhere in the $(m,T)$ plane. Starting from the grey-shaded region in Fig.\ \ref{fig:M1PhaseDiagram} where both the SYK and HO-like solutions exist, and performing the analytic continuation of the two solutions to the high-temperature regime along a path that goes around the critical point $(\mc,\Tc)$ on its right, we deduce that two distinct solutions must exist in the weakly coupled, high $T$ at fixed $m$ regime. Moreover, these two distinct solutions must have the same UV behaviour $G_{k}\sim i/\nu_{k}$ given by \eqref{GkUV}, since this behaviour cannot change under analytic continuation. But this is impossible. At high temperatures, all the Matsubara frequencies are in the UV regime $\nu_{k}\gg\la$ and thus any solution with the standard UV behaviour will necessarily be very near the perturbative solution $G(\tau)\simeq \frac{1}{2}\sign(\tau)$. A unique series expansion for any such solution can be generated from the Schwinger-Dyson equations; this series coincides, of course, with the high-temperature perturbative series. It converges and yields a unique solution, thus proving that the bound $m_{+}(T)$ necessarily exists.

The numerical results suggest that for the SYK-like phase the lines $Q(m)$ at fixed temperature $T<\Tc$ all terminate at a strictly positive charge $Q_{\text{min}}(T)>0$. This minimum possible charge $Q_{\text{min}}(T)$ corresponds to the endpoints of the upper dashed lines in Fig.\ \ref{fig:M1MQPhaseDiagram}, and can be computed at any given temperature evaluating $Q(m)$ at $m=m_{+}(T)$. For instance, we find at zero temperature $Q_{\text{min}}(T=0) = 0.150_2$. Note that $Q_{\text{min}}$ is of course less than the value of the charge below which the solution becomes thermodynamically unstable; at $T=0$, this latter charge corresponds to the upper left endpoint of the black line in Fig.\ \ref{fig:M1MQPhaseDiagram} and is found to be approximately $Q_* = Q(T = 0, m = m_*) = 0.243_1$. Our numerical results seem to indicate that $Q_{\text{min}}(T)$ is a decreasing function of $T$ which goes to the value $Q_{\text c}=0.050_{1}$ at the critical point when $T=\Tc$. 

However, it is also conceivable that our numerical algorithm may fail to converge for a small interval of masses near $m_{+}(T=0)$, for which the slope $|\partial Q/\partial m|$ is large; one may then imagine that the zero-temperature line $Q(m)$ (the red dashed line in Fig.\ \ref{fig:M1MQPhaseDiagram}) might extend all the way down to zero charge. On the other hand, at non-zero temperature it seems very unnatural to obtain solutions with zero charge and finite chemical potential. Then $Q_{\text{min}}(T)$ should be strictly positive, at least for all $T>0$, while going to $\Qc$ when $T=\Tc$ (since at the critical point the SYK-like solution merges with the HO-like solution).

\subsubsection{\label{Pha1basic}The physics at zero temperature}

\begin{figure}[h!]
\centering
\def\svgwidth{4in}
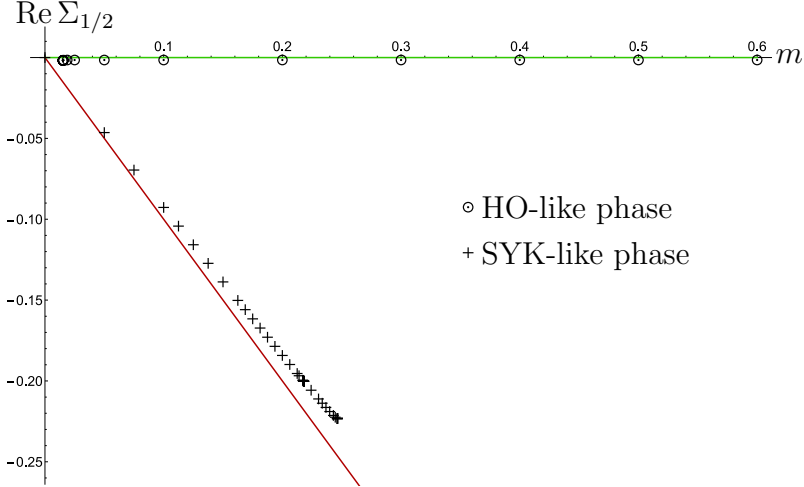
\caption{\label{fig:M1SigmaZero}Values of $\re\Sigma_{1/2} \simeq \tilde\Sigma(0)$ as a function of the mass $m$, for the \modone\ model at $T = 10^{-3}$ in natural units. Datapoints for the SYK-like phase (crosses) closely follow the constraint \eqref{Sigtildeconst}, $\tilde\Sigma(0) = m$  (red or dark gray line), whereas datapoints for the HO-like phase (circles) are compatible with $\tilde\Sigma(0) = 0$ (green or light gray line).}
\end{figure}

It is certainly very interesting to study the phase diagram at zero temperature, but the numerical algorithm used to solve the Schwinger-Dyson equations operates inherently at finite-temperature. Of course, the identification of the two distinct solutions in the \modone\ model as SYK- and HO-like already constitutes a statement about the low-temperature regime. However, this identification is possible using only very elementary data, such as the value of the entropy (see Fig.~\ref{fig:Hysteresis}) or the rough criterion associated to the constraint \eqref{Sigtildeconst}. Indeed, as shown in Fig.\ \ref{fig:M1SigmaZero} the value of $\re\Sigma_{1/2}$ at a small but finite temperature $T = 10^{-3}$ is approximately equal to the mass for the analytic continuation of the $m = 0$ solution, whereas it essentially vanishes for the analytic continuation of the large mass solution.

\begin{figure}[h!]
\vskip 0.2cm
\centering
\begin{tabular}{cc}
\def\svgwidth{7cm}
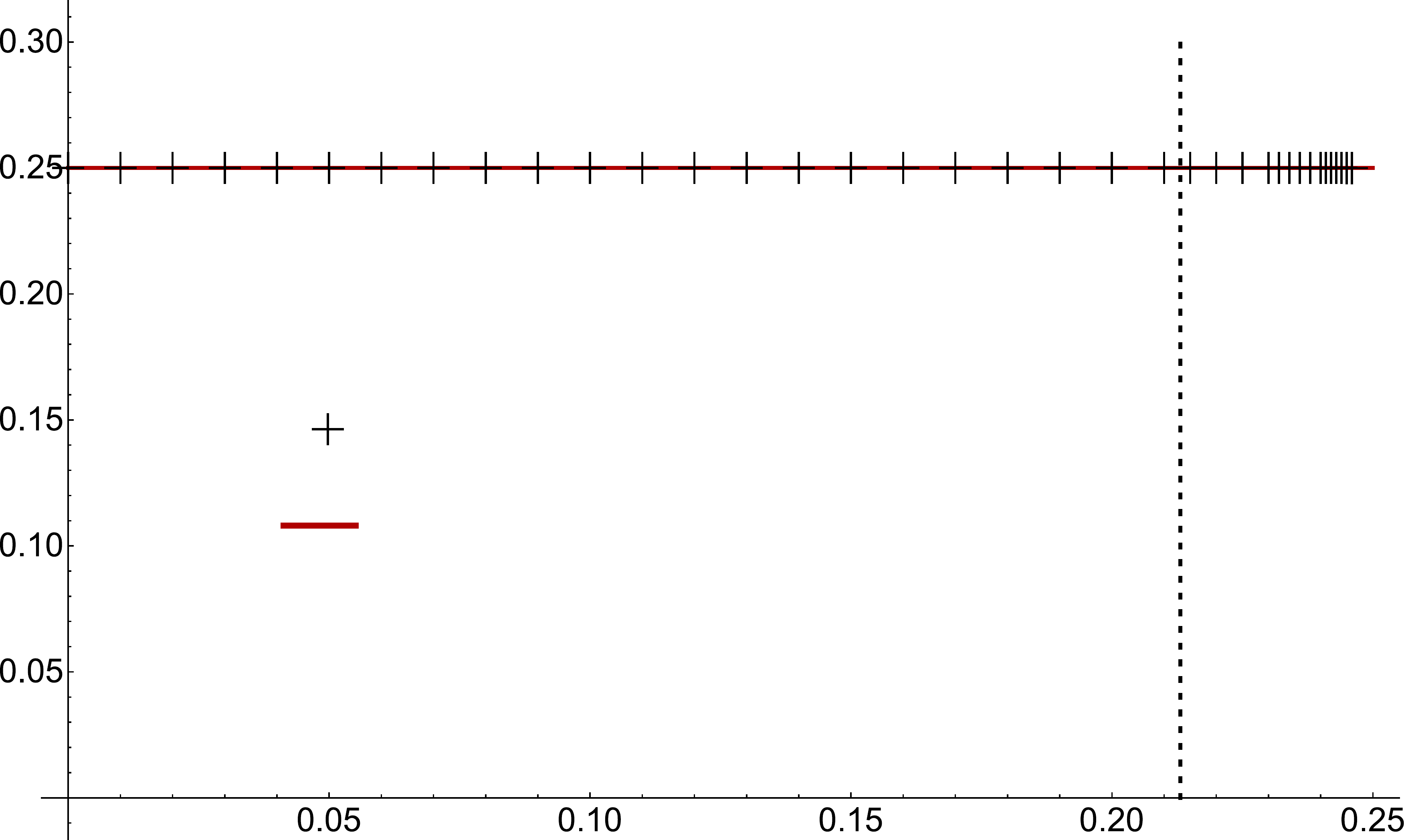 &
\def\svgwidth{7cm}
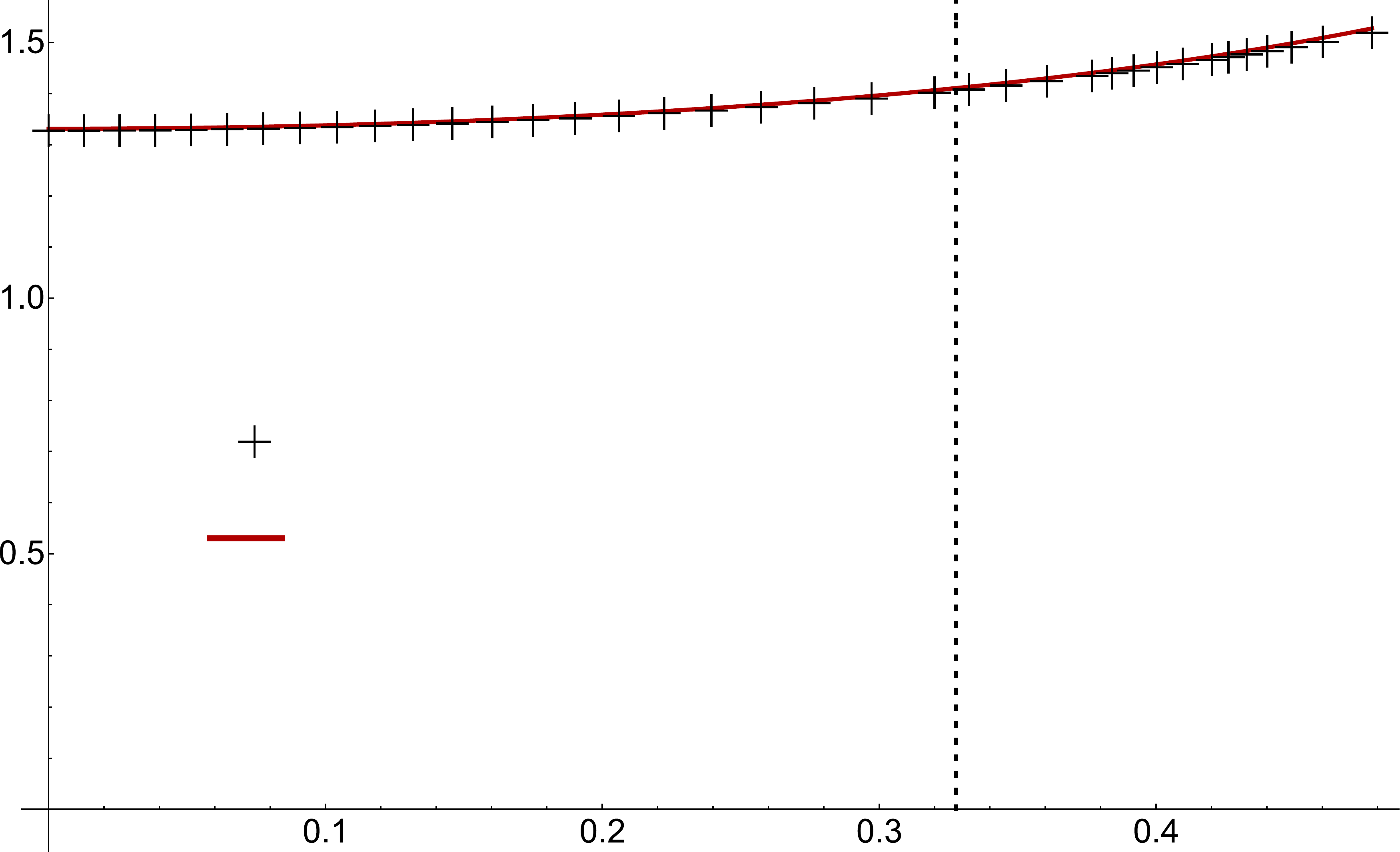 \\
(a) & (b)
\end{tabular}
\caption{\label{fig:M1ZeroTDelta}$T\to0$ extrapolation of the scaling dimension $\Delta$ (a) and the IR parameter $C$ (b), as functions of the mass and the IR parameter $\theta$, respectively. Numerical values were computed using the finite-temperature fitting procedure illustrated in Fig.~\ref{fig:M1DeltaFiniteT}, with linear extrapolations to $T\to0$ (represented with crosses) performed using $\sim20$ points with $\beta \in [500, 1000]$. The vertical dotted lines mark the zero-temperature phase transition values $m_* = \mStarOne$ and $\theta_* = \theta(m_*) = 0.328_2$, for (a) and (b) respectively. The scaling dimension matches the $\Delta = 1/4$ value of \eqref{Deltaform1}, whereas the IR parameters $C$ and $\theta$ defined by \eqref{Cthetadef} satisfy the constraint $C^4 \cos(2\theta) = \pi$ of \eqref{Cthetarel} (red or dark gray lines in each plot).}
\end{figure}

In order to provide further details about the zero-temperature physics, a careful analysis involving more delicate extrapolations is required. We have already shown in Fig.~\ref{fig:M1DeltaFiniteT} an example for how such a procedure may be carried out for some small but non-vanishing mass, its results matching the expectations coming from the low-temperature ansatz, \eqref{Deltaform1} and \eqref{Cthetarel}. A more global picture is provided in Fig.~\ref{fig:M1ZeroTDelta} by systematically repeating that process for various masses. We clearly see that the agreement is very good all along the range of masses for which the SYK-like phase actually exists.

\begin{figure}[h!]
\centering
\begin{tabular}{cc}
\def\svgwidth{7cm}
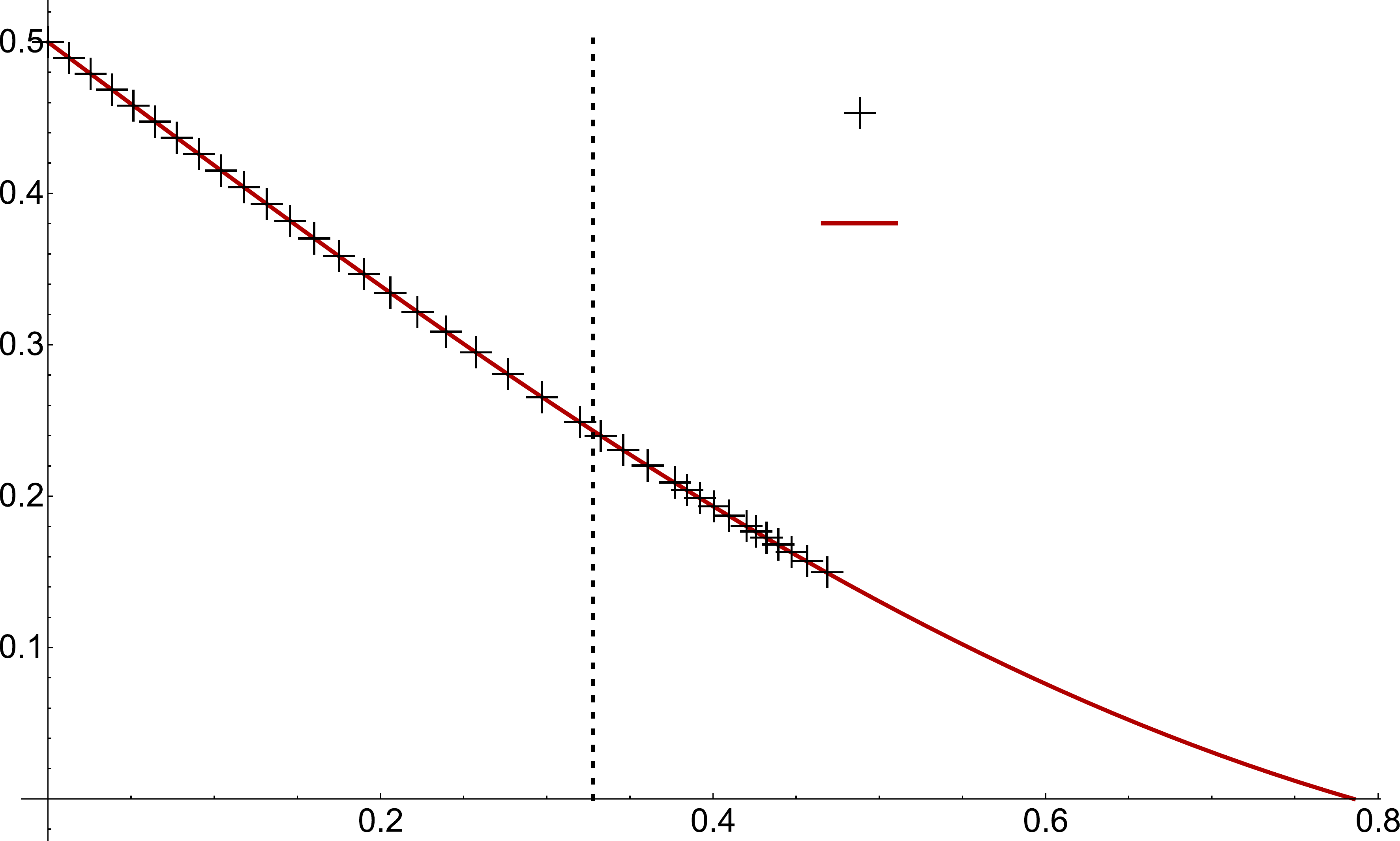 &
\def\svgwidth{7cm}
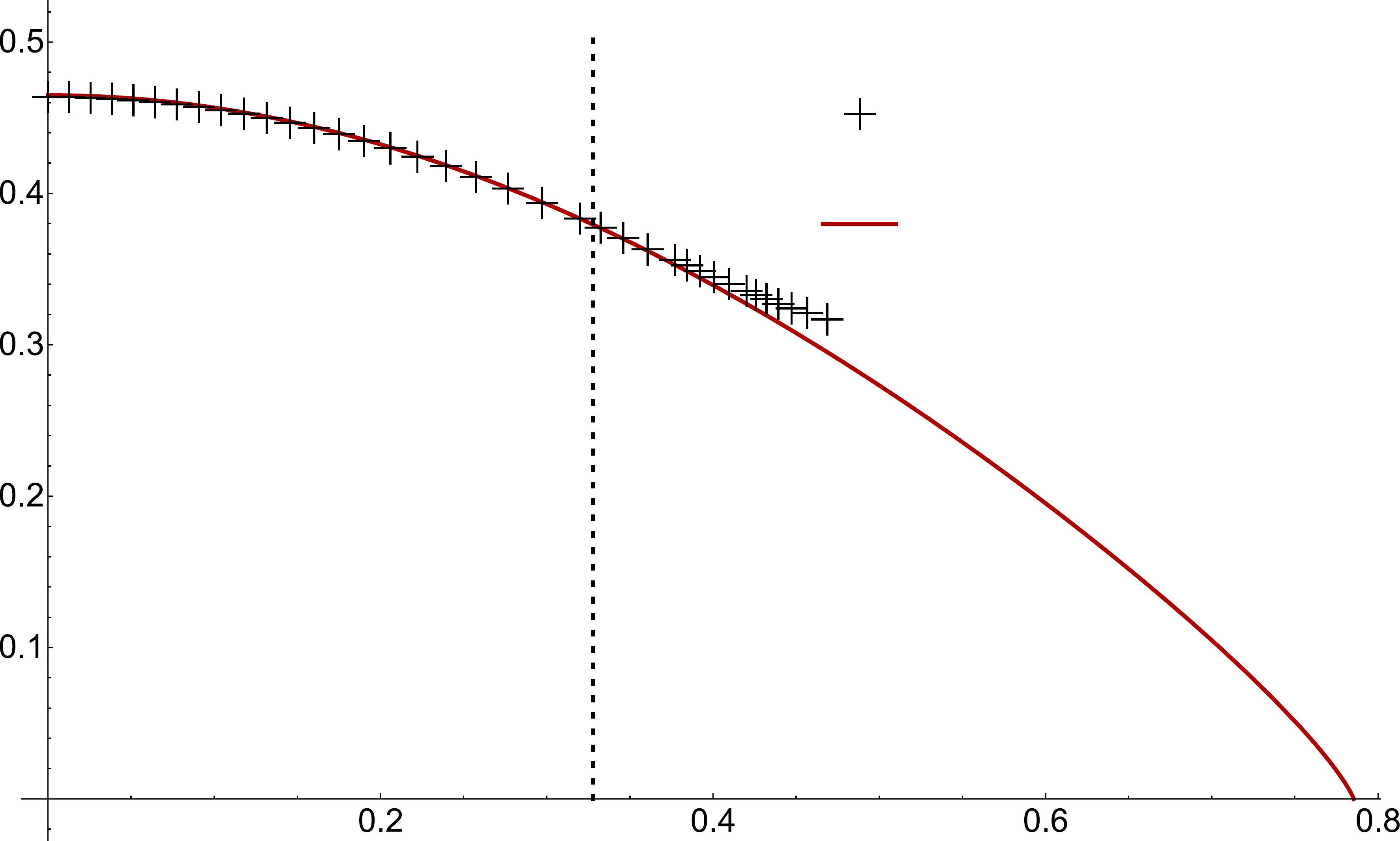 \\
(a) & (b)
\end{tabular}
\caption{\label{fig:M1ZeroTTheta}Zero-temperature charge (a) and entropy (b) as functions of the IR parameter $\theta$ defined by \eqref{Cthetadef}. Numerical values are represented with crosses, solid red (dark gray) lines  correspond to the formulas \eqref{thetaQrel} and \eqref{zeroTentropy2} for (a) and (b), respectively. Extrapolations were performed as explained in Fig.~\ref{fig:M1ZeroTDelta}, but taking $\beta \in [50,100]$ for $S/N$. The vertical dotted line marks the value $\theta_* = \theta(m_*) = 0.328_2$.
}
\end{figure}

A further non-trivial check of the extrapolation procedure is made possible by the exact $T = 0$ results for the SYK-like phase reviewed in Sec.~\ref{EucformSec}. In Fig.~\ref{fig:M1ZeroTTheta} we plot the $T\to0$ extrapolations of the charge and entropy as functions of the IR parameter $\theta$, along with the exact results \eqref{thetaQrel} and \eqref{zeroTentropy2} for comparison. The agreement between numerical and theoretical values is very good for masses up to $m = m_*$ and a bit higher, but starts exhibiting deviations in the entropy for larger masses. We can attribute these discrepancies to numerical artifacts becoming more and more relevant as the mass increases, the entropy being very sensitive to the truncation implemented by the frequency-cutoff of the numerical algorithm (see App.~\ref{NumericalErrorsApp}). In any case, these demonstrations of the precision of the numerical extrapolation used to implement the $T\to0$ limit give us confidence that our results are very robust at least up to $m \sim m_*$.

\begin{figure}[h!]
\centering
\begin{tabular}{cc}
\def\svgwidth{7cm}
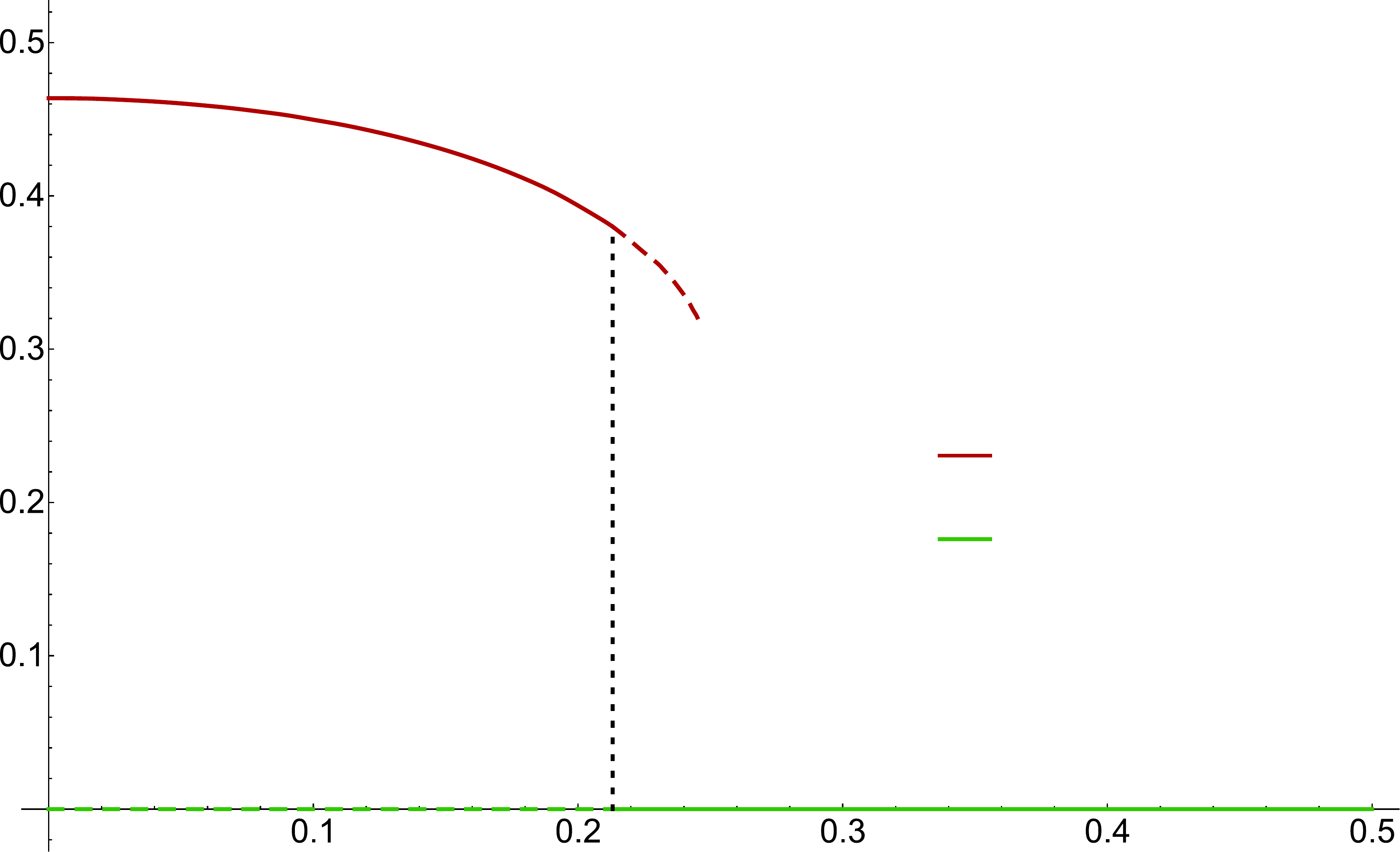 &
\def\svgwidth{7cm}
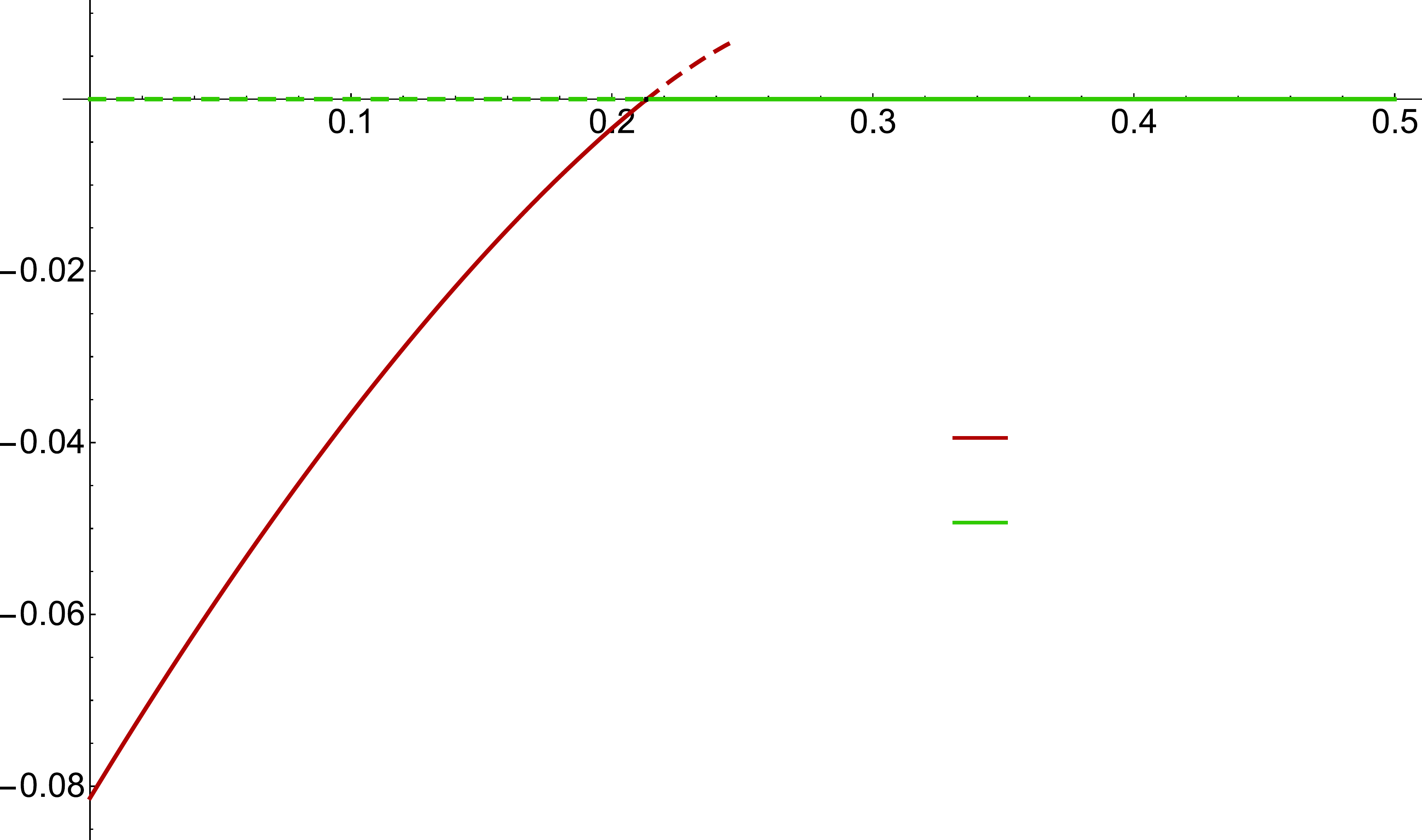 \\
(a) & (b)\\[0.5cm]
\def\svgwidth{7cm}
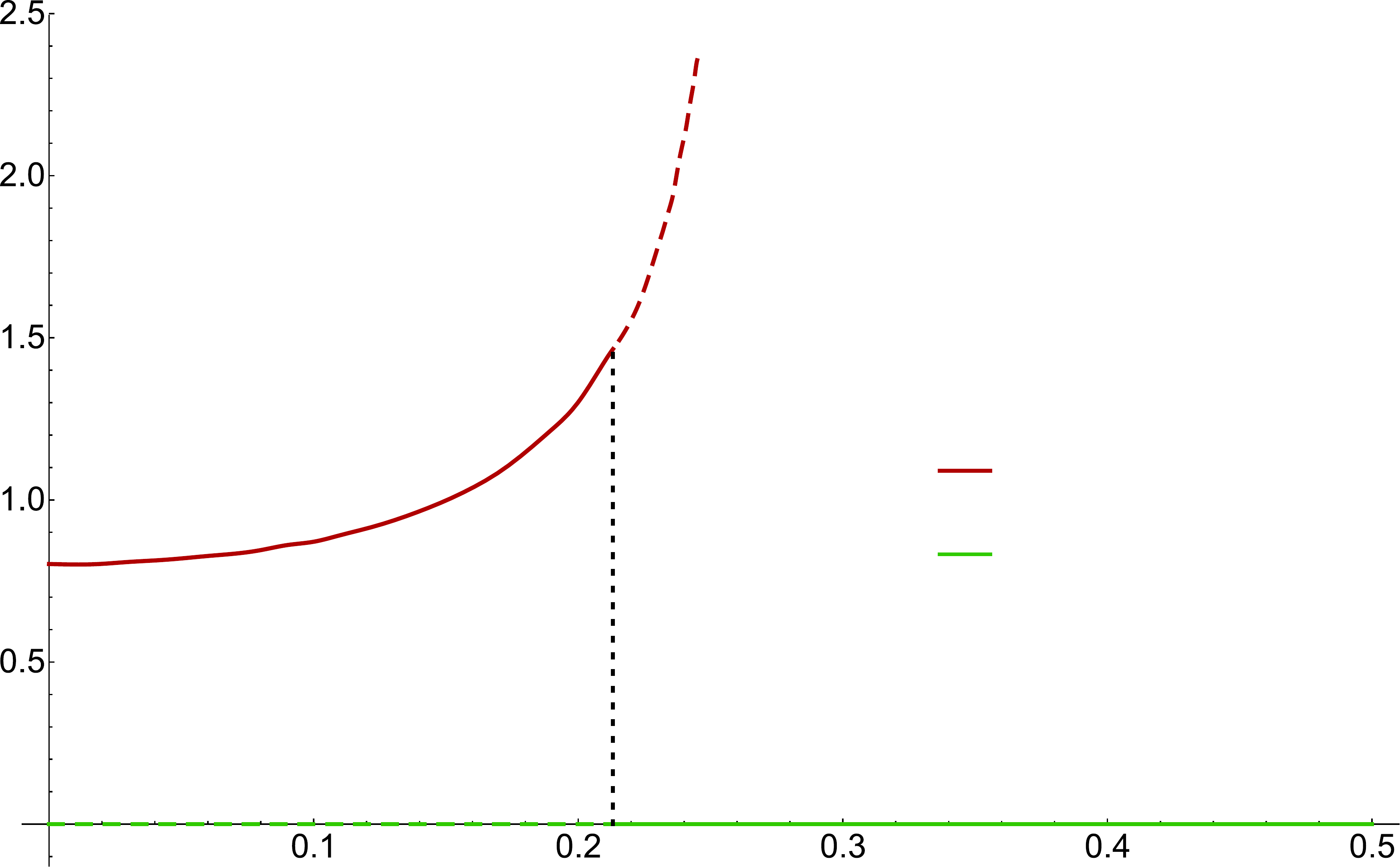 &
\def\svgwidth{7cm}
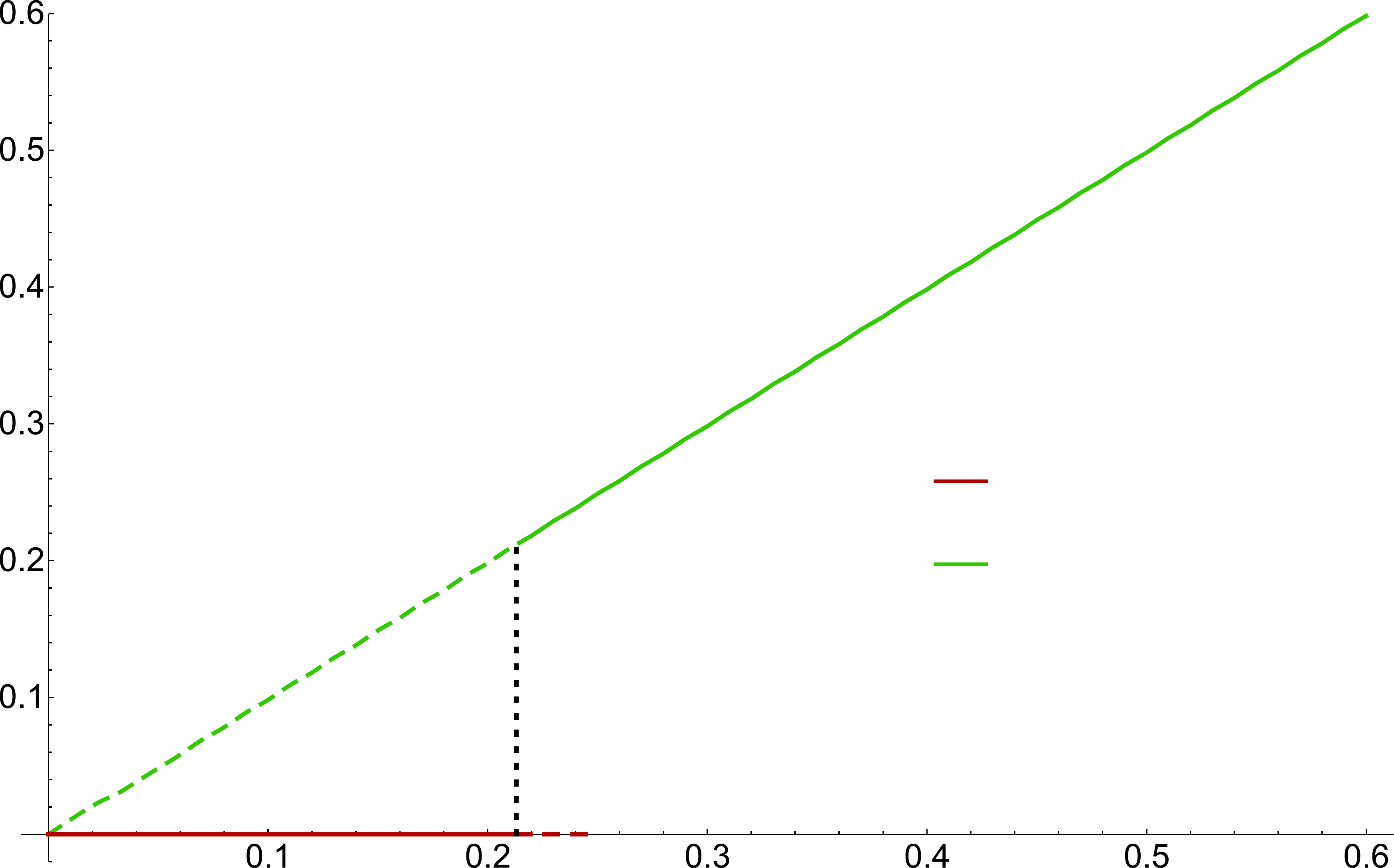 \\
(c) & (d)
\end{tabular}
\caption{\label{fig:M1ZeroTThermo}Plots of the zero-temperature entropy $\sigma$ (a), energy $\epsilon$ (b), specific heat coefficient $c$ defined in \eqref{zeroTdef} (c) and effective mass $\meff$ (d), as functions of the mass. Solid lines correspond to the thermodynamically stable solution, which is the SYK-like solution for $m<m_{*}$ (red, dark gray) and the HO-like solution for $m>m_{*}$ (green, light gray). Values for metastable solutions (the SYK-like solution for $m>m_*$, the HO-like one for $m<m_*$) are represented with dashed lines. Vertical black dotted lines denote a finite jump in the corresponding functions. All extrapolations are performed as in Fig.~\ref{fig:M1ZeroTDelta} and Fig.~\ref{fig:M1ZeroTTheta}.
}
\end{figure}

Moving on, in Fig.\ \ref{fig:M1ZeroTThermo}, we plot the zero-temperature energy, entropy and specific heat divided by $T$,
\be\label{zeroTdef} \epsilon = \frac{1}{N}\lim_{T\rightarrow 0}E\, ,\quad
\sigma = \frac{1}{N}\lim_{T\rightarrow 0}S\, ,\quad
c =\frac{1}{N}\lim_{T\rightarrow 0}\frac{C}{T}= \frac{1}{N}\lim_{T\rightarrow 0}\frac{\partial S}{\partial T}\, \cvp\ee
as functions of the mass, for the SYK-like solution and the most stable solution. Note that all these quantities are trivially vanishing for the HO-like solution, since it coincides with the tree-level result at zero temperature. The thermodynamical potentials can be expanded in terms of $\epsilon$, $\sigma$ and $c$ at low temperature as
\be\label{lowTpotentials} \frac{E}{N} = \epsilon + \frac{1}{2} cT^{2} + o(T^{2})\, ,\quad \frac{S}{N} = \sigma + c T + o(T)\, ,\quad \frac{F}{N} = \epsilon - \sigma
T -\frac{1}{2} c T^{2}+o(T^{2})\, .\ee
The results are consistent with the general discussion in Sec.~\ref{phasemotivSec}. The ground state jumps from the Fock vacuum at $m>m_{*} = \mStarOne$ to a highly degenerate state at $m<m_{*}$. The heat capacity $c$ grows from its $c(m=0) = 0.80_1$ value matching the result reported for SYK \cite{Maldacena:2016hyu}, and experiences a finite jump to zero at $m = m_*$. While the extrapolation for this quantity should certainly experience similar numerical artifacts to those present in the entropy (see Fig.~\ref{fig:M1ZeroTTheta}b), it seems plausible that $c$ will diverge when $m\to m_{+}(T=0)$. We also plot in Fig.~\ref{fig:M1ZeroTThermo}d the effective mass obtained from the exponential decay of the HO-like solution, extrapolated to zero temperature, finding perfect agreement with the $m_{\rm eff} = m$ prediction. This result will serve as a baseline reference when studying the \modtwo\ model in the following.

\noindent\emph{Remark}: There is no contradiction between the fact that the mass gap $\meff$ jumps discontinuously from $\meff=m$ for $m>m_{*}$ to $\meff=0$ for $m<m_{*}$ (Fig.\ \ref{fig:M1ZeroTThermo}d) and the fact that the zero-temperature energy is a continuous function of $m$ at $m=m_{*}$ (Fig.\ \ref{fig:M1ZeroTThermo}b). This is a large $\dof$ effect: the eigenstates of the Hamiltonian responsible for the small negative energy $\epsilon$ when $m$ is just below $m_{*}$ do not contribute to the large $\dof$ two-point function when $m>m_{*}$.

\subsubsection{\label{Pha1exponentsqSec}The $q$-generalizations and critical exponents}

\begin{figure}
\centering
\def\svgwidth{4.5in}
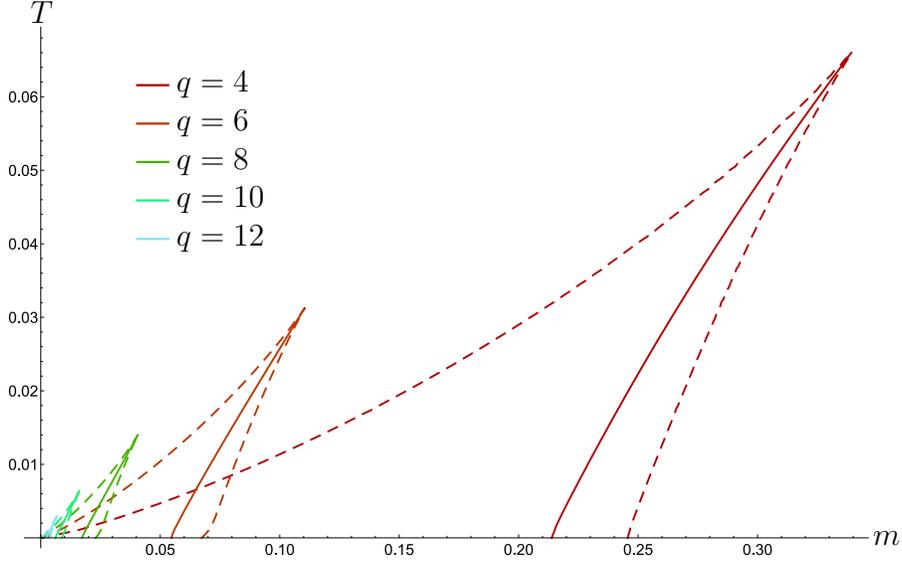
\caption{\label{fig:PhaseDiagramGenQ}Superimposition of the phase diagram of the $q$-generalizations of the \modone\ model in the $(m,T)$-plane, in the units $\la_{q}=1$, for $q=4,6,8,10,12$. Solid lines correspond to first order phase transitions, dashed lines delimit the regions of coexistence of the SYK-like and HO-like solutions in each model.}
\end{figure}

The \modone\ model admits a straightforward generalization described in App.\ \ref{qGenApp}, for which the quartic interaction term in the Hamiltonian is replaced by a sum of terms of even degrees. In this subsection we restrict our attention to the case where only one such term is included, with degree $q\geq 4$. The qualitative physics of the models is then essentially the same for all values of $q$, and the discussion presented in the previous subsections for the case $q=4$ can be straightforwardly generalized. This is illustrated in Fig.\ \ref{fig:PhaseDiagramGenQ}, in which the phase diagram for the original $q=4$ model, taken from Fig.\ \ref{fig:M1PhaseDiagram}, is superimposed to those of the $q=6,8,10,12$ generalizations. In each case, there is a line of first order phase transitions that terminates at a critical point $(\mc^{(q)},\Tc^{(q)})$, with the zero-temperature phase transition occurring at $m = \smash{m_*^{(q)}}$. The phase diagrams are all very similar, with the transition line shrinking to an ever smaller region in the $(m,T)$-plane as $q$ is increased (in the natural units $\la_{q}=1$). Values of $\smash{\mc^{(q)}}$, $\smash{\Tc^{(q)}}$ and $\smash{m_*^{(q)}}$ for the first few $q$'s are presented in Table~\ref{tab:PhaseDiagramParameters}, along with the charge at the critical point $\Qc^{(q)}$.

\begin{table}[h!]
\centering
\begin{tabular}{|c|c|c|c|c|}
\hline
$q$ & $m_*^{(q)}$ & $\mc^{(q)}$ & $\Tc^{(q)}$ & $\Qc^{(q)}$\\
\hline
4 & $\mStarOne$ & $\mcValue$ & $\TcValue$ & $0.050_1$\\
\hline
6 & $0.055_1$ & $0.111_2$ & $0.0317_5$ & $0.0996_1$\\
\hline
8 & $0.017_1$ & $0.0412_6$ & $0.0143_4$ & $0.1343_2$\\
\hline
10 & $0.006_1$ & $0.0163_2$ & $0.0065_2$ & $0.1612_2$\\
\hline
12 & $0.0022_5$ & $0.0068_2$ & $0.0030_1$ & $0.1837_7$\\
\hline
\end{tabular}
\caption{\label{tab:PhaseDiagramParameters}Values of the phase diagram parameters for the $q$-generalizations of the \modone\ model, for the first few values of $q$, in the units $\la_{q}=1$. The first order phase transition line starts at $T = 0$ with $m = \smash{m_*^{(q)}}$ and terminates at the critical point $\smash{(\mc^{(q)},\Tc^{(q)})}$ with charge $\Qc^{(q)}$.}
\end{table}

The $q$-body interacting generalizations of the SYK model and closely related models have been considered in the literature (see \emph{e.g.}\ \cite{Maldacena:2016hyu,Maldacena:2018lmt,qSYK}) because the large $q$ limit can be solved analytically and captures interesting aspects of the physics of the models at finite $q$. However, as we have already mentioned in our comments on the Maldacena-Qi model in Sec.\ \ref{PhaseMod1Sec}, the large $q$ limit of the \modone\ model turns out to be rather trivial when defined in the standard way. In particular it is not able to capture the line of first order phase transitions or the associated critical point, which are the most interesting features of the model. Trying to understand better the large $q$ behaviour in this model, as well as possible $q$-body generalizations of the \modtwo\ model, could be an interesting direction for future research.

For our purposes, the main interest of considering the $q$-generalizations of the \modone\ model introduced in App.~\ref{qGenApp} is that we can study the properties of the critical point for different values of $q$. There are various natural observables one may consider; we choose to focus on the following quantities:
\begin{enumerate}[i)]
\item The heat capacity
\be\label{Cdefexp}
C = T\frac{\partial S}{\partial T}\,,
\ee
as a function of temperature at $m=\mc^{(q)}$. We shall compute the right and left critical exponents $\alpha_{\pm}^{(q)}$ defined by
\be\label{crit1def} C\bigl(T,\mc^{(q)}\bigr)\underset{T\rightarrow \Tc^{(q)^{\pm}}}{\sim} \frac{C_{\pm}^{(q)}}{\bigl|T-\Tc^{(q)}\bigr|^{\alpha_{\pm}^{(q)}}}\,\cvp\ee
which govern the singular behaviour of $C$ when $T$ approaches the critical value $\Tc^{(q)}$ from above ($\alpha^{(q)}_+$ for $T>\Tc^{(q)}$) or from below ($\alpha^{(q)}_-$ for $T<\Tc^{(q)}$).
\item The susceptibility
\be\label{chidefexp}
\chi = \frac{\partial Q}{\partial m}\,,
\ee
as a function of temperature at $m=\mc^{(q)}$, with right and left critical exponents $\gamma_{\pm}^{(q)}$ defined by
\be\label{crit2def} \chi\bigl(T,\mc^{(q)}\bigr)\underset{T\rightarrow \Tc^{(q)^{\pm}}}{\sim} \frac{\chi_{\pm}^{(q)}}{\bigl|T-\Tc^{(q)}\bigr|^{\gamma_{\pm}^{(q)}}}\,\cdotp\ee
\item The charge $Q$ as a function of temperature at $m=\mc^{(q)}$, with right and left critical exponents $q_{\pm}^{(q)}$ defined by
\be\label{crit3def} Q\bigl(T,\mc^{(q)}\bigr)-Q\bigl(\Tc^{(q)},\mc^{(q)}\bigr)\underset{T\rightarrow \Tc^{(q)^{\pm}}}{\sim} Q_{\pm}^{(q)}\bigl|T-\Tc^{(q)}\bigr|^{q_{\pm}^{(q)}}\,\cdotp\ee
\item The charge difference $\Delta Q$ between the two phases along their coexistence line, as a function of temperature. This quantity is of course defined only for $T<\smash{\Tc^{(q)}}$, and the associated critical exponent $\smash{\beta^{(q)}}$ is defined by
\be\label{crit4def} \Delta Q(T)\underset{T\rightarrow \Tc^{(q)^{-}}}{\sim} \Delta Q_{-}^{(q)}\bigl (\Tc^{(q)}-T\bigr)^{\beta^{(q)}}\,\cdotp\ee
Note that we could also consider the same quantity as a function of mass, but since $m_{\rm t}^{(q)}(T)$ is always smooth the associated critical exponent is also $\beta^{(q)}$, a fact that we have verified numerically.
\item The entropy $S$ as a function of mass at $T=\Tc^{(q)}$, with right and left critical exponents $s_{\pm}^{(q)}$ defined by
\be\label{crit5def} S\bigl(\Tc^{(q)},m\bigr)-S\bigl(\Tc^{(q)},\mc^{(q)}\bigr)\underset{m\rightarrow \mc^{(q)^{\pm}}}{\sim} S_{\pm}^{(q)}\bigl|m-\mc^{(q)}\bigr|^{s_{\pm}^{(q)}}\,\cdotp\ee
\item The charge $Q$ as a function of mass at $T=\Tc^{(q)}$, with right and left critical exponents $\tilde q_{\pm}^{(q)}$ defined by
\be\label{crit6def} Q\bigl(\Tc^{(q)},m\bigr)-Q\bigl(\Tc^{(q)},\mc^{(q)}\bigr)\underset{m\rightarrow \mc^{(q)^{\pm}}}{\sim} \tilde Q_{\pm}^{(q)}\bigl|m-\mc^{(q)}\bigr|^{\tilde q_{\pm}^{(q)}}\,\cdotp\ee
\end{enumerate}

\begin{figure}[h!]
\centering
\begin{tabular}{cc}
\def\svgwidth{7cm}\hskip -0.4cm
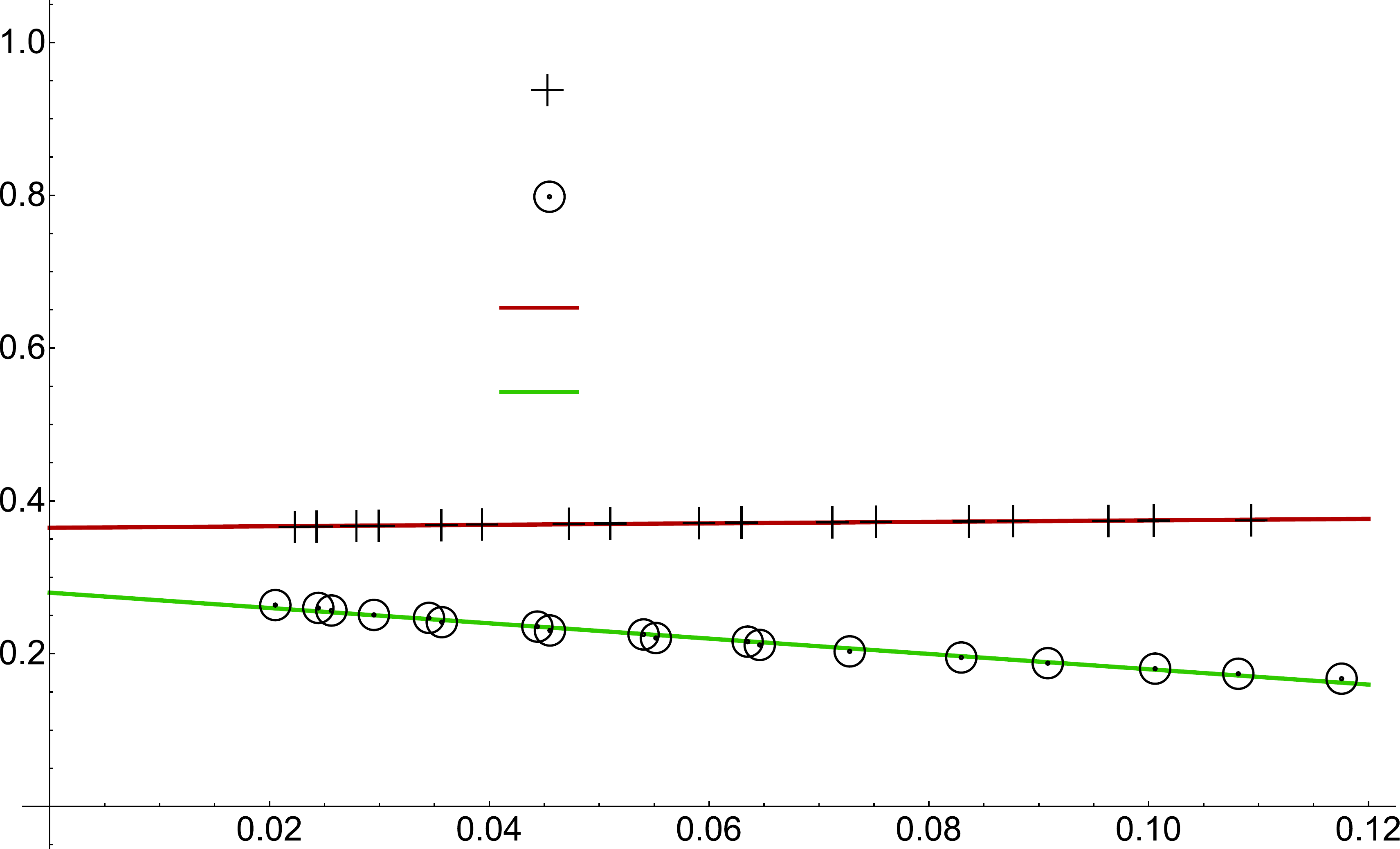 &\hskip 0.2cm
\def\svgwidth{7cm}
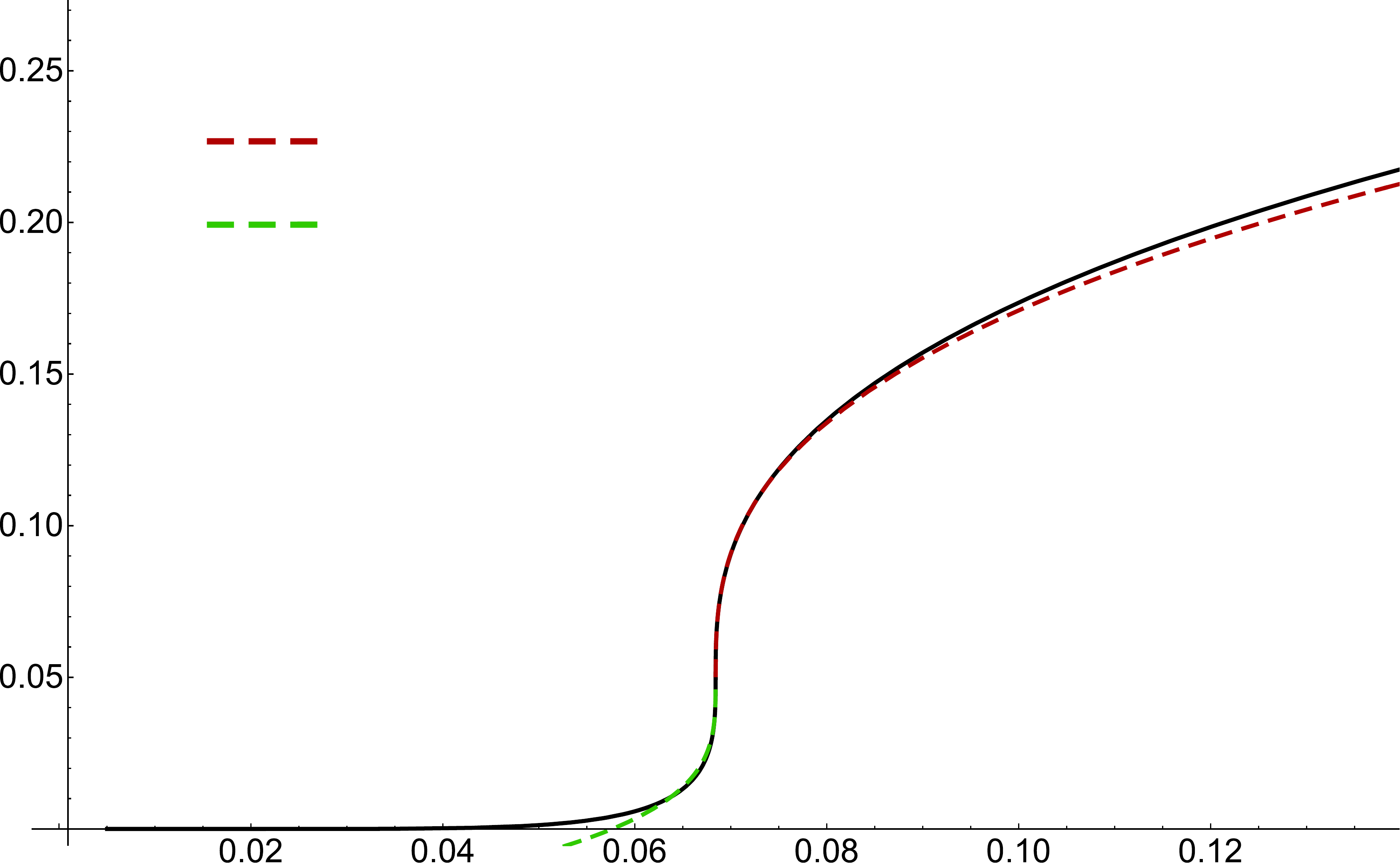 \\
(a) & (b)
\end{tabular}
\caption{\label{fig:CriticalExponent}(a) Plot of $q^{(4)}=\partial\ln |Q(T,\mc^{(4)})-Q(\Tc^{(4)},\mc^{(4)})|/\partial\ln |T-\Tc|$ as a function of $|T - \Tc|/\Tc$, for both $T>\Tc$ (crosses) and $T<\Tc$ (circles). The critical exponents $\smash{q_\pm^{(4)}}$ are defined by $\smash{q_{\pm}^{(4)}=\lim_{T\to\Tc^{\pm}}q^{(4)}(T)}$, implemented numerically through linear extrapolations (red and green, or dark and light gray lines for $T>\Tc$ and $T<\Tc$, respectively). (b) Charge $Q$ at $\smash{m=\mc^{(4)}}$, as a function of temperature, together with the best power-law fits near $\smash{T = \Tc^{(4)}}$ (red or dark gray dashed line for $T>\smash{\Tc^{(4)}}$, green or light gray dashed line for $T<\smash{\Tc^{(4)}}$).}
\end{figure}

To obtain all of these critical exponents we proceed in a way which is very similar to what we did before for the scaling dimension. For example, to extract $q_\pm^{(q)}$ we may plot
\be
q^{(q)}=\frac{\partial\ln |Q(T,\mc^{(q)})-Q(\Tc^{(q)},\mc^{(q)})|}{\partial\ln |T-\Tc^{(q)}|}
\ee
as a function of $|T - \Tc^{(q)}|$, for both $T>\Tc^{(q)}$ and $T<\Tc^{(q)}$, see Fig.~\ref{fig:CriticalExponent}a. We observe an approximately linear behaviour of $q^{(q)}$ within a fitting window where $T$ is close, but not too close, to the critical temperature $\smash{\Tc^{(q)}}$. This allows for a reliable extrapolation to $T\to\smash{\Tc^{(q)}}$, giving $\smash{q_+^{(q)}}$ and $\smash{q_-^{(q)}}$ when extrapolating $T>\smash{\Tc^{(q)}}$ and $T<\smash{\Tc^{(q)}}$, respectively. Having fixed the exponents $\smash{q_\pm^{(q)}}$ we can finally fit the $\smash{Q_\pm^{(q)}}$ coefficients in \eqref{crit3def} directly, again using an appropriate fitting window, as shown in Fig.~\ref{fig:CriticalExponent}b.

The example of this procedure illustrated in Fig.~\ref{fig:CriticalExponent} for $q = 4$ is entirely typical, with other exponents and $q$-generalizations behaving analogously. The results for $q=4,6,8,10,12$ are listed in Table \ref{tab:CriticalExponents}.
\begin{table}[h!]
\centering
\begin{tabular}{|c|c|c|c|c|c|}
\hline
 & $q = 4$ & $q = 6$ & $q = 8$ & $q = 10$ & $q = 12$\\
\hline
$\alpha_+^{(q)}$	& $0.68_2$ & $0.68_2$ & $0.67_2$ & $0.69_2$ & $0.69_2$\\
\hline
$\alpha_-^{(q)}$		& $0.66_2$ & $0.62_1$ & $0.59_1$ & $0.60_1$ & $0.59_1$\\
\hline
$\gamma_+^{(q)}$	& $0.51_1$ & $0.59_1$ & $0.54_1$ & $0.57_2$ & $0.57_1$\\
\hline
$\gamma_-^{(q)}$	& $0.80_1$ & $0.74_2$ & $0.86_2$ & $0.80_1$ & $0.83_1$\\
\hline
$q_+^{(q)}$			& $0.36_1$ & $0.36_1$ & $0.37_1$ & $0.36_1$ & $0.35_1$\\
\hline
$q_-^{(q)}$			& $0.28_2$ & $0.32_2$ & $0.36_1$ & $0.36_1$ & $0.36_1$\\
\hline
$\beta^{(q)}$		& $0.50_2$ & $0.52_2$ & $0.54_2$ & $0.53_2$ & $0.53_2$\\
\hline
$s_+^{(q)}$			& $0.28_2$ & $0.33_1$ & $0.37_1$ & $0.37_1$ & $0.37_1$\\
\hline
$s_-^{(q)}$			& $0.36_1$ & $0.34_1$ & $0.34_1$ & $0.33_1$ & $0.32_1$\\
\hline
$\tilde{q}_+^{(q)}$	& $0.22_2$ & $0.27_2$ & $0.30_1$ & $0.32_2$ & $0.32_2$\\
\hline
$\tilde{q}_-^{(q)}$	& $0.43_1$ & $0.40_1$ & $0.40_1$ & $0.38_1$ & $0.37_2$\\
\hline
\end{tabular}
\caption{\label{tab:CriticalExponents}Various critical exponents in the $q$-generalizations of the \modone\ model, for $q=4,6,8,10,12$. For the purposes of calculating these values, we have taken $\smash{(\mc^{(4)},\Tc^{(4)})} = (0.3445,0.0685)$, $\smash{(\mc^{(6)},\Tc^{(6)})} = (0.112,0.0323)$, $\smash{(\mc^{(8)},\Tc^{(8)})} = (0.0417,0.0147)$, $\smash{(\mc^{(10)},\Tc^{(10)})} = (0.0164,0.0066)$ and $\smash{(\mc^{(12)},\Tc^{(12)})} = (0.0068,0.0030)$. Reported error bars correspond to typical variations when adjusting the fitting window, see App.~\ref{NumericalErrorsApp}. Further variations can be expected when approaching the critical points more precisely, but general properties and trends should however be robust.
}
\end{table}

Let us emphasize three important properties of these results.

First, the critical exponents that we find are not mean-field. In other words, the critical points are non-trivial and strongly coupled. This may appear as a surprise, since the whole formalism in which the solution of the models is discussed is mean-field-like: all the thermodynamical properties are derived from the classical effective actions that we extremize, \eqref{Seffmat1} and its $q$-body generalization \eqref{Seffmatq}. The crucial difference with the standard mean-field theory is that these effective actions depend on \emph{functions} and not simply on a finite number of variables. We see that this functional version of mean-field theory yields a much richer theory of critical phenomena than Landau's, including the possibility for non-trivial critical exponents.

Second, the critical exponents are asymmetric, for instance $\alpha_{+}^{(q)} \neq \alpha_{-}^{(q)}$, $\tilde{q}_+^{(q)} \neq \tilde{q}_-^{(q)}$ \emph{etc}. This is an unusual property \cite{Leonard:2015wyg}. In ordinary local quantum field theory, standard renormalization group arguments imply scaling forms near the critical points which ensure that the critical exponents must be symmetric (the constants of proportionality, like $C_{\pm}^{(q)}$ in \eqref{Cdefexp}, may however be asymmetric). This general result does not contradict our findings, since the models we are dealing with here have no notion of space or locality.

Third, the critical exponents are $q$-dependent, especially for lower values of $q$. We thus obtain an infinite series of inequivalent critical points. We shall argue in the next subsection that these points actually belong to a continuous line of inequivalent critical points. 

Clearly, we are here just scratching the surface of a very non-trivial new phenomenon in large $\dof$ quantum mechanical models. The critical points we are dealing with remain to a large extent very mysterious, and further study is needed to better understand their properties. In particular, we do not know if some version of the renormalization group formalism can be applied in the present context.

\subsubsection{\label{Pha1continuousexpSec}Multi-$q$ models, domain walls and a line of critical points}

When turning on several couplings $\la_{q}$ at the same time, one expects to find an even richer structure of critical and possibly multi-critical points in the corresponding phase diagrams. A detailed study is outside the scope of the present work, but we want to display here two interesting features.

\paragraph{RG flow and domain walls}

One way to visualize the renormalization group flow from the asymptotically free regime $\omega \gg \lambda_{q}$ to the conformal region $\omega \ll \lambda_{q}$ at low masses is to plot in log-log scale the imaginary part of the Matsubara coefficients $\im G_{k}$ as a function of the Matsubara frequencies $\nu_k$.

\begin{figure}[h!]
\centering
\vskip 0.3cm
\begin{tabular}{cc}
\def\svgwidth{7cm}
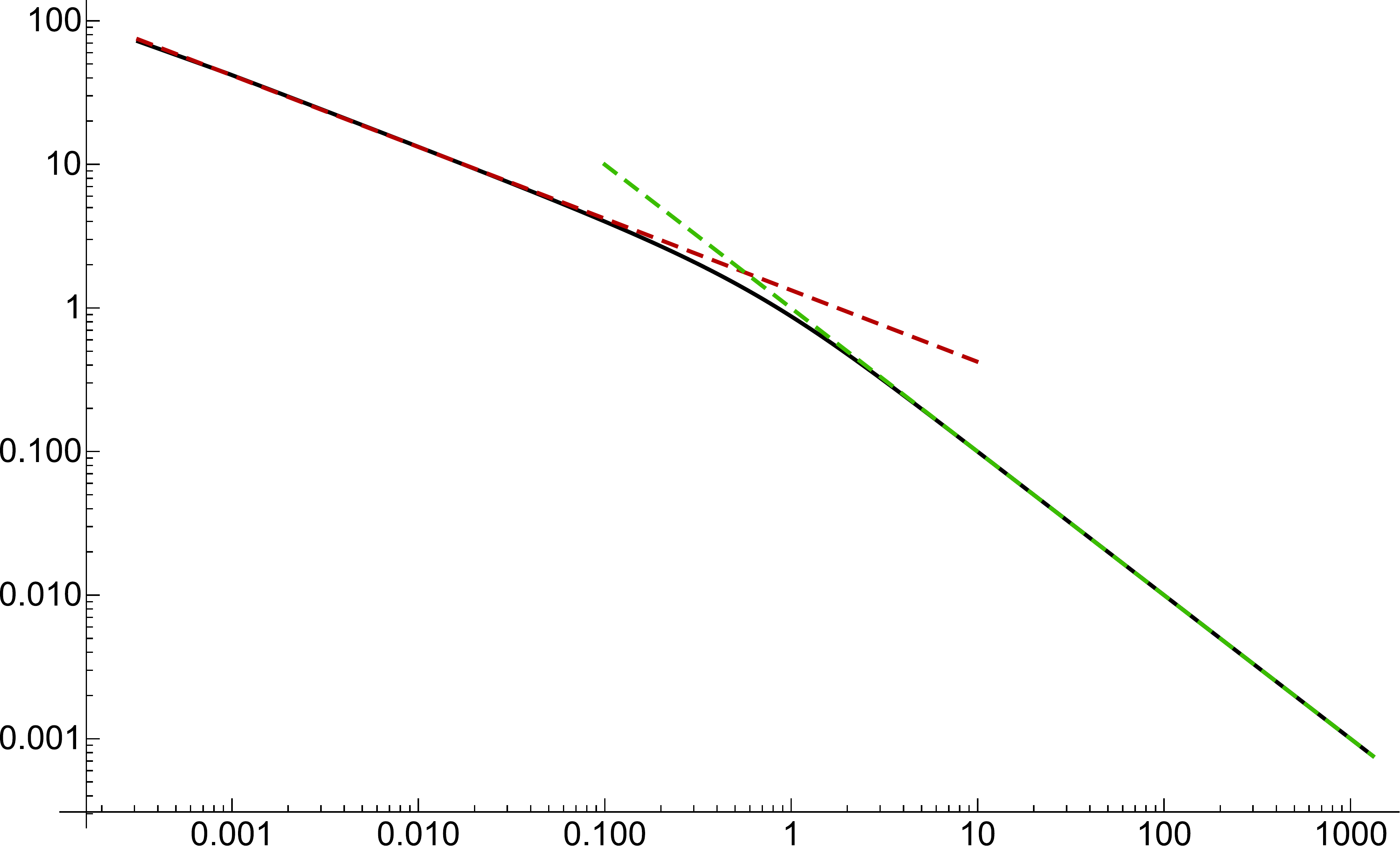 &
\def\svgwidth{7cm}
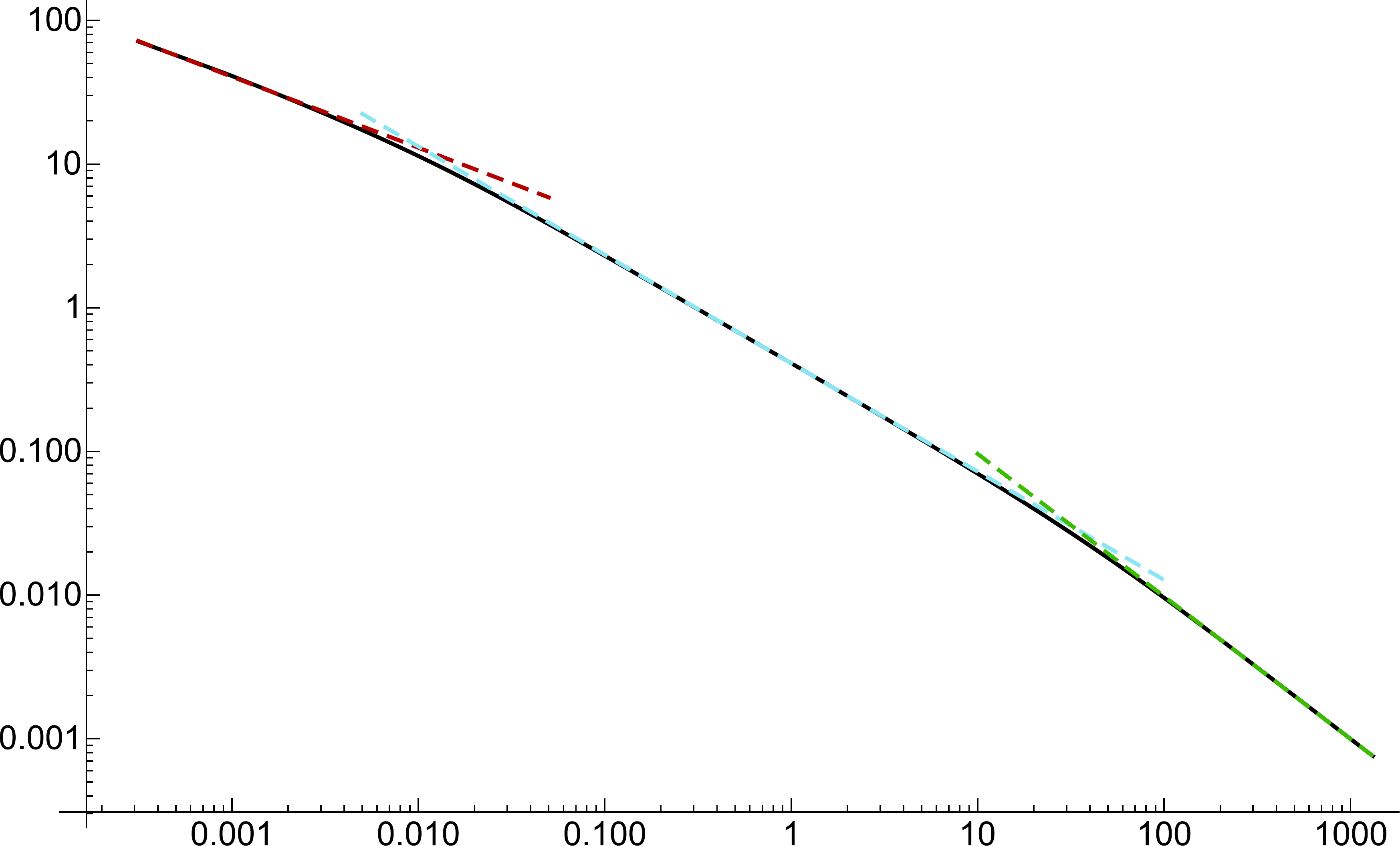 \\
(a) & (b)
\end{tabular}
\caption[RGFlowExample]{\label{fig:RGFlow}Log-log plot of $\im G(\omega)$ as a function of the frequency $\omega$ at $T=10^{-4}$ and $m=0$ in the \modone\ model (a) and in the model with both $\la_{4}=1$ and $\la_{8}=200$ turned on (b). The RG flow from the asymptotically free region at high frequencies ($-1$ slope, green or gray) to the conformal behaviour dominated by the most relevant quartic coupling at low frequencies ($-1/2$ slope, red or dark gray) is clearly visible in both cases. In (b) the transition occurs through an intermediate regime dominated by the $q=8$ interaction term ($-0.75$ slope, blue or light gray).\footnotemark}
\end{figure}

This is illustrated in Fig.~\ref{fig:RGFlow}a for the standard case of the \modone\ model, where only $\la_{4}=1$ is turned on and we take $m = 0$ for simplicity. The structure of the RG flow is then clearly visible: we have at large frequencies a $-1$ slope characteristic of the asymptotic regime \eqref{GkUV}, $G_{k}\sim i/\nu_{k}$; and a slope of $2\Delta -1=-1/2$ determined by \eqref{FourierGle} in the conformal region at low frequencies. The transition between the two regimes occurs at frequencies of order $\omega\sim\la_4 = 1$ and is very visible at the small but finite temperature $T = 10^{-4}$ used here.

Now consider the case shown in Fig.~\ref{fig:RGFlow}b, where an additional coupling, chosen to be $\la_{8}=200$ for illustrative purposes, is turned on. There is a deep-UV region $\nu_{k} \gg \la_{8}$ again governed by asymptotic freedom, as well as a deep-IR region $\nu_{k} \ll \la_{4}$ governed by the conformal behaviour associated with the most relevant coupling $\la_{4}$. Moreover, between them now lies an intermediate region $\la_{4}\ll\nu_{k}\ll\la_{8}$, which is approximately conformal with a slope $2\Delta - 1\simeq -0.75$, corresponding to the dimension associated with the $q=8$ interaction term, $\Delta\simeq 1/8$.

Note that we are using a model with $\la_{4}$ and $\la_{8}$ turned on, instead of, for example, $\la_{4}$ and $\la_{6}$. This is because the larger difference in the associated exponents $\Delta$ makes the intermediate approximately conformal region easier to visualize when one works at small yet finite temperature and ratio $\la_{4}/\la_{8}$, a limitation inherent to the numerical approach. More generally, one may turn on many couplings $\la_{q_{1}},\ldots,\la_{q_{r}}$, organized hierarchically with $q_{i-1}< q_{i}$ and $\la_{q_{i-1}}\ll\la_{q_{i}}$. In the frequency regime $\la_{q_{i-1}}\ll\omega\ll\la_{q_{i}}$ the model is then in an approximate conformal phase dominated by the coupling $\la_{q_{i}}$ and characterized by an exponent $\Delta = 1/q_{i}$. In a holographic picture, along the lines of \cite{holomod}, this behaviour should be mapped to interesting domain wall solutions interpolating between different $\text{AdS}_{2}$ spaces. To the best of our knowledge, these solutions have not yet been discussed in the literature. \footnotetext{Both plots in Fig.~\ref{fig:RGFlow} were made using solutions computed with a weighting factor $\alpha=0.1$ and $\maxk=2^{21}$, \emph{i.e.}\ approximately $4\times10^6$ frequencies.}

\paragraph{A line of inequivalent critical points}

\begin{figure}[h!]
\centering
\def\svgwidth{4.5in}
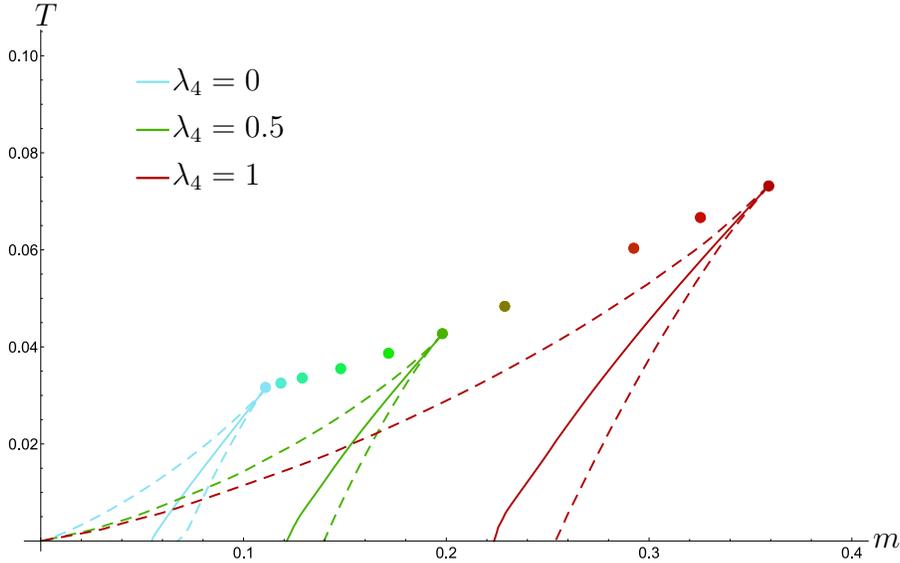
\caption{\label{fig:CriticalLine}Phase diagram of the model with two couplings $\la_{4}$ and $\la_{6}=1$ turned on, for three different values of $\la_{4}$. The critical points for $\lambda_4 = 0$, $0.1$, $0.2$, $0.3$, $0.4$, $0.5$, $0.6$, $0.7$, $0.8$, $0.9$, $1$ are marked with colored circles from blue (light) to red (dark gray), illustrating the continuous nature of the phase diagram dependence on the $\lambda_4$ coupling.}
\end{figure}
\begin{figure}[h!]
\centering
\vskip 0.5cm
\def\svgwidth{4.5in}
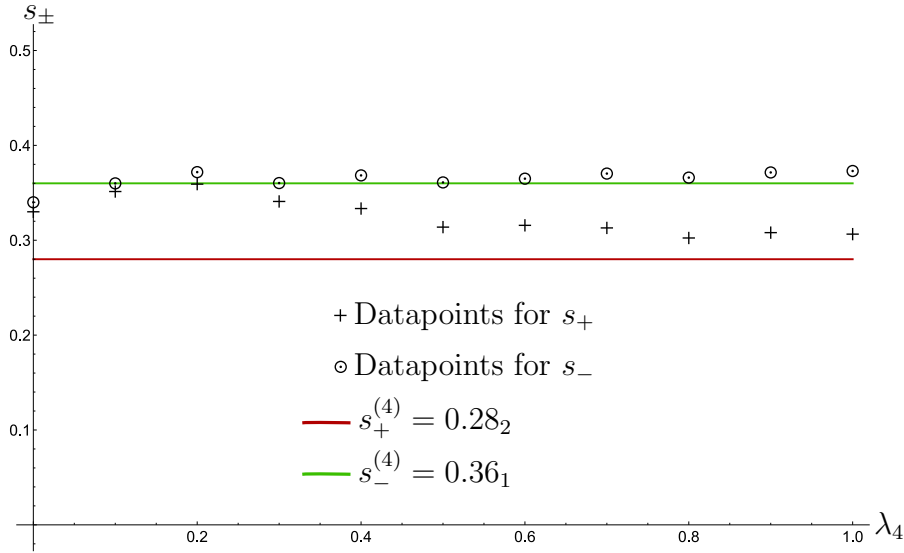
\caption{\label{fig:DoubleCriticalExponents}Critical exponents $s_{\pm}$ for the entropy as a function of mass at $T = \Tc$, for various values of $\la_4$ in the model where both $\la_4$ and $\la_6=1$ are turned on. The continuous variation of the exponents implies that we have a line of inequivalent critical points with marginal coupling $\la_4$. Solid lines mark the values corresponding to $\lambda_4 \gg \lambda_6$ (red or dark gray for $s_+$, green or light gray for $s_-$).}
\end{figure}

In multi-$q$ models we expect that the phase diagram smoothly depends on the various couplings, so that we have in fact critical lines, surfaces or hyper-surfaces when considering models with two, three or more interaction terms, respectively.

One of the simplest cases to analyze corresponds to the model in which both $\la_4$ and $\la_6$ are turned on, with $\lambda_6 = 1$ setting the units. Its phase diagram is depicted in Fig.~\ref{fig:CriticalLine}, projecting different values of $\la_4$ to the $(m,T)$-plane. We see that there is a critical line interpolating between the critical point of the $q=6$ model when $\la_4=0$ and the critical point of the $q=4$ model when $\la_{4}\gg\la_6$. This is nicely confirmed by our analysis of the critical exponents. For example, in Fig.~\ref{fig:DoubleCriticalExponents} we show the entropy exponents $s_{\pm}$, computed as explained for $q_\pm$ in Fig.~\ref{fig:CriticalExponent}, as a function of $\la_4$. We observe a continuous variation between the values $\smash{s_+^{(6)}=0.33_1}$, $\smash{s_-^{(6)}=0.34_1}$ and $\smash{s_+^{(4)}=0.28_2}$, $\smash{s_-^{(4)}=0.36_1}$ (see Table \ref{tab:CriticalExponents}), with the gap $s_- - s_+$ smoothly widening as $\lambda_4$ becomes larger. This provides convincing evidence that we have a continuous line of inequivalent critical points, for which the coupling $\la_4$ is exactly marginal. 

\subsection{\label{PhaseMod2Sec}Phase diagram of the \modtwo\ model}
\begin{figure}[h!]
\centering
\def\svgwidth{4.5in}
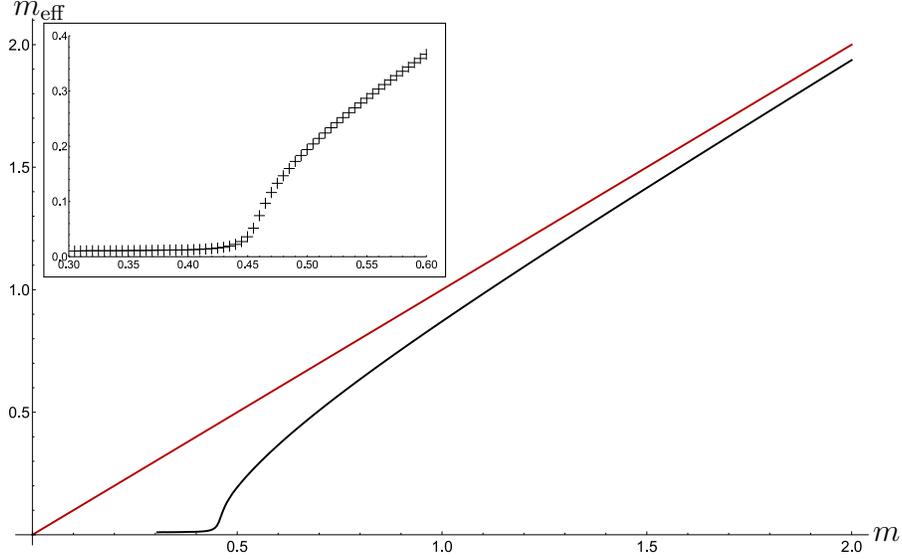
\caption{\label{fig:M2MassGap}Zero-temperature extrapolation of the mass gap in the \modtwo\ model (black) as a function of the mass parameter $m$. Numerical values were computed fitting the log-plots of finite-temperature Euclidean two-point functions, linearly extrapolating results for $\sim10$ points in the range $\beta \in [50, 100]$. The mass gap decreases from its asymptotic value $\meff\simeq m$ at large mass (red or dark gray) to zero for a finite critical value $m = m_{*} = \mStarTwo$ (detail shown in inset, with numerical values represented with crosses).}
\end{figure}

Unlike the case of the \modone\ model that we have discussed in the previous Section, the thermodynamics of the \modtwo\ model is governed by a unique solution to the Schwinger-Dyson equations that can be analytically continued for all values of $m$ and $T$ without undergoing monodromies. In particular, we do not find first order phase transitions or non-trivial critical points at finite temperature in the \modtwo\ model. All the interesting physics occurs at zero, or near zero, temperature. As discussed in Sec.~\ref{phasemotivSec}, this important difference with the \modone\ model is made possible by the fact that the zero-temperature Euclidean two-point function gets non-trivial corrections even at relatively large masses.

More precisely, at large enough mass the system is weakly coupled. The ground state must then coincide with the Fock vacuum $|0\rangle$ defined by the conditions $\psi_{\mu}|0\rangle = 0$, with a mass gap $\meff$ of order $m$. The spectral function \eqref{specdef} then has a support only for $\omega\geq\meff>0$ at zero temperature, and the spectral decomposition formula \eqref{Gspecdec} implies that
\begin{align}\label{Gmod2zeroT1} G(\tau;T=0) &= 0 && \hspace{-2cm}\text{for }\tau<0\,,\\
\label{Gmod2zeroT2} G(\tau;T=0) &\propto e^{-\meff\tau} && \hspace{-2cm}\text{for }\tau\rightarrow +\infty\,.
\end{align}
In Fig.~\ref{fig:M2MassGap} we plot the mass gap $\meff$ as a function of the mass parameter $m$, evaluated numerically by matching finite-temperature solutions with the exponential behaviour \eqref{Gmod2zeroT2} and then extrapolating $\meff$ to $T\to0$. The essential feature to note here is that the gap closes for a finite value $m = m_{*} = \mStarTwo$.

\begin{figure}[h!]
\centering
\def\svgwidth{4.5in}
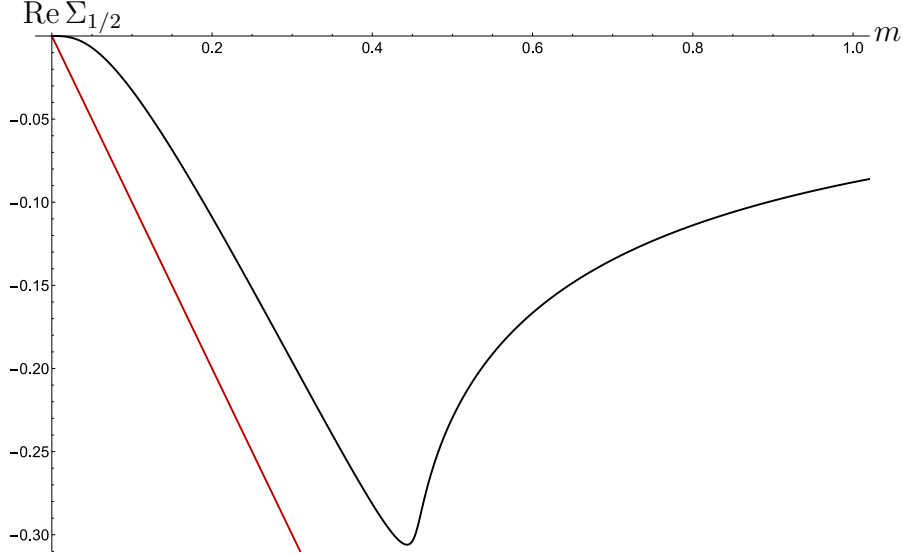
\caption{\label{fig:M2SigmaZero}Plot of $\re\Sigma_{1/2} \simeq \tilde\Sigma(0)$ as a function of the mass (black), for the \modtwo\ model at $T = 0.01$ in natural units. We observe a clear crossover from a gapless phase, which must have $\tilde\Sigma(0) = m$ at zero temperature (red or dark gray), as in \eqref{Sigtildeconst}, to a gapped phase, for which $\tilde\Sigma(0) \simeq 0$ at very large masses. The transition occurs at $m = m_* = \mStarTwo$, consistently with the appearance of a non-vanishing effective mass, see Fig.~\ref{fig:M2MassGap}.}
\end{figure}
\begin{figure}[t!]
\centering
\begin{tabular}{cc}
\def\svgwidth{7cm}
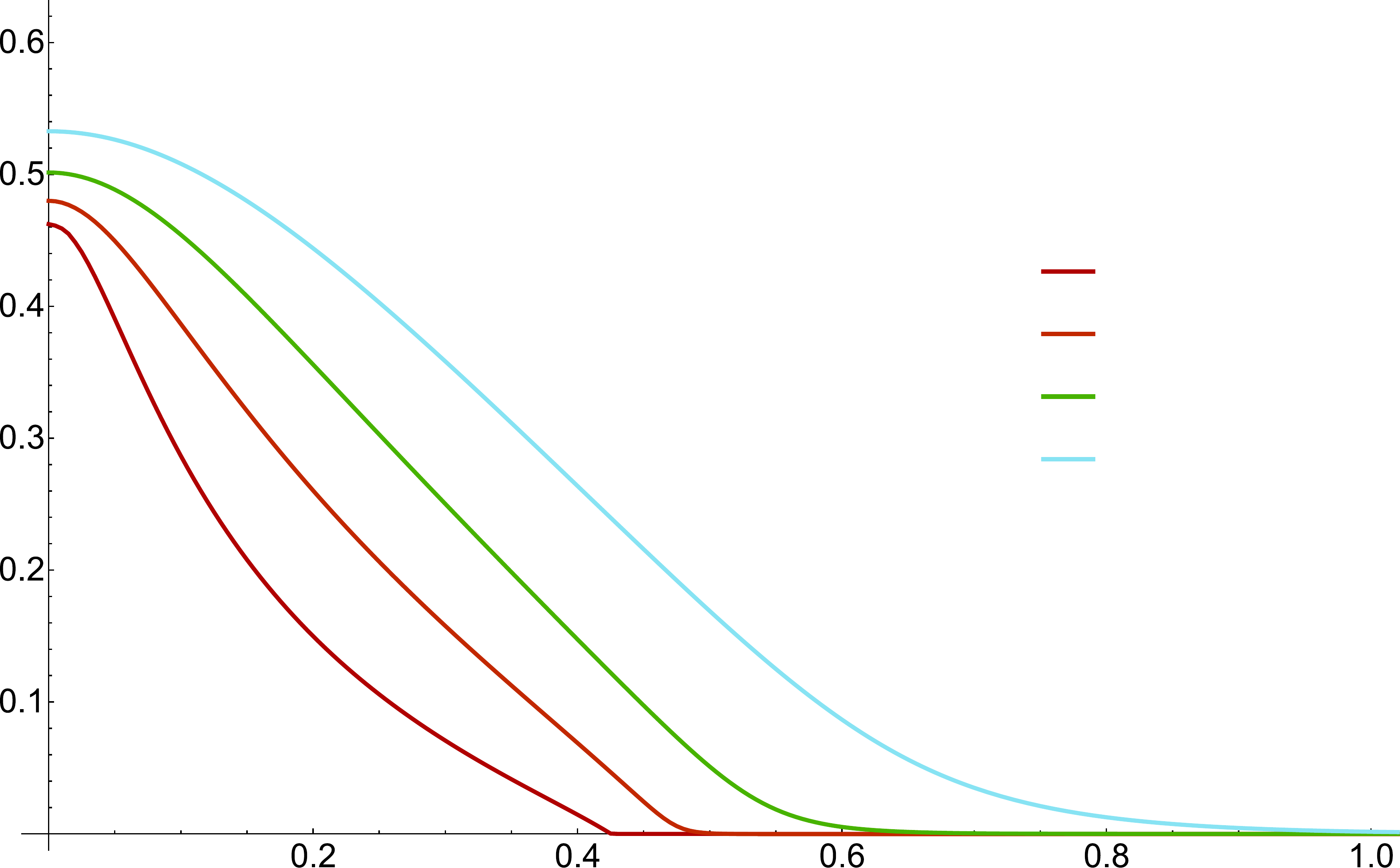 &
\def\svgwidth{7cm}
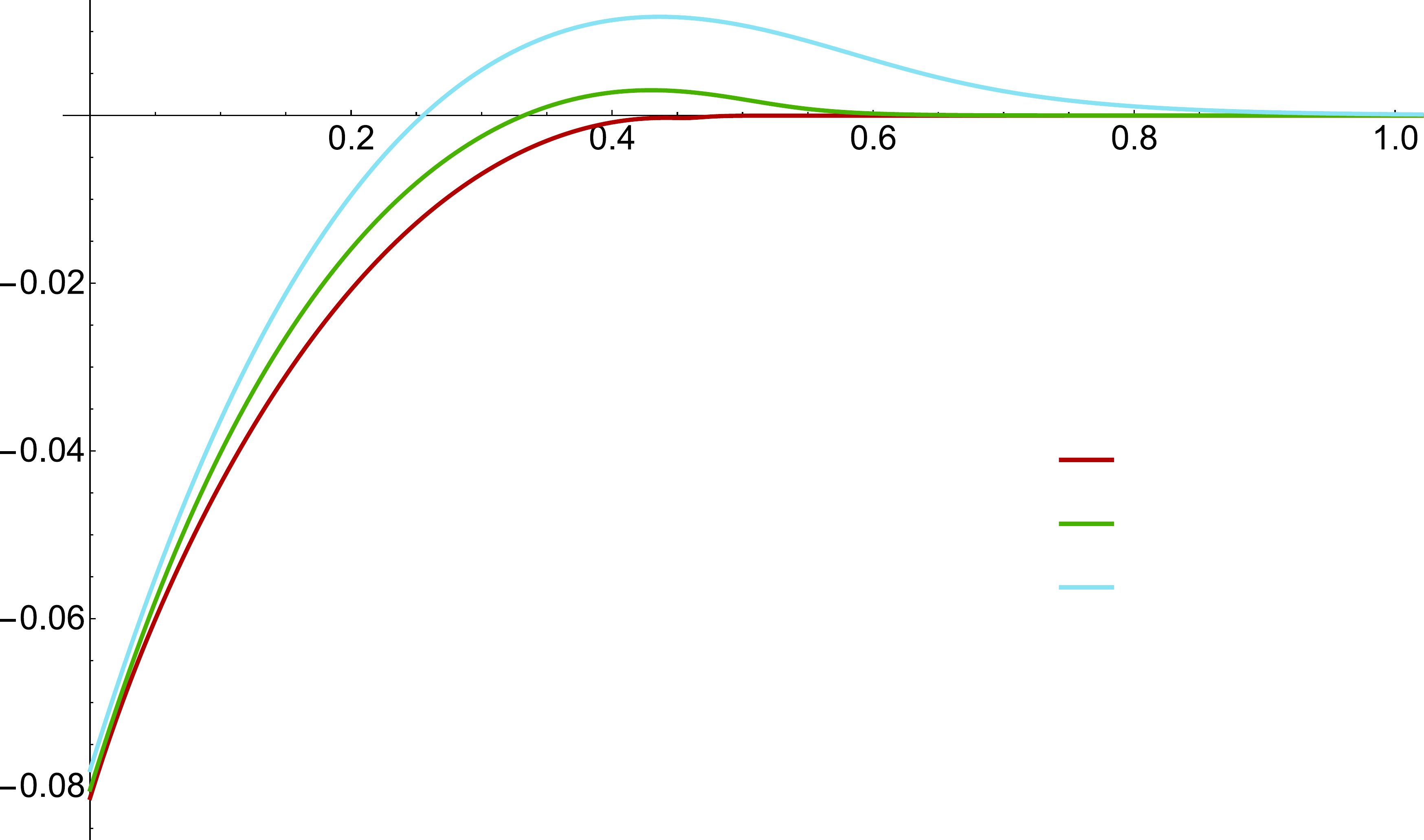 \\
(a) & (b)\\[0.5cm]
\multicolumn{2}{c}{\def\svgwidth{7cm}
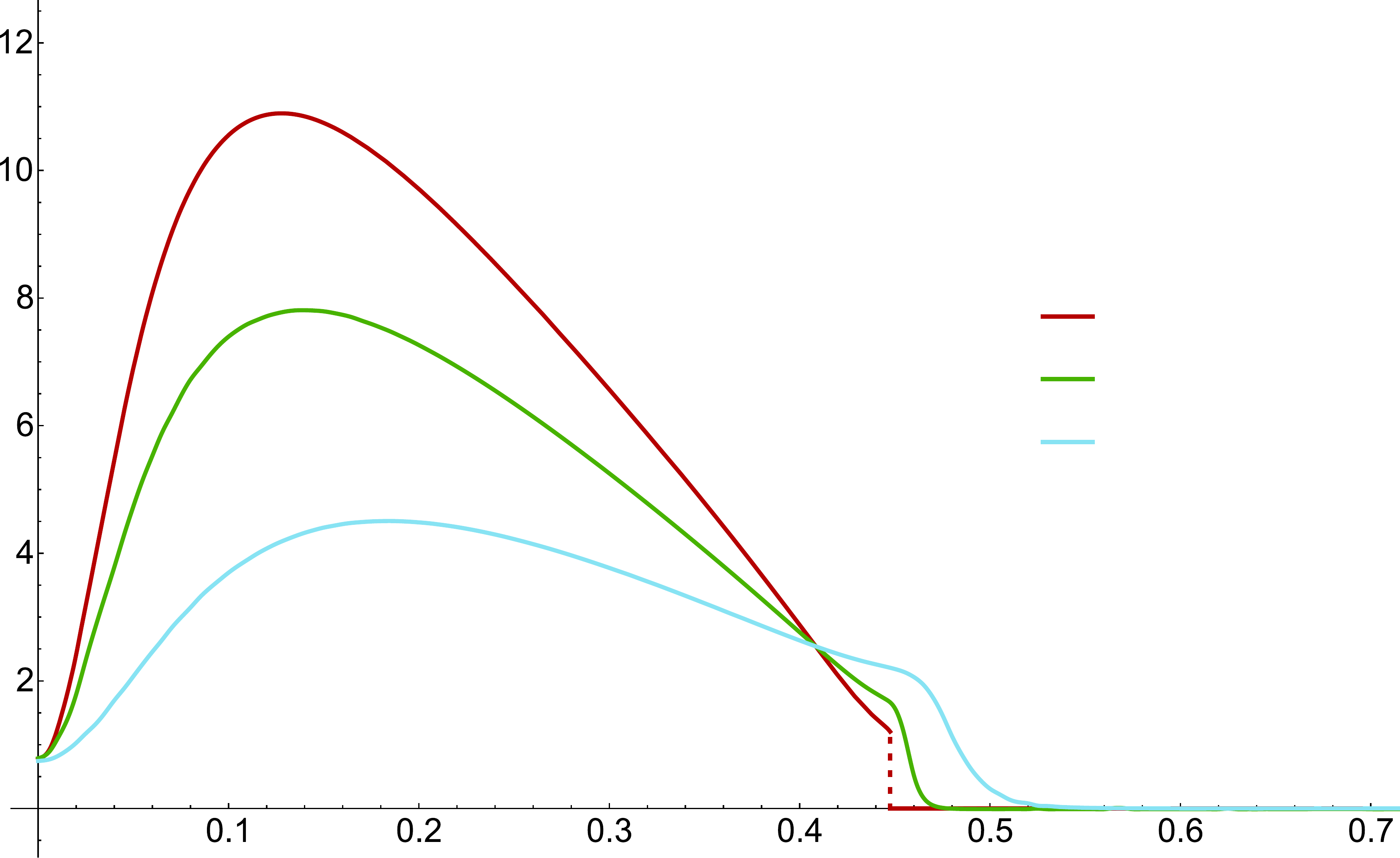}\\
\multicolumn{2}{c}{(c)}
\end{tabular}
\caption{\label{fig:M2Thermo}Plots of the entropy $S/\dof$ (a), energy $E/\dof$ (b) and heat capacity coefficient $\tfrac{C}{\dof T} = \tfrac{1}{\dof} \tfrac{\partial S}{\partial T}$ (c), as functions of the mass in the \modtwo\ model. $T\to0$ extrapolations were performed using linear fits of $\sim10$ finite-temperature solutions with $\beta \in [50, 100]$. At zero temperature all quantities become non-trivial below the quantum critical point $m=m_{*} = \mStarTwo$, where the heat capacity coefficient experiences a finite jump (red or dark gray dotted line).}
\end{figure}

A complementary point of view is provided by the constraint \eqref{Sigtildeconst}. In Fig.~\ref{fig:M2SigmaZero} we plot $\re\Sigma_{1/2}$ as a function of the mass at $T = 0.01$ in natural units. We observe here a clear crossover between the regime at low masses approximately satisfying \eqref{Sigtildeconst}, and the regime at large masses for which $\tilde\Sigma(0) \simeq 0$. The turning point is at $m=m_{*}=\mStarTwo$, perfectly consistent with the appearance of the effective mass discussed above.

The point $m=m_{*}$ is a new kind of large $\dof$, strongly-coupled quantum critical point.  For $m<m_{*}$ the zero-temperature Euclidean two-point function behaves as \eqref{Gasympnew}, with asymmetric exponents $\Delta_{+}\not = \Delta_{-}$. This feature was already discussed in detail in Sec.\ \ref{EucSolmod2Sec}, and is in sharp contrast with the IR behaviour of the two-point function in the \modone\ model. As pointed out in Sec.~\ref{NewansatzSec}, the $\slR$ subgroup of the low energy reparameterization symmetry \eqref{diffeoG1}-\eqref{diffeoS1}, including the scale transformations $\tau\mapsto\alpha\tau$ for $\alpha>0$, is now spontaneously broken, even though the phase is gapless. It is thus a phenomenon of spontaneous scale symmetry breaking without the generation of a mass gap.

Plots of the energy, entropy and heat capacity divided by $T$ are provided in Fig.~\ref{fig:M2Thermo}. At zero temperature these quantities are all vanishing for $m>m_{*}$, but become non-zero for $m<m_*$, very much like in the \modone\ model (compare with Fig.~\ref{fig:M1ZeroTThermo}). The low-temperature properties of the \modone\ and \modtwo\ models are thus quite similar as far as the thermodynamical potentials are concerned. For example, they both have non-vanishing zero-temperature entropy and a heat capacity vanishing linearly with the temperature in the gapless phase $m < m_*$. A basic difference with the \modone\ model is that the entropy is now continuous. Its derivative is discontinuous, however, and this translates into a finite jump for $C/T$ at $m = m_*$.

We conclude this Section with a few comments on some subtleties, both conceptual and numerical, associated with the new IR behaviour found in the \modtwo\ model for $0<m<m_{*}$.

We have argued that the deep IR behaviour changes discontinuously from the standard SYK behaviour at $m=0$, with symmetric exponents $\Delta_{+}=\Delta_{-}=1/4$, to the new asymmetric behaviour with $\Delta_{+}\not = \Delta_{-}$ for any $m>0$, even if arbitrarily small. In this sense, the mass operator is relevant at $m=0$ (and marginal for all $0<m<m_{*}$), even though no mass gap is created: as soon as it is turned on, the IR fixed point of the model jumps from one gapless solution to another very different gapless solution. The fact that the jump must be discontinuous can be understood \emph{e.g.}\ as a consequence of the formulas \eqref{model2sol1} and \eqref{model2sol2} for $b_{+}^{4}$, which show that this coefficient is discontinuous when one goes from $\Delta_{+}=\Delta_{-}$ to $\Delta_{+}< \Delta_{-}$, even if the difference between $\Delta_{+}$ and $\Delta_{-}$ is very small. 

This discontinuity at $m=0$ contrasts with the fact that all the thermodynamical quantities are found to be perfectly continuous when $m\rightarrow 0$, including at $T=0$, as shown in Fig.~\ref{fig:M2Thermo}. At $m=0$, they match with the results for the \modone\ model, which is essentially just SYK. But there is no contradiction here; the discontinuous change in the IR behaviour occurs only for a range of frequencies $|\omega|\ll m$ that shrinks to zero size when $m\rightarrow 0$. At very small $m\ll|\xi|$, the model thus has an interesting RG flow: from the asymptotically free regime at frequencies $\omega\gg|\xi|$, it flows to an approximate conformal phase \emph{\`a la} SYK with $\Delta=1/4$ in the regime $m\ll\omega\ll |\xi|$ before eventually flowing to the non-standard deep IR behaviour with $\Delta_{+}<\Delta_{-}$ when $\omega\ll m$.

The fact that the correct IR behaviour can emerge only for $\omega\ll m$ implies that the numerical characterization of the deep IR regime at small mass is particularly difficult. To do this, one should probe temperatures satisfying $\beta m\gg1$ as well as $\beta|\xi|\gg1$, but the former condition is much stronger than the latter in the regime $m\ll|\xi|$.

\begin{figure}[h!]
\centering
\def\svgwidth{4.5in}
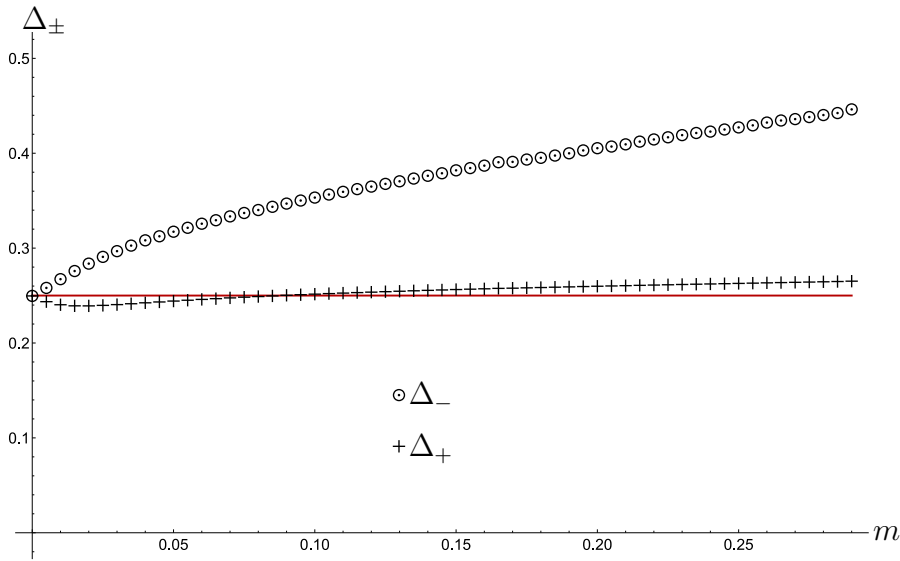
\caption{\label{fig:M2MassDelta}Numerical scaling dimensions $\Delta_\pm$ of the \modtwo\ model as functions of the mass, for small and intermediate values of $m$. Each of the datapoints for $\Delta_+$ (crosses) and $\Delta_-$ (circles) was computed by linearly extrapolating results obtained from fitting finite-temperature two-point functions, as exemplified in Fig.~\ref{fig:M2DeltaBpmExtrapolationa}. While $\Delta_+$ remains close to the value $1/4$ (red or dark gray line), as required by the generalized ansatz \eqref{model2sol2}, the numerical evaluation of $\Delta_-$ exhibits mass dependence and quickly departs from this value, in clear violation of the standard ansatz \eqref{model2sol1}. See the main text for a detailed discussion.}
\end{figure}

To illustrate this difficulty, we plot in Fig.~\ref{fig:M2MassDelta} the exponents $\Delta_{+}$ and $\Delta_{-}$ for various values of $m$, evaluated numerically by following exactly the same procedure as in Fig.\ \ref{fig:M2DeltaBpmExtrapolationa} of Sec.\ \ref{EucSolmod2Sec}. We see that $\Delta_+$ stays reasonably close to the value $\Delta_+ = 1/4$, as required by \eqref{model2sol2}, with small variations that may be attributed to numerical artifacts surviving the $T\to 0$ extrapolation. This is not surprising, since the value $\Delta_+ = 1/4$ coincides for both the standard SYK conformal phase and the new non-standard phase with $\Delta_{+}<\Delta_{-}$. A precise determination of $\Delta_{+}$ is thus expected as soon as $\beta|\xi|\gg1$, even if $m\ll|\xi|$. The situation is very different for $\Delta_{-}$. The numerical values we find exhibit a marked mass dependence, starting from $\Delta_{-}=0.250_1$ at $m=0$ and increasing up to $\Delta_{-}=0.446_2$ at $m=0.29$.

The fact that the numerical result for $\Delta_{-}$ varies smoothly from its $m=0$ value is an automatic consequence of the discussion above: at extremely small masses the lowest temperature we can study numerically will always be in the near-SYK regime $1/|\xi|\ll\beta\ll1/m$ and the genuine deep IR regime $\beta\gg 1/m$ is not probed. It is therefore not possible to decide, based on the numerical analysis alone, whether the actual value of $\Delta_{-}$ in the phase $\Delta_{+}\not = \Delta_{-}$ does depend continuously on $m$ or not. Two possibilities thus remain open:
\begin{enumerate}[i)]
	\item The deep IR value of $\Delta_{-}$ depends smoothly on $m$ and our numerical analysis can actually provide a reasonable approximation to the actual value, possibly even for small masses.
	\item The deep IR value of $\Delta_{-}>\Delta_{+}$ is fixed and does not depend on $m$, but our numerical analysis cannot reliably determine it for small $m$. The mass dependence of $\Delta_{-}$ in Fig.\ \ref{fig:M2MassDelta} may then be interpreted as mimicking a RG flow, since the deep IR is probed better and better when $m$ increases (as explained below, this is no longer true when the mass is too close to $m_{*}$).
\end{enumerate}
Note that, in either case, the fact that $\Delta_+<\Delta_-$ when $m>0$ is unambiguously established. As will be explained elsewhere \cite{Ferranew}, the possibility ii) seems to be strongly favoured.

Finally, let us mention that obtaining numerical solutions at very low temperatures becomes increasingly harder as the mass approaches the critical value $m_* = \mStarTwo$, which explains why the highest mass appearing in Fig.\ \ref{fig:M2MassDelta} is $m=0.29$. Indeed, the solutions near the quantum critical point are quite subtle. They must be very near the low entropy, perturbative-like solutions above some effective mass scale that goes to zero when $m\rightarrow m_{*}$ from below, but at the same time should reproduce the non-standard IR behaviour with $\Delta_{+}<\Delta_{-}$ below this small effective mass scale.

\section{\label{RealTimeSec}Real time physics}
\subsection{\label{rttwoptSec}Real time two-point functions}

The goal of this subsection is to explain in detail how to compute real time two-point functions. After recalling standard definitions and results in Sec.~\ref{defrealtimeSec}, we present a simple derivation of the real time version of the Schwinger-Dyson equations in Sec.~\ref{realSDSec}. The argument can be used in any melonic quantum mechanics. The resulting equations are the real time counterparts of the Euclidean equations \eqref{Sigmadef}, \eqref{SDone}, \eqref{SDtwo}. They can be solved numerically using a method outlined in Sec.~\ref{realnum2ptSec}. Note that our results generalize and are consistent with known formulas, presented for example in \cite{Maldacena:2016hyu} for the SYK model; the only true novelty here may be the rigorous justification of the derivations, especially in Sec.~\ref{realSDSec}, and the pedagogical presentation.

Real time two-point functions can be used to compute quasi-normal frequencies. The results are presented in Sec.~\ref{quasifreqSec}. We find a very interesting behaviour near the finite temperature critical point in the \modone\ model. Real time two-point functions will also serve as a crucial building block in the analysis of the four-point functions in Sec.~\ref{real4ptSec}, allowing the calculation of Lyapunov exponents in the following subsections.

\subsubsection{\label{defrealtimeSec}General definitions and properties}

For any operator $\mathscr O$ and complex ``time'' $u$ we define
\be\label{cplxetimeevol}
\mathscr O(u) = e^{H u}\mathscr O e^{-Hu}\,,
\ee
denoting
\be\label{udectaut} \tau_{u}=\re u\, ,\quad t_{u}=\im u\, .\ee
If $u=\tau_{u}$ is real \eqref{cplxetimeevol} is equivalent to the Euclidean time evolution \eqref{Euctimeevolution}. If $u=it_{u}$ is purely imaginary, on the other hand, \eqref{cplxetimeevol} yields the standard real time Heisenberg evolution. We can then define a generalized two-point function for complex time arguments $u$ and $v$ satisfying $0<|\tau_{u}-\tau_{v}|<\beta$,
\be\label{Gcplxedef} G(u,v) =
\begin{cases}
\frac{1}{\N}\bigl\langle\text{T}\tr\psi_{\mu}(u)\psi_{\mu}^{\dagger}(v)\bigr\rangle_{\beta} & \text{(matrix models)}\\
\bigl\langle\text{T}\chi^{i}(u)\chi_{i}^{\dagger}(v)\bigr\rangle_{\beta} & \text{(disordered models)}\, ,
\end{cases}
\ee
where T indicates an ordering with respect to the real parts of $u$ and $v$, \emph{i.e.}\ with respect to the Euclidean time. This is an important point: it is the Euclidean time that always dictates the ordering of operators in a correlator. Clearly $G(u,v)$ depends on $u-v$ only, and we shall thus also note $G(u,v)=G(u-v)$. The KMS condition
\be\label{KMS2} G(u+\beta)=-G(u)\ee
can be used to extend $G(u)$ to all values of $\tau_{u}$. It is sometimes convenient to decompose the generalized two-point function in terms of Wightman correlators,
\be\label{GdecGpGmdef} G(u) = \Theta(\tau_{u})G_{+}(u) + \Theta(-\tau_{u})G_{-}(u)\, ,\ee
which are defined by\footnote{We stick to the matrix model notation from now on.}
\be\label{GpGmdef} G_{+}(u) = \tfrac1\N \bigl\langle\tr\psi_{\mu}(u)\psi_{\mu}^{\dagger}\bigr\rangle_{\beta}\, \quad\text{and}\quad G_{-}(u) = -\tfrac1\N \bigl\langle\tr\psi_{\mu}^{\dagger}\psi_{\mu}(u)\bigr\rangle_{\beta}\, .\ee

At tree level,
\be\label{Gtreecplxe}
G^{(0)}(u;m) =
\begin{cases} 
\nF(m)e^{m(\beta - u)} &\quad\text{if}\quad 0<\tau_{u}<\beta\\
-\nF(m)e^{-m u} &\quad\text{if}\quad -\beta<\tau_{u}<0\, ,
\end{cases}
\ee
generalizing \eqref{Gtree}. The Euclidean spectral decomposition \eqref{Gspecdec} is generalized to
\be\label{Gcplxespecdec} G(u) = \int_{-\infty}^{+\infty}G^{(0)}(u;\omega)\rho(\omega)\,\d\omega\, ,\ee
where the spectral function $\rho$ is defined in \eqref{specdef}. We can also use the Fourier decomposition
\be\label{Fouriercplxe} G(\tau + i t ) = \frac{1}{2\pi\beta}\sum_{k\in\mathbb Z+\frac{1}{2}}\int_{-\infty}^{+\infty}\tilde G_{k}(\omega)\,e^{-i\omega t-i\nu_{k}t}\, \d\omega\, ,\ee
with
\be\label{tildeGdef} \tilde G_{k}(\omega) = \frac{2\pi\rho(\omega)}{\omega - i\nu_{k}}\,\cdotp\ee
It is important to note that the correct analytic continuation from Euclidean time to complex time cannot be obtained by replacing $\tau \mapsto \tau +it$ in the Euclidean Fourier expansion \eqref{GFourier}. The analytic continuation to complex time is simple in correlators for which the operator ordering is fixed, because the complex time operator evolution \eqref{cplxetimeevol} is a simple analytic continuation of the Euclidean time evolution \eqref{Euctimeevolution}, but is non-trivial in time-ordered correlators.

Purely real time correlators are obtained from the complex time correlators by letting $\tau$ go to zero, the ordering of operators being fixed by the way the limit is taken, \emph{i.e.}\ $\tau\to0^+$ or $\tau\to0^-$. Eq.\ \eqref{Gcplxespecdec} implies that all the real time two-point functions, for any ordering of the operators, can be expressed in terms of the spectral function $\rho$. For instance, 
\begin{align}\label{Spdef} &S_{+}(t) = \frac{1}{2n}\bigl\langle\tr [\psi_{\mu}(it),\psi^{\dagger}_{\mu}]\bigr\rangle_{\beta} =\frac{1}{2}\bigl(G_{+}(it) + G_{-}(it)\bigr)= \frac{1}{2\pi}\int_{-\infty}^{+\infty}\tilde S_{+}(\omega)e^{-i\omega t}\,\d\omega\,,\\
\label{Smdef} &S_{-}(t)  =\frac{1}{2n}\bigl\langle\tr \{\psi_{\mu}(it),\psi^{\dagger}_{\mu}\}\bigr\rangle_{\beta} =\frac{1}{2}\bigl(G_{+}(it) - G_{-}(it)\bigr)= \frac{1}{2\pi}\int_{-\infty}^{+\infty}\tilde S_{-}(\omega)e^{-i\omega t}\,\d\omega
\end{align}
are given by
\be\label{tildeSpm} \tilde S_{+}(\omega) = \pi\frac{e^{\beta\omega}-1}{e^{\beta\omega}+1}\,\rho(\omega)\, \quad\text{and}\quad \tilde S_{-}(\omega) = \pi\rho(\omega)\, .\ee
It is also convenient to introduce the retarded and advanced two-point functions
\begin{align}\label{Grdef} &\Gr(t) = 2i\Theta(t)S_{-}(t) = i\Theta(t) \bigl(G_{+}(it) - G_{-}(it)\bigr) =  \frac{1}{2\pi}\int_{-\infty}^{+\infty} \chir(\omega)e^{-i\omega t}\,\d\omega\,,\\\label{Gadef}  &\Ga(t) = -2i\Theta(-t)S_{-}(t) = -i\Theta(-t) \bigl(G_{+}(it) - G_{-}(it)\bigr) =  \frac{1}{2\pi}\int_{-\infty}^{+\infty} \chia(\omega)e^{-i\omega t}\,\d\omega\, ,
\end{align}
for which
\be \chir(\omega) = -\int_{-\infty}^{+\infty}\frac{\rho(\omega')}{\omega-\omega'+i0^{+}}\,\d\omega'\, ,\quad
\chia(\omega) = -\int_{-\infty}^{+\infty}\frac{\rho(\omega')}{\omega-\omega'-i0^{+}}\,\d\omega'\, .\ee
From the definition of the resolvent in \eqref{resolventdef}, we see that
\be\label{chiRrel} \chir(\omega) = - R(\omega+i0^{+})\, \quad\text{and}\quad \chia(\omega) = - R(\omega-i0^{+})\, .\ee
The retarded and advanced Fourier transforms $\chir$ and $\chia$ can thus be analytically continued to holomorphic functions on the upper and lower half-planes respectively, an important and well-known property. Of course, they are not independent since the reality of the spectral function implies that
\be\label{GaGrrel} \Ga(t)^{*}=\Gr(-t)\, ,\quad \chia(z)^{*}=\chir(z^{*})\, .\ee

Real time two-point functions are in this way completely encoded in the spectral function $\rho$, whereas the Euclidean two-point function is encoded in the Fourier-Matsubara coefficients $G_{k}$. The real time and Euclidean time data thus look very different: frequencies in real time are always continuous, whereas their Euclidean counterparts are discrete. There is a powerful theorem due to Carlson \cite{Rubel} that shows that these two different-looking sets of data are nevertheless strictly equivalent: the real time data determines uniquely the Euclidean time data and vice versa. That the real time data determines the Euclidean time data is easy to understand: from the spectral function $\rho$ one can get the $G_{k}$'s from \eqref{resolventdef} and \eqref{GkRrel}. The converse is much more subtle. The knowledge of the Matsubara-Fourier coefficients fixes the values \eqref{GkRrel} of the resolvent on a discrete set of points $z=i\nu_{k}$ on the imaginary axis. Moreover, we know that the resolvent is a holomorphic function for $\im z>0$ and $\im z<0$ and has the simple asymptotic behaviour \eqref{Rasymp}. Carlson's theorem then implies that $R(z)$ is determined uniquely for all complex values of $z$.

The theorem itself does not provide an explicit procedure to reconstruct $\rho$ from the $G_{k}$'s, but such a procedure was discovered in \cite{Cuniberti} and generalized in \cite{ferraRG}. The explicit formulas in \cite{ferraRG} actually allow to reconstruct the resolvent from the knowledge of the Euclidean data above an arbitrary cut-off $k_{0}$ in frequency, \emph{i.e.}\ the set of Matsubara-Fourier coefficients $\{G_{k}, k\geq k_{0}>0\}$ . A startling consequence of this fact is the existence in any quantum mechanical theory of a rigorous version of the Renormalization Group, as a consequence solely of analyticity properties \cite{ferraRG}.\footnote{Ref.\ \cite{ferraRG} focuses on the bosonic case, but the fermionic case can be treated in exactly the same way.} The explicit reconstruction formulas are thus very interesting conceptually, but in practice their use is limited by the fact that a reliable reconstruction requires the knowledge of the $G_{k}$'s to an enormous precision. They are therefore impractical except in simple cases where $\rho$ is known to be very smooth, without sharp features. Since we do not want to make any such \emph{a priori} assumptions, we prefer to base our analysis on the real time version of the Schwinger-Dyson equations derived in the next subsection. Interestingly, the derivation of these equations that we now present is going to use Carlson's theorem in a central way.

\subsubsection{\label{realSDSec}Melonic real time two-point functions}

\begin{figure}[h!]
\centering
\def\svgwidth{4.5in}
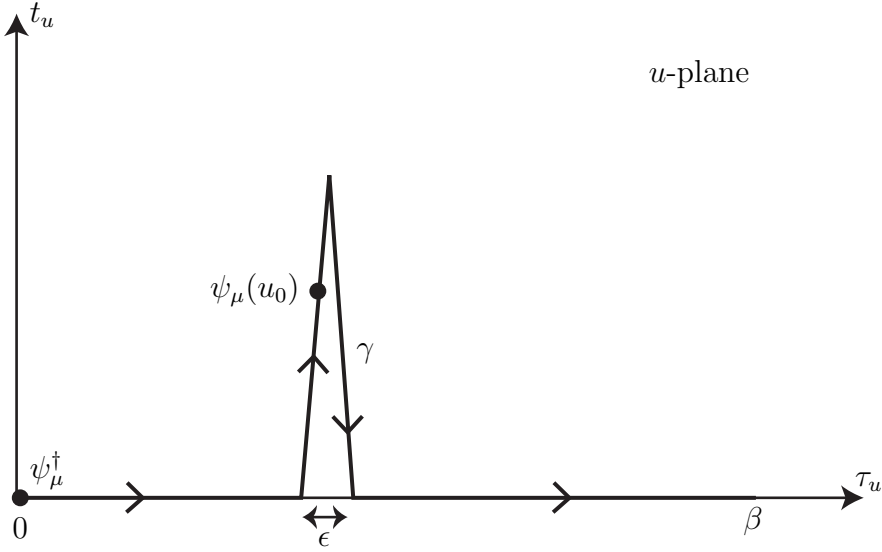
\caption{Keldysh contour $\gamma$ (thick line) used to compute the two-point functions at an arbitrary complex time $u_{0}$. The insertions of the operators $\psi_{\mu}$ at $u=0$ and $\psi_{\mu}^{\dagger}$ at $u=u_{0}$ are represented by bullets along the contour. The width $\epsilon$ of the triangle-shaped piece of the contour may be taken to zero. Note that the contour may be chosen such that the insertion point at $u=u_{0}$ sits at the top of the triangle, but this is not required.
\label{SKcontour1}}
\end{figure}

The standard general approach to real time correlation functions is based on the Schwinger-Keldysh formalism. This formalism is a direct consequence of the path integral formulation, suitably adapted to cases where operators are inserted at arbitrary complex times, as in \eqref{cplxetimeevol}. One introduces an oriented closed Keldysh contour $\gamma$ on the complex time cylinder where $u$ and $u+\beta$ are identified. The contour starts at $u=0$, ends at $u=\beta$ and goes through all the operator insertion points, say $\mathscr O_{1}(u_{1}),\ldots,\mathscr O_{r}(u_{r})$. The correlator associated with such a contour is 
\be\label{Keldyshcorr} G_{\gamma}(u_{1},\ldots, u_{r}) = \bigl\langle\text{T}_{\gamma}\mathscr O_{1}(u_{1})\cdots\mathscr O_{r}(u_{r})\bigr\rangle_{\beta}\,,\ee
where the symbol $\text{T}_{\gamma}$ orders the operators according to their position along the contour. For instance, to compute the two-point function $G(u_{0})$, Eq.\ \eqref{Gcplxedef}, we can use the contour depicted in Fig.\ \ref{SKcontour1}, with an insertion of $\psi_{\mu}^{\dagger}$ at $u=0$ and an insertion of $\psi_{\mu}$ at $u=u_{0}$. The standard Euclidean formalism corresponds to the special case where all the insertions are on the Euclidean time axis $\im u=0$, so that we can then use the contour $\im u=0$. In practice, it is convenient to use contours along which $\re u$ is monotonically increasing, so that the contour ordering coincides with the usual Euclidean time ordering. This is why the real time piece of $\gamma$ in Fig.~\ref{SKcontour1} was chosen to have a triangular shape.

The advantage of this approach is that the Feynman rules take a familiar form. The only differences with the Euclidean case is that integrals over $\tau$ are replaced by contour integrals along $\gamma$, and propagators are associated with the complex time tree-level two-point function \eqref{Gtreecplxe}. The arguments leading to the Euclidean Schwinger-Dyson equations \eqref{SDone} and \eqref{SDtwo}, in their path integral or diagrammatic versions for the disordered and matrix models respectively, can be repeated without change in the present, more general context. The self-energy for the contour of Fig.\ \ref{SKcontour1} is defined by the equation
\be\label{Sigmadefgamma} \Bigl(\frac{\d}{\d u} + m\Bigr)G(u-u') +\int_{\gamma}\Sigma(u-v)G(v-u')\,\d v = \delta_{\gamma}(u-u')\, ,\ee
generalizing \eqref{Sigmadef2}. Eqs.\ \eqref{SDone} and \eqref{SDtwo} are now replaced by
\begin{align}
\label{SDone2} \Sigma(u) &=\la^{2}G(u)^{2}G(-u) &&\text{for the \modone\ model}\,,\\
\label{SDtwo2} \Sigma(u) &= -\frac{|\xi|^{2}}{4}\bigl(G(u)^{3}+3 G(u)G(-u)^{2}\bigr) &&\text{for the \modtwo\ model}\,.
\end{align}
Traditionally, one proceeds by using the decomposition \eqref{GdecGpGmdef} to rewrite \eqref{Sigmadefgamma} in the form of the so-called Kadanov-Baym equations; see \emph{e.g.}\ \cite{KBeqs} for a pedagogical introduction. The Kadanov-Baym equations can then be integrated numerically using \eqref{SDone2} or \eqref{SDtwo2}.\footnote{For a recent interesting use of the Kadanov-Baym equations in the context of the SYK model, see \emph{e.g.}\ \cite{SachdevKB}.}

We are not going to use this rather tedious route here. Instead, we present a simple argument based on Carlson's theorem, leading to a very convenient expression of the solution which entirely avoids having to deal with the details of the Kadanov-Baym equations.

Let us start by defining
\be\label{Rsigmadef} R_{\Sigma}(z) = \frac{1}{R(z)}+m-z\,,\ee
so that Eqs.\ \eqref{GkRrel} and \eqref{Sigmadef} imply
\be\label{Rsigmk} R_{\Sigma}(i\nu_{k}) = -\Sigma_{k}\, .\ee
Using \eqref{resolventdef} and denoting $x=\re z$ and $y=\im z$, we get
\be\label{imRformula} \im R(x+iy) = -y\int_{-\infty}^{+\infty}\frac{\rho(\omega)}{(x-\omega)^{2}+y^{2}}\,\d\omega\, .\ee
From the positivity of the spectral function $\rho$, which is manifest from its definition \eqref{specdef}, we deduce that $\im R(z)\not = 0$ away from the real axis, and thus also $R(z)\not = 0$ there. Since $R$ is holomorphic and has no zeros for $\im z>0$ or $\im z<0$, Eq.\ \eqref{imRformula} implies that $R_{\Sigma}$ is also holomorphic for $\im z>0$ and $\im z<0$. Moreover, the expansion \eqref{Rasymp} shows that
\be\label{RSigmaasymp} R_{\Sigma}(z)\underset{z\rightarrow\infty}{=}\mathcal{O}(1/z)\, .\ee
All the conditions for the applicability of Carlson's theorem to the function $R_{\Sigma}$ are thus met: if we are able to find some function $\tilde R_{\Sigma}(z)$, holomorphic for $\im z>0$ and $\im z<0$, behaving as $\mathcal{O}(1/z)$ when $z\rightarrow\infty$ and such that $\tilde R_{\Sigma}$ and $R_{\Sigma}$ match for all $z=i\nu_{k}$, then the theorem states that $\tilde R_{\Sigma}= R_{\Sigma}$ for all $z$:
\be\label{RsRsid} \tilde R_{\Sigma}(i\nu_{k}) =  R_{\Sigma}(i\nu_{k})\quad \forall k\in\mathbb Z+\frac{1}{2}\implies R_{\Sigma}(z) =  R_{\Sigma}(z)\quad \forall z \in \mathbb C\, .
\ee
We can combine the equations \eqref{SDone2} or \eqref{SDtwo2} with the spectral decomposition \eqref{Gcplxespecdec} to guess a formula for $R_{\Sigma}(z)$. We shall then check that the above criteria are satisfied.

In the case of the \modone\ model, we consider
\be\label{tildeRmodone} \tilde R_{\Sigma}(z) = -\la^{2}\int_{-\infty}^{+\infty}
\prod_{j=1}^{3}\bigl[\d\omega_{j}\rho(\omega_{j})\nF(\omega_{j})\bigr]
\frac{e^{\beta (\omega_{1}+\omega_{2})}}{\nF(\omega_{3}-\omega_{1}-\omega_{2})}
\frac{1}{z+\omega_{3}-\omega_{1}-\omega_{2}}\,\cdotp\ee
The convergence of the integral is trivial to check. The function $\tilde R_{\Sigma}$ so defined is manifestly holomorphic for $\im z>0$ and $\im z<0$ and is $\mathcal{O}(1/z)$ at large $z$. Moreover, if we evaluate $\Sigma_{k}$ by taking the Fourier transform of \eqref{SDone2}, in which we use the spectral decomposition formula \eqref{Gcplxespecdec}  for $G$ and the explicit form \eqref{Gtreecplxe} for $G^{(0)}$, we immediately find that $\Sigma_{k}=-\tilde R_{\Sigma}(i\nu_{k})$ is satisfied. From \eqref{Rsigmk} we conclude that $\tilde R_{\Sigma}(i\nu_{k}) = R_{\Sigma}(i\nu_{k})$ and thus from \eqref{RsRsid} $\tilde R_{\Sigma} = R_{\Sigma}$. 

Exactly the same reasoning can be applied to the \modtwo\ model starting from
\begin{align}\label{tildeRmodtwo}&\tilde R_{\Sigma}(z) = -\frac{|\xi^{2}|}{4}
\int_{-\infty}^{+\infty}
\prod_{j=1}^{3}\bigl[\d\omega_{j}\rho(\omega_{j})\nF(\omega_{j})\bigr]\\
&\quad\times\biggl[\frac{1}{\nF(\omega_{1}+\omega_{2}+\omega_{3})}
\frac{1}{z-\omega_{1}-\omega_{2}-\omega_{3}}+
\frac{3 e^{\beta (\omega_{2}+\omega_{3})}}{\nF(\omega_{1}-\omega_{2}-\omega_{3})}
\frac{1}{z+\omega_{2}+\omega_{3}-\omega_{1}}\biggr]\,,\nonumber
\end{align}
and then showing that $\tilde R_{\Sigma}(i\nu_{k})=R_{\Sigma}(i\nu_{k})$ and thus $\tilde R_{\Sigma}=R_{\Sigma}$.

We should stress that it is crucial to understand that the property $\tilde R_{\Sigma}(i\nu_{k}) = R_{\Sigma}(i\nu_{k})=-\Sigma_{k}$ is \emph{not} sufficient in itself to conclude $\tilde R_{\Sigma}(z) = R_{\Sigma}(z)$. For example, a naive way to do the analytic continuation from Euclidean time to real time in the \modone\ model is to start from the Euclidean equation \eqref{SDone} written in terms of the Matsubara-Fourier coefficients,
\be\label{SigGFourierex} \Sigma_{k}=\frac{\la^{2}}{\beta^{2}}\sum_{k_{1},k_{2}\in\mathbb Z+\frac{1}{2}}G_{k_{1}}G_{k_{2}}G_{k_{1}+k_{2}-k}
=-\frac{\la^{2}}{\beta^{2}}\sum_{k_{1},k_{2}\in\mathbb Z+\frac{1}{2}} G_{k_{1}}G_{k_{2}}R(i\nu_{k_{1}+k_{2}}-i\nu_{k})\, .\ee
This might induce us to propose
\be\label{wrongpostSigma} R_{\Sigma}(z)\overset{?}{=}\frac{\la^{2}}{\beta^{2}}\sum_{k_{1},k_{2}\in\mathbb Z+\frac{1}{2}} G_{k_{1}}G_{k_{2}}R(i\nu_{k_{1}+k_{2}}-z)\, .\ee
Although this expression satisfies by construction $R_{\Sigma}(i\nu_{k})=-\Sigma_{k}$, it does not have the required analytic properties; indeed, the right-hand side of \eqref{wrongpostSigma} is holomorphic only within bands $ni<\im z<(n+1)i$ where $n\in\mathbb Z$, but not on the whole half-planes $\im z>0$ and $\im z<0$ as required.

\begin{figure}[h!]
\centering
\def\svgwidth{3.5in}
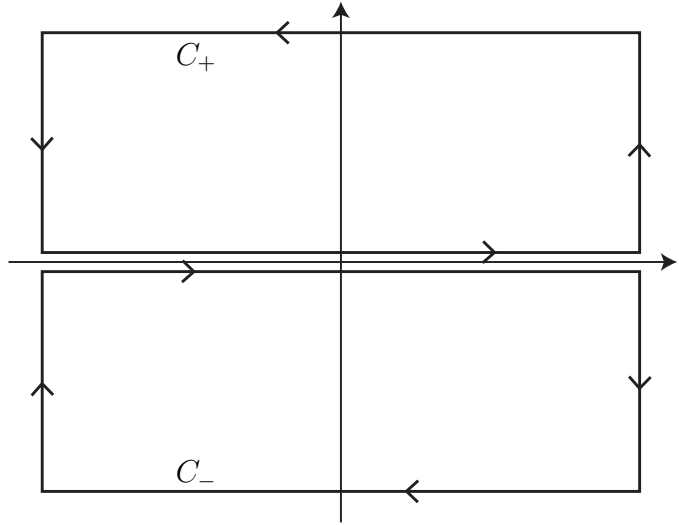
\caption{Contours $C_{+}$ and $C_{-}$ in the upper and lower $w$-plane used in the main text for computing \eqref{Iint1def} and similar integrals. The limit for which the rectangles have an infinite size is taken.
\label{intcontourfig}}
\end{figure}

For the purpose of improving the efficiency of the numerical algorithm presented in the next subsection, it is useful to perform explicitly one of the integrals over the frequencies in \eqref{tildeRmodone} and \eqref{tildeRmodtwo}. For example, in the \modone\ model we write
\be\label{RSigmod1int} R_{\Sigma}(z) = -\la^{2}\int_{-\infty}^{+\infty}\!\d\omega_{1}\d\omega_{2}\,\rho(\omega_{1})\nF(\omega_{1})\rho(\omega_{2})\nF(\omega_{2})e^{\beta (\omega_{1}+\omega_{2})}\, \mathcal I(\omega_{1}+\omega_{2};z)\,,\ee
where
\be\label{Iint1def} \mathcal I (\Omega;z) = \int_{-\infty}^{+\infty}\!\d\omega\,\rho(\omega)\frac{\nF(\omega)}{\nF(\omega-\Omega)}\frac{1}{z+\omega-\Omega}\,\cdotp\ee
We can then use \eqref{rhofromR} and the well-behaved properties of the integrand at infinity to get
\be\label{Iint1calc1} \mathcal I (\Omega;z) =\frac{i}{2\pi} \int_{C_{+}\cup C_{-}} \frac{\nF(w)}{\nF(w-\Omega)}\frac{R(w)}{z+w-\Omega}\d w\, \cvp\ee
where $C_{+}$ and $C_{-}$ are semi-infinite rectangles on the upper and lower half complex $w$-plane respectively, see Fig.\ \ref{intcontourfig}. Picking up the poles at $w=i\nu_{k}$ for all $k\in\mathbb Z+\frac{1}{2}$, as well as at $w=\Omega-z$, yields
\be\label{Iint1final} \mathcal I (\Omega;z) =\frac{e^{-\beta\Omega}-1}{\beta}\sum_{k\in\mathbb Z+\frac{1}{2}}\frac{G_{k}}{z+i\nu_{k}-\Omega}-\bigl(1+e^{-\beta z}\bigr)\frac{R(\Omega-z)}{e^{\beta (\Omega-z)}+1}\,\cdotp\ee
Overall, we thus obtain for the \modone\ model
\begin{align}\label{RSigmod1result} R_{\Sigma}(z) &= -\la^{2}\int_{-\infty}^{+\infty}\!\d\omega_{1}\d\omega_{2}\,\rho(\omega_{1})\nF(\omega_{1})\rho(\omega_{2})
\nF(\omega_{2})\\
&\qquad\times\Biggl[ \frac{1-e^{\beta (\omega_{1}+\omega_{2})}}{\beta}\sum_{k\in\mathbb Z+\frac{1}{2}}\frac{G_{k}}{z+i\nu_{k}-\omega_{1}-\omega_{2}}
-\bigl(1+e^{\beta z}\bigr)\frac{R(\omega_{1}+\omega_{2}-z)}{e^{\beta (z-\omega_{1}-\omega_{2})}+1}\Biggr].\nonumber
\end{align}
In the \modtwo\ model, similar manipulations yield
\begin{align}R_{\Sigma}(z) &= -\frac{|\xi|^{2}}{4}\int_{-\infty}^{+\infty}\!\d\omega_{1}\d\omega_{2}\,\rho(\omega_{1})\nF(\omega_{1})\rho(\omega_{2})
\nF(\omega_{2})\nonumber\\
\label{RSigmod2result} &\qquad\times\Biggl[ \frac{2\bigl(e^{\beta (\omega_{1}+\omega_{2}\bigr)}-1)}{\beta}\sum_{k\in\mathbb Z+\frac{1}{2}}\frac{i\nu_{k}+2(\omega_{1}+\omega_{2})-z}{(z-i\nu_{k})^{2}-(\omega_{1}+\omega_{2})^{2}}\,G_{k}\\
&\qquad\qquad+\bigl(e^{\beta z}+1\bigr)\biggl(\frac{R(z-\omega_{1}-\omega_{2})}{e^{\beta (z-\omega_{1}-\omega_{2})}+1}+3e^{-\beta z}\frac{R(z+\omega_{1}+\omega_{2})}{e^{-\beta (z+\omega_{1}+\omega_{2})}+1}\biggr)\Biggr].\nonumber
\end{align}
Equations \eqref{Rsigmadef} and \eqref{RSigmod1result} (for the \modone\ model) or \eqref{RSigmod1result} (for the \modtwo\ model) are the precise real time counterparts of the Euclidean equations \eqref{Sigmadef} and \eqref{SDone} or \eqref{SDtwo}, respectively, and completely determine the spectral function $\rho$.

Let us conclude this derivation with the following remark. The Kadanov-Baym equations obtained from \eqref{Sigmadefgamma} must of course be satisfied by our solution, and we have checked this explicitly. This is an interesting albeit  somewhat tedious exercise in contour integration. One must consider several cases depending on the position of $u$ and $u'$ along the Keldysh contour $\gamma$. It is convenient to use double Fourier decompositions of the form \eqref{Fouriercplxe} for both $G(u)$ and $\Sigma(u)$. All the required identities follow from the analytic properties of $R$ and $R_{\Sigma}$ discussed above.

\subsubsection{\label{realnum2ptSec}Numerical solution of real time equations}

As already mentioned, all the real time two-point functions can be straightforwardly derived from the spectral function $\rho(\omega)$. The basic formula relating $\rho$ to the resolvent $R$ is \eqref{rhofromR}, and noting that for any real $\epsilon$
\be\label{Rcc} R(\omega+i\epsilon)^{*}=R(\omega-i\epsilon)\,,\ee
we can rewrite this equation as
\be\label{rhoRnumrel} \rho(\omega) = -\frac{1}{\pi}\im R(\omega+ i0^{+})\, .\ee
We will therefore focus on the computation of the resolvent through equations \eqref{Rsigmadef} and \eqref{RSigmod1result} or \eqref{RSigmod2result}. In practice, we evaluate $R(\omega+ i\epsilon)$ for a small positive $\epsilon$, and assume throughout
\be\label{rhoepsapprox} \rho(\omega) \simeq \rho_\epsilon(\omega) = -\frac{1}{\pi}\im R(\omega+ i\epsilon)\, .\ee

The real time Schwinger-Dyson equations can be solved numerically using an iterative algorithm that is very similar to the one applied in the Euclidean case, described in detail in Sec.~\ref{NumSolEucSec}. There are however a few new subtleties associated with the real time case.

The algorithm starts from an approximate solution $\smash{R^{[0]}(\omega+i\epsilon)}$, and progressively refines it in successive iterations. Denoting $R^{[i]}(\omega+i\epsilon)$ the approximate solution at step $i$, and $\rho_\epsilon^{[i]}(\omega)$ the spectral function associated to it through \eqref{rhoepsapprox}, we first compute $R^{[i+1]}_{\Sigma}(\omega + i\epsilon)$ by plugging these in the right-hand side of \eqref{RSigmod1result} or \eqref{RSigmod2result}, according to the model under study. To complete the iteration, we define the resolvent at step $i+1$ to be
\be\label{Rrealiterative} R^{[i+1]}(\omega+i\epsilon) = (1-\alpha)R^{[i]}(\omega+i\epsilon) + \frac{\alpha}{\omega+i\epsilon - m +R_{\Sigma}^{[i+1]}(\omega+i\epsilon)}\,\cvp\ee
where $\alpha$ is an \emph{a priori} arbitrary real weighting factor.

Because equations \eqref{RSigmod1result} and \eqref{RSigmod2result} involve integrals over $\omega$, in practice we need to choose a finite number of discrete frequencies over which the resolvent is evaluated,
\be\label{discretefreqchoice}
\omega_{p}= \delta\omega \, p \qquad\text{for}\qquad p = p_i, p_i+1, \dots p_f-1, p_f \,.
\ee
This defines a unidimensional grid on a finite interval $[\omega_{\min}, \omega_{\max}] = [\delta\omega \, p_i, \delta\omega \, p_f ]$, outside which we use asymptotic freedom to identify the solution with the tree-level result \eqref{Rstep0tree}, thus transforming all integrals into finite sums. The discrete step $\delta\omega$ is chosen consistently with the parameter $\epsilon>0$, and sets an intrinsic upper limit on the resolution with which $\rho$ can be reconstructed from \eqref{rhoepsapprox}. 

Note that the Euclidean two-point function also appears on the right-hand side of equations \eqref{RSigmod1result} and \eqref{RSigmod2result} through its Matsubara-Fourier coefficients $G_{k}$. We will therefore assume that a solution for these is already known, possibly by having run the Euclidean algorithm beforehand, so that they are simply constant coefficients in this context.

The iteration of the algorithm is repeated until the equations are solved up to a prescribed tolerance $\varepsilon$, \emph{i.e.}
\be
\frac{1}{P}\sum_{p=p_i}^{p_f} \biggl|R^{[i]}(\omega_p + i \epsilon) - \frac{1}{\omega + i\epsilon - m + R_\Sigma^{[i]}(\omega_p + i\epsilon)}\biggr| < \varepsilon \quad\text{with}\quad P = p_f-p_i+1\,.
\ee
As in the Euclidean case, we will take $\varepsilon = 10^{-12}$ (in natural units $\lambda = |\xi| = 1$), ensuring that the solution is found with a very good accuracy while staying safely above machine-precision issues.

Before moving on, let us make a few further comments on the algorithm described above. First, the choice of initial approximation is very important for its convergence. We may use the tree-level result corresponding to $R^{(0)}_\Sigma(\omega+i\epsilon) = 0$,
\be\label{Rstep0tree}
R^{(0)}(\omega+i\epsilon) = \frac{1}{\omega+i\epsilon-m}\,\cvp
\ee
or any previously found solution deemed to be close to the actual solution we are seeking. As in the Euclidean case, this allows us to map the parameter space by taking small steps in either the $m$ or $T$ directions to ``trace'' solutions.

Another common element with the Euclidean algorithm is the presence of an arbitrary weighting factor $\alpha$. This comes about for the same reasons discussed in Sec.~\ref{NumSolEucSec}, having a similar impact on convergence and performance as was seen there, and again we find that $\alpha = 0.75$ works well for most cases.

A new ingredient in the real time algorithm is the arbitrary frequency-domain discretization \eqref{discretefreqchoice}, which was naturally given by the Matsubara frequencies in the Euclidean setting. While as already mentioned $\delta\omega$ is related to the resolution of the resulting spectral function, its support is determined by $p_i$ and $p_f$. Taking Eq.~\eqref{rhoone} into account, these parameters need to be chosen such that the interval $[\omega_{\min}, \omega_{\max}]$ is approximately centered around $\omega = m$, and extends to either side at least up to the regime where asymptotic freedom takes over.

More generally, the choice \eqref{discretefreqchoice} used here has the advantage that it is additive and symmetric about the origin, so that it satisfies the convenient properties
\be
\omega_{p_1} + \omega_{p_2} = \omega_{p_1+p_2} \quad\text{and}\quad -\omega_p = \omega_{-p}\,.
\ee
These properties simplify the practical implementation of the algorithm and improve its performance by completely avoiding costly interpolations. Furthermore, they allow for the efficient pre-calculation of the Matsubara-Fourier coefficient sums in \eqref{RSigmod1result} and \eqref{RSigmod2result}, which after truncation of the Euclidean data can be done in $\mathcal{O}(P)$ memory and $\mathcal{O}(P \times \maxk)$ time, instead of the naive $\mathcal{O}(P^3)$ and $\mathcal{O}(P^3 \times \maxk)$, respectively. It is for this reason that we chose to use the formulas \eqref{RSigmod1result} and \eqref{RSigmod2result}, with two integrals over frequencies, instead of \eqref{tildeRmodone} and \eqref{tildeRmodtwo}, with three integrals over frequencies.\footnote{One should note that precision is also improved by this choice, since Euclidean data is generally of better quality than real time one.}

There is no guarantee that the algorithm described here terminates, but if it does so in $I_\alpha(\varepsilon)$ steps then the total running time is $\mathcal{O}(I_\alpha(\varepsilon) \times P^3 + P\times\maxk)$. Additional pre-processing in $\mathcal{O}(P^2)$ time and memory is useful in practice to reduce constant factors and the incidence of numerical errors, but this is not relevant asymptotically. For our {\tt C++} implementation running on a desktop computer one may then take $P = \mathcal{O}(10^3)$, which makes the pre-processing subleading for typical values of $\maxk = \mathcal{O}(10^5)$. Depending on the model and values of the mass, this is enough to reach temperatures of order $T = \mathcal{O}(10^{-2})$ in natural units, but quickly becomes insufficient below that point.

\medskip

\begin{figure}[h!]
\centering
\begin{tabular}{cc}
\def\svgwidth{7cm}
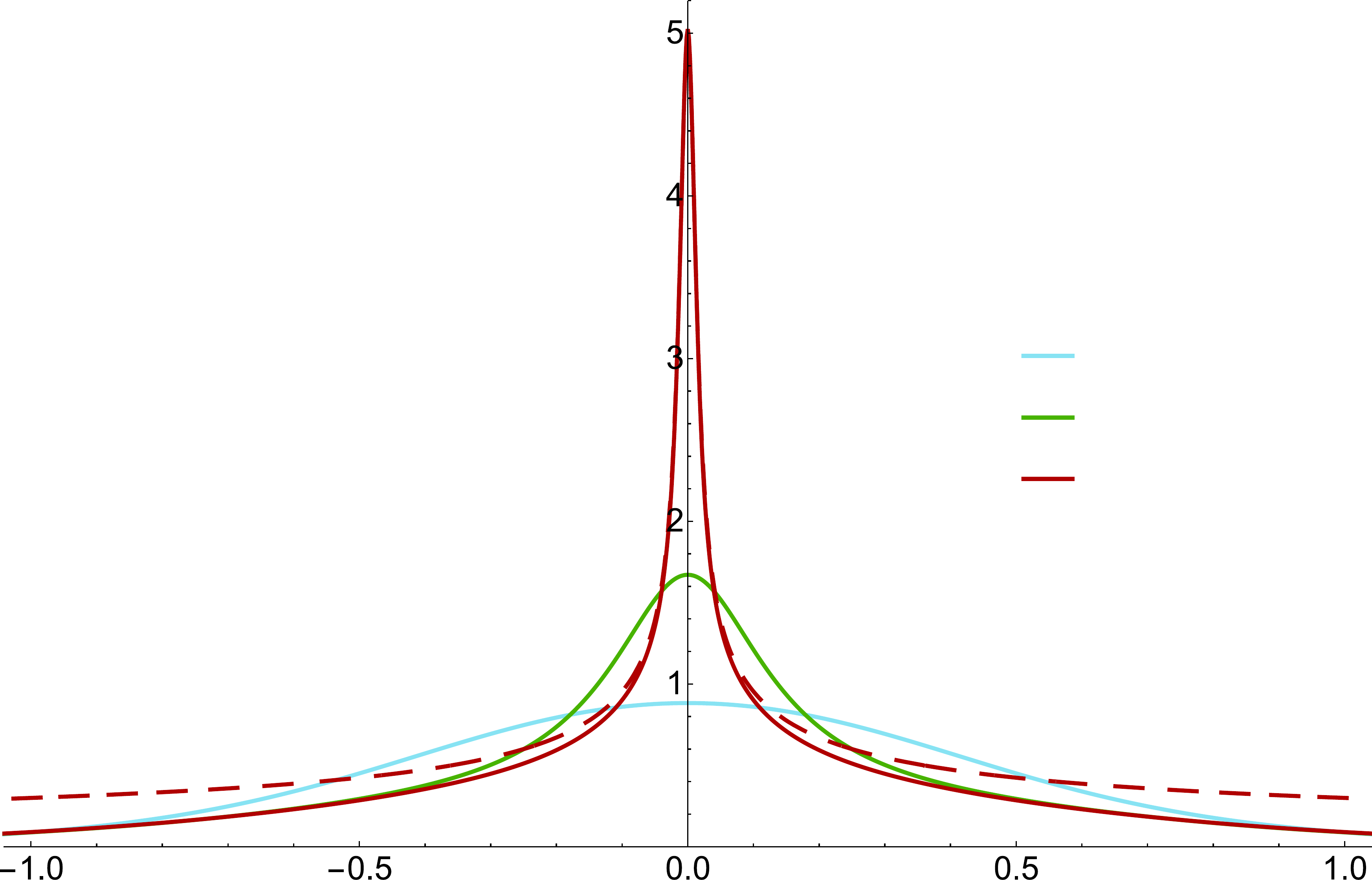 &
\def\svgwidth{7cm}
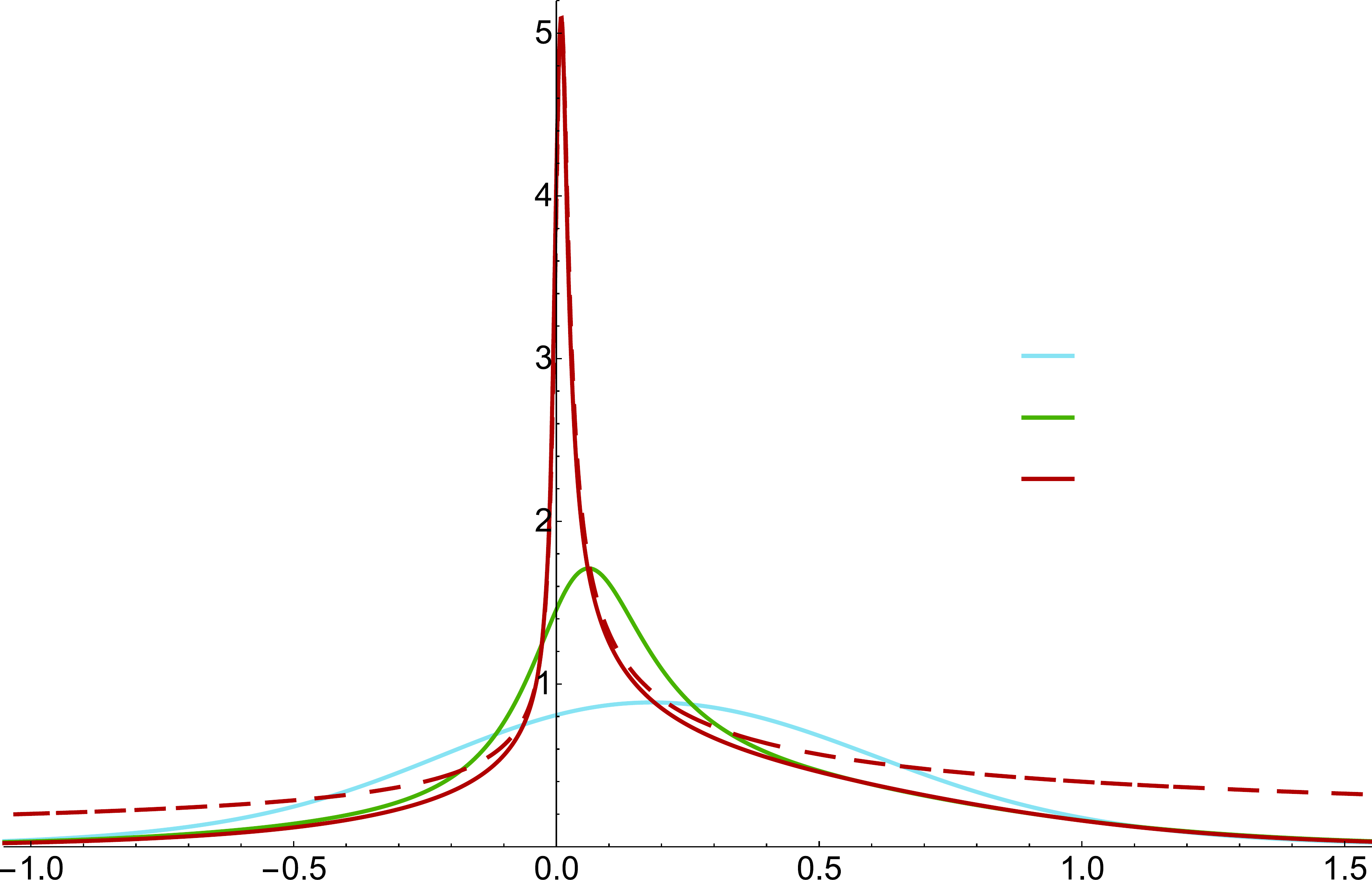 \\
(a) & (b)\\[5pt]
\multicolumn{2}{c}{\def\svgwidth{7cm}
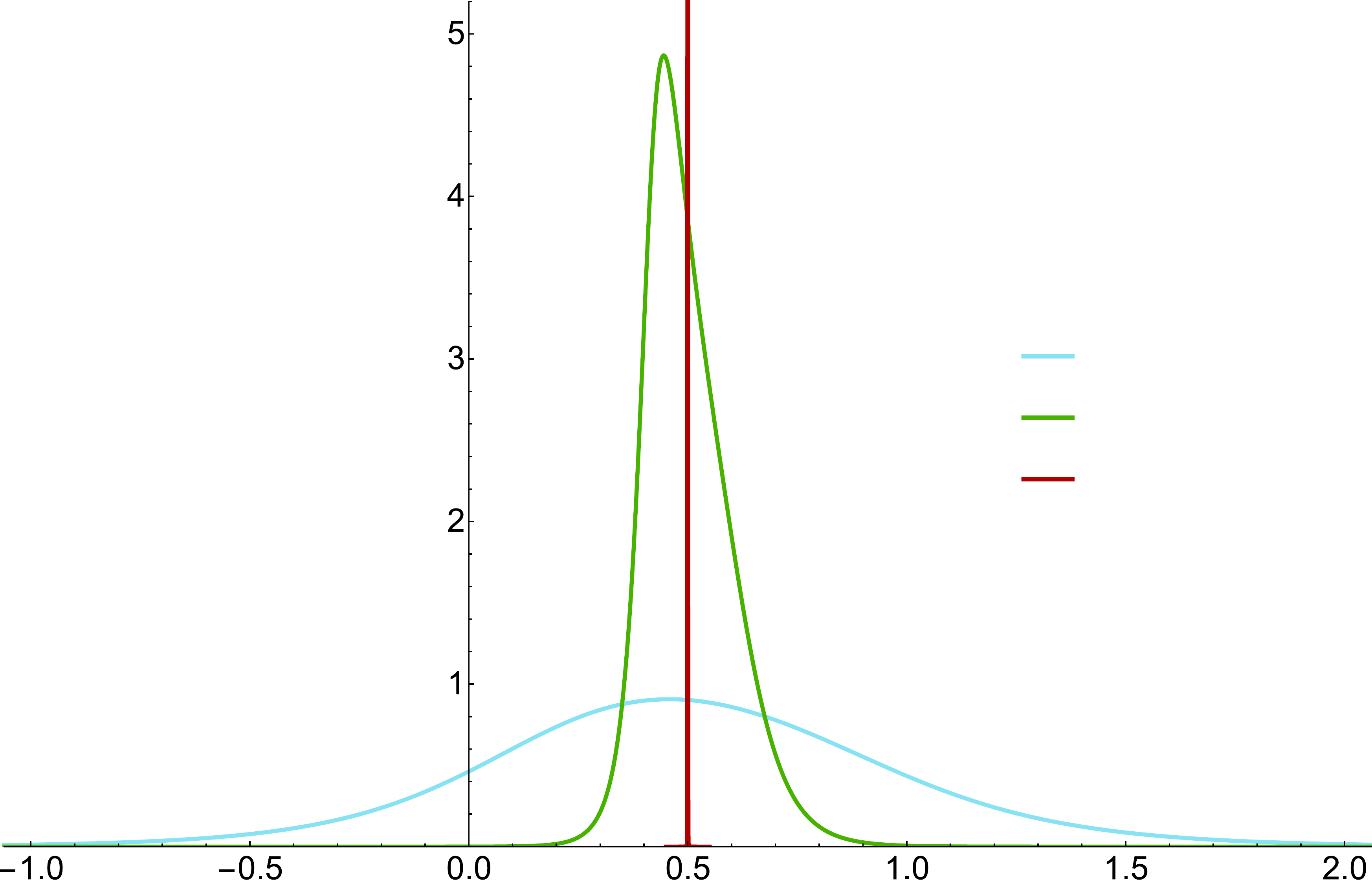}\\
\multicolumn{2}{c}{(c)}
\end{tabular}
\caption{\label{fig:M1SpectralFunction}The spectral function $\rho(\omega)$ in the \modone\ model, for $m = 0$ (a), $m = 0.2$ (b) and $m = 0.5$ (c), at various temperatures in natural units. Plots (a) and (b) correspond to the SYK-like phase, in which the spectral function approaches the conformal form at low temperatures for $|\omega| \ll \lambda = 1$ (red or dark gray dashed lines show Eq.~\eqref{rhoscale} with parameters $\theta$ and $a$ fitted as in Fig.~\ref{fig:M1DeltaFiniteT} for the solution at $T = 0.01$). Plot (c) corresponds to the HO-like phase, where the low-temperature limit $\rho(\omega)\to\delta(\omega-m)$ is already evident for $T = 0.03$ (vertical axis is truncated for clarity). All solutions were obtained with $P = 1001$, $\Delta\omega = \delta\omega(p_f-p_i) = 10$ and $\epsilon = 10^{-5}$, except for the lowest temperature in (c) for which $\Delta\omega = 0.1$ and $\epsilon = 10^{-7}$.}
\end{figure}

To illustrate the results obtained with this algorithm, we show in Fig.~\ref{fig:M1SpectralFunction} examples of the spectral function computed in the \modone\ model, for various masses and temperatures. At $m=0$ it is an even function as a consequence of the particle-hole symmetry, and coincides with the spectral function of the \modtwo\ model. These properties are lost when $m>0$. At small masses and temperatures, $\rho(\omega)$ approaches the conformal form \eqref{rhoscale}, whereas for large masses it converges to the perturbative result $\delta (\omega-m)$.

\begin{figure}[h!]
\centering
\begin{tabular}{cc}
\def\svgwidth{7cm}
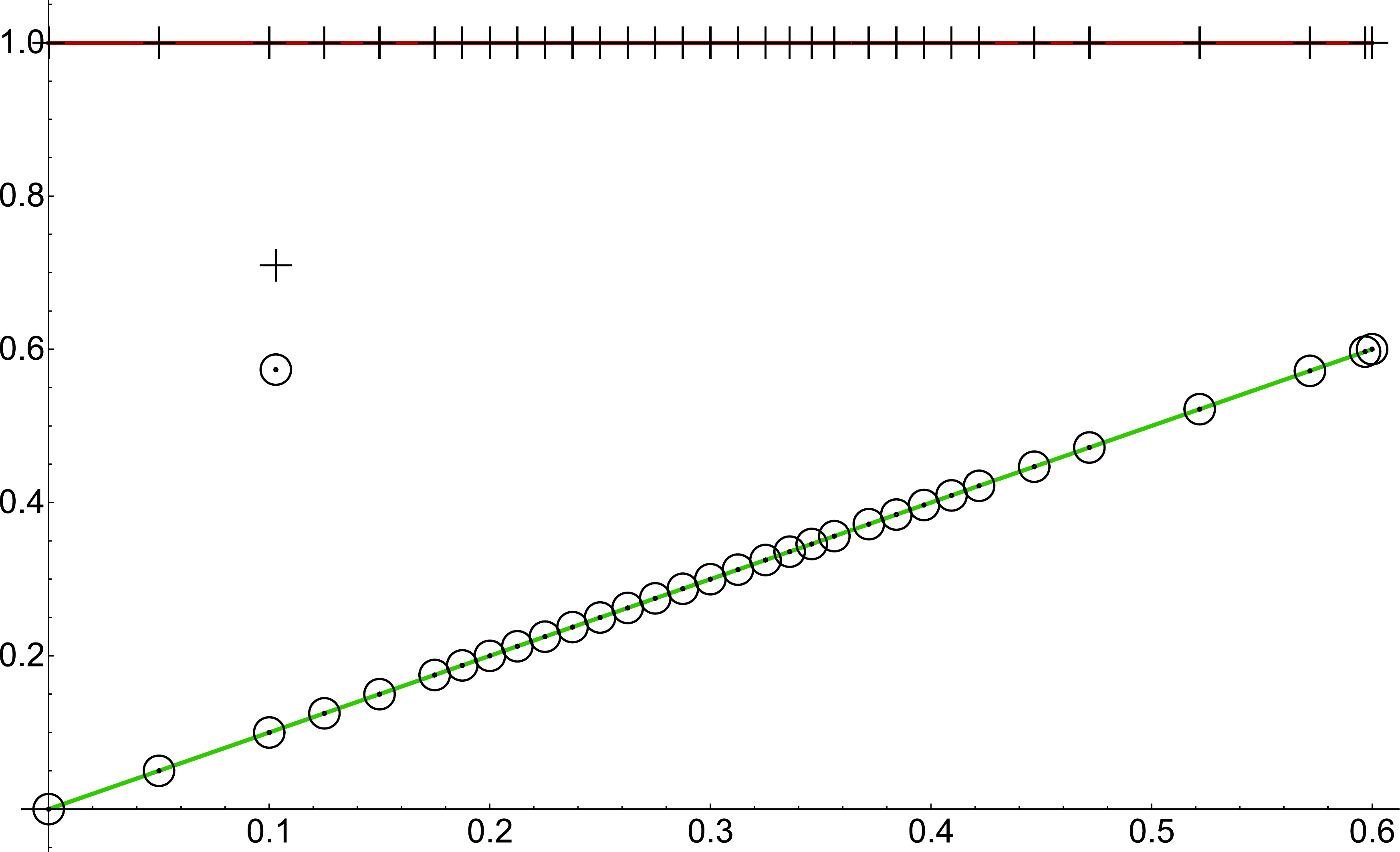 &
\def\svgwidth{7cm}
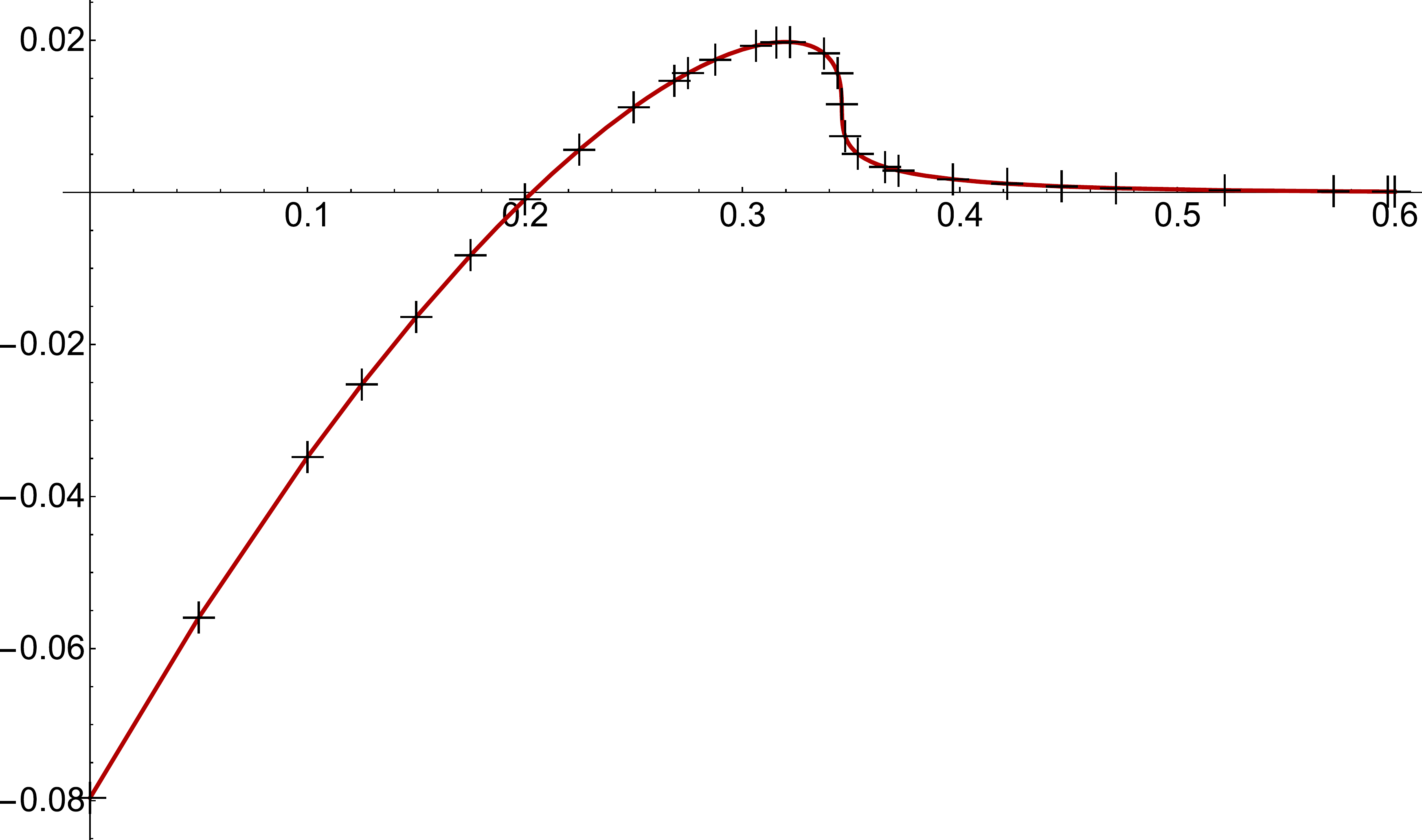 \\
(a) & (b)
\end{tabular}
\caption{\label{fig:M1SpectralFunctionCheck}(a) First and second moments $\rho_{0}$ (crosses) and $\rho_{1}$ (circles) of the spectral function of the \modone\ model, for various masses at $T = \Tc$; solid lines correspond to the exact values $\rho_{0}=1$ (red or dark gray) and $\rho_{1}=m$ (green or light gray). (b) Energy of the \modone\ model, as calculated from the spectral function through Eq.~\eqref{Erhorel}, for various masses at $T = \Tc$ (datapoints shown with crosses); the solid red (dark gray) line corresponds to the numerical values computed independently from purely Euclidean data, through the formula \eqref{Energy2}. Consistency between the real time numerics and the analytic and Euclidean results is excellent. Note that $T = \Tc$ is used here for illustrative purposes and similar results are obtained for other temperatures.}
\end{figure}

In order to check the precision of our numerical methods, we show in Fig.~\ref{fig:M1SpectralFunctionCheck}a the values of the first two moments of the spectral function, $\rho_{0}$ and $\rho_{1}$ defined in Eq.\ \eqref{rhomomentdef}, which are to be compared with the exact results \eqref{rhozero} and \eqref{rhoone}, \emph{i.e.}\ $\rho_{0}=1$ and $\rho_{1}=m$. Another check is provided in Fig.~\ref{fig:M1SpectralFunctionCheck}b, where we plot the energy computed from the real time data via Eq.~\eqref{Erhorel}, and compare with the results obtained from the Euclidean formula \eqref{Energy2}. Similar results are obtained for the \modone\ model at temperatures other than $T = \Tc$, used here for illustrative purposes, as well as for other thermodynamic quantities such as the fermion number (computed from Euclidean or real time data through \eqref{QFourier2} and \eqref{Qrhorel}, respectively).

\begin{figure}[h!]
\centering
\vskip 0.3cm
\begin{tabular}{cc}
\def\svgwidth{7cm}
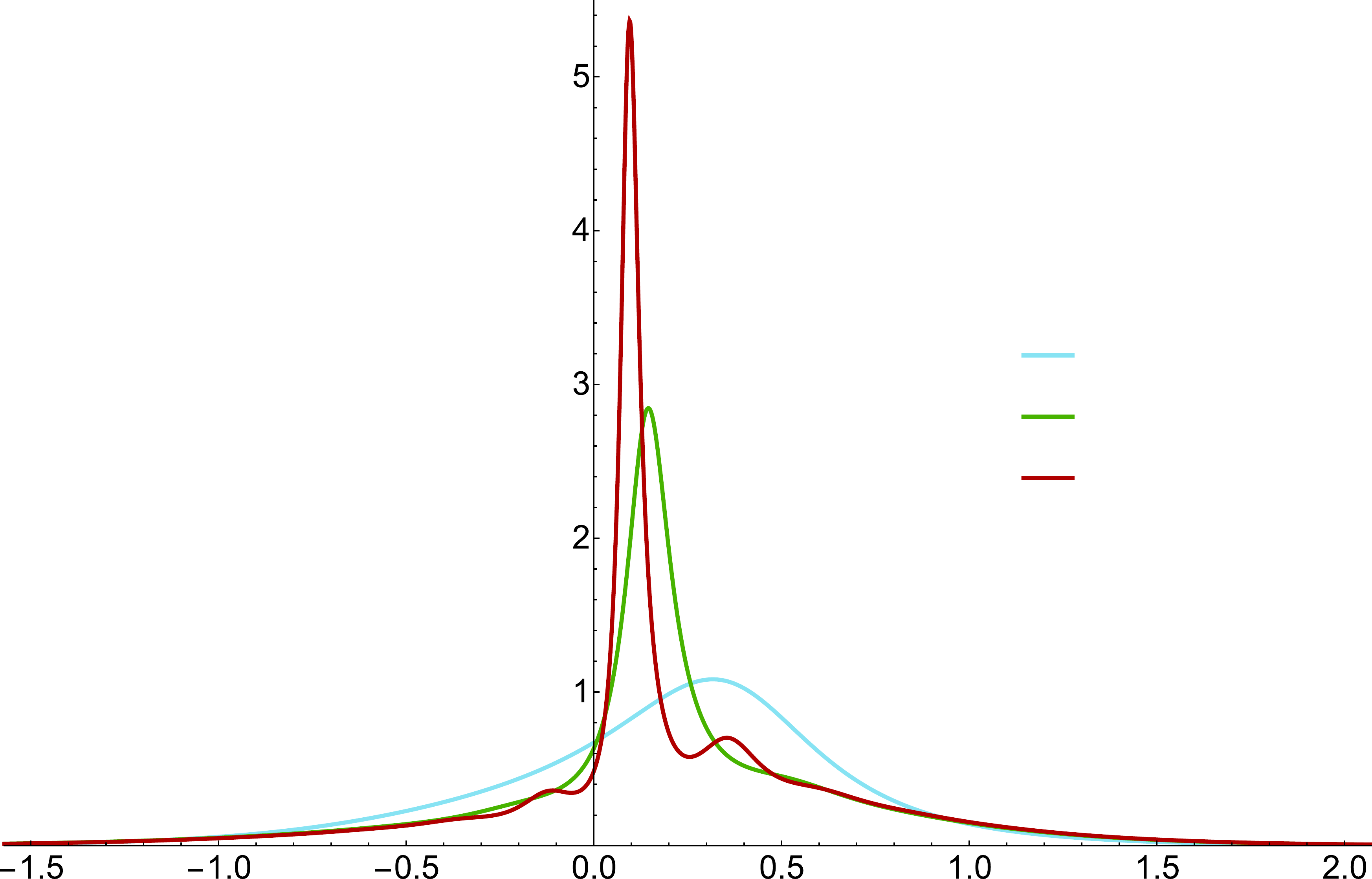 &
\def\svgwidth{7cm}
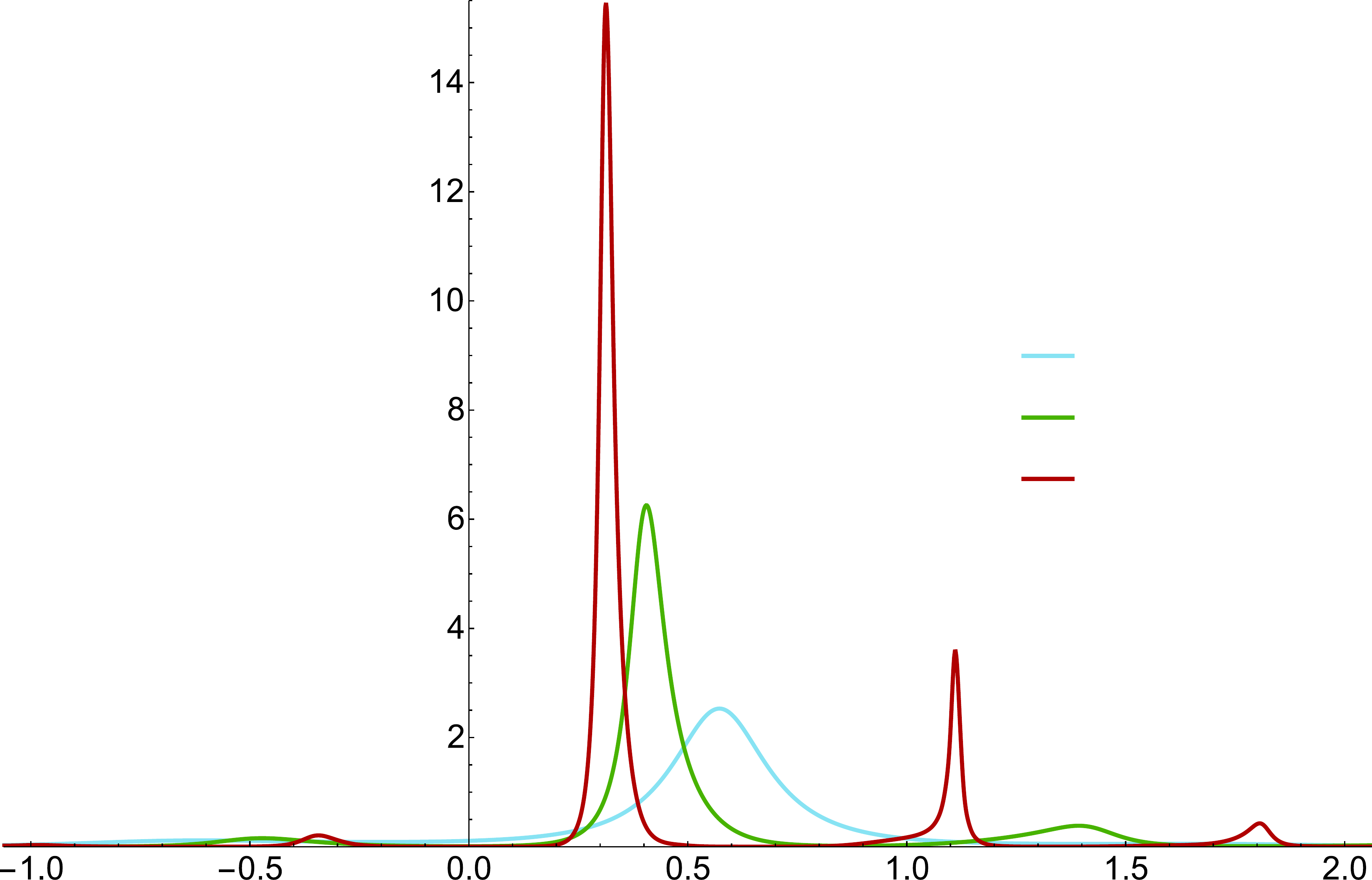 \\
(a) & (b)
\end{tabular}
\caption{\label{fig:M2SpectralFunction}The spectral function $\rho(\omega)$ in the \modtwo\ model, for $m = 0.2$ (a) and $m = 0.5$ (b), at various temperatures in natural units. All solutions were obtained with $P = 1001$, $\Delta\omega = \delta\omega(p_f-p_i) = 10$ and $\epsilon = 10^{-5}$.}
\end{figure}

While Fig.~\ref{fig:M1SpectralFunction} and \ref{fig:M1SpectralFunctionCheck} correspond to the \modone\ model, very similar results are also obtained for the \modtwo\ model. The main difference in this case resides in the fact that the spectral function has a more involved structure for $m>0$, presenting multiple peaks instead of just one, see Fig.~\ref{fig:M2SpectralFunction}. While this is not intrinsically problematic, these peaks become sharper as the temperature is lowered, thus increasing our resolution requirements and therefore raising the lower bound on the temperatures we can probe numerically in a reasonable amount of time.

\subsubsection{\label{quasifreqSec}Application: quasi-normal frequencies}

Quasi-normal frequencies $\nuqn^{(p)}$, $p=0,1,2,\ldots$, govern the way a system responds to a small perturbation. Linear response theory relates these frequencies to the long time behaviour of the retarded two-point function $\Gr$, with frequency $\smash{\nuqn^{(p)}}$ contributing a term proportional to $\smash{\exp(-i\nuqn^{(p)}t)}$. Equivalently, they correspond to poles of $\chir$ on the lower-half complex frequency plane.\footnote{Or to poles of $\chia$ on the upper-half plane, see \eqref{GaGrrel}.} A non-zero negative imaginary part in $\nuqn^{(p)}$ indicates an exponential decay at long times of the associated mode, with time scale
\be\label{tqnpdef} \tqn^{(p)}=-\frac{1}{\im\nuqn^{(p)}}\, \cdotp\ee
In the context of black holes, having $\im\nuqn^{(p)}< 0$ is a landmark feature of horizon physics, signaling the absorption of any small external perturbation by the black hole and its subsequent return to equilibrium. 

For example, for the \modone\ model at low temperatures in the SYK-like phase, we can use \eqref{Gscaling} and \eqref{Cthetadef} to write down the retarded function in the regime $\beta\la\gg 1$, $t\gg1/\la$, 
\be\label{GrlowTmod1} \Gr (t) = i\Theta(t)\frac{C}{\sqrt{\beta}}\frac{e^{-i(\theta + 2\pi a t/\beta)}}{\sqrt{\sinh\bigl(\pi t/\beta\bigr)}}\,\cdotp\ee
As expected, the large $t$ asymptotic expansion of this function is a sum of decreasing exponentials of the form $\smash{\exp(-i\nuqn^{(p)}t)}$, with
\be\label{qnzeroTmod1} \nuqn^{(p)} \underset{\beta\rightarrow\infty}{=}\frac{2\pi}{\beta}\Bigl(a  - \frac{i}{4}\bigl(1+4p\bigr)\Bigr)\, ,\quad p = 0, 1, 2, \dots\, .\ee

Amongst all the quasi-normal frequencies $\nuqn^{(q)}$, the one having the largest time scale \eqref{tqnpdef} dominates at long times. It is called the leading quasi-normal frequency and will be denoted simply by $\nuqn$,
\be\label{GRlargetasump} \Gr(t)\underset{t\rightarrow +\infty}{\propto} e^{-i\nuqn t}\, .\ee
The associated time scale
\be\label{tqnscale} \tqn = -\frac{1}{\im\nuqn}\ee
is often referred to in the literature as the thermalization or dissipation time scale.

\begin{figure}[h!]
\centering
\begin{tabular}{cc}
\def\svgwidth{7cm}
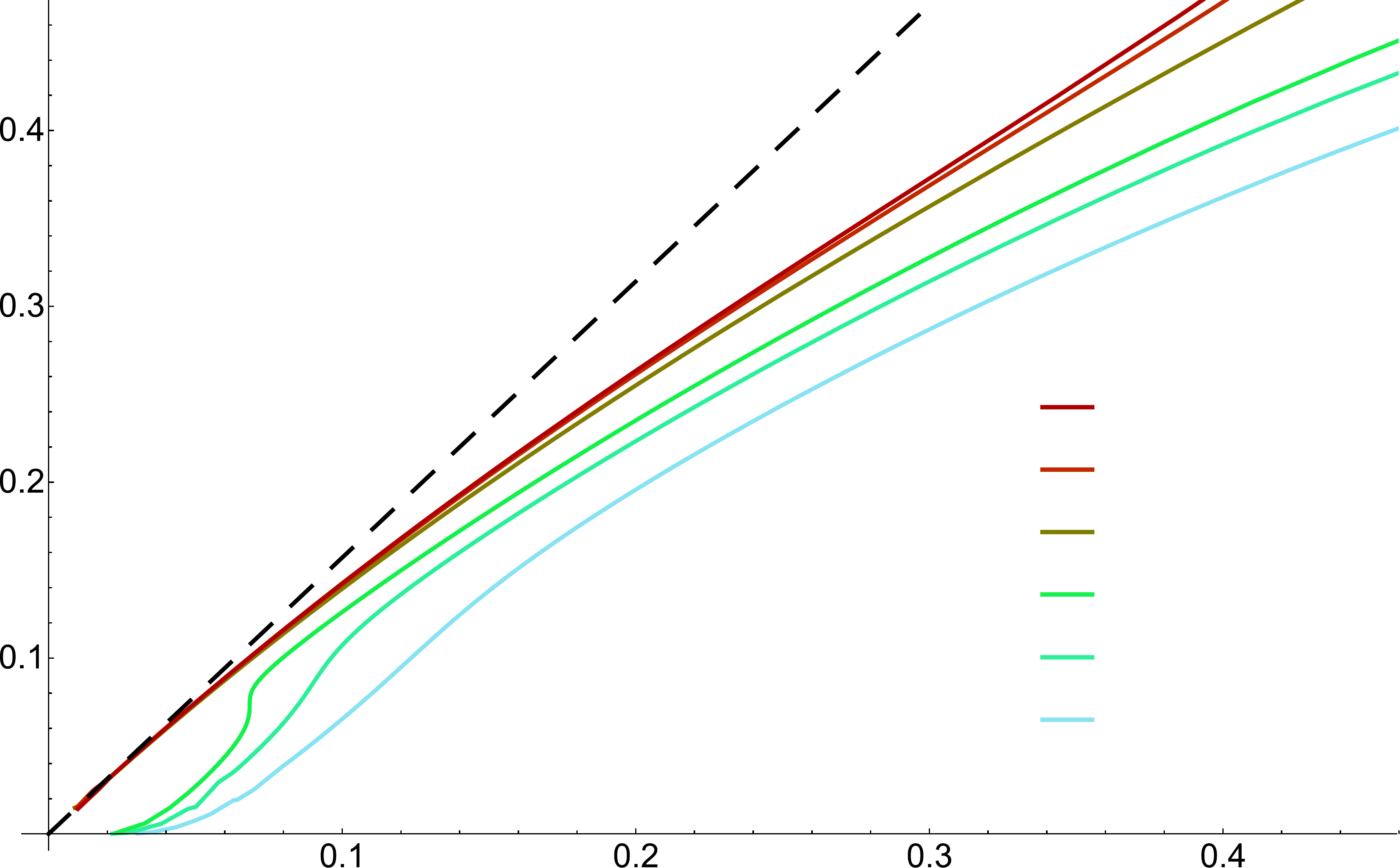 &
\def\svgwidth{7cm}
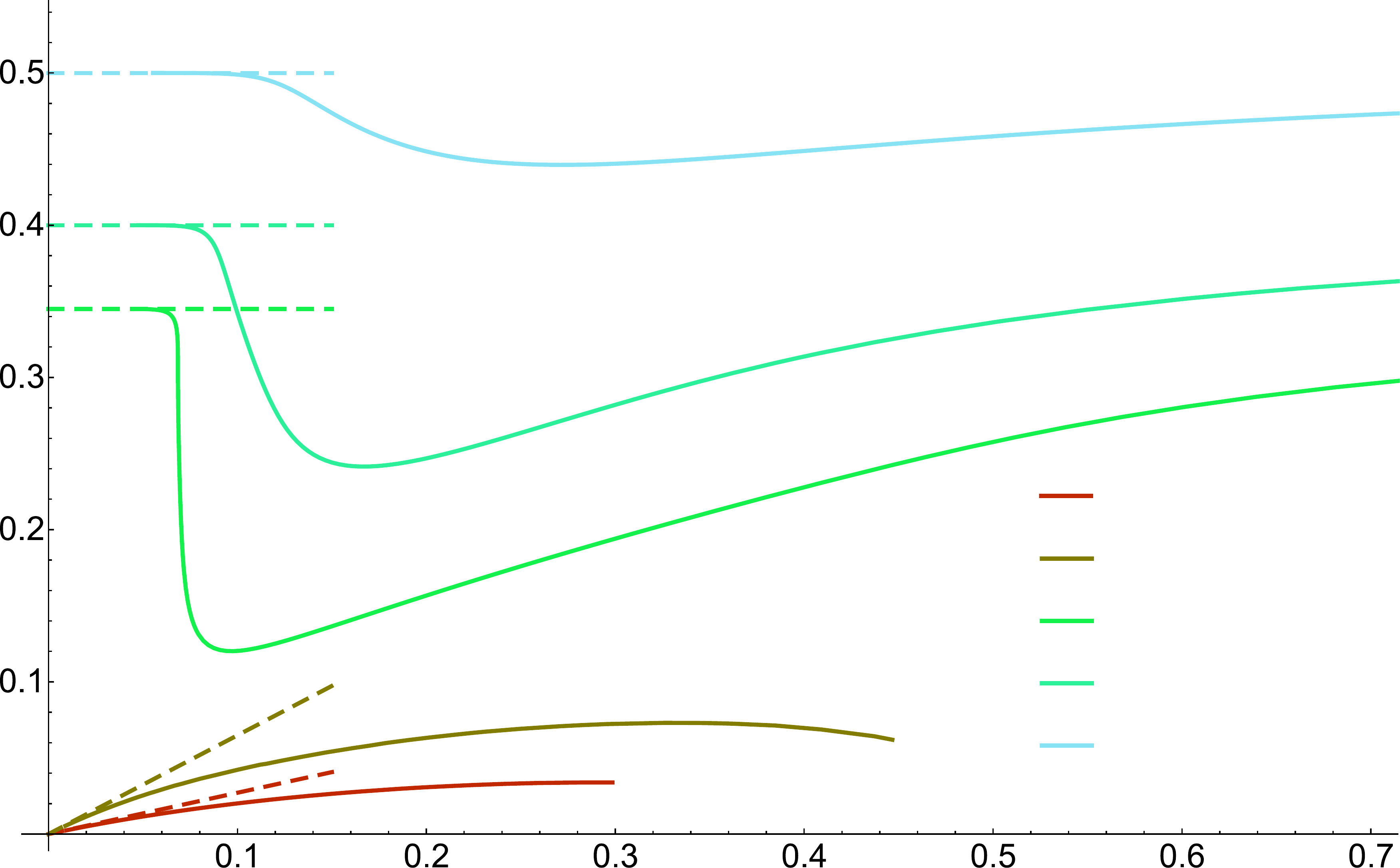 \\
(a) & (b)
\end{tabular}
\caption{\label{fig:M1QNF}The leading quasi-normal frequency $\nuqn$ in the \modone\ model, as a function of the temperature and for a few fixed masses. The low-temperature behaviour for the SYK-like phase with $m<m_* = \mStarOne$ is given by Eq.~\eqref{qnzeroTmod1}. (a) Imaginary part $-\im\nuqn$, matching $-\im\nuqn = \pi T/2$ for the SYK-like phase at small $T$ (black dashed line). (b) Real part $\re\nuqn$, matching $\re\nuqn = 2\pi a T$ for the SYK-like phase and $\re\nuqn = m$ for the HO-like phase at small $T$ (dashed lines, with the parameter $a$ extrapolated from the finite-temperature fits of $\sim10$ solutions with $\beta\in[50,100]$).}
\end{figure}

In Fig.\ \ref{fig:M1QNF}, we plot $-\im\nuqn$ and $\re\nuqn$ as functions of the temperature, for various values of the mass in the \modone\ model. As explained in Sec.\ \ref{PhaseSec}, for masses $m<m_{*} = \mStarOne$ the IR physics is dominated by the SYK-like solution; our numerical results are then consistent with the asymptotic values  \eqref{qnzeroTmod1} for $p=0$. For masses $m>m_{*}$, on the other hand, we find that the behaviour of $-\im\nuqn$ is $o(T)$.

\begin{figure}[h!]
\centering
\vskip0.3cm
\begin{tabular}{cc}
\def\svgwidth{7cm}
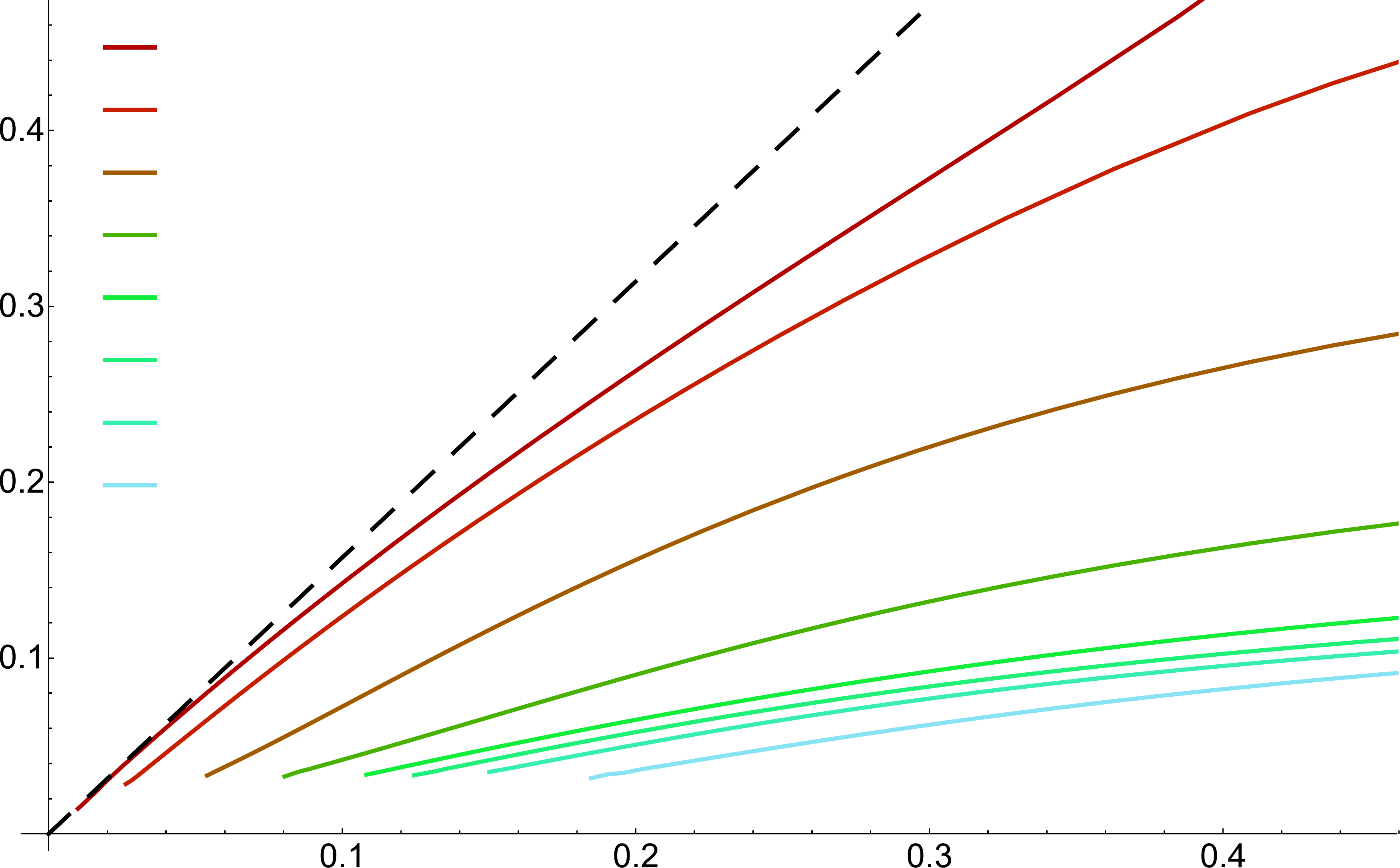 &
\def\svgwidth{7cm}
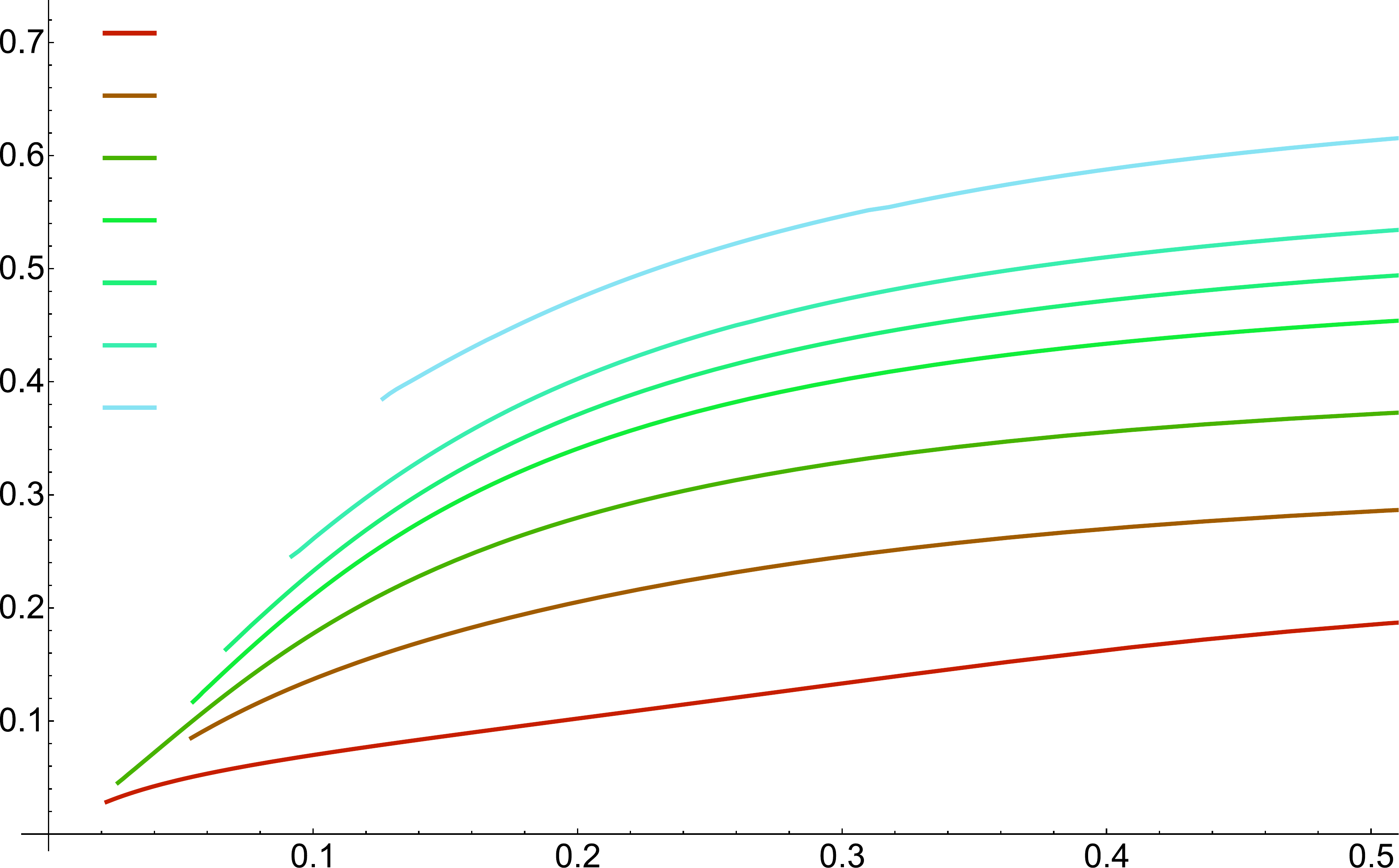 \\
(a) & (b)
\end{tabular}
\caption{\label{fig:M2QNF}The leading quasi-normal frequency $\nuqn$ in the \modtwo\ model, as a function of the temperature and for a few fixed masses. (a) Imaginary part $-\im\nuqn$; (b) Real part $\re\nuqn$. In the gapless phase, the low-temperature behaviour is linear with seemingly mass-dependent slopes, matching $-\im\nuqn = \pi T/2$ only for $m = 0$ (black dashed line).}
\end{figure}

The situation in the \modtwo\ model is illustrated in Fig.~\ref{fig:M2QNF}. For low enough masses, in the gapless phase, the low-temperature behaviour appears to be linear for both the real and imaginary parts of $\nuqn$, but the slope is seemingly dependent of the mass. Note that we have no analytical results to contrast with this observation. For larger masses, in the gapped phase, numerics do not allow us to reliably probe the very low temperature regime.

\begin{figure}[h!]
\centering
\begin{tabular}{cc}
\def\svgwidth{7cm}
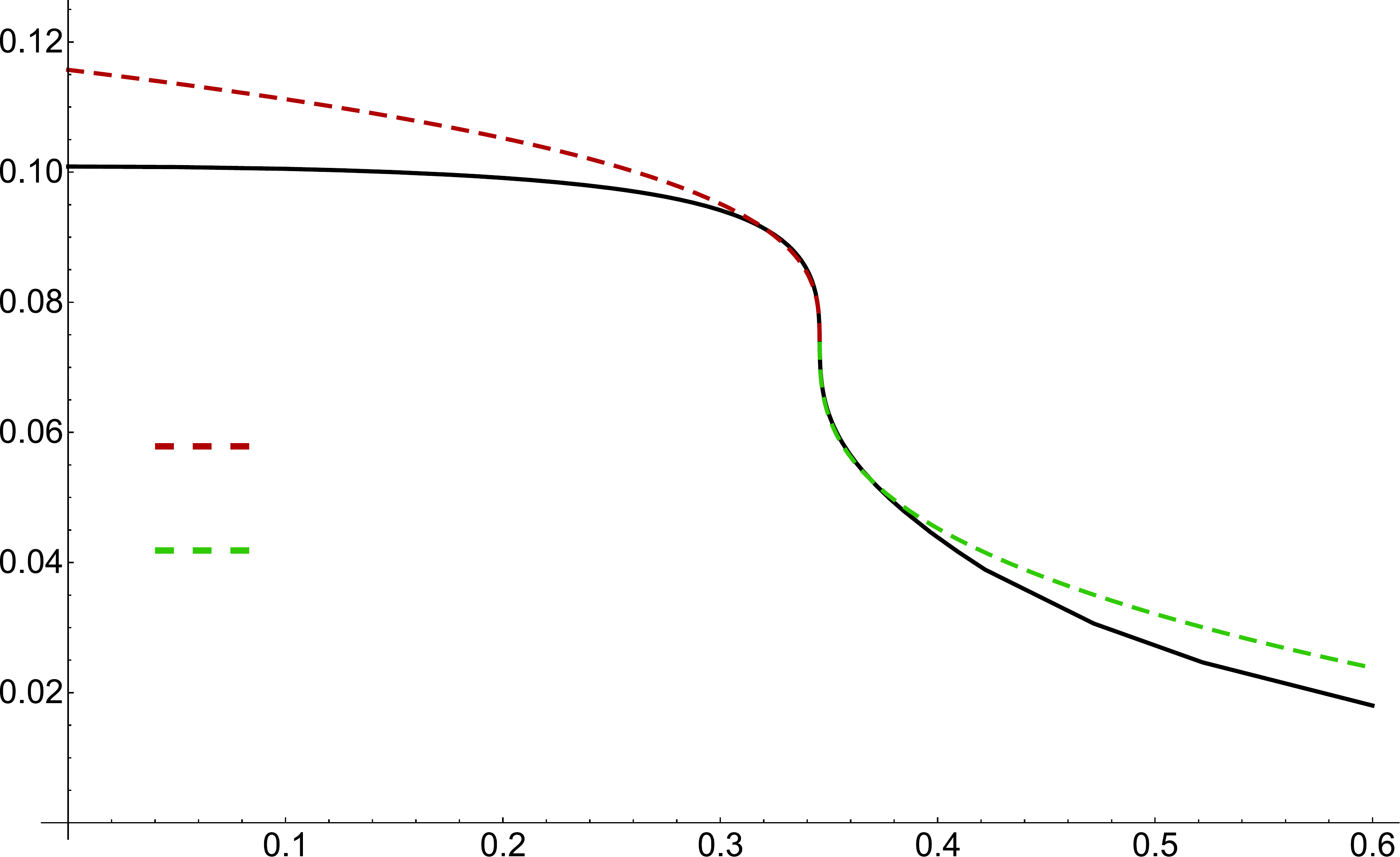 &
\def\svgwidth{7cm}
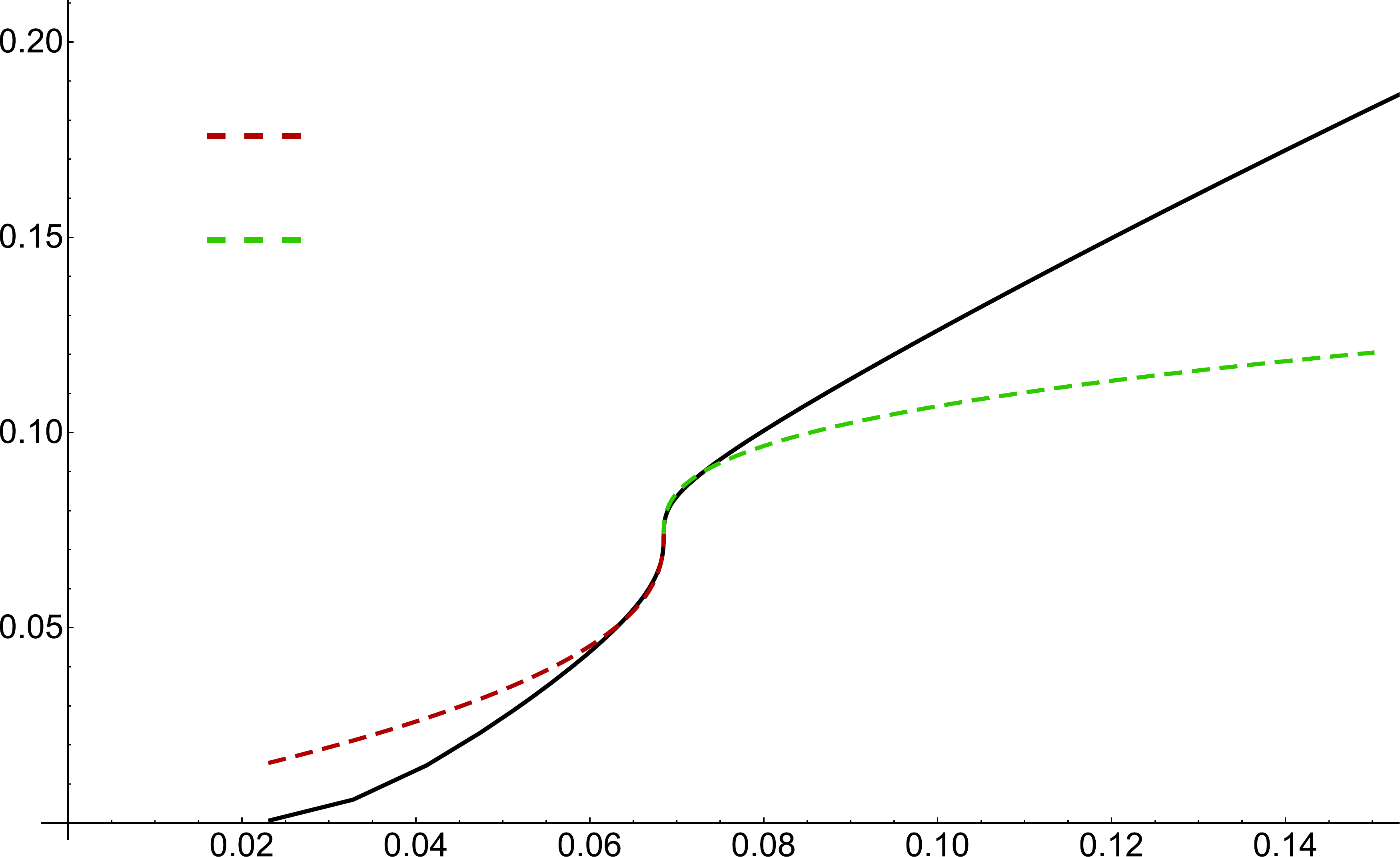 \\
(a) & (b)
\end{tabular}
\caption{\label{fig:M1QNFCritExp}Leading quasi-normal frequency $-\im\nuqn$ as a function of mass at $T=\Tc$ (a) and as a function of temperature at $m=\mc$ (b), together with the best power-law fits near the critical point $(\mc,\Tc)$ (dashed lines). Power-law fits were performed as in Sec.~\ref{Pha1exponentsqSec}, see \emph{e.g.}\ Fig.~\ref{fig:CriticalExponent}.}
\end{figure}

Just as with the thermodynamic quantities studied in Sec.~\ref{Pha1exponentsqSec}, the behaviour of the leading quasi-normal frequency near the critical point of the \modone\ model turns out to be very interesting. In Fig.~\ref{fig:M1QNFCritExp} we plot $-\im\nuqn$ as a function of the mass at $T=\Tc$ and as a function of the temperature at $m=\mc$. The results are consistent with a critical power-law behaviour,
\be\label{critqndef1} \frac{\partial\im\nuqn}{\partial m}\bigl(m,\Tc\bigr) \underset{m\rightarrow\mc^{\pm}}{\sim}\frac{\nu_{\pm}}{\bigl|m-\mc\bigr|^{f_{\pm}}}\,\cvp\quad
\frac{\partial\im\nuqn}{\partial T}\bigl(\mc,T\bigr) \underset{T\rightarrow\Tc^{\pm}}{\sim}\frac{\tilde\nu_{\pm}}{\bigl|T-\Tc\bigr|^{\tilde f_{\pm}}}\,\cvp\ee
with critical exponents found to be
\be\label{fexpvalues} f_{+}=0.64_1\, ,\quad f_{-}=0.67_1\, ,\quad \tilde f_{+} = 0.64_2\, ,\quad \tilde f_{-} = 0.57_2\, .
\ee
\subsection{\label{real4ptSec}Real time four-point functions}

This subsection is devoted to the computation of real time four-point functions, in particular in the out-of-time-ordered (OTOC) case which is relevant to probe chaos \cite{Larkin,Kitaev,chaosrefs,Maldacena:2015waa}. We review basic properties in Sec.~\ref{real4ptgenSec} and the Schwinger-Keldysh set-up in Sec.~\ref{realSD4ptSec}. We introduce the Lyapunov exponent $\lyap$ in Sec.~\ref{OTOCSec} and explain in Sec.~\ref{OTOCSec2} how the exact formulas derived in Sec.~\ref{realSD4ptSec} can be simplified if one wishes to compute $\lyap$ only, instead of the detailed time evolution of the four-point function. The simple numerical algorithms presented in Sec.~\ref{LyapnumSec} are then used in Sec.~\ref{LyapAppSec} to compute $\lyap$ in various cases. We uncover in the process new and interesting features. In particular, we find that the chaos bound of \cite{Maldacena:2015waa} is saturated in the SYK-like phase of the \modone\ model, but find evidence that this is not so in the \modtwo\ model except at strictly $m=0$. We also find that the Lyapunov exponent has a singular behaviour near the critical point of the \modone\ model, governed by non-trivial critical exponents.

\subsubsection{\label{real4ptgenSec}General definitions and properties}

A four-point function for complex time arguments $u_{i}$ satisfying $0<\tau_{u_{i}}<\beta$, generalizing \eqref{PhiEucliddef}, can be defined by
\be\label{Phirealdef}
\mathcal \phi(u_{1},u_{2},u_{3},u_{4}) = 
\begin{cases}\displaystyle
\frac{1}{n^{2}}\bigl\langle\text{T}\tr\psi_{\mu}(u_{1})\psi^{\dagger}_{\mu}(u_{2})\tr\psi_{\nu}(u_{3})\psi^{\dagger}_{\nu}(u_{4})\bigr\rangle_{\beta}&\text{(matrix models)}\\\displaystyle
\bigl\langle\text{T}\chi^{i}(u_{1})\chi^{\dagger}_{i}(u_{2})\chi^{j}(u_{3})\chi^{\dagger}_{j}(u_{4})\bigr\rangle_{\beta}&\text{(disordered models)}\, ,
\end{cases}
\ee
where as usual T indicates the ordering with respect to the Euclidean times $\tau_{u_{i}}=\re u_{i}$. This definition can be extended to all values of $\tau_{u_{i}}$ using the KMS condition
\be\label{phiantireal} \phi(u_{1}+n_{1}\beta,u_{2}+n_{2}\beta,u_{3}+n_{3}\beta,u_{4}+n_{4}\beta) = (-1)^{n_{1}+n_{2}+n_{3}+n_{4}}\phi (u_{1},u_{2},u_{3},u_{4})\, ,\ n_{i}\in\mathbb Z\, .\ee
This four-point function satisfies the properties
\begin{align}
\label{fourptrealsym1} & \phi (u_{1},u_{2},u_{3},u_{4})^{*}=\phi(-u_{4}^{*},-u_{3}^{*},-u_{2}^{*},-u_{1}^{*})\,,\\
\label{fourptrealsym2} & \phi (u_{1},u_{2},u_{3},u_{4}) = \phi (u_{3},u_{4},u_{1},u_{2})\,,\\
\label{fourptrealsym3} & \phi (u_{1},u_{2},u_{3},u_{4})^{*}=\phi (u_{1}^{*},u_{2}^{*},u_{3}^{*},u_{4}^{*})\, ,
\end{align}
generalizing the similar conditions \eqref{fourptsym1}, \eqref{fourptsym2} and \eqref{fourptsym3} in the Euclidean setting. One may decompose
\be\label{phidecperm} \phi(u_{1},u_{2},u_{3},u_{4}) = \sum_{\sigma\in\mathfrak S_{4}}\Theta(\tau_{u_{\sigma(1)}},\tau_{u_{\sigma(2)}},\tau_{u_{\sigma(3)}},\tau_{u_{\sigma(4)}}) \phi_{\sigma}(u_{1},u_{2},u_{3},u_{4})\, ,\ee
where the sum is over all the elements of the group $\mathfrak S_{4}$ of permutations of four elements,
\be\label{Thetagendef} \Theta (\tau_{1},\tau_{2},\tau_{3},\tau_{4}) = \Theta (\tau_{1}-\tau_{2})\Theta (\tau_{2}-\tau_{3})\Theta (\tau_{3}-\tau_{4})\ee
orders the Euclidean times and $\phi_{\sigma}$ is one of the $4!=24$ possible four-point correlators with a fixed given ordering of the operators. Note that, by construction, the Euclidean times appearing as the arguments of $\phi_{\sigma}$ are always ordered according to the permutation $\sigma$, with $\smash{\tau_{u_{\sigma(1)}}>\tau_{u_{\sigma(2)}}>\tau_{u_{\sigma(3)}}>\tau_{u_{\sigma(4)}}}$ and $\smash{0<\tau_{u_{\sigma(1)}}-\tau_{u_{\sigma(4)}}<\beta}$. The real times (imaginary parts of the $u_{i}$), on the other hand, are unconstrained. Any purely real time four-point function can be obtained by taking the Euclidean times to zero while keeping their ordering fixed. Spectral decomposition formulas read
\be\label{specdec4pt} \phi_{\sigma}(u_{1},u_{2},u_{3},u_{4}) = 
\int_{-\infty}^{+\infty}\prod_{j=1}^{4}\bigl[\d\omega_{j}\bigr]\delta\Bigl(\sum_{j=1}^{4}\omega_{j}\Bigr) \mathfrak r_{\sigma}(\omega_{1},\omega_{2},
\omega_{3},\omega_{4})\, e^{-\sum_{j=1}^{4}\omega_{j}u_{j}}\,,\ee
with spectral functions that can be expressed in terms of sums over the eigenstates $|p\rangle$ and eigenvalues $E_{p}$ of the Hamiltonian,
\begin{multline}\label{rsigmaform}\delta\Bigl(\sum_{j=1}^{4}\omega_{j}\Bigr) \mathfrak r_{\sigma}(\omega_{1},\omega_{2},
\omega_{3},\omega_{4}) =
\delta\Bigl(\sum_{j=1}^{4}\omega_{j}\Bigr)\frac{1}{Z}\sum_{p,q,r,s}e^{-\beta E_{p}}\alpha_{pqrs}^{\sigma}\\\delta(\omega_{\sigma(1)}+E_{p}-E_{q})\delta(\omega_{\sigma(2)}+E_{q}-E_{r})\delta(\omega_{\sigma(3)}+E_{r}-E_{s})\, .
\end{multline}
The coefficients $\smash{\alpha_{pqrs}^{\sigma}}$ depend on the matrix elements of the operators and on the permutation $\sigma$ in a natural way. If we use the explicit notation
\be\label{sigexplicit} \sigma = [\sigma(1)\sigma(2)\sigma(3)\sigma(4)]\,,\ee
we have for example in the matrix model notation
\begin{align}\label{alphagenex1} &\alpha_{pqrs}^{[1234]} = \langle p|\psi^{a}_{\mu\, b}|q\rangle
\langle q|\psi^{\dagger b}_{\mu\, a}|r\rangle\langle r|\psi^{c}_{\nu\, d}|s\rangle
\langle s|\psi^{\dagger d}_{\nu\, c}|p\rangle\,,\\\label{alphagenex2} 
& \alpha_{pqrs}^{[2134]} = -\langle p|\psi^{\dagger b}_{\mu\, a}|q\rangle
\langle q|\psi^{a}_{\mu\, b}|r\rangle\langle r|\psi^{c}_{\nu\, d}|s\rangle
\langle s|\psi^{\dagger d}_{\nu\, c}|p\rangle\, ,\ \text{\emph{etc}.}
\end{align}

Correlators with the same cyclic ordering of the operators are related by the KMS conditions \eqref{phiantireal}. These conditions follow, for instance, from the cyclicity property of the coefficients $\smash{\alpha_{pqrs}^{\sigma}}$,
\be\label{KMScoeffrel}\alpha^{[\sigma(1)\sigma(2)\sigma(3)\sigma(4)]}_{pqrs} = -
\alpha^{[\sigma(4)\sigma(1)\sigma(2)\sigma(3)]}_{spqr}\, .\ee
In terms of the spectral functions \eqref{rsigmaform}, they read
\be\label{KMSonspec} \mathfrak r_{[\sigma(4)\sigma(1)\sigma(2)\sigma(3)]}(\omega_{1},\omega_{2},
\omega_{3},\omega_{4}) =-e^{\beta\omega_{\sigma(4)}} \mathfrak r_{[\sigma(1)\sigma(2)\sigma(3)\sigma(4)]}(\omega_{1},\omega_{2},
\omega_{3},\omega_{4})\, .\ee
The KMS conditions thus reduce from $4!=24$ to $3!=6$ the number of independent spectral functions or correlators. Moreover, in our particular case one may check that the additional symmetry \eqref{fourptrealsym2} leaves only four independent spectral functions or correlators,
\begin{align}
\label{phi1234} &\phi_{[1234]}(u_{1},u_{2},u_{3},u_{4}) = \bigl\langle\psi^{a}_{\mu\, b}(u_{1})\psi^{\dagger b}_{\mu\, a}(u_{2})\psi^{c}_{\nu\, d}(u_{3})\psi^{\dagger d}_{\nu\, c}(u_{4})\bigr\rangle_{\beta}\,,\\
\label{phi2134} &\phi_{[2134]}(u_{1},u_{2},u_{3},u_{4}) = -\bigl\langle\psi^{\dagger b}_{\mu\, a}(u_{2})\psi^{a}_{\mu\, b}(u_{1})\psi^{c}_{\nu\, d}(u_{3})\psi^{\dagger d}_{\nu\, c}(u_{4})\bigr\rangle_{\beta}\,,\\
\label{phi1324} &\phi_{[1324]}(u_{1},u_{2},u_{3},u_{4}) = -\bigl\langle\psi^{a}_{\mu\, b}(u_{1})\psi^{c}_{\nu\, d}(u_{3})\psi^{\dagger b}_{\mu\, a}(u_{2})\psi^{\dagger d}_{\nu\, c}(u_{4})\bigr\rangle_{\beta}\,,\\
\label{phi3214} &\phi_{[3214]}(u_{1},u_{2},u_{3},u_{4}) = -\bigl\langle\psi^{c}_{\nu\, d}(u_{3})\psi^{\dagger b}_{\mu\, a}(u_{2})\psi^{a}_{\mu\, b}(u_{1})\psi^{\dagger d}_{\nu\, c}(u_{4})\bigr\rangle_{\beta}\, .
\end{align}
The other two \emph{a priori} independent correlators are given by
\begin{align}\label{phi2314} & \phi_{[2314]}(u_{1},u_{2},u_{3},u_{4}) = -\phi_{[1324]}(u_{3},u_{4},u_{1},u_{2}-\beta)\,,\\\label{phi3124}
& \phi_{[3124]}(u_{1},u_{2},u_{3},u_{4}) = -\phi_{[2134]}(u_{3},u_{4}+\beta,u_{1},u_{2})\, .
\end{align}

The above extends the discussion of Sec.\ \ref{defrealtimeSec} for the two-point functions, for which there is only one independent spectral function. It can be straightforwardly generalized to arbitrary $p$-point functions, with generically $(p-1)!$ independent spectral functions, see \emph{e.g.}\ \cite{gencorrref} for a thorough discussion.

\subsubsection{\label{realSD4ptSec}Melonic real time four-point functions}

\begin{figure}[h!]
\centering
\def\svgwidth{4.5in}
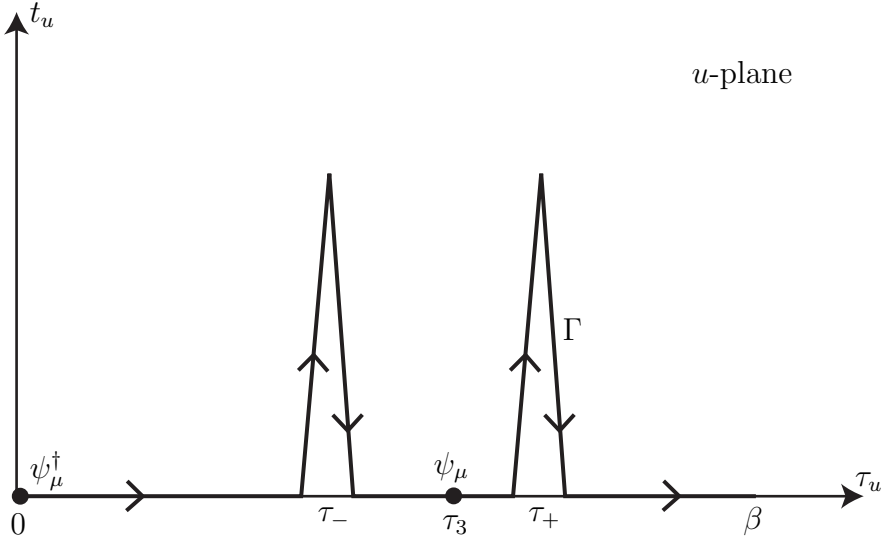
\caption{Keldysh contour $\Gamma$ (thick line) used to compute the four-point functions with a choice of triplet $(\tau_{-},\tau_{3},\tau_{+})$ for the Euclidean part of their time arguments. The triangle-shaped pieces of the contour are centered on the Euclidean times $\tau_{-}$ and $\tau_{+}$. Their width $\epsilon$ is taken to zero and their height to infinity. Fixed insertions at $u=0$ and $u=\tau_{3}$ are indicated.\label{SKcontour2}}
\end{figure}

By time translation invariance, we can restrict our attention to $\phi(u_{1},u_{2},u_{3},u_{4}=0)$. For the purpose of studying chaos, along the lines of \cite{Kitaev,Maldacena:2015waa}, we can further limit ourselves to a real
\be\label{u3real} u_{3}=\tau_{3}\, .\ee
We also fix the real parts of $u_{1}$ and $u_{2}$ to be $\tau_{+}$ and $\tau_{-}$ (or $\tau_{-}$ and $\tau_{+}$) such that
\be\label{choicetaupm}
0<\tau_{-}<\tau_{3}<\tau_{+}<\beta\,,
\ee
and we note
\be\label{t1t2def} t_{1}=\im u_{1}\, ,\quad t_{2}=\im u_{2}\, .\ee
The associated Keldysh contour $\Gamma$ for these choices of insertion points has two triangle-shaped pieces and is depicted in Fig.\ \ref{SKcontour2}. The height of the triangles must be at least $\max (t_{1},t_{2})$ but may conveniently be taken to be infinite.

As already emphasized in Sec.\ \ref{realSDSec}, the advantage of the Keldysh contour formalism is that the Feynman rules and diagrammatics take the familiar form, with only simple and obvious modifications with respect to the standard Euclidean case. In particular, the arguments presented in Sec.\ \ref{EuclidefourSec} go through. The leading Feynman graphs are still the ladders depicted in Fig.\ \ref{figladder}, \ref{figladder1} and \ref{figladder2}, but now the positions of the vertices have to be integrated over the Keldysh contour $\Gamma$ instead of the Euclidean interval $[0,\beta]$. If we decompose
\be\label{PhiFreal} \phi(u_{1},u_{2},u_{3},u_{4})= G(u_{1},u_{2})G(u_{3},u_{4}) + \frac{1}{N}\mathcal F(u_{1},u_{2},u_{3},u_{4})\,,\ee
as in \eqref{PhiFref}, $\mathcal F$ is determined by the equation
\be\label{SDreal4pteq} \mathcal F(u_{1},u_{2},\tau_{3},0) = \mathcal F_{0} (u_{1},u_{2},\tau_{3},0)+ \int_{\Gamma\times\Gamma}\mathcal K(u_{1},u_{2},u,u')\mathcal F(u,u',\tau_{3},0)\, \d u \d u' .\ee
Generalizing the matrix multiplication rule \eqref{matrixnot} to
\be\label{matrixnotreal} (\mathbb A\cdot\mathbb B)(u_{1},u_{2},u_{3},u_{4}) = \int_{\Gamma\times\Gamma}\mathcal A(u_{1},u_{2},u,u')\mathcal B(u,u',u_{3},u_{4})\,\d u\d u'\, ,\ee
with integrals taken along $\Gamma$, \eqref{SDreal4pteq} is solved by resumming a geometric series or equivalently inverting the matrix $\mathbb I-\mathbb K$,
\be\label{Ffromkernel2} \mathbb F = \sum_{n\geq 0}\mathbb K^{n}\cdot\mathbb F_{0}=(\mathbb I-\mathbb K)^{-1}\cdot\mathbb F_{0}\, ,\ee
exactly as in \eqref{Ffromkernel}. The function $\mathcal F_{0}$ and the kernel $\mathcal K$ are given by the same formulas as in the Euclidean case, \eqref{F0def} and \eqref{Kone} (in the \modone\ model) or \eqref{Ktwo} (in the \modtwo\ model), with arguments $(\tau_{1},\tau_{2},\tau_{3},\tau_{4})$ replaced by $(u_{1},u_{2},u_{3},u_{4})$ and the Euclidean two-point function $G(\tau_{1},\tau_{2})$ replaced by the more general complex time two-point function $G(u_{1},u_{2})$ defined in \eqref{Gcplxedef}.

The points $u_{1}$ and $u_{2}$ in \eqref{SDreal4pteq} may be chosen to be anywhere on the contour $\Gamma$. The real time OTOCs correspond to insertions on the two triangles, \emph{i.e.}
\be\label{u1u2relevant} u_{1} = \tau_{+}+i t_{1}\, ,\ u_{2} = \tau_{-}+i t_{2}\quad\text{or}\quad u_{1} = \tau_{-}+i t_{1}\, ,\ u_{2} = \tau_{+}+i t_{2}\, ,\ee
which yield
\begin{align}\label{OTOC1} 
\begin{split}\mathscr C_{1}(t_{1},t_{2}) &=-\phi_{[1324]}(\tau_{+}+i t_{1},\tau_{-}+i t_{2},\tau_{3},0)\\&
=\bigl\langle 
\psi^{a}_{\mu\, b}(\tau_{+}+it_{1})\psi^{c}_{\nu\, d}(\tau_{3})\psi^{\dagger b}_{\mu\, a}(\tau_{-}+it_{2})\psi^{\dagger d}_{\mu\, c}\bigr\rangle_{\beta}\,,
\end{split}\\
\label{OTOC2}
\begin{split}
\mathscr C_{2}(t_{1},t_{2}) &= \phi_{[2314]}(\tau_{-}+i t_{1},\tau_{+}+i t_{2},\tau_{3},0)\\ &=\bigl\langle 
\psi^{\dagger b}_{\mu\, a}(\tau_{+}+it_{2})\psi^{c}_{\nu\, d}(\tau_{3})
\psi^{a}_{\mu\, b}(\tau_{-}+it_{1})\psi^{\dagger d}_{\mu\, c}\bigr\rangle_{\beta}\, .
\end{split}
\end{align}
These correlators are decomposed, following \eqref{PhiFreal}, as
\begin{align}\label{C1dec} &\mathscr C_{1}(t_{1},t_{2}) = - G\bigl(\tau_{+}-\tau_{-}+i(t_{1}-t_{2})\bigr)G\bigl(\tau_{3}\bigr) + \frac{1}{N} C_{1}(t_{1},t_{2})\,,\\\label{C2dec} &\mathscr C_{2}(t_{1},t_{2}) = G\bigl(\tau_{-}-\tau_{+}+i(t_{1}-t_{2})\bigr)G\bigl(\tau_{3}\bigr) + \frac{1}{N} C_{2}(t_{1},t_{2})\,,
\end{align}
with
\begin{align}\label{C1Frel} & C_{1}(t_{1},t_{2}) = -\mathcal F(\tau_{+}+i t_{1},\tau_{-}+i t_{2},\tau_{3},0)\,,\\\label{C2Frel}&
C_{2}(t_{1},t_{2}) = \mathcal F(\tau_{-}+i t_{1},\tau_{+}+i t_{2},\tau_{3},0)\, .
\end{align}
The two connected OTOCs \eqref{C1Frel} and \eqref{C2Frel} can be inserted on the left-hand side of Eq.\ \eqref{SDreal4pteq} by setting $u_{1}$ and $u_{2}$ to the appropriate values \eqref{u1u2relevant}. However, it is important to realize that several other types of correlators then necessarily appear on the right-hand side of \eqref{SDreal4pteq} when one performs the integrals over $\Gamma$. This includes, for example, correlators like
\be\label{tocs}\bigl\langle 
\psi^{a}_{\mu\, b}(\tau_{+}+it_{1})\psi^{\dagger b}_{\mu\, a}(\tau_{+}+it_{2})\psi^{c}_{\nu\, d}(\tau_{3})\psi^{\dagger d}_{\mu\, c}\bigr\rangle_{\beta}\, ,\, \bigl\langle \psi^{\dagger b}_{\mu\, a}(\tau_{+}+it_{2})\psi^{a}_{\mu\, b}(\tau_{+}+it_{1})\psi^{c}_{\nu\, d}(\tau_{3})\psi^{\dagger d}_{\mu\, c}\bigr\rangle_{\beta}\, ,\ee
when $u$ and $u'$ are both on the triangle at $\re u=\re u' =\tau_{+}$, similar correlators when $u$ and $u'$ are both on the triangle at $\re u=\re u' =\tau_{-}$, and also correlators with only one real time turned on or purely Euclidean correlators when both $u$ and $u'$ are on the Euclidean segments of $\Gamma$. 

It is thus impossible to write an exact closed-form system of integral equations similar to \eqref{SDreal4pteq} involving the OTOCs alone.\footnote{This point seems to be overlooked in part of the existing literature.} However, we shall explain in Sec.~\ref{OTOCSec2} that such an \emph{approximate} system can be used if we are only concerned with evaluating the asymptotic, long time behaviour. It is this behaviour which is relevant to compute the chaos Lyapunov exponents, which we discuss in the following.

\subsubsection{\label{OTOCSec}On Lyapunov exponents}

Let us start by summarizing the standard lore on the late real time behaviour of four-point functions such as $\phi(u_{1},u_{2},\tau_{3},0)$ in models like the ones we are studying. Recall that $t_{1}$ and $t_{2}$ are defined in \eqref{t1t2def}; we want to discuss the regime where $t=t_{1}=t_{2}$ is very large.

We assume that the number $\dof$ of degrees of freedom is large but finite. This is important, because time scales of the order of $\ln N$ play a role in the discussion.  Much larger time scales, like the Poincar\'e recurrence time scale, which are exponentially large with $N$, are irrelevant for our purposes and can be considered to be infinite.

Depending on the ordering of the operators in the correlators, \emph{i.e.}\ on the positions of $u_{1}$ and $u_{2}$ on the Keldysh contour of Fig.\ \ref{SKcontour2}, two qualitatively distinct behaviours are expected:
\begin{enumerate}[i)]
	\item If $u_{1}$ and $u_{2}$ are on the same triangle-shaped piece of $\Gamma$, then the correlator will factorize and approach a constant value as soon as $t=t_{1}=t_{2}$ is much larger than the dissipation time scale defined in \eqref{tqnscale},
\be\label{phifactor1} \phi(u_{1},u_{2},\tau_{3},0)\underset{t=t_{1}=t_{2}\gg\tqn}\simeq G_{\pm}(0)G(\tau_{3})\, .\ee
The function $G_{+}$ or $G_{-}$ appears on the right-hand side of \eqref{phifactor1} depending on the relative position of $u_{1}$ and $u_{2}$ along $\Gamma$. Note that the result \eqref{phifactor1} is consistent with the large $N$ expansion, since the right-hand side of this equation matches with the first term on the right-hand side of \eqref{PhiFreal}. In other words, the long time and large $N$ limits commute for this class of correlators.\footnote{As indicated above, we do not consider time scales larger than $\ln N$ in our discussion.} 
	\item If $u_{1}$ and $u_{2}$ are on distinct triangle-shaped pieces of $\Gamma$, which is the case corresponding to the OTOCs \eqref{OTOC1} and \eqref{OTOC2}, it is expected that the correlator will go to zero when $t=t_{1}=t_{2}\rightarrow\infty$,
\be\label{phizerolimit} \lim_{t=t_{1}=t_{2}\rightarrow +\infty}\phi(u_{1},u_{2},\tau_{3},0)= 0\, .\ee
Intuitively, this is because the insertions at $u_{1}$ and $u_{2}$ are infinitely far apart on the Keldysh contour $\Gamma$ in the limit $t\rightarrow\infty$. The result \eqref{phizerolimit} can actually be used as a good criterion to detect chaos in a strongly coupled quantum system \cite{Maldacena:2015waa}. Indeed, \eqref{phizerolimit} is inconsistent with the first term on the right-hand side of \eqref{PhiFreal}, which predicts \eqref{phifactor1} at $\dof\to\infty$. In other words, the long time and large $\dof$ limits do not commute for the OTOCs. The $\mathcal F/\dof$ term, which naively is a small correction at large $\dof$, must eventually become of order one to cancel out the constant $GG$ term. More precisely, it is expected that
\be\label{phiLyap} \mathcal F(u_{1},u_{2},\tau_{3},0)\propto e^{\lyap t}\ee
as soon as $t=t_{1}=t_{2}\gg\tqn$, the exponential growth of the ``small'' correction $\mathcal F/\dof$ signaling the onset of chaos. The constant $\lyap$ is called the Lyapunov exponent. Eventually, the OTOCs should vanish on a time scale, called the scrambling time scale, that can be estimated by setting $\mathcal F/\dof\sim 1$ in \eqref{phiLyap},
\be\label{scrambling} t_{\text{scr}} \approx \frac{\ln N}{\lyap}\,\cdotp\ee
\end{enumerate}

A few clarifying remarks may be made at this stage. First, in the strict $N\rightarrow\infty$ limit studied in the present paper the scrambling time $t_{\text{scr}}\sim\ln \dof$ is not seen and the vanishing of the OTOCs cannot be rigorously derived. This is a non-perturbative effect in the large $\dof$ expansion.

On the other hand, the behaviour \eqref{phiLyap} can in principle be derived by solving the exact equation \eqref{SDreal4pteq}. In practice, this can be done only in the low-temperature limit and in the conformal phases of the models. The method, presented in \cite{Maldacena:2016hyu} for SYK (see also \cite{Murugan:2017eto,Peng:2017spg,Bulycheva:2017uqj,Yoon:2017nig}), does not use \eqref{SDreal4pteq} explicitly. Instead, one first works in the Euclidean setting and exploits the conformal symmetry to explicitly invert the operator $\mathbb I-\mathbb K$ in \eqref{Ffromkernel}, and then performs the appropriate analytic continuations to real time on the exact solution. The result is the famous formula
\be\label{LyaplowT} \lim_{T\rightarrow 0}\frac{\lyap}{T} = 2\pi\ee
for the Lyapunov exponent in the zero-temperature limit, which matches with the expectations from black hole physics and holography \cite{chaosrefs}. The universality of the result \eqref{LyaplowT} can be understood as being the consequence of the small explicit breaking of the low energy reparameterization invariance \eqref{diffeoG1}-\eqref{diffeoS1} by UV effects, which yields a universal Schwarzian action for the reparameterization zero modes from which the leading behaviour \eqref{phiLyap} can be obtained.

At non-zero temperature one can in principle derive the behaviour \eqref{phiLyap} by solving \eqref{SDreal4pteq} numerically. We first compute the kernel by using the solution for the two-point functions presented in Sec.\ \ref{rttwoptSec}, and then invert the operator $\mathbb I-\mathbb K$ to get $\mathcal F$ from \eqref{Ffromkernel2}. This procedure is in principle straightforward and allows the computation of the OTOCs from the very short time scales $0\leq t_{1},t_{2}\ll\tqn$ to the chaos regime $t_{1},t_{2}\gg\tqn$. However, the size of the matrix $\mathbb I-\mathbb K$ that one needs to invert to obtain reliable results when $t_{1},t_{2}\gg\tqn$ is very large. To the best of our knowledge, this has never been done and we let this interesting challenge for the future.

Now, if the detailed time evolution of the OTOCs is not required, but one is only interested in the regime $t\gg\tqn$, the exact equation \eqref{SDreal4pteq} can be simplified. This is sufficient to extract the Lyapunov exponents, and is explained in the next subsection. The drawback of this approach is that it relies on the standard expectations concerning the behaviour of the real time correlators reviewed above, discarding in a somewhat heuristic way many terms on the right-hand side of \eqref{SDreal4pteq}. It is therefore not a first principles derivation strictly proving that this standard lore is indeed correct. However, the results obtained in this way are fully consistent with the original hypothesis and we believe that they are very reliable, see Sec.\ \ref{LyapnumSec} for examples.

An important and well-motivated conjecture \cite{Maldacena:2015waa} about the Lyapunov exponents $\lyap$ is the famous upper bound
\be\label{Lyapbound} \lyap\leq 2\pi T\, .\ee
This bound is supposed to be valid in a large class of models for which a large hierarchy exists between the dissipation time scale and the scrambling time scale, $t_{\text{scr}}\gg t_{\text d}$ \cite{Maldacena:2015waa}. It is saturated by black holes and, as Eq.\ \eqref{LyaplowT} shows, by the SYK model and many generalizations thereof, including the \modone\ model in the standard conformal phase (as emphasized in Sec.\ \ref{PhaseSec}, the standard conformal phase is not realized in the \modtwo\ model; this will be further discussed in Sec.\ \ref{LyapnumSec}).

The OTOCs we are considering are labeled by an arbitrary triplet $(\tau_{-},\tau_{3},\tau_{+})$ constrained by \eqref{choicetaupm}. The most natural choice, which yields physical, real time finite temperature correlation functions, corresponds to taking the limits
\be\label{physicalcase} \tau_{+}\rightarrow 0\, ,\quad\tau_{3}\rightarrow 0\, ,\quad\tau_{-}\rightarrow 0\, ,\ee
keeping the ordering \eqref{choicetaupm} fixed. We shall call the associated OTOCs the ``physical OTOCs.'' The existing literature has instead mainly focused on a different choice, which is singled out in \cite{Maldacena:2015waa} and corresponds to 
\be\label{maldacase}\tau_{-}=\beta/4\, ,\quad\tau_{3}=\beta/2\, ,\quad\tau_{+}=3\beta/4\, .\ee
We will refer to these as the ``standard'' correlators, which can be interpreted as particular expectation values in a model for which we take two copies of the original Hilbert space and in a state which is a deformed version of the standard thermofield double state \cite{Maldacena:2015waa}.

It is not obvious whether the bound \eqref{Lyapbound} is valid independently of the choice of triplet $(\tau_{-},\tau_{3},\tau_{+})$, even though we believe it is very likely to be so in melonic models. We shall compute Lyapunov exponents for different choices of triplets $(\tau_{-},\tau_{3},\tau_{+})$ in Sec.\ \ref{LyapnumSec} and our results support this claim. However, it is important to stress that the proof presented in \cite{Maldacena:2015waa} does not apply to the general case of arbitrary $(\tau_{-},\tau_{3},\tau_{+})$ but only to the special choice $\tau_{-}=\beta/4$, $\tau_{3}=\beta/2$, $\tau_{+}=3\beta/4$.\footnote{We would like to thank Douglas Stanford for a discussion of this point.} Indeed, the reasoning crucially depends on the possibility to analytically continue the OTOC from the real time $t=t_{1}=t_{2}$ to a strip $-\beta/4 < \im t < \beta/4$ in the complex $t$-plane. This is indeed possible when $\tau_{-}=\beta/4$, $\tau_{3}=\beta/2$, $\tau_{+}=3\beta/4$, but, unfortunately, no such analytic continuation exists in the physical case. In general, an analytic continuation is possible in a strip $-\min (\tau_{3}-\tau_{-},\beta-\tau_{+})<\im t < \min (\tau_{-},\tau_{+}-\tau_{3})$, whose width can never exceed, but is in general strictly less than, $\beta/2$; the argument of \cite{Maldacena:2015waa} may then be used to derive an upper bound which is less strict than \eqref{Lyapbound}.\footnote{A general argument proving the bound when a Schwarzian action description is valid has also appeared very recently \cite{Yoonbound}.}

\subsubsection{\label{OTOCSec2}Computing Lyapunov exponents}

Lyapunov exponents are determined by analyzing the long time exponential growth of the connected OTOCs \eqref{C1Frel} and \eqref{C2Frel}, and comparing with \eqref{phiLyap}. We start by writing \eqref{SDreal4pteq} for the relevant values \eqref{u1u2relevant} of $u_{1}$ and $u_{2}$. Using the complex time version of \eqref{F0def}, we get
\begin{align}
\label{C1exact}
\begin{split}
C_{1}(t_{1},t_{2}) & = G_{+}(\tau_{+}+ i t_{1})G_{+}(\tau_{3}-\tau_{-}-it_{2}) \\
&\hskip 2cm - \int_{\Gamma\times\Gamma}\mathcal K (\tau_{+}+it_{1},\tau_{-}+it_{2},u,u')\mathcal F(u,u',\tau_{3},0)\,\d u\d u' \,,
\end{split}
\\ \label{C2exact}
\begin{split}
C_{2}(t_{1},t_{2}) & = -G_{+}(\tau_{-}+ i t_{1})G_{-}(\tau_{3}-\tau_{+}-it_{2}) \\ &\hskip 2cm  + \int_{\Gamma\times\Gamma}\mathcal K (\tau_{-}+it_{1},\tau_{+}+it_{2},u,u')\mathcal F(u,u',\tau_{3},0)\,\d u\d u'\, .
\end{split}
\end{align}
The idea is that we can simplify the right-hand sides of these equations, because we are interested in the regime in which the OTOCs are exponentially large. We thus keep only the terms that can develop such an exponential growth in the integrals and neglect all the others. From the discussion in Sec.~\ref{OTOCSec}, we expect that these terms correspond to the integration regions $u=\tau_{+}+ i t$ and $u'=\tau_{-}+ i t'$ or $u=\tau_{-}+ i t$ and $u'=\tau_{+}+ i t'$, for which $\mathcal F(u,u',\tau_{3},0)$ coincides with the OTOCs \eqref{C1Frel}, \eqref{C2Frel}. We are left with integrals running along the two sides of the triangles, which are non-trivial because the kernel function $\mathcal K$ can undergo discontinuities when $u$ or $u'$ goes through an insertion point at $u_{1}$ or $u_{2}$. To take into account these discontinuities, it is useful to decompose the kernel along the lines of \eqref{phidecperm}, 
\be\label{kerdecperm} \mathcal K(u_{1},u_{2},u_{3},u_{4}) = \sum_{\sigma\in\mathfrak S_{4}}\Theta(\tau_{u_{\sigma(1)}},\tau_{u_{\sigma(2)}},\tau_{u_{\sigma(3)}},\tau_{u_{\sigma(4)}}) \mathcal K_{\sigma}(u_{1},u_{2},u_{3},u_{4})\, .\ee
We get in this way a closed system of equations involving $C_{1}$ and $C_{2}$ only,
\begin{align}
\label{C1trunc}
\begin{split}
C_{1}(t_{1},t_{2}) & \approx G_{+}(\tau_{+}+ i t_{1})G_{+}(\tau_{3}-\tau_{-}-it_{2}) \\
&\hskip 0.5cm + \int_{0}^{+\infty}\Bigl[ \mathcal K_{1,1} (t_{1},t_{2},t,t') C_{1}(t,t') + \mathcal K_{1,2} (t_{1},t_{2},t,t') C_{2}(t,t')\Bigr]\,\d t\d t' \,,
\end{split}
\\ \label{C2trunc}
\begin{split}
C_{2}(t_{1},t_{2}) & \approx -G_{+}(\tau_{-}+ i t_{1})G_{-}(\tau_{3}-\tau_{+}-it_{2}) \\
&\hskip 0.5cm + \int_{0}^{+\infty}\Bigl[ \mathcal K_{2,1} (t_{1},t_{2},t,t') C_{1}(t,t') + \mathcal K_{2,2} (t_{1},t_{2},t,t') C_{2}(t,t')\Bigr]\,\d t\d t' \, ,
\end{split}
\end{align}
with
\be\label{Kij}
\begin{aligned}
\begin{split} & \mathcal K_{1,1}(t_{1},t_{2},t_{3},t_{4}) = -\Theta(t_{1}-t_{3})\Theta(t_{2}-t_{4})\\
&\hskip 0.85cm \bigl[\mathcal K_{[1324]} - \mathcal K_{[1342]} - \mathcal K_{[3124]}+\mathcal K_{[3142]}\bigr]\bigl(\tau_{+}+ i t_{1},\tau_{-}+i t_{2},\tau_{+}+it_{3},\tau_{-}+it_{4}\bigr)\,,
\end{split}
\\
\begin{split} & \mathcal K_{1,2}(t_{1},t_{2},t_{3},t_{4}) = \Theta(t_{2}-t_{3})\Theta(t_{1}-t_{4})\\
&\hskip 0.85cm \bigl[\mathcal K_{[1423]} - \mathcal K_{[4123]} - \mathcal K_{[1432]}+\mathcal K_{[4132]}\bigr]\bigl(\tau_{+}+ i t_{1},\tau_{-}+i t_{2},\tau_{-}+it_{3},\tau_{+}+it_{4}\bigr)\,,
\end{split}
\\
\begin{split} & \mathcal K_{2,1}(t_{1},t_{2},t_{3},t_{4}) = \Theta(t_{2}-t_{3})\Theta(t_{1}-t_{4})\\
&\hskip 0.85cm \bigl[\mathcal K_{[2314]} - \mathcal K_{[2341]} - \mathcal K_{[3214]}+\mathcal K_{[3241]}\bigr]\bigl(\tau_{-}+ i t_{1},\tau_{+}+i t_{2},\tau_{+}+it_{3},\tau_{-}+it_{4}\bigr)\,,
\end{split}
\\
\begin{split} & \mathcal K_{2,2}(t_{1},t_{2},t_{3},t_{4}) = -\Theta(t_{1}-t_{3})\Theta(t_{2}-t_{4})\\
&\hskip 0.85cm \bigl[\mathcal K_{[2413]} - \mathcal K_{[4213]} - \mathcal K_{[2431]}+\mathcal K_{[4231]}\bigr]\bigl(\tau_{-}+ i t_{1},\tau_{+}+i t_{2},\tau_{-}+it_{3},\tau_{+}+it_{4}\bigr)\, .
\end{split}
\end{aligned}
\ee
The $\approx$ signs in \eqref{C1trunc} and \eqref{C2trunc} emphasize the fact that this is a truncated version of the exact equations, to be used only to evaluate the late time behaviour of the OTOCs. These equations can be presented in a convenient matrix form
\be\label{matrixformtrunc}
\begin{pmatrix}
\mathbb C_{1}\\ \mathbb C_{2}
\end{pmatrix}
\approx
\begin{pmatrix}
\mathbb C_{1}^{(0)}\\ \mathbb C_{2}^{(0)}
\end{pmatrix}
+
\begin{pmatrix} \mathbb K_{1,1} & \mathbb K_{1,2}\\
\mathbb K_{2,1} & \mathbb K_{2,2}\end{pmatrix}
\star
\begin{pmatrix}
\mathbb C_{1}\\ \mathbb C_{2}
\end{pmatrix}\,,
\ee
where
\begin{align}\label{C0def1} & C_{1}^{(0)}(t_{1},t_{2}) = G_{+}(\tau_{+}+ i t_{1})G_{+}(\tau_{3}-\tau_{-}-it_{2})\,,\\\label{C0def2}&
C_{2}^{(0)}(t_{1},t_{2})= -G_{+}(\tau_{-}+ i t_{1})G_{-}(\tau_{3}-\tau_{+}-it_{2})\,,
\end{align}
and we define the $\star$-product between components in \eqref{matrixformtrunc} as
\be\label{matrixstar} (\mathbb K\star\mathbb C)(t_{1},t_{2}) = \int_{0}^{+\infty}\!\mathcal K(t_{1},t_{2},t,t') C(t,t')\,\d t\d t'\, .\ee

The above formulas can be applied in any melonic model with ladder kernel $\mathcal K$. For the special cases of the \modone\ and \modtwo\ models \eqref{Kone} and \eqref{Ktwo}, the right-hand sides of \eqref{Kij} can be conveniently expressed in terms of the retarded and advanced functions defined in \eqref{Grdef} and \eqref{Gadef}, as well as the particular Wightman two-point correlators
\be\label{GWpmdef}\GWp(t) = G_{+}(\tau_{+}-\tau_{-}+ it)\, ,\quad
\GWm(t) = G_{-}(\tau_{-}-\tau_{+}+ it)\, .\ee
We find
\be\label{Kijmod1}
\begin{aligned}
\mathcal K^{\modelone}_{1,1}(t_{1},t_{2},t_{3},t_{4}) & =-2\la^{2}\Gr(t_{1}-t_{3})\Ga(t_{4}-t_{2})\GWp(t_{3}-t_{4})\GWm(t_{4}-t_{3})\,,\\
\mathcal K^{\modelone}_{1,2}(t_{1},t_{2},t_{3},t_{4}) & =\la^{2}\Gr(t_{1}-t_{4})\Ga(t_{3}-t_{2})\GWp(t_{4}-t_{3})^{2}\,,\\
\mathcal K^{\modelone}_{2,1}(t_{1},t_{2},t_{3},t_{4}) & =\la^{2}\Gr(t_{1}-t_{4})\Ga(t_{3}-t_{2})\GWm(t_{4}-t_{3})^{2}\,,\\
\mathcal K^{\modelone}_{2,2}(t_{1},t_{2},t_{3},t_{4}) & =-2\la^{2}\Gr(t_{1}-t_{3})\Ga(t_{4}-t_{2})\GWp(t_{4}-t_{3})\GWm(t_{3}-t_{4})\,,
\end{aligned}
\ee
and
\be\label{Kijmod2}
\begin{aligned}
\mathcal K^{\modeltwo}_{1,1}(t_{1},t_{2},t_{3},t_{4}) & =\frac{3}{4}|\xi|^{2}\Gr(t_{1}-t_{3})\Ga(t_{4}-t_{2})\bigl[\GWp(t_{3}-t_{4})^{2}+\GWm(t_{4}-t_{3})^{2}\bigr]\,,\\
\mathcal K^{\modeltwo}_{1,2}(t_{1},t_{2},t_{3},t_{4}) & =
-\frac{3}{2}|\xi|^{2}\Gr(t_{1}-t_{4})\Ga(t_{3}-t_{2})\GWp(t_{4}-t_{3})\GWm(t_{3}-t_{4})\,,\\
\mathcal K^{\modeltwo}_{2,1}(t_{1},t_{2},t_{3},t_{4}) & =
-\frac{3}{2}|\xi|^{2}\Gr(t_{1}-t_{4})\Ga(t_{3}-t_{2})\GWp(t_{3}-t_{4})\GWm(t_{4}-t_{3})\,,\\
\mathcal K^{\modeltwo}_{2,2}(t_{1},t_{2},t_{3},t_{4}) & =\frac{3}{4}|\xi|^{2}\Gr(t_{1}-t_{3})\Ga(t_{4}-t_{2})\bigl[\GWp(t_{4}-t_{3})^{2}+\GWm(t_{3}-t_{4})^{2}\bigr]\, .
\end{aligned}
\ee
\subsubsection{\label{LyapnumSec}Numerical implementation}

The equations \eqref{C1trunc} and \eqref{C2trunc}, or equivalently \eqref{matrixformtrunc}, can be used to compute $C_{1}(t_{1},t_{2})$ and $C_{2}(t_{1},t_{2})$ numerically by introducing a cutoff $t_{\max}$, discretizing the real time variables $(t_1, t_2) \in [0, t_{\max}]^2$, and inverting the resulting finite size matrix. Note that the integrals effectively extend from zero to $t_{1}$ or $t_{2}$ due to the presence of the $\Theta$ functions in \eqref{Kij}, so that this procedure allows us to obtain $C_{1}(t_{1},t_{2})$ and $C_{2}(t_{1},t_{2})$ for all $t_1,t_2 < t_{\max}$ without further truncations. The choice of the cutoff is also related to the support of the kernels \eqref{Kijmod1} or \eqref{Kijmod2}, which can be estimated at low temperatures using the conformal forms derived from \eqref{Gcplxespecdec} with $\rho(\omega)$ given by \eqref{rhoscale}, or alternatively determined beforehand from the explicit numerical solution for the real time two-point functions.

In order to see the exponential growth of the OTOCs we need to reliably evaluate $C_{1}$ and $C_{2}$ up to long times. To accurately approximate the integrals and achieve an appropriate resolution we may therefore need a significant number of points $P$ in the $[0,t_{\max}]$ grid. The runtime of the algorithm is dominated by the $P^2 \times P^2$ matrix inversion, which for the generic matrices we are dealing with here can be done in $\mathcal{O}(P^{2\delta})$ for some $2 < \delta < 3$. This is the main reason why the truncated system \eqref{C1trunc}-\eqref{C2trunc} is much more convenient than the exact equation \eqref{SDreal4pteq}, since the latter involves more correlators and integrals thus enlarging the final size of the matrix to be inverted.

In practice, our Mathematica implementation allows for $P\sim10^2$ points in the interval $[0,t_{\max}]$, resulting in the inversion of $\sim 10^4 \times 10^4$ matrices performed in a few seconds with a desktop computer.\footnote{Much improvement can be expected from a GPU implementation, which might enable the numerical solution of the full set of equations \eqref{SDreal4pteq}. We found however that this was not necessary for our purposes here.} The typical time scale beyond which the exponential behaviour of $C_{1}$ and $C_{2}$ is seen is of the order $\tqn$, and we have shown in Figs.~\ref{fig:M1QNF}a and \ref{fig:M2QNF}a that $\tqn^{-1} = -\im\nuqn$ behaves linearly with $T$ at low-temperatures in the SYK-like phase of the \modone\  and \modtwo\ models. We are thus required to take $t_{\max} \sim \beta$, so that the resolution requirements then impose a lower bound on the temperatures we can reach. We find that this bound is of the same order of that imposed by the real time algorithm used for the computation of the two-point functions themselves, so it is only occasionally a limitation.

\begin{figure}[h!]
\centering
\def\svgwidth{4.5in}
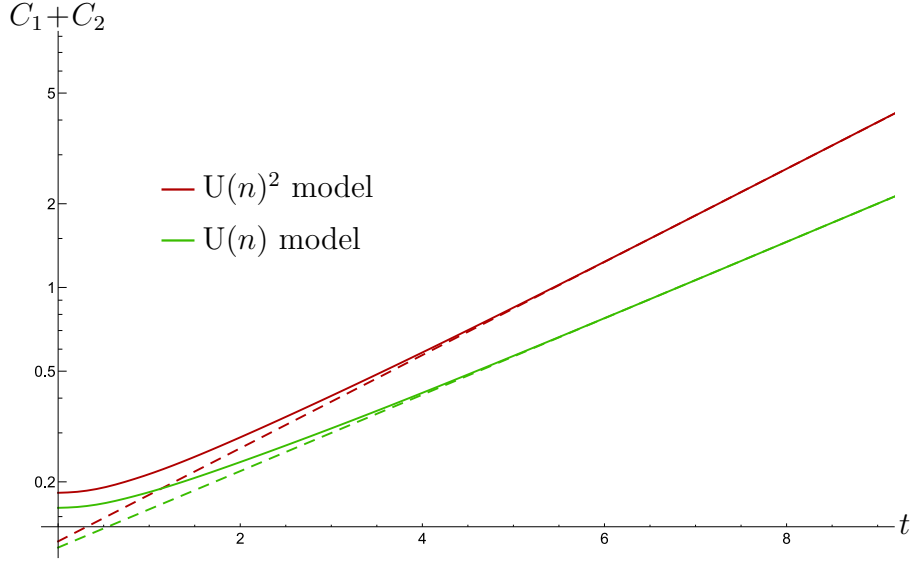 
\caption{\label{fig:F4ptStdDirect}Log-plot of the numerical evaluation of $C_{1}(t,t)+C_{2}(t,t)$, as obtained from \eqref{matrixformtrunc}, for the standard choice of correlators \eqref{maldacase} at $m=0.2$ and $T=0.1$ in natural units (\modone\ model in red or dark gray; \modtwo\ model in green or light gray). The exponential growth is clearly seen at long times, with associated Lyapunov exponents $\tfrac{\lyap}{2\pi T} = 0.614_5$ and $\tfrac{\lyap}{2\pi T} = 0.503_5$ obtained  for the \modone\ and \modtwo\ models, respectively, through linear fits in log-scale (shown with dashed lines). These results were calculated using $P = 100$ points and a cutoff of $t_{\max} = 10.5$.}
\end{figure}

Finally, we should note that while the physical correlators defined by \eqref{physicalcase} are better motivated, there is a practical advantage associated to the standard choice corresponding to  \eqref{maldacase}. Indeed, in the latter case the system \eqref{matrixformtrunc} decouples when taking the combinations $C_{1}\pm C_{2}$, further reducing the size of the problem. Moreover, $C_{1} + C_{2}$ is real and displays exponential growth, whereas we find that $|C_1-C_2|$ is bounded and need not be computed if only Lyapunov exponents are sought. We show in Fig.~\ref{fig:F4ptStdDirect} the combination $C_{1} + C_{2}$, obtained from \eqref{matrixformtrunc} as explained above, for typical values of the parameters $m$ and $T$.\footnote{The truncated system \eqref{matrixformtrunc} can be used only to compute the exponentially growing contributions to the correlators and thus does not yield a reliable evaluation of the combination $C_{1}-C_{2}$. All we can say in this case is that $|C_{1}-C_{2}|$ is not exponentially growing.} The exponential behaviour is clearly seen at long times and yields a precise determination of the associated Lyapunov exponents through linear fitting in log-scale. Note that the values at short times are not reliable, since we use the truncated system \eqref{matrixformtrunc} and not the exact equations \eqref{SDreal4pteq}.

\medskip

We can also use an alternative method to obtain the Lyapunov exponents, due to Kitaev, that can be justified in the following way. We start from the truncated system \eqref{C1trunc}-\eqref{C2trunc} in which we plug the following ansatz,
\be\label{Cansatz} C_{j}(t_{1}+t_{0},t_{2}+t_{0}) = e^{\frac{\lyap}{2}(t_{1}+t_{2})+\lyap t_{0}} D_{j}\Bigl(\frac{t_{1}-t_{2}}{2}\Bigr) \quad \text{for }j=1,2\, ,\ee
for some \emph{a priori} unknown functions $D_{1}$ and $D_{2}$. In the limit of very large $t_{0}$, the first term on the right-hand sides of \eqref{C1trunc} and \eqref{C2trunc} drops. It is convenient to use the new variables
\be\label{TXtx} T=\frac{t_{1}+t_{2}}{2}\,\cvp\quad X=\frac{t_{1}-t_{2}}{2}\,\cvp\quad T'=\frac{t+t'}{2}\,\cvp\quad X'=\frac{t-t'}{2}\,,\ee
to get
\begin{multline}\label{tCeq1} 
e^{\lyap T}D_{j}(X)  = 2\int_{-\infty}^{+\infty}\!\!\d T'\d X'\, e^{\lyap T'}\Bigl[\mathcal K_{j,1}(T+X,T-X,T'+X',T'-X') D_{1}(X')\\ +
\mathcal K_{j,2}(T+X,T-X,T'+X',T'-X') D_{2}(X')\Bigr]\, , \quad j=1,2\, .
\end{multline}
The integrals over $T'$ can be performed and we are left with a new linear system of the form
\be\label{matrixformKit}
\begin{pmatrix}
\mathbb D_{1}\\ \mathbb D_{2}
\end{pmatrix}
=
\begin{pmatrix} \mathbb M_{1,1} & \mathbb M_{1,2}\\
\mathbb M_{2,1} & \mathbb M_{2,2}\end{pmatrix}
*
\begin{pmatrix}
\mathbb D_{1}\\ \mathbb D_{2}
\end{pmatrix}\,,
\ee
where the $*$-product between components now corresponds to
\be\label{matrixstar2} (\mathbb M*\mathbb D)(X) = \int_{-\infty}^{+\infty}\!\mathcal M(X,X') D(X')\,\d X'\, .\ee
More explicitly, if
\be\label{adefu} a(x) = \int_{|x|}^{+\infty}\Gr(x+T)\Ga(u-T)e^{-\lyap T}\,\d T\, ,\ee
we find for the \modone\ and \modtwo\ models respectively,
\be\label{Mijmod1}
\begin{aligned}
\mathcal M^{\modelone}_{1,1}(X,X') & =-4\la^{2}\GWp(2X')\GWm(-2X')a(X-X')\,,\\
\mathcal M^{\modelone}_{1,2}(X,X') & =2\la^{2}\GWp(-2X')^{2}a(X+X')\,,\\
\mathcal M^{\modelone}_{2,1}(X,X') & =2\la^{2}\GWm(-2X')^{2}a(X+X')\,,\\
\mathcal M^{\modelone}_{2,2}(X,X') & =-4\la^{2}\GWp(-2X')\GWm(2X')a(X-X')\,,
\end{aligned}
\ee
and
\be\label{Mijmod2}
\begin{aligned}
\mathcal M^{\modeltwo}_{1,1}(X,X') & =\frac{3}{2}|\xi|^{2}\bigl[\GWp(2X')^{2}+\GWm(-2X')^{2}\bigr]a(X-X')\,,\\
\mathcal M^{\modeltwo}_{1,2}(X,X') & =-3|\xi|^{2}\GWm(2X')\GWp(-2X')a(X+X')\,,\\
\mathcal M^{\modeltwo}_{2,1}(X,X') & =-3|\xi|^{2}\GWm(-2X')\GWp(2X')a(X+X')\,,\\
\mathcal M^{\modeltwo}_{2,2}(X,X') & =\frac{3}{2}|\xi|^{2}\bigl[\GWp(-2X')^{2}+\GWm(2X')^{2}\bigr]a(X-X')\, .\end{aligned}
\ee
Equation \eqref{matrixformKit} shows that the ansatz \eqref{Cansatz} is consistent as long as the matrix $\bigl(\begin{smallmatrix} \mathbb M_{1,1} & \mathbb M_{1,2}\\\mathbb M_{2,1} & \mathbb M_{2,2}\end{smallmatrix}\bigr)$ has an eigenvalue one. The Lyapunov exponent will correspond to the largest $\lyap$ for which the matrix does have this eigenvalue, since such a value will obviously dominate at long times. Note that this largest $\lyap$ always exists; indeed, \eqref{adefu} shows that the matrix elements go to zero when $\lyap\rightarrow\infty$. We thus get $\lyap$ by shooting for the largest eigenvalue of $\bigl(\begin{smallmatrix} \mathbb M_{1,1} & \mathbb M_{1,2}\\\mathbb M_{2,1} & \mathbb M_{2,2}\end{smallmatrix}\bigr)$ to be one. Of course, the drawback is that this approach uses the exponential ansatz \eqref{Cansatz} \emph{a priori}, whereas the exponential growth of the correlators emerges on its own from the solution of \eqref{matrixformtrunc}. 

The method described above can be efficiently implemented numerically. The dimensions of the matrix $\smash{\bigl(\begin{smallmatrix} \mathbb M_{1,1} & \mathbb M_{1,2}\\\mathbb M_{2,1} & \mathbb M_{2,2}\end{smallmatrix}\bigr)}$ are linear in the number $P$ of discretization points used to perform the integrals, instead of quadratic as in \eqref{matrixformtrunc}. This allows for greater values of $t_{\max}$, which then results in higher matrix sparsity. Specialized algorithms such as Arnoldi iteration can then be used to compute the largest eigenvalue of the matrix faster than its inverse. At this point, it also becomes important to avoid the $\mathcal{O}(P^2)$ naive computation of the one-dimensional kernel $a(x)$. Instead, one may rewrite \eqref{adefu} as a convolution to be computed in Fourier space in $\mathcal{O}(P\log P)$. At the end of the day, taking into account the binary search used to shoot $\lyap$ for the maximal eigenvalue equal to one, the algorithm runs in $\mathcal{O}\left(\log\tfrac{1}{\varepsilon} \times (P\log P + P^{\delta'})\right)$, where $\varepsilon$ is the desired precision of $\lyap$ and $\delta'$ characterizes the diagonalization method used. In practice, we find $P \sim 10^3$ and $\varepsilon = 10^{-5}$ to be feasible on a desktop computer within a few seconds.

We used both methods in all our evaluations of the Lyapunov exponents below, and the results turn out to be perfectly consistent.

\noindent\emph{Remark}: Kitaev's method is sometimes justified by taking the large time $t=t_{1}=t_{2}$ limit directly in \eqref{C1trunc} and \eqref{C2trunc}, or \eqref{matrixformtrunc}, and using the quasi-normal behaviour of the two point functions to neglect the first term on the right-hand sides of these equations. This is not correct. In particular, the matrix $\bigl(\begin{smallmatrix} \mathbb K_{1,1} & \mathbb K_{1,2}\\\mathbb K_{2,1} & \mathbb K_{2,2}\end{smallmatrix}\bigr)$ does \emph{not} have an eigenvalue one. If it had one, the system \eqref{matrixformtrunc} would not be invertible! Note also that the integrals in \eqref{C1trunc}, \eqref{C2trunc} or \eqref{matrixstar} run from zero to infinity, unlike the integrals in \eqref{tCeq1} or \eqref{matrixstar2} for which one integrates from minus infinity to plus infinity. 

\subsubsection{\label{LyapAppSec}Applications}

\paragraph{Zero-temperature limit and chaos bound}

\begin{figure}[h!]
\centering
\vskip 0.3cm
\def\svgwidth{4.5in}
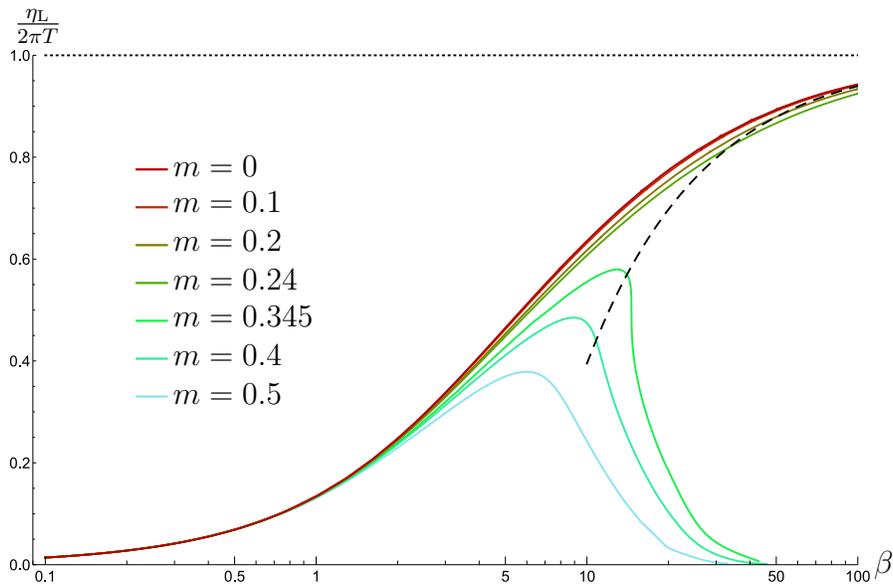
\caption{\label{fig:LyapunovExponentsa}Normalized Lyapunov exponent $\lyap/(2\pi T)$ of the \modone\ model, as a function of the inverse temperature $\beta$ (in logarithmic scale) and for various masses. The bound $\lyap/(2\pi T) = 1$ is marked by a horizontal dotted line, with the dashed black curve corresponding to the asymptotic result $\lyap \simeq \tfrac{2\pi}{\beta}(1 - \tfrac{6.05}{\beta})$ computed for SYK in \cite{Maldacena:2016hyu}. The bound is saturated in the SYK-like phase in the zero-temperature limit.}
\end{figure}

\begin{figure}[h!]
\centering
\def\svgwidth{4.5in}
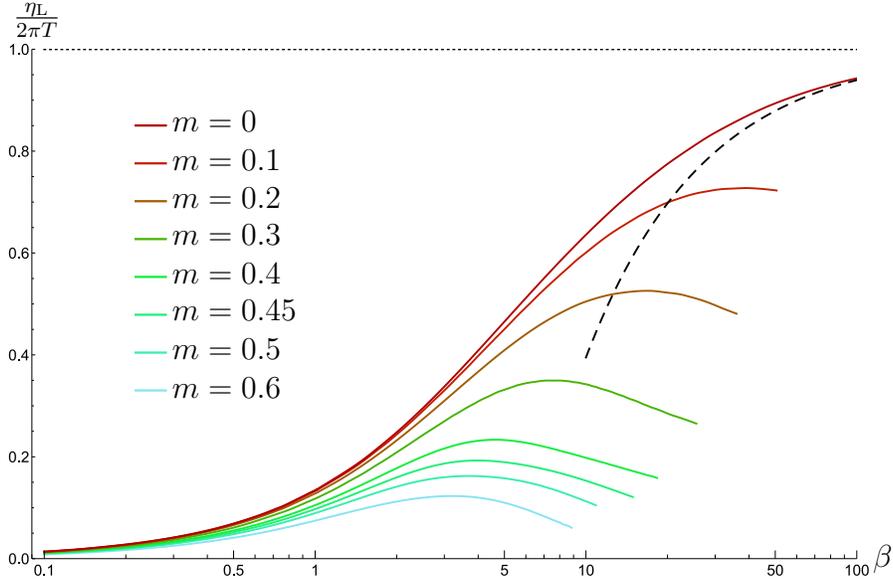 
\caption{\label{fig:LyapunovExponentsb}Normalized Lyapunov exponent $\lyap/(2\pi T)$ of the \modtwo\ model, as a function of the inverse temperature $\beta$ (in logarithmic scale) and for various masses. The bound $\lyap/(2\pi T) = 1$ is marked by a horizontal dotted line, with the dashed black curve corresponding to the asymptotic result $\lyap \simeq \tfrac{2\pi}{\beta}(1 - \tfrac{6.05}{\beta})$ computed for SYK in \cite{Maldacena:2016hyu}. Saturation of the bound is only clearly seen for $m = 0$ and seems to be violated for all $m>0$, including in the gapless phase $0<m<m_{*}$.
}
\end{figure}

\begin{figure}[h!]
\centering
\def\svgwidth{4.5in}
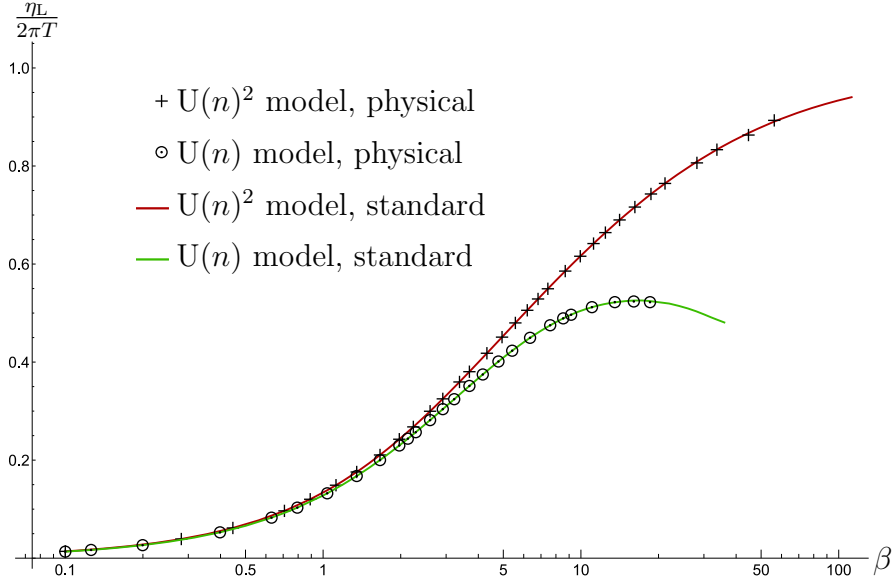
\caption{\label{fig:LyapunovComparison}Lyapunov exponents of the \modone\ and \modtwo\ models, as functions of inverse temperature $\beta$ (in logarithmic scale), for $m = 0.2$ and in natural units. Red and green solid lines (dark and light gray, respectively) correspond to the standard correlators \eqref{maldacase}, with values for the physical case \eqref{physicalcase} represented with crosses and circles, for the \modone\ and \modtwo\ model respectively. All results were computed using Kitaev's method with $P = 1001$ and $\varepsilon = 10^{-5}$. As discussed in Sec.~\ref{OTOCSec2} equations \eqref{matrixformtrunc} can be decoupled in the standard case, leading to a faster and more precise solution allowing for slightly lower temperatures. An equally good agreement is observed between Lyapunov exponents for standard and physical correlators for other masses, both in SYK-like and HO-like phases.
}
\end{figure}

In Fig.~\ref{fig:LyapunovExponentsa} and \ref{fig:LyapunovExponentsb}, we plot $\lyap/(2\pi T)$ as a function of inverse temperature $\beta$, for various values of the mass parameter in the \modone\ and \modtwo\ models. These results were computed for the standard case \eqref{maldacase}, but complete agreement is seen with the Lyapunov exponents of the physical one \eqref{physicalcase}, as exemplified for $m =0.2$ in Fig.~\ref{fig:LyapunovComparison}.


In the case of the \modone\ model, the results are as expected. We find that the bound $\lyap \leq 2\pi T$ of \eqref{Lyapbound} is saturated for the SYK-like phase in the zero-temperature limit, including for $m > m_*$ where this solution is metastable (\emph{e.g.}\ for $m = 0.24$ in Fig.~\ref{fig:LyapunovExponentsa}).

The situation seems to be very different in the case of the \modtwo\ model. As soon as $m>0$, we find that the curve for $\lyap/(2\pi T)$ reaches a maximum and then starts to decrease as the temperature is lowered. Our numerics do not allow us to probe the deep IR regime, so we must remain cautious about the conclusions we may draw for the strict zero-temperature limit. However, the behaviour is clearly qualitatively different from the \modone\ model and suggests that the bound \eqref{Lyapbound} is not saturated already in the phase where the non-standard ansatz discussed in Sec.\ \ref{NewansatzSec} is valid. The fact that the $\slR$ subgroup of the reparameterization symmetry \eqref{diffeoG1}-\eqref{diffeoS1} is broken, including the scale invariance, is certainly crucial to understand this property. This seemingly very different behaviour of the Lyapunov exponent is contrasting with other properties that we have discussed previously, like the existence of a non-vanishing zero-temperature entropy, that are similar to the SYK-like phase of the \modone\ model. Clearly, this raises interesting questions with respect to the gravitational interpretation of the \modtwo\ model, in particular in relation with a possible description in terms of a suitable generalization of the Schwarzian action, and deserves further study.

\noindent\emph{Remark}: The non-saturation of the chaos bound discussed above in the gapless phase of the $\uN$ model must not be confused with the usual and expected non-saturation of the bound found in the gapped phase, for $m>m_{*}$. The gapped phases of both the \modone\ and \modtwo\ models are similar to what is found in the SYK model with a random mass term or in the Maldacena-Qi model, where non-saturation of the chaos bound was already observed \cite{Garcia-Garcia:2017bkg,Maldacena:2018lmt,Garcia-Garcia:2019poj}.

\paragraph{Critical properties of the Lyapunov exponent}

\begin{figure}[h!]
\centering
\begin{tabular}{cc}
\def\svgwidth{7cm}
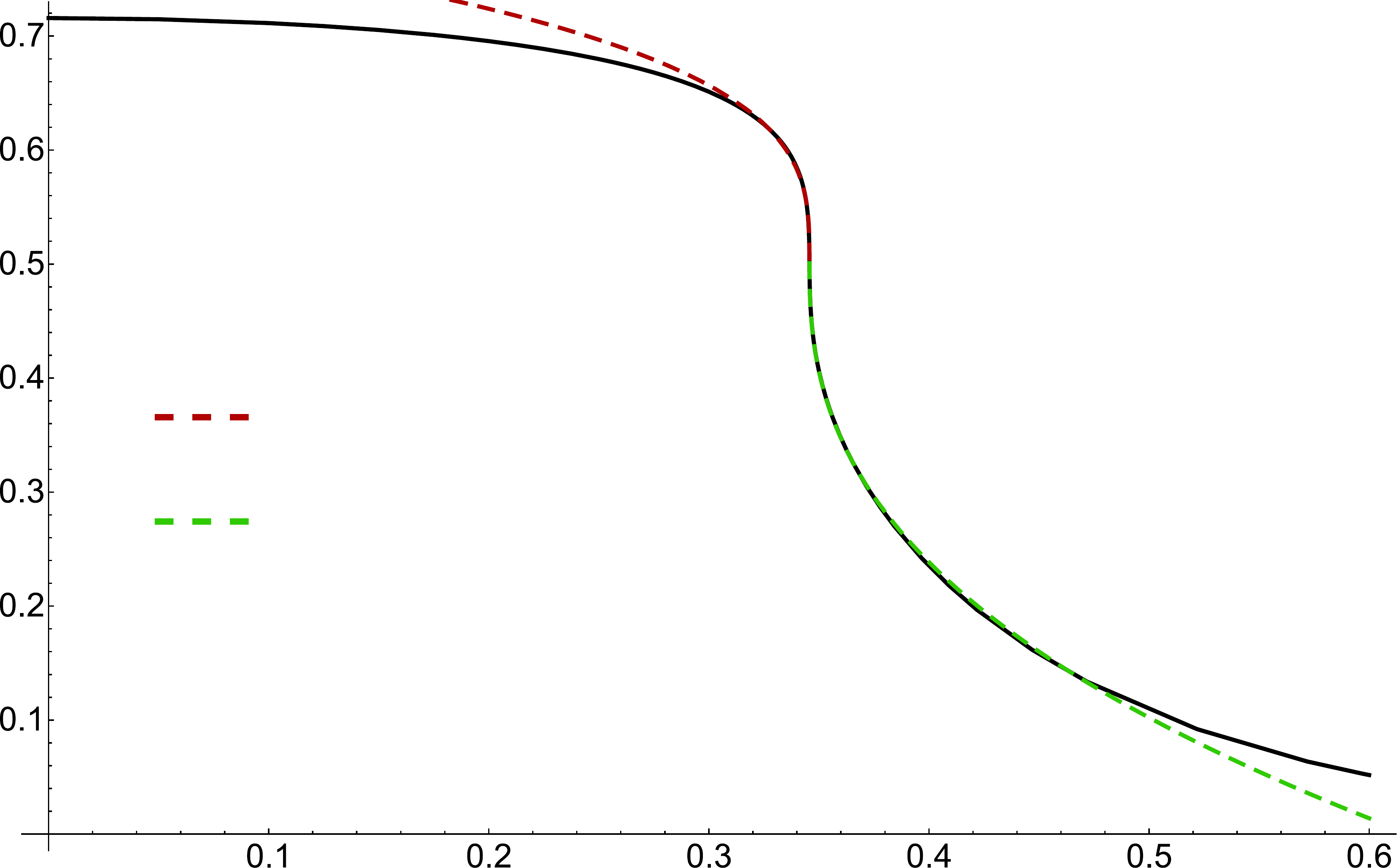 &
\def\svgwidth{7cm}
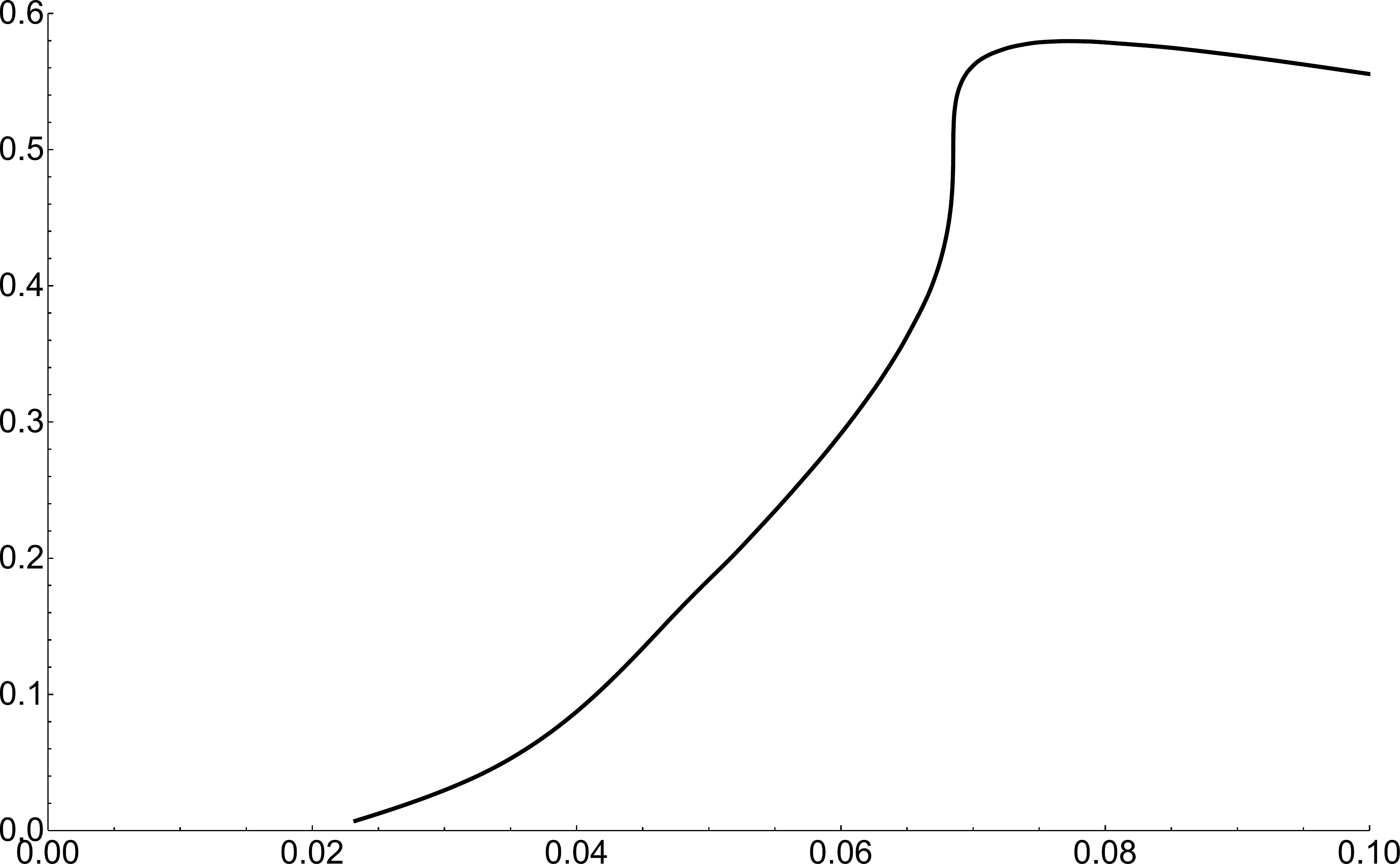 \\
(a) & (b)
\end{tabular}
\caption{\label{fig:LyapunovCriticalExp}Lyapunov exponent for the \modone\ model, as a function of mass at fixed temperature $T = \Tc$ (a) and as a function of temperature at fixed mass $m = \mc$ (b). For (a) critical exponents were obtained as in Sec.~\ref{Pha1exponentsqSec}, see \emph{e.g.}\ Fig.~\ref{fig:CriticalExponent}. In (b) a singular behaviour is clearly seen when $T\to\Tc$, but numerical data is not precise enough to extract critical exponents in a reliable way.}
\end{figure}

We close our analysis with a discussion of the very interesting behaviour of the Lyapunov exponent near the critical point $(\mc, \Tc)$ of the \modone\ model. In Fig.\ \ref{fig:LyapunovCriticalExp}a, $\lyap/(2\pi T)$ is plotted as a function of $m$ for $T=\Tc$. We find a singular behaviour near $m=\mc$, consistent with 
\be
\frac{\partial\lyap}{\partial m}\bigl(m,\Tc\bigr) \underset{m\rightarrow\mc^{\pm}}{\sim}\frac{\lyap^{\pm}}{\bigl|m-\mc\bigr|^{g_{\pm}}}\,,
\ee
critical exponents being
\be
g_+ = 0.69_1 \qquad,\quad g_- = 0.60_1\,.
\ee
We have also studied $\lyap/(2\pi T)$ as a function of $T$ at $m=\mc$, see Fig.\ \ref{fig:LyapunovCriticalExp}b. It is very plausible that critical exponents can be defined in this case as well, but our data is not precise enough to extract them in a reliable way.

\section*{Acknowledgements}

We would like to thank Igor Klebanov, Juan Maldacena, Subir Sachdev, Douglas Stanford and Herman Verlinde for useful discussions and comments. F.\ S.\ M.\ acknowledges affiliation to the IBS Center for Theoretical Physics of the Universe (Seoul, South Korea) during the initial stages of this project. This research is supported in part by the Belgian Fonds National de la Recherche Scientifique FNRS (convention IISN 4.4503.15) and the F\'ed\'eration Wallonie-Bruxelles (Advanced ARC project ``Holography, Gauge Theories and Quantum Gravity''). This work was performed in part at the Aspen Center for Physics, which is supported by the National Science Foundation grant PHY-1607611, and was partially supported by a grant from the Simons Foundation.

\appendix\clearpage

\section{\label{qGenApp}$q$-body interacting Hamiltonians}

One can consider the following extension of the \modone\ model, which has interesting features discussed in Sec.~\ref{PhaseMod1Sec}. We generalize the random Hamiltonian \eqref{Honerandom} to
\be\label{Hqhamilton} H_{\mathbf q}=Nm\chi_{i}^{\dagger}\chi^{i} +\sqrt{N} \sum_{q\in\mathbf q}
\la_{j_{1}\cdots j_{q/2}}^{i_{1}\cdots i_{q/2}}\chi^{\dagger}_{i_{1}}\cdots\chi^{\dagger}_{i_{q/2}}\chi^{j_{1}}\cdots\chi^{j_{q/2}}\,,
\ee
where the sum is over even integers $q\in\mathbf q = \{q_{1},\ldots,q_{p}\}$, $4\leq q_{1}<q_{2}<\cdots <q_{p}$, and the $\smash{\la_{j_{1}\cdots j_{q/2}}^{i_{1}\cdots i_{q/2}}}$ are random Gaussian couplings satisfying constraints similar to the $\smash{\la_{ij}^{kl}}$ in \eqref{rancouplingcond1} and \eqref{rancouplingcond2}. One can also provide a matrix model version of the model, using the complete interaction vertex studied in \cite{ferra2}. The effective action \eqref{Seffmat1} is generalized to
\begin{multline}
\label{Seffmatq}
\frac{1}{n^{2}d}\sSeff^{\modelone}[\sG,\sS] =-\ln\bigl(1+e^{-\beta m}\bigr) -\ln\frac{\det\mathscr O_{\sS}}{\det\mathscr O_{0}} + \int_{[0,\beta]^{2}}\Bigl[\sS(\tau_{2},\tau_{1})\sG(\tau_{1},\tau_{2})\\
-\sum_{q\in\mathbf q}\frac{(-1)^{q/2}}{q}\la_q^{2}\sG(\tau_{1},\tau_{2})^{q/2}\sG(\tau_{2},\tau_{1})^{q/2}\Bigr]\d \tau_{1}\d \tau_{2}\, ,
\end{multline}
and the Schwinger-Dyson equation \eqref{SDone} is replaced by
\be\label{SDq} \Sigma(\tau) = \sum_{q\in\mathbf q}(-1)^{q/2}\la_{q}^{2}G(\tau)^{q/2}G(-\tau)^{q/2-1}\, .\ee
At low energy and when the mass parameter is below a critical value (see Sec.~\ref{PhaseSec}), the models flow to a conformal phase. In this regime, only the $q_1$-body interaction term contributing in \eqref{SDq} is relevant. The ansatz  \eqref{Gasymp} and Eq.\ \eqref{FourierGle}, \eqref{rhozeroTmodelone} are still valid. The generalizations of \eqref{Cthetadef}, \eqref{Deltaform1}, \eqref{Cthetarel}, \eqref{Gscaling} and \eqref{aform} read
\begin{align}\label{Cthetaqrel} b_{+} &= \frac{C}{\Gamma(1-2\Delta)}\frac{\sin (\pi\Delta+\theta)}{\sin (2\pi\Delta)}\, \cvp\quad
b_{-}=\frac{C}{\Gamma(1-2\Delta)}\frac{\sin (\pi\Delta-\theta )}{\sin (2\pi\Delta)}\, \cvp\\
\label{Deltaformq}\Delta &= \frac{1}{q_1}\, \cvp\\
\label{Cthetarelq} C^{q_1} &= \frac{(1-2\Delta)\bigl(\Gamma(1-2\Delta)\bigr)^{q_1}}{\pi\la_{q_1}^{2}}\frac{\bigl(\sin(2\pi\Delta)\bigr)^{q_1-1}}{\bigl(\sin(\pi\Delta+\theta)\sin(\pi\Delta-\theta)\bigr)^{\frac{q_1}{2}-1}}\,\cvp\\
\label{Gscalingq}G(\tau) &= \Bigl(\frac{\pi}{\beta}\Bigr)^{2\Delta}\frac{e^{-2\pi a \tau/\beta}}{\bigl(\sin (\pi |\tau|/\beta )\bigr)^{2\Delta}}\Bigl(b_{+}\Theta(\tau) - b_{-}\Theta(-\tau)\Bigr)\, ,\\
\label{aformq} a &= \frac{1}{2\pi}\ln\frac{b_{+}}{b_{-}}=\frac{1}{2\pi}\ln\frac{\sin(\pi\Delta+\theta)}{\sin(\pi\Delta-\theta)}\,\cdotp
\end{align}

In Sec.~\ref{PhaseMod1Sec} we study the thermodynamics of the models. The formulas \eqref{QFourier} or \eqref{QFourier2} for the fermion number are still valid, but the formulas for the free energy \eqref{FreeEnergy} and the energy \eqref{Energy} or \eqref{Energy2} are generalized to
\begin{multline}\label{FreeEnergyq}
\frac{F}{\dof} = m - T\ln\bigl(1+e^{\beta m}\bigr)
+T\sum_{k\in\mathbb N+\frac{1}{2}}\ln\biggl|\frac{G_{k}}{G_{k}^{(0)}}\biggr|^2\\
+\sum_{q\in\mathbf q}\frac{q-1}{q}(-1)^{q/2}\la_{q}^{2}\int_{0}^{\beta}G(\tau)^{q/2}G(-\tau)^{q/2}\,\d\tau\, ,
\end{multline}
\begin{multline}
\label{Energyq} \frac{E}{N} =m\,\nF(m) -\frac{2m}{\beta}\sum_{k\in\mathbb N+\frac{1}{2}}
\re\bigl( G_{k} - G_{k}^{0}\bigr)\\-2\sum_{q\in\mathbf q}\frac{(-1)^{q}}{q}\la_{q}^{2}
\int_{0}^{\beta}G(\tau)^{q/2}G(-\tau)^{q/2}\,\d\tau\, .
\end{multline}
If only one coupling $\la_{q}$ is turned on, these formulas simplify to
\begin{align}\label{FreeEnergyq2}
\frac{F}{\dof} &= m - T\ln\bigl(1+e^{\beta m}\bigr)
+T\sum_{k\in\mathbb N+\frac{1}{2}}\biggl[\ln\biggl|\frac{G_{k}}{G_{k}^{(0)}}\biggr|^2 + \frac{2(q-1)}{q}\Bigl(1-\re\frac{G_k}{G_k^{(0)}}\Bigr)\biggr]\, ,\\
\label{Energyq2} \frac{E}{\dof} 
&=m\,\nF(m) -\frac{1}{\beta}\sum_{k\in\mathbb N+\frac{1}{2}}
\biggl[2m\re\bigl( G_{k} - G_{k}^{0}\bigr) +\frac{4}{q}\Bigl( 1 - \re\frac{G_k}{G_k^{(0)}}\Bigr)\biggr]\, .\end{align}
\section{\label{algApp}Algebraic derivation of the kernel formula}

The goal of this Appendix is to provide all the elementary details of the algebraic derivation of the formulas \eqref{Ffromkernel} and \eqref{Kalggen} given in Sec.\ \ref{EuclidefourSec} for the connected four-point function.

We start from the equations \eqref{fourptfuncder2}, \eqref{SeffJ} and \eqref{ZJform}. In order to simplify the notation we write functions or operators like $G(t_{1},t_{2})$ or $\mathscr O_{\sS}$ as $G_{ij}$ or $(\mathscr O_{\sS})_{ij}$, integrals over $t$ are replaced by sums and functional derivatives by partial derivatives. In these notations, and using \eqref{ZJform} to express the free energy in terms of the on-shell effective action, the four-point function \eqref{fourptfuncder2} is given by
\be\label{fptapp1}
\mathcal F_{ijkl} = \left.-\frac{1}{N}\frac{\partial^{2}\Shat}{\partial J_{ji}\partial J_{lk}}\right|_{J=0}\,,
\ee
and the equations \eqref{Ffromkernel} and \eqref{Kalggen} that we want to prove are equivalent to
\begin{align}\label{fapSD} & \mathcal F_{ijkl}=-G_{il}G_{kj} + \sum_{p,q}\mathcal K_{ijpq}\mathcal F_{pqkl}\, ,\\\label{Kapder} &
\mathcal K_{ijkl} = \sum_{p,q}G_{ip}G_{qj}\frac{\partial^{2}s}{\partial G_{qp}\partial G_{kl}}\,\cdotp
\end{align}

From the explicit expressions \eqref{Seffalggen} and \eqref{SeffJ}, we get
\begin{align}\label{dSdS}\frac{1}{N}\frac{\partial\sShat}{\partial\sS_{ji}}&=
-\bigl(\mathscr O^{-1}_{\sS+J}\bigr)_{ij}+ \sG_{ij}\, ,\\\label{dSdG}
\frac{1}{N}\frac{\partial\sShat}{\partial\sG_{ji}}&=\sS_{ij}+\frac{\partial s}{\partial \sG_{ji}}\bigl[\sG\bigr]\, ,\\
\label{dSdJ}\frac{1}{N}\frac{\partial\sShat}{\partial J_{ji}} &= -\bigl(\mathscr O^{-1}_{\sS+J}\bigr)_{ij}\, .
\end{align}
In particular, the on-shell values $G[J]$ and $\Sigma[J]$ satisfy the saddle-point equations
\begin{align}\label{saddlealg1} G_{ij}&=\bigl(\mathscr O^{-1}_{\Sigma+J}\bigr)_{ij}\, ,
\\ \label{saddlealg2}
\Sigma_{ij}&= -\frac{\partial s}{\partial G_{ji}}\bigl[G\bigr]\, \cdotp
\end{align}
Taking the derivative with respect to $J$ of the definition of the on-shell action,
\be\label{Sonshelldef}
\Shat[J]=\sShat\bigl[G[J],\Sigma[J],J\bigr]\,,
\ee
and using the saddle-point equations together with \eqref{dSdJ}, we get
\be\label{dSonshell}\frac{1}{N} \frac{\partial\Shat}{\partial J_{ji}}=\frac{1}{N} \frac{\partial\sShat}{\partial J_{ji}}\Bigl[G[J],\Sigma[J],J\Bigr] = -G_{ij}\, .\ee
The definition \eqref{fptapp1} thus implies that
\be\label{FdGrel} \mathcal F_{ijkl} = \frac{\partial G_{ij}}{\partial J_{lk}}\,\cdotp\ee
Let us now differentiate the saddle-point equations \eqref{saddlealg1} and \eqref{saddlealg2} with respect to $J$. We get
\begin{align}\label{SDdiffJ1} \frac{\partial G_{ij}}{\partial J_{lk}} &=
-G_{il}G_{kj} - \sum_{r,s}G_{ir}G_{sj}\frac{\partial\Sigma_{rs}}{\partial J_{lk}}\, \cvp
\\
\label{SDdiffJ2} \frac{\partial\Sigma_{rs}}{\partial J_{lk}} &= -\sum_{p,q}\frac{\partial^{2}s}{\partial G_{sr}\partial G_{pq}}
\frac{\partial G_{pq}}{\partial J_{lk}}\,\cvp
\end{align}
so that plugging \eqref{SDdiffJ2} into \eqref{SDdiffJ1} we find that \eqref{FdGrel} yields
\be\label{Falgfinal} \mathcal F_{ijkl}=-G_{il}G_{kj}+\sum_{p,q,r,s}
G_{ir}G_{sj}
\frac{\partial^{2}s}{\partial G_{sr}\partial G_{pq}} \mathcal F_{pqkl}\, .\ee
This is precisely the equation \eqref{fapSD} we wanted to prove, with the definition \eqref{Kapder} for the kernel.

For the models we are studying, the relevant $s[\sG]$ functionals are \eqref{s1alg} and \eqref{s2alg}, or in the notation of this Appendix
\begin{align}
s_{\modelone}[\sG] &= -\frac{\la^{2}}{4} \sum_{i,j} \sG_{ij}^2 \sG_{ji}^2\, ,\\ 
s_{\modeltwo}[\sG] &= \frac{|\xi|^{2}}{4} \sum_{i,j} \sG_{ij}^3 \sG_{ji}\, .
\end{align}
Then 
\begin{align}
\label{Konealg} (\mathcal{K}^{\modelone})_{ijkl} &= -\la^{2}\Bigl[G_{il}G_{kj}G_{lk}^{2}+2G_{ik}G_{lj}G_{lk}G_{kl}\Bigr]\,,\\
\label{Ktwoalg} (\mathcal{K}^{\modeltwo})_{ijkl} &= \frac{3}{4}|\xi|^{2}\Bigl[\bigl(G_{lk}^{2}+G_{kl}^{2}\bigr)G_{ik}G_{lj}+ 2 G_{kl}G_{lk}G_{il}G_{kj}\Bigr]\,.
\end{align}
\section{\label{NumericalErrorsApp}On the numerical errors}

Many of the results presented in this paper are numerical in nature, consisting either of specific values for various relevant quantities (critical temperatures, masses, exponents, \emph{etc}.) or observations concerning the solutions of Schwinger-Dyson equations, such as \emph{e.g.}\ the existence of monodromies in the space of parameters of the \modone\ model. It is therefore important to understand the limitations of the numerical methods employed, which we briefly discuss in this Appendix.

Lacking statistical fluctuations in our results given the exclusive use of deterministic algorithms, there are two main classes of numerical errors or uncertainties to consider:
\begin{enumerate}[i)]
	\item Systematic errors are incurred in whenever imposing a cutoff to truncate an infinite set of equations or realize a limiting procedure, such as in \eqref{SDcomplete} for the Euclidean setting and \eqref{rhoepsapprox} for the real time one; or introducing a discretization to approximate continuous quantities, as in \eqref{discretefreqchoice} for the real time case. This is essentially unavoidable, and completely trumps the numerical errors inherent to the machine-precision floating-point arithmetic used during all computations.\footnote{Proper care needs to be taken to ensure numerical stability in the real time case, given the potentially large exponential factors in the coefficients of \eqref{RSigmod1result} and \eqref{RSigmod2result}.}
	\item Methodological uncertainties are present whenever fitting numerical results to analytical models, either for parameter extraction (scaling dimensions, critical exponents, \emph{etc}.)\ or for zero-temperature extrapolations. They also naturally appear whenever using binary search to determine quantities such as critical temperatures and masses, or Lyapunov exponents as explained in Sec.~\ref{LyapnumSec}.
\end{enumerate}

As already mentioned in Sec.~\ref{EucformSec} and \ref{NumSolEucSec}, systematic errors translate into any and all quantities computed using the numerical solutions we obtain. While we have not rigorously bounded them in terms of the various parameters involved, they can be roughly estimated by comparing with exact analytical results whenever possible. For example, to estimate the precision of the thermodynamic functions computed for the \modone\ model at $T=10^{-3}$ in natural units $\lambda = 1$, we may choose a large enough mass to make sure that the only solution to the Schwinger-Dyson equations is the perturbative one, very close to \eqref{Gtree} at this temperature. In this case taking $m = 1$ is certainly enough, so that using a cutoff of $\maxk = 2^{16}$ we find $F/\dof \simeq 6 \times 10^{-5}$ instead of $F/\dof = 0$ as expected for \eqref{Gtree} at $T = 0$. We may then assume that a systematic error of $\sim 10^{-4}$ is present when computing free energies at $T = 10^{-3}$ with a cutoff $\maxk = 2^{16}$, also for other large masses in the \modone\ model. An analogous procedure can be carried out for the entropy at small masses comparing with the exact formula \eqref{zeroTentropy2}, yielding an estimate of $\sim0.1$ for the systematic error in the numerical results when using a cutoff $\maxk = 2^{16}$ at $T = 10^{-3}$. This is consistent with the $\beta = 10^{3}$ factor relating the entropy and free energy in \eqref{entropy}, so that the systematic errors are observed to be approximately mass-independent in this case, at least for large enough cutoffs and low enough temperatures.

The methodological uncertainties, on the other hand, originate in most cases both in the fitting models and techniques we use, as well as in the actual data we choose to analyze. The effect of the latter is generally the most relevant one, with the statistical error of the fitting procedures being considerably smaller than the typical fluctuations observed when modifying the underlying data. While we have not attempted here a rigorous treatment of these issues, we can again roughly estimate their effects. For example, when computing scaling dimensions at small temperatures by fitting the zero-temperature ans\"atze \eqref{Gasymp} or \eqref{Gasympnew}, as shown in Fig.~\ref{fig:M1DeltaZeroTa} and \ref{fig:M2DeltaZeroTa}, we need to choose a fitting interval satisfying $1 \ll \tau \ll \beta$ in natural units $\lambda = |\xi| = 1$. One simple choice is to take a $10\%$ section of the full Euclidean time interval, say $\tau/\beta \in [0.2,0.12]$ for some large $\beta$, and keep track of the fluctuations of the fitted scaling dimensions when considering $10\%$ fluctuations of its endpoints, \emph{i.e.}\ when fitting the ranges $\tau/\beta \in [0.1,0.11]$ and $\tau/\beta \in [0.3,0.13]$. Similar considerations apply to the calculation of critical exponents, where there is an additional uncertainty in the critical temperatures and masses to be used, so that fluctuations in these parameters need to be accounted for in an analogous manner.

Other error estimation methods are of course possible, but the important observation here is that results show certain universality: for all reasonable choices, even if arbitrary, the numerical fluctuations we encounter are consistent. The methodological error bar obtained in this way is then fairly robust. Moreover, when the method above is applied systematically it serves the additional purpose of providing an indicator for relative certainty in these determinations. It is in this sense that we may begin to quantify the naked-eye impression that the determination of $\Delta_-$ in Fig.~\ref{fig:M2DeltaZeroTa} is less precise than that of $\Delta_+$.

\medskip

In practice, both types of numerical errors and uncertainties described above are intertwined. For example, the systematic error in numerically computed thermodynamic functions prevents us from distinguishing branches of solutions to the Schwinger-Dyson equations with arbitrary precision, thereby forcing us to prematurely interrupt binary searches for critical temperatures (thus prompting a methodological uncertainty). Alternatively, we may use extrapolation procedures as shown in Sec.~\ref{Pha1basic} to account for sytematic errors, obtaining numerical results that have better precision than one could naively expect, see \emph{e.g.}\ Fig.~\ref{fig:M1ZeroTDelta}.

The error bars we report in the various quantities discussed in this work account for the methodological uncertainties, estimated  as explained above, and for systematic errors only as far as they are encompassed by the former. They are therefore not exhaustive, and subject to vary at least slightly for different data and methodologies. It is for these reasons we do not present them in figures, but provide them only for the most important values explicitly referenced in the main text. In any case, the general features and conclusions we derive from these values never depend on the fine-grained details. Indeed, wherever this could be an issue, such as in the detailed characterization of the low-temperature conformal phase of the \modtwo\ model, we opt for the more conservative route and only provide definite assertions if the numerical results are conclusive beyond reasonable doubt.

\end{document}